\documentclass[12pt]{article}
\usepackage{a4wide}
\usepackage{amsmath}
\usepackage{amssymb}
\usepackage{booktabs}
\usepackage{graphicx}
\usepackage{url}

\usepackage{listings}
\usepackage{xcolor}

\definecolor{codegreen}{rgb}{0,0.6,0}
\definecolor{codegray}{rgb}{0.5,0.5,0.5}
\definecolor{codepurple}{rgb}{0.58,0,0.82}
\definecolor{backcolour}{rgb}{0.95,0.95,0.92}

\lstdefinestyle{mystyle}{
    backgroundcolor=\color{backcolour},   
    commentstyle=\color{codegreen},
    keywordstyle=\color{magenta},
    numberstyle=\tiny\color{codegray},
    stringstyle=\color{codepurple},
    basicstyle=\ttfamily\footnotesize,
    breakatwhitespace=false,         
    breaklines=true,                 
    captionpos=b,                    
    keepspaces=true,                 
    numbers=left,                    
    numbersep=5pt,                  
    showspaces=false,                
    showstringspaces=false,
    showtabs=false,                  
    tabsize=2
}

\usepackage{color}
\definecolor{comment}{rgb}{0,0.3,0}
\definecolor{identifier}{rgb}{0.0,0,0.3}
\usepackage{listings}
\usepackage{hyperref}

\newcommand{\etal}{\textit{et al}.}

\begin{document}
  
  \title{\sf MPLAPACK version 2.0.1 user manual}
  \author{NAKATA Maho$^1$\\
  \normalsize
  $^1$RIKEN Cluster for Pioneering Research, 2-1 Hirosawa, Wako-City, Saitama 351-0198, JAPAN}

\date{}

\maketitle

\begin{abstract}
The MPLAPACK (formerly MPACK) is a multiple-precision version of LAPACK \\
(\url{https://www.netlib.org/lapack/}).
MPLAPACK version 2.0.1 is based on LAPACK version 3.9.1 and translated from Fortran 90 to C++ using FABLE, a Fortran to C++ source-to-source conversion tool (\url{https://github.com/cctbx/cctbx\_project/tree/master/fable/}). 
MPLAPACK version 2.0.1 provides the real and complex version of MPBLAS, and the real and complex versions of MPLAPACK support all LAPACK features: solvers for systems of simultaneous linear equations, least-squares solutions of linear systems of equations, eigenvalue problems, and singular value problems, and related matrix factorizations except for mixed-precision routines. The MPLAPACK defines an API for numerical linear algebra, similar to LAPACK. It is easy to port legacy C/C++ numerical codes using MPLAPACK. MPLAPACK supports binary64, binary128, FP80 (extended double), MPFR, GMP, and QD libraries (double-double and quad-double). Users can choose MPFR or GMP for arbitrary accurate calculations, double-double or quad-double for fast 32 or 64-decimal calculations. We can consider the binary64 version as the C++ version of LAPACK.
Moreover, it comes with an OpenMP accelerated version of MPBLAS for some routines and CUDA (A100 and V100 support) for double-double versions of {\tt Rgemm} and {\tt Rsyrk}. The peak performances of the OpenMP version are almost proportional to the number of cores, and the performances of the CUDA version are impressive, and approximately 400-600 GFlops. MPLAPACK is available at GitHub (\url{https://github.com/nakatamaho/mplapack/}) under the 2-clause BSD license.
\end{abstract}

\section{Release note for version 2.0.1}
\begin{itemize}
    \item Version 2.0.1 supports all the complex LAPACK functions.
    \item Version 2.0.1 supports Rectangular Full Packed (RFP) Format.
    \item Version 2.0.1 comes with an acceleration of double-double {\tt Rgemm} and {\tt Rsyrk} routines on NVIDIA Tesla V100 and A100.
    \item Many small bug fixes and improvements in usability.
    \item The quality assurance results and benchmark results on Intel CPU and Arm CPU are available.
    \item Mixed precision version is not supported yet.
    \item Version 2.0.1 was released on 2022-09-12.
\end{itemize}

\section{Introduction}

Numerical linear algebra aims to solve mathematical problems, such as simultaneous linear equations, eigenvalue problems, and least-squares methods, using arithmetic with finite precision on a computer~\cite{GoluVanl96}. We can formulate many problems as numerical linear algebra in various fields, such as natural science, social science, and engineering. Therefore, its importance is magnificent. 

The numerical linear algebra package standards are BLAS~\cite{10.1145/567806.567807} and LAPACK~\cite{laug}. The BLAS library defines how computers should perform vector, matrix-vector, and matrix-matrix operations in FORTRAN77. Almost all other vector and matrix arithmetic libraries are compatible with it or have very similar interfaces. Furthermore, LAPACK solves linear problems such as solving linear equations, singular value problems, eigenvalue problems, and least-square fitting problems using BLAS as a building block.

The main focus in numerical linear algebra has been on the speed and the size of solving the problem at approximately 8 or 16 decimal digits (binary32 or 64) using highly optimized BLAS and LAPACK~\cite{4610935,10.1145/1356052.1356053,6413635,8519659,cublas,tdb10}. The use of numbers with higher precision than binary64 is not common.

However, there are some problems in numerical linear algebra that require higher precision operations. In particular, when we solve ill-conditioned problems, large-scale simulations, and compute inversions of matrices, we usually need functions of higher precision numbers~\cite{BAILEY201210106,math3020337,high:ASNA2}.

Solving positive semidefinite programming (SDP) is another example requiring multi-precision computation. Solving this problem using binary64 usually yields values up to eight decimal digits. The accumulation of numerical error occurs because the matrices' condition number at the optimal solution usually becomes infinite. As a result, Cholesky factorization fails near the optimal solution; the approximate solution diverges toward the optimal solution using the primal and dual interior point method~\cite{SDP}. Therefore, we require multiple precision calculations if we need more than eight decimal digits for optimal solutions. For this reason, we have developed SDPA-GMP~\cite{JCP2008,SDPA-GMP,SDPA}, the GNU MP version of semidefinite programming solver based on SDPA~\cite{SDPA}, which is one of the fastest SDP solvers.

We have been developing MPLAPACK as a drop-in replacement for BLAS and LAPACK in SDPA since SDPA performs Cholesky decomposition and solves symmetric eigenvalue problems via 50 BLAS and LAPACK routines~\cite{sdpa-gmpgithub}.

The features of MPLAPACK are:
\begin{itemize}
\item Provides Application Programming Interface (API) numerical algebra, similar to LAPACK and BLAS. 
\item Like BLAS and LAPACK, we can implement an optimized version of MPBLAS and MPLAPACK; we provide a simple OpenMP version of some MPBLAS routines and two CUDA routines (Rgemm, Rsyrk dd version) as proof of concept.
\item Completely rewrote LAPACK and BLAS in C++ using FABLE and f2c.
\item C style programming. We do not introduce new matrix and vector classes. 
\item Supports seven floating-point formats by precision independent programming; binary64 (16 decimal digits), binary128 (32 decimal digits), FP80 (extended double; 19 decimal digits), double-double (32 decimal digits), quad-double (64 decimal digits), GMP, and MPFR (arbitrary precision, and the default is 153 decimal digits).
\item MPLAPACK 2.0.1 is based on LAPACK 3.9.1.
\item Version 2.0.1 supports all real and complex versions of BLAS and LAPACK functions; simultaneous linear equations, least-squares solutions of linear systems of equations, eigenvalue problems, singular value problems, and related matrix factorization and full packed matrix form except for mixed-precision version.
\item Reliability: We extended the original test programs to handle multiple-precision numbers, and most MPLAPCK routines have passed the test.
\item Runs on Linux/Windows/Mac.
\item Released at \url{https://github.com/nakatamaho/mplapack/} under 2-BSD clause license. 
\end{itemize}

Unless otherwise noted, this paper gives examples with Ubuntu 20.04 amd64 inside Docker as the reference environment.

The rest of the sections are organized as follows: Section~\ref{sec:supportedcpu} describes supported CPUs, OSes, and compilers. Section~\ref{sec:howtoinstall} describes how to install MPLAPACK. Section~\ref{sec:floatingpointformat} describes the supported floating-point format. Section~\ref{sec:nameconventions} illustrates LAPACK and BLAS standard naming conventions and available MPBLAS and MPLAPACK routines; section~\ref{sec:howtouse} describes how to use MPBLAS and MPLAPACK. We use the Docker environment to try these examples to avoid over-complicating the notation with different environments.
Section~\ref{sec:howtotest} describes how we tested MPBLAS and MPLAPACK.
Section~\ref{sec:howtoconvert} describes how we rewrote Fortran90 codes to C++ and extended them to multiple precision versions. Next, section ~\ref{sec:benchmarks} describes benchmarking MPBLAS routines. Next, section ~\ref{sec:history} describes the history, and section~\ref{sec:relatedworks} describes related works. Finally, section~\ref{sec:futureplans} describes future plans for MPLAPACK. 

\section{Supported CPUs, OSes, and compilers}
\label{sec:supportedcpu}
Only 64-bit CPUs are supported. The following OSes are supported:

\begin{itemize}
\item CentOS 7 (amd64, aarch64)
\item CentOS 8 (amd64, aarch64)
\item Ubuntu 22.04 (amd64, aarch64)
\item Ubuntu 20.04 (amd64, aarch64)
\item Ubuntu 18.04 (amd64)
\item Windows 10 (amd64)
\item MacOS (Intel)
\end{itemize}

We support the following compilers:
\begin{itemize}
\item GCC (GNU Compiler Collection) 9 and later
\item Intel One API. (you need {\tt -fp-model precise} to compile dd and qd)
\end{itemize}
Note that we use GCC by MacPorts on macOS, which is NOT the default compiler. However, we need to use GCC (Apple Clang) to compile GMP~\cite{macosgccbug}.
\clearpage

We support the following GPUs to accelerate matrix-matrix multiplication. See section~\ref{sec:nameconventions} for CUDA-enabled routines. 
\begin{itemize}
    \item NVIDIA A100
    \item NVIDIA V100
\end{itemize}

We use the mingw64 and Wine64 environments on Linux to build and test the Windows version. On different configurations, MPLAPACK may build and work without problems. We welcome reports or patches from the community.

\section{Installation}
\label{sec:howtoinstall}
\subsection{Using Docker (recommended for Linux environment), a simple demo}
The easiest way to install MPLAPACK is to build inside Docker~\cite{merkel2014docker} and use it inside Docker. 
The following command will build inside docker on Ubuntu amd64 (also known as x86\_64) or aarch64 (also known as arm64), and {\bf we use Docker ubuntu 20.04 amd64 environment for the reference in this paper and showing examples}.

\begin{verbatim}
$ git clone https://github.com/nakatamaho/mplapack/
$ cd mplapack
$ /usr/bin/time docker build -t mplapack:ubuntu2004 \
   -f Dockerfile_ubuntu20.04 . 2>&1 | tee log.ubuntu2004
\end{verbatim}
It will take a while, depending on CPU cores and OSes. For example, a Docker build took 35 minutes on a Ryzen 3970X (3.7GHz, 32 cores) Ubuntu 20.04 machine, 
 a Docker build took 1 hour 50 minutes on Xeon E5-2623 v3 (3.0GHz, 2 CPUs 8 cores) Ubuntu 20.04 machine,
 a Docker build took 14 hours 40 minutes on Raspberry Pi 4 (Cortex A72 1.5GHz, four cores) Ubuntu 20.04 machine,
Furthermore, the build of MPLAPACK on the Mac mini (2018, Core i5-8500B, 3.0GHz and six cores) took about 3.5 hours.

To run a simple demo (matrix-matrix multiplication in binary128) can be done as follows:
{\footnotesize
\begin{verbatim}
$ docker run -it mplapack:ubuntu20.04 /bin/bash
docker@2cb33bf4c36f:~$ ls
MPLAPACK  mplapack-2.0.1  mplapack-2.0.1.tar.xz
docker@2cb33bf4c36f:~ cd MPLAPACK/share/examples/mpblas/
docker@2cb33bf4c36f:~/MPLAPACK/share/examples/mpblas$ make -f Makefile.linux
c++ -c -O2 -fopenmp -I/home/docker/MPLAPACK/include -I/home/docker/MPLAPACK/include/mplapack \
-I/home/docker/MPLAPACK/include/qd Rgemm_mpfr.cpp
...
docker@2cb33bf4c36f:~/MPLAPACK/share/examples/mpblas$ ./Rgemm__Float128
# Rgemm demo...
a =[ [ +1.00000000000000000000000000000000000e+00, +8.00000000000000000000000000000000000e+00,
+3.00000000000000000000000000000000000e+00]; [ +2.00000000000000000000000
...
ans =[ [ +2.10000000000000000000000000000000000e+01, -1.92000000000000000000000000000000000e+02,
+2.28000000000000000000000000000000000e+02]; [ -6.40000000000000000000000000000000000e+01,
-1.46000000000000000000000000000000000e+02, +2.66000000000000000000000000000000000e+02]; 
[ +2.10000000000000000000000000000000000e+02, +3.61000000000000000000000000000000000e+02,
-3.80000000000000000000000000000000000e+01] ]
#please check by Matlab or Octave following and ans above
alpha * a * b + beta * c
\end{verbatim}
}
If it fails, it is a bug. Please report a problem via GitHub issue.

In Table~\ref{dockerfiles}, we list corresponding Dockerfiles for CPUs, GPUs, OSes, and compilers. 
When the build is finished, all the files will be installed under {\tt /home/docker/MPLAPACK\_CUDA} for the CUDA version,
{\tt /home/docker/MPLAPACK\_MINGW} for Windows version, {\tt /home/docker/MPLAPACK\_INTELONEAPI} for Intel One API version,
and others are under {\tt /home/docker/MPLAPACK}.
We can build corresponding environments (other OSes, GPUs, CPUs), and you can choose the Dockerfile appropriately 
listed on~\ref{dockerfiles}.

\begin{table}
\caption{The name of the Dockerfile and the corresponding OS and CPU}\label{dockerfiles}
\begin{center}
\begin{tabular}{l|c|c|c}
Docker filename                      & CPU          & OS            & Compiler \\ \hline
\tt Dockerfile\_CentOS7              & amd64        & CentOS 7      & GCC \\
\tt Dockerfile\_CentOS7\_AArch64     & aarch64 only & CentOS 7      & GCC\\
\tt Dockerfile\_CentOS8              & all CPUs     & CentOS 8      & GCC \\
\tt Dockerfile\_ubuntu18.04          & amd64 only   & Ubuntu 18.04  & GCC  \\
\tt Dockerfile\_ubuntu20.04          & all CPUs     & Ubuntu 20.04  & GCC\\
\tt Dockerfile\_ubuntu20.04\_inteloneapi & amd64 only   & Ubuntu 20.04  & Intel oneAPI \\
\tt Dockerfile\_ubuntu20.04\_mingw64 & amd64 only   & Ubuntu 20.04     & GCC \\ 
\tt Dockerfile\_ubuntu20.04\_cuda    & NVIDIA A100, V100 & Ubuntu 20.04     & GCC \\ 
\tt Dockerfile\_ubuntu22.04          & all cpus     & Ubuntu 22.04     & GCC \\ 
\tt Dockerfile\_debian\_bullseye     & all cpus     & Debian(bullseye) & GCC \\ \hline
\end{tabular}
\end{center}
\end{table}

\subsection{Compiling from the source}
We list prerequisites for compiling from the source in Table~\ref{prerequisites}; users can satisfy these prerequisites using MacPorts on macOS. Homebrew may be used as an alternative.

{\footnotesize
\begin{verbatim}
$ sudo port install gcc10 coreutils git ccache
$ wget https://github.com/nakatamaho/mplapack/releases/download/v2.0.1/mplapack-2.0.1.tar.xz
$ tar xvfz mplapack-2.0.1.tar.xz
$ cd mplapack-2.0.1
$ CXX="g++-mp-10" ; export CXX
$ CC="gcc-mp-10" ; export CC
$ FC="gfortran-mp-10"; export FC
$ ./configure --prefix=/usr/local --enable-gmp=yes --enable-mpfr=yes \
--enable-_Float128=yes --enable-qd=yes --enable-dd=yes --enable-double=yes \
--enable-_Float64x=yes --enable-test=yes
...
$ make -j6
...
$ sudo make install
\end{verbatim}
}

\begin{table}
\caption{Prerequisites for compiling from source}
\begin{center}\label{prerequisites}
\begin{tabular}{lc}
Package & Version \\ \hline
GCC & 9 or later \\
gmake & 4.3 or later \\
git & 2.33.0 or later\\
autotools & 2.71 \\
automake & 1.16 \\ 
GNU sed & 4.1 or later \\ 
\hline
\end{tabular}
\end{center}
\end{table}

\section{Supported Floating point formats}
\label{sec:floatingpointformat}

We support binary64, FP80 (extended double), binary128, double-double, quad-double (QD library), GMP, and MPFR floating point arithmetics in MPLAPACK, and floating point types in C++ are summarized in table~\cite{mplapackformats}.

Binary64~\cite{4610935} is the so-called double precision: the number of significant digits in decimal is about 16, and CPUs perform arithmetic processing in hardware. As a result, the CPU can perform binary64 operations very fast.

FP80~\cite{30711} is the so-called extended double precision: the number of significant digits in decimal is about 19, and Intel and AMD CPUs perform arithmetic processing in hardware. ISO defined C real floating types~\cite{18661-3} as {\tt \_Float64x}. Thus We always use {\tt \_Float64x} as a type for FP80 numbers in MPLAPACK. However, no SIMD support, and due to its architecture, FP80 operations are more than ten times slower than binary64 operations. Besides, FP80 was removed since IEEE 754-2008~\cite{4610935}.

Binary128, sometimes called quadruple precision, has been defined since IEEE754-2008, \cite{4610935} and this format has approximately 33 significant decimal digits. Usually, binary128 arithmetic is done by software. Therefore, operations are prolonged. We know that only IBM z processors have been the only commercial platform supporting quadruple precision \cite{IBMz13}. Some processors like aarch64, Sparc, RISCV64, and MIPS64 define instructions for quadruple-precision arithmetic. These processors emulate binary128 instructions by software. ISO defined C real floating types~\cite{18661-3} as {\tt \_Float128} but not yet a standard of C++. GCC has already provided a type for binary128 as {\tt \_\_float128} since GCC 4.6. Intel one API also provides binary128 as {\tt \_Quad}. We always use {\tt \_Float128} as a type for binary128 numbers in MPLAPACK. This type may be the same as {\tt long double} or {\tt \_\_float128} depends on the environment, we typedef appropriate type to {\tt \_Float128}.

The double-double casts two binary64 numbers as one number and has approximately 32 decimal significant digits, and the quad-double casts four binary64 numbers as one number and has approximately 64 decimal significant digits. We use the QD library \cite{hida2000} to support double-double and quad-double precision. The double-double precision and quad-double precision use Kunth and Dekker's algorithm \cite{Knuth1997,Dekker1971}, which can evaluate the addition and multiplication of two binary64 numbers rigorously. Then, we can define the addition and multiplication of double-double numbers. The pros of using these formats are that they are speedy. Since all arithmetic can be done by binary64 and accelerated by hardware, calculation speed is approximately ten times faster than software implemented binary128. The cons of using these formats are programs written for expecting that the IEEE754 feature might not work with these precisions. Historically, IBM XL compilers for PowerPCs and GCC targeted to PowerPCs and PowerMacs, ``long double" has been equivalent to the double-double~\cite{ibmdd}. However, other environments do not always support ``long double." Therefore, we use the QD library for MPLAPACK, and we guess that is why Hida \etal{} developed the library.

GMP~\cite{Granlund12} is a C library for arbitrary precision arithmetic, operating on signed integers, rational numbers, and floating-point numbers. We can perform calculations to any accuracy, as long as the machine's resources allow. GMP comes with C++ binding, and we use this {\tt mpf\_class} as an arbitrary floating-point number. In addition, we support complex numbers by preparing ${\tt mpc\_class}$. 

MPFR~\cite{10.1145/1236463.1236468} is a C library for multiple-precision floating-point computations with correct rounding based on GMP. Unlike GMP, MPFR does not come with C++ binding, and we use {\tt mpreal}~\cite{mpreal} as an arbitrary floating-point number. We support complex numbers using MPC, a library for multiple-precision complex arithmetic with correct rounding~\cite{MPC} via ${\tt mpcomplex}$ class (more precisely, the final LGPL version with our customization). Both libraries provide almost the same functionalities, but MPFR and MPC are smooth extensions to IEEE 754 and further support trigonometric functions necessary for cosine-sine decomposition, elementary and special functions. Therefore, we will drop GMP support in the future.

{\tt long double} is no longer supported in MPLAPACK since version 1.0 because the situation regarding {\tt long double} is very different and confusing by  CPUs and OSes. E.g., on Intel CPUs, {\tt long double} is equivalent to {\tt \_Float64x} on Linux; however, on Windows, {\tt long double} is equivalent to {\tt double}. Similar confusion happens in the AArch64 environment. The official ARM ABI defines {\tt long double} to be binary128~\cite{arm64abi}. Nevertheless, Apple ABI overwrites {\tt long double} to double on the OS side. On IBM PowerPC or Power Macs, {\tt long double} have been double-double, as described above. When long double transit to {\tt \_\_float128} on PowerPC in the future, double-double will also be supported as {\tt \_\_ibm128}~\cite{ibmdd} as the GNU extension.

\begin{table}
\caption{The floating-point formats used in MPLAPACK, their type names, and accuracy in decimal digits.}\label{mplapackformats}
\begin{center}
\begin{tabular}{c|c|c} \hline
Library or format & type name & accuracy in decimal digits \\ \hline
GMP             & {\tt mpf\_class, mpc\_class}       & 154 (default) and arbitrary\\
MPFR            & {\tt mpreal, mpcomplex}            & 154 (default) and arbitrary \\
double-double   & {\tt dd\_real, dd\_complex}        & 32\\
quad-double     & {\tt qd\_real, qd\_complex}        & 64 \\
binary64        & {\tt double, std::complex<double>} & 16 \\
extended double & {\tt \_Float64x, std::complex<\_Float64x>} & 19  \\
binary128       & {\tt \_Float128, std::complex<\_Float128>} & 33 \\ 
integer         & mplapackint & (32bit on Win, 64bit on Linux and macOS) \\ \hline
\end{tabular}
\end{center}
\end{table}

\section{LAPACK and BLAS Routine Naming Conventions and available routines}
\label{sec:nameconventions}
BLAS and LAPACK prefix routine names with ``s" and ``d" for real single and double precision real numbers and ``c" and ``z" for single and double precision complex numbers. FORTRAN77 and Fortran90 are not case-sensitive in function and subroutine names, while C++ is case-sensitive in function names.

The prefix for real and complex routine names in MPLAPACK is an uppercase ``R" for real numbers and ``C" for complex numbers. All other letters in the routines are lowercase.
Also, while we do not distinguish function names by floating-point class (e.g. {\tt mpf\_class, \_Float128})~\ref{mplapackformats}, we use the same function name for different floating-point classes to take advantage of function overloading (e.g., no matter which floating-point class is adopted, program calls {\tt Raxpy} appropriately). The floating-point class is added after the routine name for routines that take no arguments or only integers. Otherwise, we add the prefix ``M" or insert ``M" after ``i."

For example, \begin{itemize}
    \item {\tt daxpy, zaxpy $\rightarrow$ Raxpy, Caxpy}
    \item {\tt dgemm, zgemm $\rightarrow$ Rgemm, Cgemm}
    \item {\tt dsterf, dsyev $\rightarrow$ Rsterf, Rsyev}
    \item {\tt dzabs1, dzasum $\rightarrow$ RCabs1, RCasum}
    \item {\tt lsame $\rightarrow$ Mlsame\_mpfr,  Mlsame\_gmp, Mlsame\_\_Float128} $\cdots$ etc.
    \item {\tt dlamch $\rightarrow$ Rlamch\_mpfr,  Rlamch\_gmp, Rlamch\_\_Float128} $\cdots$ etc.
    \item {\tt ilaenv $\rightarrow$ iMlaenv\_mpfr,  iMlaenv\_gmp, iMlaenv\_\_Float128} $\cdots$ etc.
\end{itemize}

In table~\ref{mpblasroutines}, we show all supported MPBLAS routines.
The prototype definitions of these routines can be found in the following headers.
\begin{itemize} 
\item {\tt /home/docker/MPLAPACK/include/mplapack/mpblas\_\_Float128.h}
\item {\tt /home/docker/MPLAPACK/include/mplapack/mpblas\_\_Float64x.h}
\item {\tt /home/docker/MPLAPACK/include/mplapack/mpblas\_dd.h}
\item {\tt /home/docker/MPLAPACK/include/mplapack/mpblas\_double.h}
\item {\tt /home/docker/MPLAPACK/include/mplapack/mpblas\_gmp.h}
\item {\tt /home/docker/MPLAPACK/include/mplapack/mpblas\_mpfr.h}
\item {\tt /home/docker/MPLAPACK/include/mplapack/mpblas\_qd.h}
\end{itemize}
A simple OpenMP version of MPBLAS is available.
In table~\ref{mpblasroutinesopt}, we show all OpenMP accelerated MPBLAS routines.
In table~\ref{mpblasroutines_cuda_dd}, we show all CUDA accelerated MPBLAS (double-double) routines. 

In table~\ref{mplapackdriver_real}, we show all supported MPLAPACK real driver routines.
In table~\ref{mplapackcomp_real}, we show all supported MPLAPACK real computational routines.
In table~\ref{mplapackdriver_complex}, we show all supported MPLAPACK complex driver routines.
In table~\ref{mplapackcomp_complex}, we show all supported MPLAPACK complex computational routines.

We do not list them here, but there are also many MPLPACK auxiliary routines.
We usually use MPLAPACK driver routines and MPBLAS routines directly. However, the driver routines also implicitly use computational routines and auxiliary routines. 

The prototype definitions of the MLAPACK routines can be found in the following headers.
\begin{itemize}
\item {\tt /home/docker/MPLAPACK/include/mplapack/mplapack\_\_Float128.h}
\item {\tt /home/docker/MPLAPACK/include/mplapack/mplapack\_\_Float64x.h}
\item {\tt /home/docker/MPLAPACK/include/mplapack/mplapack\_dd.h}
\item {\tt /home/docker/MPLAPACK/include/mplapack/mplapack\_double.h}
\item {\tt /home/docker/MPLAPACK/include/mplapack/mplapack\_gmp.h}
\item {\tt /home/docker/MPLAPACK/include/mplapack/mplapack\_mpfr.h}
\item {\tt /home/docker/MPLAPACK/include/mplapack/mplapack\_qd.h}
\end{itemize}

\begin{table}
\caption{Available MPBLAS routines}\label{mpblasroutines}
\begin{center}
{\tt
\begin{tabular}{cccccccccc} \hline
Crotg & Cscal & Rrotg & Rrot & Rrotm & CRrot & Cswap & Rswap & CRscal & Rscal \\
Ccopy & Rcopy & Caxpy & Raxpy & Rdot & Cdotc  & Cdotu & RCnrm2 & Rnrm2 & Rasum \\
iCasum & iRamax & RCabs1& Mlsame & Mxerbla \\ \hline
Cgemv & Rgemv & Cgbmv & Rgbmv & Chemv & Chbmv & Chpmv & Rsymv & Rsbmv & Ctrmv \\
Cgemv & Rgemv & Cgbmv & Rgemv & Chemv & Chbmv & Chpmv & Rsymv & Rsbmv & Rspmv \\
Ctrmv & Rtrmv & Ctbmv & Ctpmv & Rtpmv & Ctrsv & Rtrsv & Ctbsv & Rtbsv & Ctpsv \\
Rger  & Cgeru & Cgerc & Cher  & Chpr & Cher2  & Chpr2  & Rsyr  & Rspr & Rsyr2 \\
Rspr2 \\ \hline
Cgemm & Rgemm & Csymm & Rsymm & Chemm & Csyrk  & Rsyrk & Cherk & Csyr2k& Rsyr2k \\
Cher2k& Ctrmm & Rtrmm & Ctrsm & Rtrsm\\\hline
\end{tabular}
}
\end{center}
\end{table}

\begin{table}
\caption{Available OpenMP version of MPBLAS routines}\label{mpblasroutinesopt}
\begin{center}
{\tt
\begin{tabular}{cccccccccc} \hline
Raxpy & Rcopy & Rdot & Rgemm \\ \hline
\end{tabular}
}
\end{center}
\end{table}

\begin{table}
\caption{Available CUDA version of MPBLAS (double-double) routines}\label{mpblasroutines_cuda_dd}
\begin{center}
{\tt
\begin{tabular}{cccccccccc} \hline
Rgemm & Rsyrk \\ \hline
\end{tabular}
}
\end{center}
\end{table}

\begin{table}
\caption{Available MPLAPACK Real Driver routines (2.0.1)}\label{mplapackdriver_real}
\begin{center}
{\tt 
\begin{tabular}{cccccccccc} \hline
Rgesv  & Rgesvx & Rgbsv  & Rgbsvx & Rgtsv & Rgtsvx & Rposv  & Rposvx & Rppsv  & Rppsvx \\
Rpbsv  & Rpvsvx & Rptsv  & Rptsvx & Rsysv & Rsysvx & Rpspv  & Rspsvx \\ \hline %%https://netlib.org/lapack/lug/node26.html
Rgels  & Rgelsy & Rgelss & Rgelsd \\ \hline %%https://netlib.org/lapack/lug/node27.html
Rgglse & Rggglm \\ \hline %%https://netlib.org/lapack/lug/node28.html
Rsyev & Rsyevd & Rsyevx & Rsyevr & Rspev & Rspevd & Rspevx & Rsbev & Rsbevd & Rsbevx \\
Rstev & Rstevd & Rstevx & Rstevr & Rgees & Rgeesx & Rgeev & Rgeevx & Rgesvd & Rgesdd \\ \hline %%https://netlib.org/lapack/lug/node32.html
Rsygv & Rsygvd & Rsygvx & Rspgv & Rspgvd & Rspgvx & Rsbgv & Rsbgv & Rsbgvx & Rgges \\
Rggesx & Rggev & Rggevx & Rggsvd \\ \hline %%https://netlib.org/lapack/lug/node36.html
\end{tabular}
}
\end{center}
\end{table}

\begin{table}
\caption{Available MPLAPACK Complex Driver routines (2.0.1)}\label{mplapackdriver_complex}
\begin{center}
{\tt 
\begin{tabular}{cccccccccc} \hline
Cgesv & Cgesvx & Cgbsv & Cgbsvx & Cgtsv & Cgtsvx & Cposv & Cposvx & Cppsv & Cppsvx \\
Cpbsv & Cpbsvx & Cptsv & Cptsvx & Chesv & Chesvx & Csysv & Csysvx & Chpsv & Chpsvx \\
Cspsv & Cspsvx \\ \hline %%https://netlib.org/lapack/lug/node26.html
Cgels & Cgelsy & Cgelss & Cgelsd  \\ \hline %%https://netlib.org/lapack/lug/node27.html
Cgglse & Cggglm \\ \hline %%https://netlib.org/lapack/lug/node28.html
Cheev & Cheevd & Cheevx & Cheevr & Chpev & Chpevd & Chpevx & Chbev & Chbevd & Chbevx \\
Cgees & Cgeesx & Cgeev & Cgeevx & Cgesvd & Cgesdd \\ \hline %%https://netlib.org/lapack/lug/node32.html
Chegv & Chegvd & Chegvx & Chpgv & Chpgvd & Chpgvx & Chbgv & Chbgvd & Chbgvx & Cgges \\ %%https://netlib.org/lapack/lug/node36.html
Cggesx & Cggev & Cggevx & Cggsvd \\ \hline 
\end{tabular}
}
\end{center}
\end{table}

\begin{table}
\caption{Available MPLAPACK Computational Real routines (2.0.1)}\label{mplapackcomp_real}
\begin{center}
{\tt
\begin{tabular}{cccccccccccc} \hline
Rgetrf & Rgetrs & Rgecon & Rgerfs & Rgetri & Rgeequ & Rgbtrf & Rgbtrs & Rgbcon & Rgbrfs \\
Rgbequ & Rgttrf & Rgttrs & Rgtcon & Rgtrfs & Rpotrf & Rpotrs & Rpocon & Rporfs & Rpotri \\
Rpoequ & Rpptrf & Rpptrs & Rppcon & Rpprfs & Rpptri & Rppequ & Rpbtrf & Rpbtrs & Rpbcon \\
Rpbrfs & Rpbequ & Rpttrf & Rpttrs & Rptcon & Rptrfs & Rsytrf & Rsytrs & Rsycon & Rsyrfs \\
Rsytri & Rsptrf & Rsptrs & Rspcon & Rsprfs & Rsptri & Rtrtrs & Rtrcon & Rtrrfs & Rtrtri \\
Rtptrs & Rtpcon & Rtprfs & Rtptri & Rtbtrs & Rtbcon & Rtbrfs \\ \hline
Rgeqp3 & Rgeqrf & Rorgqr & Rormqr & Rgelqf & Rorglq & Rormlq & Rgeqlf & Rorgql & Rormql \\
Rgerqf & Rorgrq & Rormrq & Rtzrzf & Rormrz \\ \hline
Rsytrd & Rsptrd & Rsbtrd & Rorgtr & Rormtr & Ropgtr & Ropmtr & Rsteqr & Rsterf & Rstedc \\
Rstegr & Rstebz & Rstein & Rpteqr \\ \hline
Rgehrd & Rgebal & Rgebak & Rorghr & Rormhr & Rhseqr & Rhsein & Rtrevc & Rtrexc & Rtrsyl \\
Rtrsna & Rtrsen \\ \hline
Rgebrd & Rgbbrd & Rorgbr & Rormbr & Rbdsqr & Rbdsdc \\ \hline
Rsygst & Rspgst & Rpbstf & Rsbgst \\ \hline
Rgghrd & Rggbal & Rggbak & Rhgeqz & Rtgevc & Rtgexc & Rtgsyl & Rtgsna & Rtgsen \\ \hline
Rggsvp & Rtgsja \\
\hline
\end{tabular}
}
\end{center}
\end{table}

\begin{table}
\caption{Available MPLAPACK Computational Complex routines (2.0.1)}\label{mplapackcomp_complex}
\begin{center}
{\tt
\begin{tabular}{cccccccccccc} \hline
Cgetrf & Cgetrs & Cgecon & Cgerfs & Cgetri & Cgeequ & Cgbtrf & Cgbtrs & Cgbcon & Cgbrfs \\ %%https://netlib.org/lapack/lug/node38.html
Cgbequ & Cgttrf & Cgttrs & Cgtcon & Cgtrfs & Cpotrf & Cpotrs & Cpocon & Cporfs & Cpotri \\
Cpoequ & Cpptrf & Cpptrs & Cppcon & Cpprfs & Cpptri & Cppequ & Cpbtrf & Cpbtrs & Cpbcon \\
Cpbrfs & Cpbequ & Cpttrf & Cpttrs & Cptcon & Cptrfs & Chetrf & Chetrs & Checon & Cherfs \\
Chetri & Csytrf & Csytrs & Csycon & Csyrfs & Csytri & Chptrf & Chptrs & Chpcon & Chprfs \\
Chptri & Csptrf & Csptrs & Cspcon & Csprfs & Csptri & Ctrtrs & Ctrcon & Ctrrfs & Ctrtri \\
Ctptrs & Ctpcon & Ctprfs & Ctptri & Ctbtrs & Ctbcon & Ctbrfs \\ \hline
Cgeqp3 & Cgeqrf & Cungqr & Cunmqr & Cgelqf & Cunglq & Cunmlq & Cgeqlf & Cungql & Cunmql \\ %%https://netlib.org/lapack/lug/node44.html
Cgerqf & Cungrq & Cunmrq & Ctzrzf & Cunmrz \\ \hline
Chetrd & Chptrd & Chbtrd & Cungtr & Cunmtr & Cupgtr & Cupmtr & Csteqr & Cstedc & Cstegr \\ %%https://netlib.org/lapack/lug/node48.html
Cstein & Cpteqr \\ \hline
Cgehrd & Cgebal & Cgebak & Cunghr & Cunmhr & Chseqr & Chsein & Ctrevc & Ctrexc & Ctrsyl \\ %%https://netlib.org/lapack/lug/node52.html
Ctrsna & Ctrsen \\ \hline
Cgebrd & Cgbbrd & Cungbr & Cunmbr & Cbdsqr \\ \hline %%https://netlib.org/lapack/lug/node53.html
Chegst & Chpgst & Cpbstf & Chbgst \\ \hline %%https://netlib.org/lapack/lug/node54.html
Cgghrd & Cggbal & Cggbak & Chgeqz & Ctgevc & Ctgexc & Ctgsyl & Ctgsna & Ctgsen \\ \hline %%https://netlib.org/lapack/lug/node58.html
Cggsvp & Ctgsja \\ 
\hline
\end{tabular}
}
\end{center}
\end{table}

%% how to convert
%% REAL
%% $ wget https://www.netlib.org/lapack/lug/node59.html
%% $ grep TD node59.html  | sed 's/<TD ALIGN="LEFT">//g' | grep ^D | sed 's/</ /g'  | awk '{print $1}' |  tr "[A-Z]" "[a-z]" | awk '{ print toupper(substr($0, 1, 1)) substr($0, 2, length($0) - 1) }'  | sed 's/D/R/g'  | awk '{print $1 " &" }' | tr "\n" " ";echo
%% COMPLEX
%% $ https://www.netlib.org/lapack/lug/node38.html
%% $ grep TD node38.html  | sed 's/<TD ALIGN="LEFT">//g' | grep ^Z | sed 's/</ /g'  | awk '{print $1}' |  tr "[A-Z]" "[a-z]" | awk '{ print toupper(substr($0, 1, 1)) substr($0, 2, length($0) - 1) }'  | sed 's/Z/C/g'  | awk '{print $1 " &" }' | tr "\n" " ";echo

\section{How to use MPBLAS and MPLPACK}
\label{sec:howtouse}
\subsection{Multiprecision Types}
We use seven kinds of floating-point formats. Multiple precision types are listed in the table~\ref{mplapackformats}.
For GMP, we use built-in {\tt mpf\_class} for real type. For complex type, we developed {\tt mpc\_class}.
For MPFR, we use a modified version of mpfrc++ (the final LGPL version) to treat like double or float type. For complex type,
we developed {\tt mpcomplex.h} using MPFR and MPC for complex type.
For double-double and quad-double, we use {\tt dd\_real} and {\tt qd\_real}, respectively. For complex type, we developed {\tt dd\_complex} and {\tt qd\_complex}, respectively.

For extended double, we use {\tt \_Float64x} for all Intel and AMD environments. For binary128, we use {\_Float128} for
all OSes and CPUs. For complex types for {\tt double}, {\tt \_Float64x} and {\tt \_Float128}, we use standard complex implementation of C++. Using {\tt \_\_float80}, {\tt \_\_float128}, or {\tt long double} is strongly discouraged as they are not tested and not portable.

Also, we define {\tt mplapackint} as a 64-bit signed integer to MPLAPACK that can access all the memory spaces regardless
of which data type models the environment employs (LLP64 or LP64). On Windows, {\tt mplapackint} is still a 32-bit signed integer.

\subsection{General pitfall substituting floating point numbers to multiple precision numbers}
Suppose when we substitute $1.2$ to multiple numbers {\tt alpha}
\begin{verbatim}
    alpha = 1.2;
\end{verbatim}
Such code is problematic. First, GCC translates ``1.2" into double precision, then substituting {\tt alpha}.
The compilation of this code cause exact numbers to be rounded to double precision. Thus, when we print this number
up 64 decimal digits using default precision (512 bits = 153 decimal digits) of GMP in MPBLAS, {\tt alpha} becomes
\begin{verbatim}
    -1.1999999999999999555910790149937383830547332763671875000000000000e+00
\end{verbatim}
. This may be an undesired result. To avoid this behavior, we should code as follows:
\begin{verbatim}
    alpha = "1.2";
\end{verbatim}
Alternatively, use the constructer explicitly as follows:
\begin{verbatim}
    alpha = mpf_class("1.2");
\end{verbatim}
Then, the output becomes as desired.
\begin{verbatim}
    -1.2000000000000000000000000000000000000000000000000000000000000000e+00
\end{verbatim}
For GMP, to substitute an imaginary number $1.2+1.2i$ to {\tt beta}, we should code as follows:
\begin{verbatim}
    beta = mpc_class(mpf_class("1.2"), mpf_class("1.2");
\end{verbatim}
For real numbers, we verified that all precision have string type constructor except for {\tt \_Float128} and {\tt \_Float64x} as following:
\begin{verbatim}
    alpha = "1.2";
\end{verbatim}
For complex numbers, we have to explicitly use real constructors except for {\tt \_Float128} and {\tt \_Float64x} as follows:
\begin{verbatim}
    beta = dd_complex(dd_real("1.2"), dd_real("1.2");
\end{verbatim}

\subsection{How to use MPBLAS}
\label{howtouseblas}
This subsection describes the basic usage of MPBLAS, OpenMP accelerated MPBLAS and CUDA accelerated MPBLAS.

\subsubsection{How to use reference MPBLAS}
The API of MPBLAS is very similar to the original BLAS and CBLAS. However, unlike CBLAS, we always use a one-dimensional array as a column-major type matrix in MPBLAS. 

Here, we show how to use MPBLAS by two examples; {\tt Rgemm} and {\tt Cgemm}. Of course, other routines can be used similarly. However, first, we show how to use {\tt Rgemm}, which corresponds to {\tt DGEMM} of BLAS. 

Following is the prototype definition of the multiple-precision version of matrix-matrix multiplication ({\tt Rgemm}) of the MPFR version.
\begin{verbatim}
void Rgemm(const char *transa, const char *transb, mplapackint const m, 
mplapackint const n, mplapackint const k, mpreal const alpha, mpreal *a, 
mplapackint const lda, mpreal *b, mplapackint const ldb, mpreal const beta, 
mpreal*c, mplapackint const ldc);
\end{verbatim}
Moreover, the following is the definition part of the original DGEMM:
\begin{verbatim}
SUBROUTINE DGEMM(TRANSA,TRANSB,M,N,K,ALPHA,A,LDA,B,LDB,BETA,C,LDC)
*     .. Scalar Arguments ..
      DOUBLE PRECISION ALPHA,BETA
      INTEGER K,LDA,LDB,LDC,M,N
      CHARACTER TRANSA,TRANSB
*     ..
*     .. Array Arguments ..
      DOUBLE PRECISION A(LDA,*),B(LDB,*),C(LDC,*)
\end{verbatim}
There is a clear correspondence between variables in the C++ prototype of Rgemm and variables in the header of {\tt DGEMM}.
\begin{itemize}
    \item {\tt CHARACTER} $\rightarrow$ {\tt const char *}
    \item {\tt INTEGER} $\rightarrow$ {\tt mplapackint}
    \item {\tt DOUBLE PRECISION A(LDA, *)} $\rightarrow$ {\tt mpreal *a}
\end{itemize}
We can see such correspondences for other MPBLAS and MPLAPACK routines.

Then, let us see how we multiply matrices using the MPFR version of {\tt Rgemm},
\[
 \alpha A B + \beta C \rightarrow C,
\]
where $A$, $B$ and $C$ are matrices, $\alpha$ and $\beta$ are scalars. 
Let us choose $A$, $B$, and $C$
\begin{equation}
A = \left (
\begin{array}{rrr}
1 & 8 & 3 \\
0 & 10 & 8 \\
9 & -5 & -1
\end{array}
\right ),
B = \left (
\begin{array}{rrr}
9 & 8 & 3 \\
3 & -11 & 0 \\
-8 & 6 & 1
\end{array}
\right ),
C = \left (
\begin{array}{rrr}
3 & 3 & 0 \\
8 & 4 & 8 \\
6 & 1 & -2
\end{array}
\right ), \nonumber
\end{equation}
and $\alpha = 3$, and $\beta=-2$.

The answer is:
\begin{equation}
\alpha AB + \beta C = \left (
\begin{array}{rrr}
21 & -192 & 18 \\
-118 & -194 & 8 \\
210 & 361 & 82
\end{array}
\right ). \nonumber
\end{equation}

The list is the following.
\begin{lstlisting}
#include <mpblas_mpfr.h>

//Matlab/Octave format
void printmat(int N, int M, mpreal * A, int LDA)
{
    mpreal mtmp;
    printf("[ ");
    for (int i = 0; i < N; i++) {
        printf("[ ");
        for (int j = 0; j < M; j++) {
            mtmp = A[i + j * LDA];
            mpfr_printf("%5.2Re", mpfr_ptr(mtmp));
            if (j < M - 1)
                printf(", ");
        }
        if (i < N - 1)
            printf("]; ");
        else
            printf("] ");
    }
    printf("]");
}

int main()
{
    mplapackint n = 3;
//initialization of MPFR
    int default_prec = 256;
    mpfr_set_default_prec(default_prec);

    mpreal *A = new mpreal[n * n];
    mpreal *B = new mpreal[n * n];
    mpreal *C = new mpreal[n * n];
    mpreal alpha, beta;

//setting A matrix
    A[0 + 0 * n] = 1;    A[0 + 1 * n] = 8;    A[0 + 2 * n] = 3;
    A[1 + 0 * n] = 0;    A[1 + 1 * n] = 10;   A[1 + 2 * n] = 8;
    A[2 + 0 * n] = 9;    A[2 + 1 * n] = -5;   A[2 + 2 * n] = -1;

    B[0 + 0 * n] = 9;    B[0 + 1 * n] = 8;    B[0 + 2 * n] = 3;
    B[1 + 0 * n] = 3;    B[1 + 1 * n] = -11;  B[1 + 2 * n] = 0;
    B[2 + 0 * n] = -8;   B[2 + 1 * n] = 6;    B[2 + 2 * n] = 1;

    C[0 + 0 * n] = 3;    C[0 + 1 * n] = 3;    C[0 + 2 * n] = 0;
    C[1 + 0 * n] = 8;    C[1 + 1 * n] = 4;    C[1 + 2 * n] = 8;
    C[2 + 0 * n] = 6;    C[2 + 1 * n] = 1;    C[2 + 2 * n] = -2;

    printf("# Rgemm demo...\n");

    printf("A ="); printmat(n, n, A, n); printf("\n");
    printf("B ="); printmat(n, n, B, n); printf("\n");
    printf("C ="); printmat(n, n, C, n); printf("\n");
    alpha = 3.0;
    beta = -2.0;
    Rgemm("n", "n", n, n, n, alpha, A, n, B, n, beta, C, n);

    mpfr_printf("alpha = %5.3Re\n", mpfr_ptr(alpha));
    mpfr_printf("beta  = %5.3Re\n", mpfr_ptr(beta));
    printf("ans ="); printmat(n, n, C, n); printf("\n");
    printf("#please check by Matlab or Octave following and ans above\n");
    printf("alpha * A * B + beta * C =\n");
    delete[]C;
    delete[]B;
    delete[]A;
}
\end{lstlisting}
First, we must include {\tt mpblas\_mpfr.h} to use MPFR. {\tt printmat} function prints matrix in Octave/Matlab format (lines 4 to 22). 
We set 256 bit in fraction for mpfrc++ in lines 28 to 29; MPFR Real accuracy setted to  77 decimal digits ($77.06 = \log_{10} 2^{256}$).
We allocate the matrix as a one-dimensional array (lines 31 to 33). Then we set the matrix $A$, $B$, and $C$. We always input the matrix by row-major format to an array (lines 36 to 47). Matrix-matrix multiplication {\tt Rgemm} is called in line 58. 
One can input the list and save as {\tt Rgemm\_mpfr.cpp} or you can find {\tt /home/docker/mplapack/examples/mpblas/} directory.
Then, one can compile on one own following in the Docker environment as follows:
\begin{verbatim}
$ g++ -O2 -I/home/docker/MPLAPACK/include -I/home/docker/MPLAPACK/include/mplapack \
Rgemm_mpfr.cpp -Wl,--rpath=/home/docker/MPLAPACK/lib -L/home/docker/MPLAPACK/lib \
-lmpblas_mpfr -lgmp -lmpfr -lmpc
\end{verbatim}
If the compilation is done successfully, one can run as follows:
{\fontsize{9.5pt}{0.4cm}\selectfont
\begin{verbatim}
$ ./a.out
# Rgemm demo...
A =[ [ 1.00e+00, 8.00e+00, 3.00e+00]; [ 0.00e+00, 1.00e+01, 8.00e+00]; [ 9.00e+00, -5.00e+00, -1.00e+00] ]
B =[ [ 9.00e+00, 8.00e+00, 3.00e+00]; [ 3.00e+00, -1.10e+01, 0.00e+00]; [ -8.00e+00, 6.00e+00, 1.00e+00] ]
C =[ [ 3.00e+00, 3.00e+00, 0.00e+00]; [ 8.00e+00, 4.00e+00, 8.00e+00]; [ 6.00e+00, 1.00e+00, -2.00e+00] ]
alpha = 3.000e+00
beta  = -2.000e+00
ans =[ [ 2.10e+01, -1.92e+02, 1.80e+01]; [ -1.18e+02, -1.94e+02, 8.00e+00]; [ 2.10e+02, 3.61e+02, 8.20e+01] ]
#please check by Matlab or Octave following and ans above
alpha * A * B + beta * C
\end{verbatim}
}
One can check the result by comparing the result of the octave.
{\footnotesize
\begin{verbatim}
$ ./a.out | octave
octave: X11 DISPLAY environment variable not set
octave: disabling GUI features
A =

    1    8    3
    0   10    8
    9   -5   -1

B =

    9    8    3
    3  -11    0
   -8    6    1

C =

   3   3   0
   8   4   8
   6   1  -2

alpha =  3
beta = -2
ans =

    21  -192    18
  -118  -194     8
   210   361    82

ans =

    21  -192    18
  -118  -194     8
   210   361    82
\end{verbatim}
}
In this case, we see the result below two ``{\tt ans =}'' are the same. The {\tt Rgemm} result is correct up to 16 decimal digits.

Let us see how we can multiply matrices using {\tt \_Float128} (binary128). The list is the following.
\begin{lstlisting}
#include <mpblas__Float128.h>
#include <stdio.h>
#define BUFLEN 1024

void printnum(_Float128 rtmp)
{
    int width = 42;
    char buf[BUFLEN];
#if defined ___MPLAPACK_WANT_LIBQUADMATH___
    int n = quadmath_snprintf (buf, sizeof buf, "%+-#*.35Qe", width, rtmp);
#elif defined ___MPLAPACK_LONGDOUBLE_IS_BINARY128___
    snprintf (buf, sizeof buf, "%.35Le", rtmp);
#else
    strfromf128(buf, sizeof(buf), "%.35e", rtmp);
#endif
    printf ("%s", buf);
    return;
}

//Matlab/Octave format
void printmat(int N, int M, _Float128 * A, int LDA)
{
    _Float128 mtmp;

    printf("[ ");
    for (int i = 0; i < N; i++) {
        printf("[ ");
        for (int j = 0; j < M; j++) {
            mtmp = A[i + j * LDA];
            printnum(mtmp);
            if (j < M - 1)
                printf(", ");
        }
        if (i < N - 1)
            printf("]; ");
        else
            printf("] ");
    }
    printf("]");
}

int main()
{
    mplapackint n = 3;

    _Float128 *A = new _Float128[n * n];
    _Float128 *B = new _Float128[n * n];
    _Float128 *C = new _Float128[n * n];
    _Float128 alpha, beta;

//setting A matrix
    A[0 + 0 * n] = 1;    A[0 + 1 * n] = 8;    A[0 + 2 * n] = 3;
    A[1 + 0 * n] = 2.5;  A[1 + 1 * n] = 10;   A[1 + 2 * n] = 8;
    A[2 + 0 * n] = 9;    A[2 + 1 * n] = -5;   A[2 + 2 * n] = -1;

    B[0 + 0 * n] = 9;    B[0 + 1 * n] = 8;    B[0 + 2 * n] = 3;
    B[1 + 0 * n] = 3;    B[1 + 1 * n] = -11;  B[1 + 2 * n] = 4.8;
    B[2 + 0 * n] = -8;   B[2 + 1 * n] = 6;    B[2 + 2 * n] = 1;

    C[0 + 0 * n] = 3;    C[0 + 1 * n] = 3;    C[0 + 2 * n] = 1.2;
    C[1 + 0 * n] = 8;    C[1 + 1 * n] = 4;    C[1 + 2 * n] = 8;
    C[2 + 0 * n] = 6;    C[2 + 1 * n] = 1;    C[2 + 2 * n] = -2;

    printf("# Rgemm demo...\n");

    printf("A ="); printmat(n, n, A, n); printf("\n");
    printf("B ="); printmat(n, n, B, n); printf("\n");
    printf("C ="); printmat(n, n, C, n); printf("\n");
    alpha = 3.0;
    beta = -2.0;
    Rgemm("n", "n", n, n, n, alpha, A, n, B, n, beta, C, n);

    printf("alpha = "); printnum(alpha); printf("\n");
    printf("beta  = "); printnum(beta);  printf("\n");
    printf("ans ="); printmat(n, n, C, n); printf("\n");
    printf("#please check by Matlab or Octave following and ans above\n");
    printf("alpha * A * B + beta * C \n");
    delete[]C;
    delete[]B;
    delete[]A;
}
\end{lstlisting}
We do not show the output since the output is almost the same as the MPFR version.

The {\tt \_Float128} version of the program list is almost similar to the MPFR version. However, {\tt printnum} part is a bit complicated. Since there are at least three kinds of binary128 support depending on CPUs and OSes: (i) GCC only supports {\tt \_\_float128} using {\tt libquadmath} (Windows, macOS), (ii) GCC supports {\tt \_Float128} directly and there is libc support as well (Linux amd64), (iii) {\tt long double} is already binary128, and no special support is necessary (AArch64). For (i), we internally define {\tt \_\_\_MPLAPACK\_WANT\_LIBQUADMATH\_\_\_}, for (ii) we define {\tt \_\_\_MPLAPACK\_\_FLOAT128\_ONLY\_\_\_}, and for (iii),
we define {\tt \_\_\_MPLAPACK\_LONGDOUBLE\_IS\_BINARY128\_\_\_}. Users can use these internal definitions to write a portable program.

As a summary, we show how we compile and run binary64, FP80, binary128, double-double, quad-double, GMP and MPFR versions of {\tt Rgemm} demo programs in {\tt /home/docker/mplapack/examples/mpblas} as follows:
\begin{itemize}
    \item binary64 (double) version
\begin{verbatim}
$ g++ -O2 -I/home/docker/MPLAPACK/include -I/home/docker/MPLAPACK/include/mplapack \
Rgemm_double.cpp -Wl,--rpath=/home/docker/MPLAPACK/lib \
-L/home/docker/MPLAPACK/lib -lmpblas_double
$ ./a.out
\end{verbatim}
\item FP80 (extended double) version
\begin{verbatim}
$ g++ -O2 -I/home/docker/MPLAPACK/include -I/home/docker/MPLAPACK/include/mplapack \
Rgemm__Float64x.cpp -Wl,--rpath=/home/docker/MPLAPACK/lib \
-L/home/docker/MPLAPACK/lib -lmpblas__Float64x
$ ./a.out
\end{verbatim}
\item binary128 version
\begin{verbatim}
$ g++ -O2 -I/home/docker/MPLAPACK/include -I/home/docker/MPLAPACK/include/mplapack \
Rgemm__Float128.cpp -Wl,--rpath=/home/docker/MPLAPACK/lib \
-L/home/docker/MPLAPACK/lib -lmpblas__Float128
$ ./a.out
\end{verbatim}
or (on macOS, mingw64 and CentOS7 amd64)
\begin{verbatim}
$ g++ -O2 -I/home/docker/MPLAPACK/include -I/home/docker/MPLAPACK/include/mplapack \
Rgemm__Float128.cpp -Wl,--rpath=/home/docker/MPLAPACK/lib \
-L/home/docker/MPLAPACK/lib -lmpblas__Float128 -lquadmath
$ ./a.out
\end{verbatim}
\item double-double version
\begin{verbatim}
$ g++ -O2 -I/home/docker/MPLAPACK/include -I/home/docker/MPLAPACK/include/mplapack \
Rgemm_dd.cpp -Wl,--rpath=/home/docker/MPLAPACK/lib \ 
-L/home/docker/MPLAPACK/lib -lmpblas_dd -lqd
$ ./a.out
\end{verbatim}
\item quad-double version
\begin{verbatim}
$ g++ -O2 -I/home/docker/MPLAPACK/include -I/home/docker/MPLAPACK/include/mplapack \
Rgemm_qd.cpp -Wl,--rpath=/home/docker/MPLAPACK/lib \
-L/home/docker/MPLAPACK/lib -lmpblas_qd -lqd
$ ./a.out
\end{verbatim}
\item GMP version
\begin{verbatim}
$ g++ -O2 -I/home/docker/MPLAPACK/include -I/home/docker/MPLAPACK/include/mplapack \
Rgemm_gmp.cpp -Wl,--rpath=/home/docker/MPLAPACK/lib \
-L/home/docker/MPLAPACK/lib -lmpblas_gmp -lgmpxx -lgmp
$ ./a.out
\end{verbatim}
\item MPFR version
\begin{verbatim}
$ g++ -O2 -I/home/docker/MPLAPACK/include -I/home/docker/MPLAPACK/include/mplapack \
Rgemm_mpfr.cpp -Wl,--rpath=/home/docker/MPLAPACK/lib \
-L/home/docker/MPLAPACK/lib -lmpblas_mpfr -lmpfr -lmpc -lgmp
$ ./a.out
\end{verbatim}
\end{itemize}

Next, we show how to use {\tt Cgemm} the complex version gemm. This is not just an example of how to use {\tt Cgemm}, but an example of how to input floating point numbers.

Following is the prototype definition of the multiple-precision version of matrix-matrix multiplication ({\tt Cgemm}) of the GMP version.
\begin{verbatim}
void Cgemm(const char *transa, const char *transb, mplapackint const m, 
mplapackint const n, mplapackint const k, mpc_class const alpha, mpc_class *a, 
mplapackint const lda, mpc_class *b,  mplapackint const ldb, mpc_class const beta, 
mpc_class *c, mplapackint const ldc);
\end{verbatim}
The following is the definition part of the original ZGEMM:
\begin{verbatim}
*       SUBROUTINE ZGEMM(TRANSA,TRANSB,M,N,K,ALPHA,A,LDA,B,LDB,BETA,C,LDC)
*
*       .. Scalar Arguments ..
*       COMPLEX*16 ALPHA,BETA
*       INTEGER K,LDA,LDB,LDC,M,N
*       CHARACTER TRANSA,TRANSB
*       ..
*       .. Array Arguments ..
*       COMPLEX*16 A(LDA,*),B(LDB,*),C(LDC,*)
\end{verbatim}
There is a clear correspondence between variables in the C++ prototype of {\tt Cgemm} and variables in the header of {\tt ZGEMM} like {\tt Rgemm}'s case.
\begin{itemize}
    \item {\tt CHARACTER} $\rightarrow$ {\tt const char *}
    \item {\tt INTEGER} $\rightarrow$ {\tt mplapackint}
    \item {\tt COMPLEX*16 A(LDA, *)} $\rightarrow$ {\tt mpc\_class *a}
\end{itemize}
Let us see how we can multiply matrices using GMP.
\[
 \alpha A B + \beta C \rightarrow C,
\]
where
{\footnotesize
\begin{equation}
A = \left (
\begin{array}{rrr}
1-i   & 8 + 2.2i & -10i \\
2     & 10       & 8.1+2.2i \\
-9+3i & -5 + 3i  & -1.0
\end{array}
\right ),
 B = \left (
\begin{array}{rrr}
9      & 8 -0.01i   & 3.1.001i \\
3 -8i  & -11 + 0.1i & 8+0.0000i \\
-8 + i & 6.0        & 1,1 + 1.1i
\end{array}
\right ),
 C = \left (
\begin{array}{rrr}
3+i & 3+9.99i & -9-11i \\
8-i & 4+4.44i & 8 + 9i \\
6   & -1      & -2 + 1i
\end{array}
\right ), \nonumber
\end{equation}
}
and $\alpha = 3-1.2i$, and $\beta=-2-2i$.

The answer is:
\begin{equation}
\alpha AB + \beta C = \left (
\begin{array}{rrr}
194.120 -  39.920i &-324.402 - 191.934i &  235.524 -  39.798i \\
  -182.400 - 259.700i & -118.304 +  80.140i &  295.157 - 107.686i \\
  -114.000 + 289.800i &-79.102 +   1.694i & -179.720 + 156.296i
\end{array}
\right ).
\end{equation}
Following is the program list for multiplying complex matrices using GMP.

\begin{lstlisting}[basicstyle=\ttfamily]
//public domain
#include <mpblas_gmp.h>
#include <iostream>
#include <cstring>
#include <algorithm>

#define GMP_FORMAT "%+68.64Fe"
#define GMP_SHORT_FORMAT "%+20.16Fe"

inline void printnum(mpf_class rtmp) { gmp_printf(GMP_FORMAT, rtmp.get_mpf_t()); }
inline void printnum_short(mpf_class rtmp) { gmp_printf(GMP_SHORT_FORMAT, rtmp.get_mpf_t()); }
inline void printnum(mpc_class ctmp) { gmp_printf(GMP_FORMAT GMP_FORMAT "i", ctmp.real().get_mpf_t(), ctmp.imag().get_mpf_t()); }

//Matlab/Octave format
template <class X> void printvec(X *a, int len) {
    X tmp;
    printf("[ ");
    for (int i = 0; i < len; i++) {
        tmp = a[i];
        printnum(tmp);
        if (i < len - 1)
            printf(", ");
    }
    printf("]");
}

template <class X> void printmat(int n, int m, X *a, int lda)
{
    X mtmp;

    printf("[ ");
    for (int i = 0; i < n; i++) {
        printf("[ ");
        for (int j = 0; j < m; j++) {
            mtmp = a[i + j * lda];
            printnum(mtmp);
            if (j < m - 1)
                printf(", ");
        }
        if (i < n - 1)
            printf("]; ");
        else
            printf("] ");
    }
    printf("]");
}
int main()
{
    mplapackint n = 3;

    mpc_class *a = new mpc_class[n * n];
    mpc_class *b = new mpc_class[n * n];
    mpc_class *c = new mpc_class[n * n];
    mpc_class alpha, beta;

//setting A matrix
    a[0 + 0 * n] = mpc_class(1.0,-1.0);    a[0 + 1 * n] = mpc_class(8.0, 2.2);    a[0 + 2 * n] = mpc_class(0.0, -10.0);
    a[1 + 0 * n] = mpc_class(2.0, 0.0);    a[1 + 1 * n] = mpc_class(10.0,0.0);    a[1 + 2 * n] = mpc_class(8.1, 2.2);
    a[2 + 0 * n] = mpc_class(-9.0,3.0);    a[2 + 1 * n] = mpc_class(-5.0,3.0);    a[2 + 2 * n] = mpc_class(-1.0, 0.0);

    b[0 + 0 * n] = mpc_class(9.0, 0.0);    b[0 + 1 * n] = mpc_class(8.0, -0.01);  b[0 + 2 * n] = mpc_class(3.0, 1.001);
    b[1 + 0 * n] = mpc_class(3.0, -8.0);   b[1 + 1 * n] = mpc_class(-11.0, 0.1);  b[1 + 2 * n] = mpc_class(8.0, 0.00001);
    b[2 + 0 * n] = mpc_class(-8.0, 1.0);   b[2 + 1 * n] = mpc_class(6.0, 0.0);    b[2 + 2 * n] = mpc_class(1.1, 1.0);

    c[0 + 0 * n] = mpc_class(3.0, 1.0);   c[0 + 1 * n] = mpc_class(-3.0, 9.99);   c[0 + 2 * n] = mpc_class(-9.0, -11.0);
    c[1 + 0 * n] = mpc_class(8.0, -1.0);  c[1 + 1 * n] = mpc_class(4.0, 4.44);    c[1 + 2 * n] = mpc_class(8.0, 9.0);
    c[2 + 0 * n] = mpc_class(6.0, 0.0);   c[2 + 1 * n] = mpc_class(-1.0, 0.0);    c[2 + 2 * n] = mpc_class(-2.0, 1.0);

    printf("# Cgemm demo...\n");

    printf("a ="); printmat(n, n, a, n); printf("\n");
    printf("b ="); printmat(n, n, b, n); printf("\n");
    printf("c ="); printmat(n, n, c, n); printf("\n");
    alpha = mpc_class(3.0,-1.2);
    beta = mpc_class(-2.0, -2.0);
    Cgemm("n", "n", n, n, n, alpha, a, n, b, n, beta, c, n);

    printf("alpha = "); printnum(alpha); printf("\n");
    printf("beta = "); printnum(beta); printf("\n");
    printf("ans ="); printmat(n, n, c, n); printf("\n");
    printf("#please check by Matlab or Octave following and ans above\n");
    printf("alpha * a * b + beta * c \n");
    delete[]c;
    delete[]b;
    delete[]a;
}
\end{lstlisting}
First, we must include {\tt mpblas\_gmp.h} to use GMP. {\tt printmat} function prints matrix in Octave/Matlab format (lines 7 to 46). 
We use the default GMP complex precision 153 decimal digits ($154.13 = \log_{10} 2^{512}$).
We allocate the matrix as a one-dimensional array (lines 51 to 54). We use {\tt mpc\_class} for complex values.
Then we set the matrix $A$, $B$, and $C$. We always input the matrix by row-major format to an array (lines 36 to 47). Matrix-matrix multiplication {\tt Cgemm} is called in line 58. 
One can input the list and save as {\tt Cgemm\_gmp.cpp} or you can find {\tt /home/docker/mplapack/examples/mpblas/} directory.
Then, one can compile on one own following in the Docker environment as follows:
\begin{verbatim}
$ g++ -O2 -I/home/docker/MPLAPACK/include -I/home/docker/MPLAPACK/include/mplapack \
Cgemm_gmp.cpp -Wl,--rpath=/home/docker/MPLAPACK/lib -L/home/docker/MPLAPACK/lib \
-lmpblas_gmp -lgmp
\end{verbatim}
Then, you can run as follows:
{\tiny
\begin{verbatim}
$ ./a.out
# Cgemm demo...
a =[ [ +1.0000000000000000000000000000000000000000000000000000000000000000e+00-1.0000000000000000000000000000000000000000000000000000000000000000e+00i,
+8.0000000000000000000000000000000000000000000000000000000000000000e+00+2.2000000000000001776356839400250464677810668945312500000000000000e+00i,
+0.0000000000000000000000000000000000000000000000000000000000000000e+00-1.0000000000000000000000000000000000000000000000000000000000000000e+01i];
[ +2.0000000000000000000000000000000000000000000000000000000000000000e+00+0.0000000000000000000000000000000000000000000000000000000000000000e+00i,
+1.0000000000000000000000000000000000000000000000000000000000000000e+01+0.0000000000000000000000000000000000000000000000000000000000000000e+00i,
+8.0999999999999996447286321199499070644378662109375000000000000000e+00+2.2000000000000001776356839400250464677810668945312500000000000000e+00i]; 
[ -9.0000000000000000000000000000000000000000000000000000000000000000e+00+3.0000000000000000000000000000000000000000000000000000000000000000e+00i,
-5.0000000000000000000000000000000000000000000000000000000000000000e+00+3.0000000000000000000000000000000000000000000000000000000000000000e+00i,
-1.0000000000000000000000000000000000000000000000000000000000000000e+00+0.0000000000000000000000000000000000000000000000000000000000000000e+00i] ]
b =[ [ +9.0000000000000000000000000000000000000000000000000000000000000000e+00+0.0000000000000000000000000000000000000000000000000000000000000000e+00i,
+8.0000000000000000000000000000000000000000000000000000000000000000e+00-1.0000000000000000208166817117216851329430937767028808593750000000e-02i,
+3.0000000000000000000000000000000000000000000000000000000000000000e+00+1.0009999999999998898658759571844711899757385253906250000000000000e+00i];
[ +3.0000000000000000000000000000000000000000000000000000000000000000e+00-8.0000000000000000000000000000000000000000000000000000000000000000e+00i,
-1.1000000000000000000000000000000000000000000000000000000000000000e+01+1.0000000000000000555111512312578270211815834045410156250000000000e-01i,
+8.0000000000000000000000000000000000000000000000000000000000000000e+00+1.0000000000000000818030539140313095458623138256371021270751953125e-05i];
[ -8.0000000000000000000000000000000000000000000000000000000000000000e+00+1.0000000000000000000000000000000000000000000000000000000000000000e+00i,
+6.0000000000000000000000000000000000000000000000000000000000000000e+00+0.0000000000000000000000000000000000000000000000000000000000000000e+00i,
+1.1000000000000000888178419700125232338905334472656250000000000000e+00+1.0000000000000000000000000000000000000000000000000000000000000000e+00i] ]
c =[ [ +3.0000000000000000000000000000000000000000000000000000000000000000e+00+1.0000000000000000000000000000000000000000000000000000000000000000e+00i,
-3.0000000000000000000000000000000000000000000000000000000000000000e+00+9.9900000000000002131628207280300557613372802734375000000000000000e+00i,
-9.0000000000000000000000000000000000000000000000000000000000000000e+00-1.1000000000000000000000000000000000000000000000000000000000000000e+01i]; 
[ +8.0000000000000000000000000000000000000000000000000000000000000000e+00-1.0000000000000000000000000000000000000000000000000000000000000000e+00i,
+4.0000000000000000000000000000000000000000000000000000000000000000e+00+4.4400000000000003907985046680551022291183471679687500000000000000e+00i,
+8.0000000000000000000000000000000000000000000000000000000000000000e+00+9.0000000000000000000000000000000000000000000000000000000000000000e+00i];
[ +6.0000000000000000000000000000000000000000000000000000000000000000e+00+0.0000000000000000000000000000000000000000000000000000000000000000e+00i,
-1.0000000000000000000000000000000000000000000000000000000000000000e+00+0.0000000000000000000000000000000000000000000000000000000000000000e+00i,
-2.0000000000000000000000000000000000000000000000000000000000000000e+00+1.0000000000000000000000000000000000000000000000000000000000000000e+00i] ]
alpha = +3.0000000000000000000000000000000000000000000000000000000000000000e+00-1.1999999999999999555910790149937383830547332763671875000000000000e+00i
beta = -2.0000000000000000000000000000000000000000000000000000000000000000e+00-2.0000000000000000000000000000000000000000000000000000000000000000e+00i
ans =[ [ +1.9412000000000000429878355134860610085447466221730208764814304152e+02-3.9919999999999997415400798672635510784913059003625879561714777377e+01i,
-3.2440199999999999789655757975737052220306876582624715712208640779e+02-1.9193400000000000968240765342187614821595591023640492250541356027e+02i,
+2.3552422999999999997259326805967153749369165591598754371335859085e+02-3.9797933599999995135557075099860818051923306637592336735756250096e+01i];
[ -1.8239999999999999016342400182111297296053289861415149007714347172e+02-2.5970000000000000937028232783632108284631549553475262574071520758e+02i,
-1.1830400000000000489791540658757190133827036314312018406665329523e+02+8.0140000000000003122918590392487108980814301859549537116009666066e+01i,
+2.9515651999999999967519852739003531813778936267996503629002521457e+02-1.0768569999999999661823612607559887657445872205333970782644834681e+02i];
[ -1.1400000000000000333066907387546962127089500427246093750000000000e+02+2.8979999999999999715782905695959925651550292968750000000000000000e+02i,
-7.9101999999999999661257077399056923910599691859337230298536113865e+01+1.6939999999999989081927997958132429788216147581137981441043024206e+00i,
-1.7971994999999999910681342224060695849998775453588962690525666214e+02+1.5629648599999999952610174635322244325683964957536896708915619612e+02i] ]
#please check by Matlab or Octave following and ans above
alpha * a * b + beta * c
\end{verbatim}
}
This output is tough to see, and pass to octave got more readable results.
{\footnotesize
\begin{verbatim}
$ ./a.out
a =

    1.00000 -  1.00000i    8.00000 +  2.20000i    0.00000 - 10.00000i
    2.00000 +  0.00000i   10.00000 +  0.00000i    8.10000 +  2.20000i
   -9.00000 +  3.00000i   -5.00000 +  3.00000i   -1.00000 +  0.00000i

b =

    9.0000 +  0.0000i    8.0000 -  0.0100i    3.0000 +  1.0010i
    3.0000 -  8.0000i  -11.0000 +  0.1000i    8.0000 +  0.0000i
   -8.0000 +  1.0000i    6.0000 +  0.0000i    1.1000 +  1.0000i

c =

    3.0000 +  1.0000i   -3.0000 +  9.9900i   -9.0000 - 11.0000i
    8.0000 -  1.0000i    4.0000 +  4.4400i    8.0000 +  9.0000i
    6.0000 +  0.0000i   -1.0000 +  0.0000i   -2.0000 +  1.0000i

alpha =  3.0000 - 1.2000i
beta = -2 - 2i
ans =

   194.120 -  39.920i  -324.402 - 191.934i   235.524 -  39.798i
  -182.400 - 259.700i  -118.304 +  80.140i   295.157 - 107.686i
  -114.000 + 289.800i   -79.102 +   1.694i  -179.720 + 156.296i

ans =

   194.120 -  39.920i  -324.402 - 191.934i   235.524 -  39.798i
  -182.400 - 259.700i  -118.304 +  80.140i   295.157 - 107.686i
  -114.000 + 289.800i   -79.102 +   1.694i  -179.720 + 156.296i

\end{verbatim}
}
One can see that the input and output look wrong when we see the raw output. Nevertheless, the output is correct up to 153 decimal digits.

Let us explain the reason. We input many values, but we look at $\alpha = 3-1.2i$. MPBLAS GMP output is the following:
{\tiny
\begin{verbatim}
alpha = +3.0000000000000000000000000000000000000000000000000000000000000000e+00-1.1999999999999999555910790149937383830547332763671875000000000000e+00i
\end{verbatim}
}
The imaginary part $-1.2i$ is correct up to 16 decimal digits. Again, this is correct behavior. First, compilers read {\tt alpha = mpc\_class(3.0,-1.2);}, and converts the string ``$-1.2i$" to double precision number. We cannot represent ``$-1.2$" exactly, since representing $1.2$ in binary numbers results in an infinite number of circular decimals, i.e., $1.2_{(10)} = 1.00110011001100110011..._{(2)}$. Then {\tt alpha} is initialized by {\tt mpc\_class}. Thus, input and output look wrong.
We can use a workaround for this behavior by enclosing numbers in {\tt mpf\_class("")} to indicate exact numbers up to specified precision as follows:
{\tiny
\begin{verbatim}
    alpha = mpc_class(mpf_class("3.0"), mpf_class("-1.2"));
\end{verbatim}
}
Then, the alpha becomes as follows:
{\tiny
\begin{verbatim}
alpha = +3.0000000000000000000000000000000000000000000000000000000000000000e+00-1.2000000000000000000000000000000000000000000000000000000000000000e+00i
\end{verbatim}
}
Even though ``$-1.20000...0e+01$" looks exact, there is still a tiny difference between mathematically rigorous ``$1.2i$".  

Finally, we show a program and the result fixing such trancation problem as follows:
\begin{lstlisting}[basicstyle=\ttfamily]
//public domain
//public domain
#include <mpblas_gmp.h>
#include <iostream>
#include <cstring>
#include <algorithm>

#define GMP_FORMAT "%+68.64Fe"
#define GMP_SHORT_FORMAT "%+20.16Fe"

inline void printnum(mpf_class rtmp) { gmp_printf(GMP_FORMAT, rtmp.get_mpf_t()); }
inline void printnum_short(mpf_class rtmp) { gmp_printf(GMP_SHORT_FORMAT, rtmp.get_mpf_t()); }
inline void printnum(mpc_class ctmp) { gmp_printf(GMP_FORMAT GMP_FORMAT "i", ctmp.real().get_mpf_t(), ctmp.imag().get_mpf_t()); }

//Matlab/Octave format
template <class X> void printvec(X *a, int len) {
    X tmp;
    printf("[ ");
    for (int i = 0; i < len; i++) {
        tmp = a[i];
        printnum(tmp);
        if (i < len - 1)
            printf(", ");
    }
    printf("]");
}

template <class X> void printmat(int n, int m, X *a, int lda)
{
    X mtmp;

    printf("[ ");
    for (int i = 0; i < n; i++) {
        printf("[ ");
        for (int j = 0; j < m; j++) {
            mtmp = a[i + j * lda];
            printnum(mtmp);
            if (j < m - 1)
                printf(", ");
        }
        if (i < n - 1)
            printf("]; ");
        else
            printf("] ");
    }
    printf("]");
}
int main()
{
    mplapackint n = 3;

    mpc_class *a = new mpc_class[n * n];
    mpc_class *b = new mpc_class[n * n];
    mpc_class *c = new mpc_class[n * n];
    mpc_class alpha, beta;

//setting A matrix
    a[0 + 0 * n] = mpc_class(mpf_class("1.0"),  mpf_class("-1.0"));   a[0 + 1 * n] = mpc_class(mpf_class("8.0"),  mpf_class("2.2"));
    a[0 + 2 * n] = mpc_class(mpf_class("0.0"),  mpf_class("-10.0"));
    a[1 + 0 * n] = mpc_class(mpf_class("2.0"),  mpf_class("0.0"));    a[1 + 1 * n] = mpc_class(mpf_class("10.0"), mpf_class("0.0"));
    a[1 + 2 * n] = mpc_class(mpf_class("8.1"),  mpf_class("2.2"));
    a[2 + 0 * n] = mpc_class(mpf_class("-9.0"), mpf_class("3.0"));    a[2 + 1 * n] = mpc_class(mpf_class("-5.0"), mpf_class("3.0"));
    a[2 + 2 * n] = mpc_class(mpf_class("-1.0"), mpf_class("0.0"));

    b[0 + 0 * n] = mpc_class(mpf_class("9.0"),  mpf_class("0.0"));    b[0 + 1 * n] = mpc_class(mpf_class("8.0"),  mpf_class("-0.01"));
    b[0 + 2 * n] = mpc_class(mpf_class("3.0"),  mpf_class("1.001"));
    b[1 + 0 * n] = mpc_class(mpf_class("3.0"),  mpf_class("-8.0"));   b[1 + 1 * n] = mpc_class(mpf_class("-11.0"), mpf_class("0.1"));
    b[1 + 2 * n] = mpc_class(mpf_class("8.0"),  mpf_class("0.00001"));
    b[2 + 0 * n] = mpc_class(mpf_class("-8.0"), mpf_class("1.0"));    b[2 + 1 * n] = mpc_class(mpf_class("6.0"),  mpf_class("0.0"));
    b[2 + 2 * n] = mpc_class(mpf_class("1.1"),  mpf_class("1.0"));

    c[0 + 0 * n] = mpc_class(mpf_class("3.0"),  mpf_class("1.0"));    c[0 + 1 * n] = mpc_class(mpf_class("-3.0"), mpf_class("9.99"));
    c[0 + 2 * n] = mpc_class(mpf_class("-9.0"), mpf_class("-11.0"));
    c[1 + 0 * n] = mpc_class(mpf_class("8.0"),  mpf_class("-1.0"));   c[1 + 1 * n] = mpc_class(mpf_class("4.0"), mpf_class("4.44"));
    c[1 + 2 * n] = mpc_class(mpf_class("8.0"),  mpf_class("9.0"));
    c[2 + 0 * n] = mpc_class(mpf_class("6.0"),  mpf_class("0.0"));    c[2 + 1 * n] = mpc_class(mpf_class("-1.0"), mpf_class("0.0"));
    c[2 + 2 * n] = mpc_class(mpf_class("-2.0"), mpf_class("1.0"));

    printf("# Cgemm demo...\n");

    printf("a ="); printmat(n, n, a, n); printf("\n");
    printf("b ="); printmat(n, n, b, n); printf("\n");
    printf("c ="); printmat(n, n, c, n); printf("\n");
    alpha = mpc_class(mpf_class("3.0"), mpf_class("-1.2"));
    beta = mpc_class(mpf_class("-2.0"), mpf_class("-2.0"));
    Cgemm("n", "n", n, n, n, alpha, a, n, b, n, beta, c, n);

    printf("alpha = "); printnum(alpha); printf("\n");
    printf("beta = "); printnum(beta); printf("\n");
    printf("ans ="); printmat(n, n, c, n); printf("\n");
    printf("#please check by Matlab or Octave following and ans above\n");
    printf("alpha * a * b + beta * c \n");
    delete[]c;
    delete[]b;
    delete[]a;
}
\end{lstlisting}

{\tiny
\begin{verbatim}
$ ./a.out
# Cgemm demo...
a =[ [ +1.0000000000000000000000000000000000000000000000000000000000000000e+00-1.0000000000000000000000000000000000000000000000000000000000000000e+00i,
+8.0000000000000000000000000000000000000000000000000000000000000000e+00+2.2000000000000000000000000000000000000000000000000000000000000000e+00i,
+0.0000000000000000000000000000000000000000000000000000000000000000e+00-1.0000000000000000000000000000000000000000000000000000000000000000e+01i]; 
[ +2.0000000000000000000000000000000000000000000000000000000000000000e+00+0.0000000000000000000000000000000000000000000000000000000000000000e+00i, 
+1.0000000000000000000000000000000000000000000000000000000000000000e+01+0.0000000000000000000000000000000000000000000000000000000000000000e+00i, 
+8.1000000000000000000000000000000000000000000000000000000000000000e+00+2.2000000000000000000000000000000000000000000000000000000000000000e+00i]; 
[ -9.0000000000000000000000000000000000000000000000000000000000000000e+00+3.0000000000000000000000000000000000000000000000000000000000000000e+00i,
-5.0000000000000000000000000000000000000000000000000000000000000000e+00+3.0000000000000000000000000000000000000000000000000000000000000000e+00i,
-1.0000000000000000000000000000000000000000000000000000000000000000e+00+0.0000000000000000000000000000000000000000000000000000000000000000e+00i] ]
b =[ [ +9.0000000000000000000000000000000000000000000000000000000000000000e+00+0.0000000000000000000000000000000000000000000000000000000000000000e+00i,
+8.0000000000000000000000000000000000000000000000000000000000000000e+00-1.0000000000000000000000000000000000000000000000000000000000000000e-02i,
+3.0000000000000000000000000000000000000000000000000000000000000000e+00+1.0010000000000000000000000000000000000000000000000000000000000000e+00i];
[ +3.0000000000000000000000000000000000000000000000000000000000000000e+00-8.0000000000000000000000000000000000000000000000000000000000000000e+00i,
-1.1000000000000000000000000000000000000000000000000000000000000000e+01+1.0000000000000000000000000000000000000000000000000000000000000000e-01i,
+8.0000000000000000000000000000000000000000000000000000000000000000e+00+1.0000000000000000000000000000000000000000000000000000000000000000e-05i]; [
-8.0000000000000000000000000000000000000000000000000000000000000000e+00+1.0000000000000000000000000000000000000000000000000000000000000000e+00i,
+6.0000000000000000000000000000000000000000000000000000000000000000e+00+0.0000000000000000000000000000000000000000000000000000000000000000e+00i,
+1.1000000000000000000000000000000000000000000000000000000000000000e+00+1.0000000000000000000000000000000000000000000000000000000000000000e+00i] ]
c =[ [ +3.0000000000000000000000000000000000000000000000000000000000000000e+00+1.0000000000000000000000000000000000000000000000000000000000000000e+00i,
-3.0000000000000000000000000000000000000000000000000000000000000000e+00+9.9900000000000000000000000000000000000000000000000000000000000000e+00i,
-9.0000000000000000000000000000000000000000000000000000000000000000e+00-1.1000000000000000000000000000000000000000000000000000000000000000e+01i]; 
[ +8.0000000000000000000000000000000000000000000000000000000000000000e+00-1.0000000000000000000000000000000000000000000000000000000000000000e+00i,
+4.0000000000000000000000000000000000000000000000000000000000000000e+00+4.4400000000000000000000000000000000000000000000000000000000000000e+00i
+8.0000000000000000000000000000000000000000000000000000000000000000e+00+9.0000000000000000000000000000000000000000000000000000000000000000e+00i]; 
[ +6.0000000000000000000000000000000000000000000000000000000000000000e+00+0.0000000000000000000000000000000000000000000000000000000000000000e+00i,
-1.0000000000000000000000000000000000000000000000000000000000000000e+00+0.0000000000000000000000000000000000000000000000000000000000000000e+00i 
-2.0000000000000000000000000000000000000000000000000000000000000000e+00+1.0000000000000000000000000000000000000000000000000000000000000000e+00i] ]
alpha = +3.0000000000000000000000000000000000000000000000000000000000000000e+00-1.2000000000000000000000000000000000000000000000000000000000000000e+00i
beta = -2.0000000000000000000000000000000000000000000000000000000000000000e+00-2.0000000000000000000000000000000000000000000000000000000000000000e+00i
beta = -2.0000000000000000000000000000000000000000000000000000000000000000e+00-2.0000000000000000000000000000000000000000000000000000000000000000e+00i
ans =[ [ +1.9412000000000000000000000000000000000000000000000000000000000000e+02-3.9920000000000000000000000000000000000000000000000000000000000000e+01i,
-3.2440200000000000000000000000000000000000000000000000000000000000e+02-1.9193400000000000000000000000000000000000000000000000000000000000e+02i,
+2.3552423000000000000000000000000000000000000000000000000000000000e+02-3.9797933600000000000000000000000000000000000000000000000000000000e+01i];
[ -1.8240000000000000000000000000000000000000000000000000000000000000e+02-2.5970000000000000000000000000000000000000000000000000000000000000e+02i,
-1.1830400000000000000000000000000000000000000000000000000000000000e+02+8.0140000000000000000000000000000000000000000000000000000000000000e+01i,
+2.9515652000000000000000000000000000000000000000000000000000000000e+02-1.0768570000000000000000000000000000000000000000000000000000000000e+02i]; 
[ -1.1400000000000000000000000000000000000000000000000000000000000000e+02+2.8980000000000000000000000000000000000000000000000000000000000000e+02i,
-7.9102000000000000000000000000000000000000000000000000000000000000e+01+1.6940000000000000000000000000000000000000000000000000000000000000e+00i,
-1.7971995000000000000000000000000000000000000000000000000000000000e+02+1.5629648600000000000000000000000000000000000000000000000000000000e+02i] ]
#please check by Matlab or Octave following and ans above
alpha * a * b + beta * c
\end{verbatim}
}

Unfortunately, such a workaround does not always exist. For example, {\tt \_Float128} and {\tt \_Float64x} do not allow such a string to a floating point number conversion. Therefore, we should write a program to handle high-precision inputs in such a case.

In any case, the calculations are correct up to specified precisions.

\subsubsection{OpenMP accelerated MPBLAS}
We have provided reference implementation, and a simple OpenMP accelerated version for MPBLAS, shown in Table~\ref{mpblasroutinesopt}. Even though the number of optimized routines is small; however, acceleration of {\tt Rgemm} is significant as it impacts the performance.
To change linking against the optimized version, we must add ``{\tt \_opt}" for {\tt MPBLAS} library as follows:
\begin{itemize}
    \item {\tt -lmpblas\_mpfr} $\rightarrow$ {\tt -lmpblas\_mpfr\_opt} 
    \item {\tt -lmpblas\_gmp} $\rightarrow$ {\tt -lmpblas\_gmp\_opt} 
    \item {\tt -lmpblas\_double} $\rightarrow$ {\tt -lmpblas\_double\_opt} 
    \item {\tt -lmpblas\_\_Float128} $\rightarrow$ {\tt -lmpblas\_\_Float128\_opt} 
    \item {\tt -lmpblas\_dd} $\rightarrow$ {\tt -lmpblas\_dd\_opt} 
    \item {\tt -lmpblas\_qd} $\rightarrow$ {\tt -lmpblas\_qd\_opt} 
    \item {\tt -lmpblas\_\_Float128} $\rightarrow$ {\tt -lmpblas\_\_Float128\_opt} 
    \item {\tt -lmpblas\_\_Float64x} $\rightarrow$ {\tt -lmpblas\_\_Float64x\_opt}.
\end{itemize}
Besides, we must add the ``{\tt -fopenmp}" flag when compiling and linking.

We can compile the demo program for the {\tt MPFR} version and link it against the optimized version of MPBLAS.
\begin{verbatim}
$ g++ -fopenmp -O2 -I/home/docker/MPLAPACK/include \
-I/home/docker/MPLAPACK/include/mplapack \
Rgemm_mpfr.cpp -Wl,--rpath=/home/docker/MPLAPACK/lib -L/home/docker/MPLAPACK/lib \
-lmpblas_mpfr_opt -lmpfr -lmpc -lgmp
$ ./a.out
...
\end{verbatim}
Performance comparison is presented in Section~\ref{sec:benchmarks}.

\subsubsection{CUDA accelerated MPBLAS}
These routines are usable on Ampere and Volta generations of NVIDIA Tesla GPUs. 
We provide {\tt Rgemm} and {\tt Rsyrk} in double-double routines in CUDA~\cite{6424545}. First, we build the CUDA version on Docker as follows:
\begin{verbatim}
$ docker build -f Dockerfile_ubuntu20.04_cuda -t mplapack:ubuntu2004_cuda .
\end{verbatim}
In this Dockerfile, we install MPLAPACK at {\tt /home/docker/MPLAPACK\_CUDA}.
To run a shell in the Docker enabling GPUs by:
\begin{verbatim}
$ docker run --gpus all -it mplapack:ubuntu2004_cuda /bin/bash
\end{verbatim}
Finally, to link against the CUDA version, we compile the program as follows:
\begin{verbatim}
$ g++ -O2 -I/home/docker/MPLAPACK_CUDA/include \
-I/home/docker/MPLAPACK_CUDA/include/mplapack Rgemm_dd.cpp \
-Wl,--rpath=/home/docker/MPLAPACK_CUDA/lib -L/home/docker/MPLAPACK_CUDA/lib \
-lmpblas_dd_cuda -lmpblas_dd -lqd \
-L/usr/local/cuda/lib64/ -lcudart
$ ./a.out
...
\end{verbatim}
As described in~\cite{6424545}, using GPUs significantly improves the preformance of the double-double version of {\tt Rgemm} and {\tt Rsyrk}. We provide {\tt Rsyrk} for the first time. Detailed benchmark will be given in Section~\ref{sec:benchmarks}, and for large matrices, we obtained 450-600GFlops using A100 or V100.

\subsection{How to use MPLAPACK}

In this subsection, we describe the basic usage of MPLAPACK with examples. The API of MPLAPACK is very similar to the original LAPACK. We always use a one-dimensional array as a column-major type matrix in MPLAPACK. Unfortunately, we do not have an interface like LAPACKE at the moment, and we cannot specify raw- or column-major orders in MPLAPACK.

\subsubsection{Eigenvalues and eigenvectors of a real symmetric matrix ({\tt Rsyev})}
Let us show how we diagonalize a real symmetric matrix $A$
\[
\left[\begin{array}{llll}5 & 4 & 1 & 1 \\ 4 & 5 & 1 & 1 \\ 1 & 1 & 4 & 2 \\ 1 & 1 & 2 & 4\end{array}\right]
\]
using {\tt GMP}. Eigenvalues are $\lambda_1 = 10, \lambda_2 = 5, \lambda_3 = 2, \lambda_4 = 1$ and eigenvectors are
\[
x_{1}=\left[\begin{array}{l}2 \\ 2 \\ 1 \\ 1\end{array}\right], \quad x_{2}=\left[\begin{array}{c}-1 \\ -1 \\ 2 \\ 2\end{array}\right], \quad x_{3}=\left[\begin{array}{c}0 \\ 0 \\ -1 \\ 1\end{array}\right], \quad x_{4}=\left[\begin{array}{c}-1 \\ 1 \\ 0 \\ 0\end{array}\right]
\]
and $A x_{i} = \lambda x_{i}$ for $1\leq i \leq 4$ ~\cite{Gregory1969ACO}.

Following is the prototype definition of GMP's diagonalization of a symmetric matrix ({\tt Rsyev}).
\begin{verbatim}
void Rsyev(const char *jobz, const char *uplo, mplapackint const n, 
mpf_class *a, mplapackint const lda, mpf_class *w, mpf_class *work, 
mplapackint const lwork, mplapackint &info);
\end{verbatim}
And the following is the definition of the original {\tt dsyev} 
\begin{verbatim}
       SUBROUTINE dsyev( JOBZ, UPLO, N, A, LDA, W, WORK, LWORK, INFO )
 *
 *  -- LAPACK driver routine --
 *  -- LAPACK is a software package provided by Univ. of Tennessee,    --
 *  -- Univ. of California Berkeley, Univ. of Colorado Denver and NAG Ltd..--
 *
 *     .. Scalar Arguments ..
       CHARACTER          JOBZ, UPLO
       INTEGER            INFO, LDA, LWORK, N
 *     ..
 *     .. Array Arguments ..
       DOUBLE PRECISION   A( LDA, * ), W( * ), WORK( * )
 *     ..
 *
\end{verbatim}
There are clear correspondences between variables in the C++ prototype of {\tt Rsyev} and the header of {\tt DSYEV}.
\begin{itemize}
    \item {\tt CHARACTER} $\rightarrow$ {\tt const char *}
    \item {\tt INTEGER} $\rightarrow$ {\tt mplapackint}
    \item {\tt DOUBLE PRECISION A(LDA, *)} $\rightarrow$ {\tt mpf\_class *a}
    \item {\tt DOUBLE PRECISION W(*)} $\rightarrow$ {\tt mpf\_class *w}
    \item {\tt DOUBLE PRECISION WORK(*)} $\rightarrow$ {\tt mpf\_class *work}
\end{itemize}

The list is the following.
\begin{lstlisting}
//public domain
#include <mpblas_gmp.h>
#include <mplapack_gmp.h>

#define GMP_FORMAT "%+68.64Fe"
#define GMP_SHORT_FORMAT "%+20.16Fe"

inline void printnum(mpf_class rtmp) { gmp_printf(GMP_SHORT_FORMAT, rtmp.get_mpf_t()); }
inline void printnum_short(mpf_class rtmp) { gmp_printf(GMP_SHORT_FORMAT, rtmp.get_mpf_t()); }

//Matlab/Octave format
void printvec(mpf_class *a, int len) {
    mpf_class tmp;
    printf("[ ");
    for (int i = 0; i < len; i++) {
        tmp = a[i];
        printnum(tmp);
        if (i < len - 1)
            printf(", ");
    }
    printf("]");
}

void printmat(int n, int m, mpf_class * a, int lda)
{
    mpf_class mtmp;

    printf("[ ");
    for (int i = 0; i < n; i++) {
        printf("[ ");
        for (int j = 0; j < m; j++) {
            mtmp = a[i + j * lda];
            printnum(mtmp);
            if (j < m - 1)
                printf(", ");
        }
        if (i < n - 1)
            printf("]; ");
        else
            printf("] ");
    }
    printf("]");
}
int main()
{
    mplapackint n = 4;
    mplapackint lwork, info;

    mpf_class *A = new mpf_class[n * n];
    mpf_class *w = new mpf_class[n];

//setting A matrix
    A[0 + 0 * n] = 5;    A[0 + 1 * n] = 4;    A[0 + 2 * n] = 1;    A[0 + 3 * n] = 1;
    A[1 + 0 * n] = 4;    A[1 + 1 * n] = 5;    A[1 + 2 * n] = 1;    A[1 + 3 * n] = 1;
    A[2 + 0 * n] = 1;    A[2 + 1 * n] = 1;    A[2 + 2 * n] = 4;    A[2 + 3 * n] = 2;
    A[3 + 0 * n] = 1;    A[3 + 1 * n] = 1;    A[3 + 2 * n] = 2;    A[3 + 3 * n] = 4;

    printf("A ="); printmat(n, n, A, n); printf("\n");
//work space query
    lwork = -1;
    mpf_class *work = new mpf_class[1];

    Rsyev("V", "U", n, A, n, w, work, lwork, info);
    lwork = (int) cast2double (work[0]);
    delete[]work;
    work = new mpf_class[std::max((mplapackint) 1, lwork)];
//inverse matrix
    Rsyev("V", "U", n, A, n, w, work, lwork, info);
//print out some results.
    printf("#eigenvalues \n");
    printf("w ="); printmat(n, 1, w, 1); printf("\n");

    printf("#eigenvecs \n");
    printf("U ="); printmat(n, n, A, n); printf("\n");
    printf("#you can check eigenvalues using octave/Matlab by:\n");
    printf("eig(A)\n");
    printf("#you can check eigenvectors using octave/Matlab by:\n");
    printf("U'*A*U\n");

    delete[]work;
    delete[]w;
    delete[]A;
}
\end{lstlisting}

One can input the list and save it as {\tt Rsyev\_test\_gmp.cpp}, or you can find\\
{\tt /home/docker/mplapack/examples/mplapack/03\_SymmetricEigenproblems} directory. Then, one can compile on one own following in the Docker environment as follows:
\begin{verbatim}
$ g++ -O2 -I/home/docker/MPLAPACK/include -I/home/docker/MPLAPACK/include/mplapack \
Rsyev_test_gmp.cpp -Wl,--rpath=/home/docker/MPLAPACK/lib -L/home/docker/MPLAPACK/lib \
-lmplapack_gmp -lmpblas_gmp -lgmpxx -lgmp
\end{verbatim}
Finally, you can run as follows:
{\fontsize{9.5pt}{0.4cm}\selectfont
\begin{verbatim}
$ ./a.out
A =[ [ +5.0000000000000000e+00, +4.0000000000000000e+00, +1.0000000000000000e+00, +1.0000000000000000e+00];
[ +4.0000000000000000e+00, +5.0000000000000000e+00, +1.0000000000000000e+00, +1.0000000000000000e+00];]
#eigenvalues
w =[ [ +1.0000000000000000e+00]; [ +2.0000000000000000e+00]; [ +5.0000000000000000e+00]; [ +1.0000000000000000e+01] ]
#eigenvecs
U =[ [ +7.0710678118654752e-01, -1.3882760712710340e-155, -3.1622776601683793e-01, +6.3245553203367587e-01];
[ -7.0710678118654752e-01, -1.0249769098010974e-155, -3.1622776601683793e-01, +6.3245553203367587e-01]
#you can check eigenvalues using octave/Matlab by:
eig(A)
#you can check eigenvectors using octave/Matlab by:
U'*A*U
\end{verbatim}
}
One can check the result by comparing the result of the octave.
{\footnotesize
\begin{verbatim}
$./a.out | octave
octave: X11 DISPLAY environment variable not set
octave: disabling GUI features
A =

   5   4   1   1
   4   5   1   1
   1   1   4   2
   1   1   2   4

w =

    1
    2
    5
   10

U =

   0.70711  -0.00000  -0.31623   0.63246
  -0.70711  -0.00000  -0.31623   0.63246
   0.00000  -0.70711   0.63246   0.31623
   0.00000   0.70711   0.63246   0.31623

ans =

    1.00000
    2.00000
    5.00000
   10.00000

ans =

    1.0000e+00  -2.8639e-155    2.6715e-17   -5.3429e-17
  -2.6070e-155    2.0000e+00   -4.1633e-18   -2.0817e-18
   -6.7450e-17    1.7912e-16    5.0000e+00    8.5557e-17
   -3.5824e-16   -1.3490e-16    1.1229e-16    1.0000e+01

\end{verbatim}
}

As a summary, we show how we compile and run binary64, FP80, binary128, double-double, quad-double, GMP, and MPFR versions of {\tt Rsyev} demo programs in \\
{\tt /home/docker/mplapack/examples/mplapack/03\_SymmetricEigenproblems} as follows:
\begin{itemize}
    \item binary64 (double) version
\begin{verbatim}
$ g++ -O2 -I/home/docker/MPLAPACK/include -I/home/docker/MPLAPACK/include/mplapack \
Rsyev_test_double.cpp -Wl,--rpath=/home/docker/MPLAPACK/lib \
-L/home/docker/MPLAPACK/lib -lmplapack_double -lmpblas_double
$ ./a.out
\end{verbatim}
\item FP80 (extended double) version
\begin{verbatim}
$ g++ -O2 -I/home/docker/MPLAPACK/include -I/home/docker/MPLAPACK/include/mplapack \
Rsyev_test__Float64x.cpp -Wl,--rpath=/home/docker/MPLAPACK/lib \
-L/home/docker/MPLAPACK/lib -lmplapack__Float64x -lmpblas__Float64x
$ ./a.out
\end{verbatim}
\item binary128 version
\begin{verbatim}
$ g++ -O2 -I/home/docker/MPLAPACK/include -I/home/docker/MPLAPACK/include/mplapack \
Rsyev_test__Float128.cpp -Wl,--rpath=/home/docker/MPLAPACK/lib \
-L/home/docker/MPLAPACK/lib \
-lmplapack__Float128 -lmpblas__Float128
$ ./a.out
\end{verbatim}
or (on macOS, mingw64 and CentOS7 amd64)
\begin{verbatim}
$ g++ -O2 -I/home/docker/MPLAPACK/include -I/home/docker/MPLAPACK/include/mplapack \
Rsyev_test__Float128.cpp -Wl,--rpath=/home/docker/MPLAPACK/lib \
-L/home/docker/MPLAPACK/lib \
-lmplapack__Float128 -lmpblas__Float128 -lquadmath
$ ./a.out
\end{verbatim}
\item double-double version
\begin{verbatim}
$ g++ -O2 -I/home/docker/MPLAPACK/include -I/home/docker/MPLAPACK/include/mplapack \
Rsyev_test_dd.cpp -Wl,--rpath=/home/docker/MPLAPACK/lib -L/home/docker/MPLAPACK/lib \
-lmplapack_dd -lmpblas_dd -lqd
$ ./a.out
\end{verbatim}
\item quad-double version
\begin{verbatim}
$ g++ -O2 -I/home/docker/MPLAPACK/include -I/home/docker/MPLAPACK/include/mplapack \
Rsyev_test_qd.cpp -Wl,--rpath=/home/docker/MPLAPACK/lib -L/home/docker/MPLAPACK/lib \
-lmplapack_qd -lmpblas_qd -lqd
$ ./a.out
\end{verbatim}
\item GMP version
\begin{verbatim}
$ g++ -O2 -I/home/docker/MPLAPACK/include -I/home/docker/MPLAPACK/include/mplapack \
Rsyev_test_gmp.cpp -Wl,--rpath=/home/docker/MPLAPACK/lib -L/home/docker/MPLAPACK/lib \
-lmplapack_gmp -lmpblas_gmp -lgmpxx -lgmp
$ ./a.out
\end{verbatim}
\item MPFR version
\begin{verbatim}
$ g++ -O2 -I/home/docker/MPLAPACK/include -I/home/docker/MPLAPACK/include/mplapack \
Rsyev_test_mpfr.cpp -Wl,--rpath=/home/docker/MPLAPACK/lib -L/home/docker/MPLAPACK/lib \
-lmplapack_mpfr -lmpblas_mpfr -lmpfr -lmpc -lgmp
$ ./a.out
\end{verbatim}
\end{itemize}

%%%%%%%%%%%%%%%%%%%%%%%%%%%%%%%%%%%%%%%%%%%%%%%%%%%%%%%%%%%%%%%%%%%%%%%%%%%%%
\subsubsection{Eigenvalues and eigenvectors of a complex Hermitian matrix ({\tt Cheev})}
Let us show how we diagonalize a complex hermitian matrix $A$
\[
\left[\begin{array}{lll}2 & -i & 0 \\ i & 2 & 0 \\ 0 & 0 & 3 \\ \end{array}\right]
\]
using {\tt GMP}. Eigenvalues are $\lambda_1 = 1, \lambda_2 = 3, \lambda_3 = 3$ and eigenvectors are
\[
x_{1}=\left[\begin{array}{c} -1/\sqrt{2} i \\ -1/\sqrt{2} \\ 0 \end{array}\right], \quad x_{2}=\left[\begin{array}{c}-1/\sqrt{2} i \\ 1/\sqrt{2} i \\ 0 \end{array}\right], \quad x_{3}=\left[\begin{array}{c} 0 \\ 0 \\ 1 \end{array}\right] 
\]
and $A x_{i} = \lambda x_{i}$ for $1\leq i \leq 3$.

Following is the prototype definition of GMP's diagonalization of a symmetric matrix ({\tt Cheev}).
\begin{verbatim}
void Cheev(const char *jobz, const char *uplo, mplapackint const n,
mpc_class *a, mplapackint const lda, mpf_class *w, mpc_class *work, 
mplapackint const lwork, mpf_class *rwork, mplapackint &info);
\end{verbatim}
And the following is the definition of the original {\tt dsyev} 
\begin{verbatim}
*       SUBROUTINE ZHEEV( JOBZ, UPLO, N, A, LDA, W, WORK, LWORK, RWORK,
*                         INFO )
*
*       .. Scalar Arguments ..
*       CHARACTER          JOBZ, UPLO
*       INTEGER            INFO, LDA, LWORK, N
*       ..
*       .. Array Arguments ..
*       DOUBLE PRECISION   RWORK( * ), W( * )
*       COMPLEX*16         A( LDA, * ), WORK( * )

\end{verbatim}
There are clear correspondences between variables in the C++ prototype of {\tt Cheev} and the header of {\tt ZHEEV}.
\begin{itemize}
    \item {\tt CHARACTER} $\rightarrow$ {\tt const char *}
    \item {\tt INTEGER} $\rightarrow$ {\tt mplapackint}
    \item {\tt COMPLEX*16 A(LDA, *)} $\rightarrow$ {\tt mpc\_class *a}
    \item {\tt COMPLEX*16 WORK(*)} $\rightarrow$ {\tt mpc\_class *rwork}
    \item {\tt DOUBLE PRECISION W(*)} $\rightarrow$ {\tt mpf\_class *w}
    \item {\tt DOUBLE PRECISION RWORK(*)} $\rightarrow$ {\tt mpf\_class *rwork}
\end{itemize}

The list is the following.
\begin{lstlisting}
//public domain
#include <mpblas_gmp.h>
#include <mplapack_gmp.h>
#include <iostream>
#include <cstring>
#include <algorithm>

#define GMP_FORMAT "%+68.64Fe"
#define GMP_SHORT_FORMAT "%+20.16Fe"

inline void printnum(mpf_class rtmp) { gmp_printf(GMP_FORMAT, rtmp.get_mpf_t()); }
inline void printnum_short(mpf_class rtmp) { gmp_printf(GMP_SHORT_FORMAT, rtmp.get_mpf_t()); }
inline void printnum(mpc_class ctmp) { gmp_printf(GMP_FORMAT GMP_FORMAT "i", ctmp.real().get_mpf_t(), ctmp.imag().get_mpf_t()); }

//Matlab/Octave format
template <class X> void printvec(X *a, int len) {
    X tmp;
    printf("[ ");
    for (int i = 0; i < len; i++) {
        tmp = a[i];
        printnum(tmp);
        if (i < len - 1)
            printf(", ");
    }
    printf("]");
}

template <class X> void printmat(int n, int m, X *a, int lda)
{
    X mtmp;

    printf("[ ");
    for (int i = 0; i < n; i++) {
        printf("[ ");
        for (int j = 0; j < m; j++) {
            mtmp = a[i + j * lda];
            printnum(mtmp);
            if (j < m - 1)
                printf(", ");
        }
        if (i < n - 1)
            printf("]; ");
        else
            printf("] ");
    }
    printf("]");
}
int main()
{
    mplapackint n = 3;
    mplapackint lwork, info;

    mpc_class *A = new mpc_class[n * n];
    mpf_class *w = new mpf_class[n];
    mpf_class *rwork = new mpf_class[3 * n - 1];

//setting A matrix
    A[0 + 0 * n] = 2.0; A[0 + 1 * n] = mpc_class(0.0, -1.0); A[0 + 2 * n] = 0.0;
    A[1 + 0 * n] = mpc_class(0.0, 1.0); A[1 + 1 * n] = 2.0; A[1 + 2 * n] = 0.0;
    A[2 + 0 * n] = 0.0;  A[2 + 1 * n] = 0.0; A[2 + 2 * n] = 3.0;

    printf("A ="); printmat(n, n, A, n); printf("\n");
//work space query
    lwork = -1;
    mpc_class *work = new mpc_class[1];

    Cheev("V", "U", n, A, n, w, work, lwork, rwork, info);
    lwork = (int) cast2double (work[0].real());
    delete[]work;
    work = new mpc_class[std::max((mplapackint) 1, lwork)];
//inverse matrix
    Cheev("V", "U", n, A, n, w, work, lwork, rwork, info);
//print out some results.
    printf("#eigenvalues \n");
    printf("w ="); printmat(n, 1, w, 1); printf("\n");

    printf("#eigenvecs \n");
    printf("U ="); printmat(n, n, A, n); printf("\n");
    printf("#you can check eigenvalues using octave/Matlab by:\n");
    printf("eig(A)\n");
    printf("#you can check eigenvectors using octave/Matlab by:\n");
    printf("U'*A*U\n");

    delete[]work;
    delete[]w;
    delete[]A;
}
\end{lstlisting}

One can input the list and save it as {\tt Cheev\_test\_gmp.cpp}, or you can find\\
{\tt /home/docker/mplapack/examples/mplapack/03\_SymmetricEigenproblems} directory. Then, one can compile on one own following in the Docker environment as follows:
\begin{verbatim}
$ g++ -O2 -I/home/docker/MPLAPACK/include -I/home/docker/MPLAPACK/include/mplapack \
Cheev_test_gmp.cpp -Wl,--rpath=/home/docker/MPLAPACK/lib -L/home/docker/MPLAPACK/lib \
-lmplapack_gmp -lmpblas_gmp -lgmpxx -lgmp
\end{verbatim}
Finally, you can run as follows:
{\fontsize{7.5pt}{0.4cm}\selectfont
\begin{verbatim}
A =[ [ +2.0000000000000000000000000000000000000000000000000000000000000000e+00+0.0000000000000000000000000000000000000000000000000000000000000000e+00i,
+0.0000000000000000000000000000000000000000000000000000000000000000e+00-1.0000000000000000000000000000000000000000000000000000000000000000e+00i,
+0.0000000000000000000000000000000000000000000000000000000000000000e+00+0.0000000000000000000000000000000000000000000000000000000000000000e+00i];
[ +0.0000000000000000000000000000000000000000000000000000000000000000e+00+1.0000000000000000000000000000000000000000000000000000000000000000e+00i,
+2.0000000000000000000000000000000000000000000000000000000000000000e+00+0.0000000000000000000000000000000000000000000000000000000000000000e+00i,
+0.0000000000000000000000000000000000000000000000000000000000000000e+00+0.0000000000000000000000000000000000000000000000000000000000000000e+00i];
[ +0.0000000000000000000000000000000000000000000000000000000000000000e+00+0.0000000000000000000000000000000000000000000000000000000000000000e+00i, 
+0.0000000000000000000000000000000000000000000000000000000000000000e+00+0.0000000000000000000000000000000000000000000000000000000000000000e+00i, 
+3.0000000000000000000000000000000000000000000000000000000000000000e+00+0.0000000000000000000000000000000000000000000000000000000000000000e+00i] ]
#eigenvalues
w =[ [ +1.0000000000000000000000000000000000000000000000000000000000000000e+00];
[ +3.0000000000000000000000000000000000000000000000000000000000000000e+00]; 
[ +3.0000000000000000000000000000000000000000000000000000000000000000e+00] ]
#eigenvecs
U =[ [ +0.0000000000000000000000000000000000000000000000000000000000000000e+00-7.0710678118654752440084436210484903928483593768847403658833986900e-01i, 
+0.0000000000000000000000000000000000000000000000000000000000000000e+00-7.0710678118654752440084436210484903928483593768847403658833986900e-01i, 
+0.0000000000000000000000000000000000000000000000000000000000000000e+00+0.0000000000000000000000000000000000000000000000000000000000000000e+00i];
[ -7.0710678118654752440084436210484903928483593768847403658833986900e-01+0.0000000000000000000000000000000000000000000000000000000000000000e+00i, 
+7.0710678118654752440084436210484903928483593768847403658833986900e-01+0.0000000000000000000000000000000000000000000000000000000000000000e+00i, 
+0.0000000000000000000000000000000000000000000000000000000000000000e+00+0.0000000000000000000000000000000000000000000000000000000000000000e+00i];
[ +0.0000000000000000000000000000000000000000000000000000000000000000e+00+0.0000000000000000000000000000000000000000000000000000000000000000e+00i, 
+0.0000000000000000000000000000000000000000000000000000000000000000e+00+0.0000000000000000000000000000000000000000000000000000000000000000e+00i, 
+1.0000000000000000000000000000000000000000000000000000000000000000e+00+0.0000000000000000000000000000000000000000000000000000000000000000e+00i] ]
#you can check eigenvalues using octave/Matlab by:
eig(A)
#you can check eigenvectors using octave/Matlab by:
U'*A*U
\end{verbatim}
}
One can check the result by comparing the result of the octave.
{\footnotesize
\begin{verbatim}
$ ./a.out | octave
octave: X11 DISPLAY environment variable not set
octave: disabling GUI features
A =

   2 + 0i   0 - 1i   0 + 0i
   0 + 1i   2 + 0i   0 + 0i
   0 + 0i   0 + 0i   3 + 0i

w =

   1
   3
   3

U =

   0.00000 - 0.70711i   0.00000 - 0.70711i   0.00000 + 0.00000i
  -0.70711 + 0.00000i   0.70711 + 0.00000i   0.00000 + 0.00000i
   0.00000 + 0.00000i   0.00000 + 0.00000i   1.00000 + 0.00000i

ans =

   1
   3
   3

ans =

   1.00000   0.00000   0.00000
   0.00000   3.00000   0.00000
   0.00000   0.00000   3.00000
\end{verbatim}
}

As a summary, we show how we compile and run binary64, FP80, binary128, double-double, quad-double, GMP, and MPFR versions of {\tt Cheev} demo programs in \\
{\tt /home/docker/mplapack/examples/mplapack/03\_SymmetricEigenproblems} as follows:
\begin{itemize}
    \item binary64 (double) version
\begin{verbatim}
$ g++ -O2 -I/home/docker/MPLAPACK/include -I/home/docker/MPLAPACK/include/mplapack \
Cheev_test_double.cpp -Wl,--rpath=/home/docker/MPLAPACK/lib \
-L/home/docker/MPLAPACK/lib -lmplapack_double -lmpblas_double
$ ./a.out
\end{verbatim}
\item FP80 (extended double) version
\begin{verbatim}
$ g++ -O2 -I/home/docker/MPLAPACK/include -I/home/docker/MPLAPACK/include/mplapack \
Cheev_test__Float64x.cpp -Wl,--rpath=/home/docker/MPLAPACK/lib \
-L/home/docker/MPLAPACK/lib -lmplapack__Float64x -lmpblas__Float64x
$ ./a.out
\end{verbatim}
\item binary128 version
\begin{verbatim}
$ g++ -O2 -I/home/docker/MPLAPACK/include -I/home/docker/MPLAPACK/include/mplapack \
Cheev_test__Float128.cpp -Wl,--rpath=/home/docker/MPLAPACK/lib \
-L/home/docker/MPLAPACK/lib \
-lmplapack__Float128 -lmpblas__Float128
$ ./a.out
\end{verbatim}
or (on macOS, mingw64 and CentOS7 amd64)
\begin{verbatim}
$ g++ -O2 -I/home/docker/MPLAPACK/include -I/home/docker/MPLAPACK/include/mplapack \
Cheev_test__Float128.cpp -Wl,--rpath=/home/docker/MPLAPACK/lib \
-L/home/docker/MPLAPACK/lib \
-lmplapack__Float128 -lmpblas__Float128 -lquadmath
$ ./a.out
\end{verbatim}
\item double-double version
\begin{verbatim}
$ g++ -O2 -I/home/docker/MPLAPACK/include -I/home/docker/MPLAPACK/include/mplapack \
Cheev_test_dd.cpp -Wl,--rpath=/home/docker/MPLAPACK/lib -L/home/docker/MPLAPACK/lib \
-lmplapack_dd -lmpblas_dd -lqd
$ ./a.out
\end{verbatim}
\item quad-double version
\begin{verbatim}
$ g++ -O2 -I/home/docker/MPLAPACK/include -I/home/docker/MPLAPACK/include/mplapack \
Cheev_test_qd.cpp -Wl,--rpath=/home/docker/MPLAPACK/lib -L/home/docker/MPLAPACK/lib \
-lmplapack_qd -lmpblas_qd -lqd
$ ./a.out
\end{verbatim}
\item GMP version
\begin{verbatim}
$ g++ -O2 -I/home/docker/MPLAPACK/include -I/home/docker/MPLAPACK/include/mplapack \
Cheev_test_gmp.cpp -Wl,--rpath=/home/docker/MPLAPACK/lib -L/home/docker/MPLAPACK/lib \
-lmplapack_gmp -lmpblas_gmp -lgmpxx -lgmp
$ ./a.out
\end{verbatim}
\item MPFR version
\begin{verbatim}
$ g++ -O2 -I/home/docker/MPLAPACK/include -I/home/docker/MPLAPACK/include/mplapack \
Cheev_test_mpfr.cpp -Wl,--rpath=/home/docker/MPLAPACK/lib -L/home/docker/MPLAPACK/lib \
-lmplapack_mpfr -lmpblas_mpfr -lmpfr -lmpc -lgmp
$ ./a.out
\end{verbatim}
\end{itemize}
\subsubsection{Eigenvalue problem of a real non-symmetric matrix ({\tt Rgees})}  

The following example shows how to solve a real non-symmetric eigenvalue problem. For example, let $A$ be four times four matrices:
\[
A = \left[\begin{array}{llll}-2 & 2 & 2 & 2 \\ -3 & 3 & 2 & 2 \\ -2 & 0 & 4 & 2 \\ -1 & 0 & 0 & 5\end{array}\right]
\]
with eigenvalues $\lambda_1 = 1$, $\lambda_2 = 2$, $\lambda_3 = 3$, $\lambda_4=4$~\cite{Gregory1969ACO}.
First, we solve it with {\tt dd\_real} class with {\tt Rgees}. This routine computes for an $N$-by-$N$
real nonsymmetric matrix $A$, the eigenvalues, the real Schur form $T$, and, optionally, the matrix of
Schur vectors $Z$. This gives the Schur factorization $A = Z T Z^t$. $ T$ overwrites the matrix $A$.
The prototype definition of {\tt Rgees} is the following:
\begin{verbatim}
void Rgees(const char *jobvs, const char *sort, bool (*select)(dd_real, dd_real),
mplapackint const n, dd_real *a, mplapackint const lda, mplapackint &sdim,
dd_real *wr, dd_real *wi, dd_real *vs, mplapackint const ldvs, dd_real *work,
mplapackint const lwork, bool *bwork, mplapackint &info);
\end{verbatim}
The corresponding LAPACK routine is {\tt DGEES}. We show 14 lines of {\tt DGEES} are following.
\begin{verbatim}
 *       SUBROUTINE DGEES( JOBVS, SORT, SELECT, N, A, LDA, SDIM, WR, WI,
 *                         VS, LDVS, WORK, LWORK, BWORK, INFO )
 *
 *       .. Scalar Arguments ..
 *       CHARACTER          JOBVS, SORT
 *       INTEGER            INFO, LDA, LDVS, LWORK, N, SDIM
 *       ..
 *       .. Array Arguments ..
 *       LOGICAL            BWORK( * )
 *       DOUBLE PRECISION   A( LDA, * ), VS( LDVS, * ), WI( * ), WORK( * ),
 *      $                   WR( * )
 *       ..
 *       .. Function Arguments ..
 *       LOGICAL            SELECT
 *       EXTERNAL           SELECT
\end{verbatim}

A sample program is following:
\begin{lstlisting}
//public domain
#include <iostream>
#include <string>
#include <sstream>
#include <cstring>
#include <algorithm>

#include <mpblas_dd.h>
#include <mplapack_dd.h>

#define DD_PRECISION_SHORT 16

inline void printnum(dd_real rtmp) {
    std::cout.precision(DD_PRECISION_SHORT);
    if (rtmp >= 0.0) {
        std::cout << "+" << rtmp;
    } else {
        std::cout << rtmp;
    }
    return;
}

//Matlab/Octave format
void printvec(dd_real *a, int len) {
    dd_real tmp;
    printf("[ ");
    for (int i = 0; i < len; i++) {
        tmp = a[i];
        printnum(tmp);
        if (i < len - 1)
            printf(", ");
    }
    printf("]");
}
void printmat(int n, int m, dd_real * a, int lda)
{
    dd_real mtmp;
    printf("[ ");
    for (int i = 0; i < n; i++) {
        printf("[ ");
        for (int j = 0; j < m; j++) {
            mtmp = a[i + j * lda];
            printnum(mtmp);
            if (j < m - 1)
                printf(", ");
        }
        if (i < n - 1)
            printf("]; ");
        else
            printf("] ");
    }
    printf("]");
}
bool rselect(dd_real ar, dd_real ai) {
    // sorting rule for eigenvalues.
    return false;
}

int main() {
    mplapackint n = 4;

    dd_real *a = new dd_real[n * n];
    dd_real *vs = new dd_real[n * n];
    mplapackint sdim = 0;
    mplapackint lwork = 3 * n;
    dd_real *wr = new dd_real[n];
    dd_real *wi = new dd_real[n];
    dd_real *work = new dd_real[lwork];
    bool bwork[n];
    mplapackint info;
    // setting A matrix
    a[0 + 0 * n] = -2.0; a[0 + 1 * n] = 2.0; a[0 + 2 * n] = 2.0; a[0 + 3 * n] = 2.0;
    a[1 + 0 * n] = -3.0; a[1 + 1 * n] = 3.0; a[1 + 2 * n] = 2.0; a[1 + 3 * n] = 2.0;
    a[2 + 0 * n] = -2.0; a[2 + 1 * n] = 0.0; a[2 + 2 * n] = 4.0; a[2 + 3 * n] = 2.0;
    a[3 + 0 * n] = -1.0; a[3 + 1 * n] = 0.0; a[3 + 2 * n] = 0.0; a[3 + 3 * n] = 5.0;

    printf("# octave check\n");
    printf("a ="); printmat(n, n, a, n); printf("\n");
    Rgees("V", "S", rselect, n, a, n, sdim, wr, wi, vs, n, work, lwork, bwork, info);
    printf("vs ="); printmat(n, n, vs, n); printf("\n");
    printf("t ="); printmat(n, n, a, n); printf("\n");
    printf("vs*t*vs'\n");
    printf("eig(a)\n");
    for (int i = 1; i <= n; i = i + 1) {
        printf("w_%d = ", (int)i); printnum(wr[i - 1]); printf(" "); printnum(wi[i - 1]); printf("i\n");
    }
    delete[] work;
    delete[] wr;
    delete[] wi;
    delete[] vs;
    delete[] a;
}

\end{lstlisting}
One can compile this source code by:
\begin{verbatim}
$ g++ -O2 -I/home/docker/MPLAPACK/include -I/home/docker/MPLAPACK/include/mplapack \
Rgees_test_dd.cpp -L/home/docker/MPLAPACK/lib -lmplapack_dd -lmpblas_dd -lqd
\end{verbatim}
The output of the executable is the following:
{\footnotesize
\begin{verbatim}
$ LD_LIBRARY_PATH=/home/docker/MPLAPACK/lib ./a.out
# octave check
a =[ [ -2.0000000000000000e+00, +2.0000000000000000e+00, +2.0000000000000000e+00, +2.0000000000000000e+00];
[ -3.0000000000000000e+00, +3.0000000000000000e+00, +2.0000000000000000e+00, +2.0000000000000000e+00];
[ -2.0000000000000000e+00, +0.0000000000000000e+00, +4.0000000000000000e+00, +2.0000000000000000e+00];
[ -1.0000000000000000e+00, +0.0000000000000000e+00, +0.0000000000000000e+00, +5.0000000000000000e+00] ]
vs =[ [ -7.3029674334022148e-01, -6.8313005106397323e-01, +0.0000000000000000e+00, +0.0000000000000000e+00];
[ -5.4772255750516611e-01, +5.8554004376911991e-01, +5.9761430466719682e-01, -2.7760873845656105e-33];
[ -3.6514837167011074e-01, +3.9036002917941327e-01, -7.1713716560063618e-01, -4.4721359549995794e-01];
[ -1.8257418583505537e-01, +1.9518001458970664e-01, -3.5856858280031809e-01, +8.9442719099991588e-01] ]
t =[ [ +1.0000000000000000e+00, -6.9487922897230340e+00, +2.5313275267375116e+00, -1.9595917942265425e+00];
[ +0.0000000000000000e+00, +2.0000000000000000e+00, -1.3063945294843617e+00, +7.8558440484957257e-01];
[ +0.0000000000000000e+00, +0.0000000000000000e+00, +3.0000000000000000e+00, -1.0690449676496975e+00];
[ +0.0000000000000000e+00, +0.0000000000000000e+00, +0.0000000000000000e+00, +4.0000000000000000e+00] ]
vs*t*vs'
eig(a)
w_1 = +1.0000000000000000e+00 +0.0000000000000000e+00i
w_2 = +2.0000000000000000e+00 +0.0000000000000000e+00i
w_3 = +3.0000000000000000e+00 +0.0000000000000000e+00i
w_4 = +4.0000000000000000e+00 +0.0000000000000000e+00i
\end{verbatim}
}
We see that we calculated all the eigenvalues correctly. Moreover, you can check the Schur matrix by octave.
{\footnotesize
\begin{verbatim}
$ LD_LIBRARY_PATH=/home/docker/MPLAPACK/lib ./a.out | octave
octave: X11 DISPLAY environment variable not set
octave: disabling GUI features
a =

  -2   2   2   2
  -3   3   2   2
  -2   0   4   2
  -1   0   0   5

vs =

  -0.73030  -0.68313   0.00000   0.00000
  -0.54772   0.58554   0.59761  -0.00000
  -0.36515   0.39036  -0.71714  -0.44721
  -0.18257   0.19518  -0.35857   0.89443

t =

   1.00000  -6.94879   2.53133  -1.95959
   0.00000   2.00000  -1.30639   0.78558
   0.00000   0.00000   3.00000  -1.06904
   0.00000   0.00000   0.00000   4.00000

ans =

  -2.0000e+00   2.0000e+00   2.0000e+00   2.0000e+00
  -3.0000e+00   3.0000e+00   2.0000e+00   2.0000e+00
  -2.0000e+00   5.1394e-16   4.0000e+00   2.0000e+00
  -1.0000e+00   2.5697e-16  -4.5393e-17   5.0000e+00

ans =

   1.00000
   2.00000
   3.00000
   4.00000

w_1 =  1
w_2 =  2
w_3 =  3
w_4 =  4
\end{verbatim} }
We can see that we correctly calculated all the eigenvalues, the Schur form, and the Schur vectors.
%%%%%%%%%%%%%%%%%%%%%%%%%%%%%%%%%%%%%%%%%%%%%%%%%%%%%%%%%%%%%%%%%%%%%%%%%%%%%
\subsubsection{Eigenvalue problem of a complex non-symmetric matrix ({\tt Cgees})}  

The following example shows how to solve a complex non-symmetric eigenvalue problem. For example, let $A$ be four times four matrices:
\[
A = \left[\begin{array}{cccc}
5+9 i & 5+5 i & -6-6 i & -7-7 i \\
3+3 i & 6+10 i & -5-5 i & -6-6 i \\
2+2 i & 3+3 i & -1+3 i & -5-5 i \\
1+1 i & 2+2 i & -3-3 i & 4 i
\end{array}\right]
\]

with eigenvalues $\lambda_1 = 1+5i$, $\lambda_2 = 2+6i$, $\lambda_3 = 3+7i$, $\lambda_4=4+8i$~\cite{Gregory1969ACO}.
First, we solve it with {\tt dd\_real} class with {\tt Cgees}. This routine computes for an $N$-by-$N$ 
real nonsymmetric matrix $A$, the eigenvalues, the real Schur form $T$, and, optionally, the matrix of
Schur vectors $Z$. This gives the Schur factorization $A = Z T Z^t$. $ T$ overwrites the matrix $A$.
The prototype definition of {\tt Cgees} is the following:
\begin{verbatim}
void Cgees(const char *jobvs, const char *sort, bool (*select)(dd_complex), 
mplapackint const n, dd_complex *a, mplapackint const lda, mplapackint &sdim, 
dd_complex *w, dd_complex *vs, mplapackint const ldvs, dd_complex *work, 
mplapackint const lwork, dd_real *rwork, bool *bwork, mplapackint &info);
\end{verbatim}
The corresponding LAPACK routine is {\tt ZGEES}. We show 14 lines of {\tt ZGEES} are following. 
\begin{verbatim}
*       SUBROUTINE ZGGES( JOBVSL, JOBVSR, SORT, SELCTG, N, A, LDA, B, LDB,
*                         SDIM, ALPHA, BETA, VSL, LDVSL, VSR, LDVSR, WORK,
*                         LWORK, RWORK, BWORK, INFO )
*
*       .. Scalar Arguments ..
*       CHARACTER          JOBVSL, JOBVSR, SORT
*       INTEGER            INFO, LDA, LDB, LDVSL, LDVSR, LWORK, N, SDIM
*       ..
*       .. Array Arguments ..
*       LOGICAL            BWORK( * )
*       DOUBLE PRECISION   RWORK( * )
*       COMPLEX*16         A( LDA, * ), ALPHA( * ), B( LDB, * ),
*      $                   BETA( * ), VSL( LDVSL, * ), VSR( LDVSR, * ),
*      $                   WORK( * )
*       ..
*       .. Function Arguments ..
*       LOGICAL            SELCTG
*       EXTERNAL           SELCTG

\end{verbatim}

A sample program is following:
\begin{lstlisting}
//public domain
#include <iostream>
#include <string>
#include <sstream>
#include <cstring>
#include <algorithm>

#include <mpblas_dd.h>
#include <mplapack_dd.h>

#define DD_PRECISION_SHORT 16

inline void printnum(dd_real rtmp) {
    std::cout.precision(DD_PRECISION_SHORT);
    if (rtmp >= 0.0) {
        std::cout << "+" << rtmp;
    } else {
        std::cout << rtmp;
    }
    return;
}

inline void printnum(dd_complex rtmp) {
    std::cout.precision(DD_PRECISION_SHORT);
    if (rtmp.real() >= 0.0) {
        std::cout << "+" << rtmp.real();
    } else {
        std::cout << rtmp.real();
    }
    if (rtmp.imag() >= 0.0) {
        std::cout << "+" << rtmp.imag() << "i";
    } else {
        std::cout << rtmp.imag() << "i";
    }
    return;
}

//Matlab/Octave format
template <class X> void printvec(X *a, int len) {
    X tmp;
    printf("[ ");
    for (int i = 0; i < len; i++) {
        tmp = a[i];
        printnum(tmp);
        if (i < len - 1)
            printf(", ");
    }
    printf("]");
}

template <class X> void printmat(int n, int m, X *a, int lda)
{
    X mtmp;

    printf("[ ");
    for (int i = 0; i < n; i++) {
        printf("[ ");
        for (int j = 0; j < m; j++) {
            mtmp = a[i + j * lda];
            printnum(mtmp);
            if (j < m - 1)
                printf(", ");
        }
        if (i < n - 1)
            printf("]; ");
        else
            printf("] ");
    }
    printf("]");
}
bool cselect(dd_complex a) {
    // sorting rule for eigenvalues.
    return false;
}

int main() {
    mplapackint n = 4;

    dd_complex *a = new dd_complex[n * n];
    mplapackint sdim = 0;
    mplapackint lwork = 2 * n;
    dd_complex *w = new dd_complex[n];
    dd_complex *vs = new dd_complex[n * n];
    dd_complex *work = new dd_complex[lwork];
    dd_real *rwork = new dd_real[n];
    bool bwork[n];
    mplapackint info;

    // setting A matrix
    a[0 + 0 * n] = dd_complex(5.0,  9.0); a[0 + 1 * n] = dd_complex(5.0, 5.0);   a[0 + 2 * n] = dd_complex(-6.0, -6.0); a[0 + 3 * n] = dd_complex(-7.0,-7.0);
    a[1 + 0 * n] = dd_complex(3.0,  3.0); a[1 + 1 * n] = dd_complex(6.0,10.0);   a[1 + 2 * n] = dd_complex(-5.0, -5.0); a[1 + 3 * n] = dd_complex(-6.0,-6.0);
    a[2 + 0 * n] = dd_complex(2.0,  2.0); a[2 + 1 * n] = dd_complex(3.0, 3.0);   a[2 + 2 * n] = dd_complex(-1.0, 3.0);  a[2 + 3 * n] = dd_complex(-5.0,-5.0);
    a[3 + 0 * n] = dd_complex(1.0,  1.0); a[3 + 1 * n] = dd_complex(2.0, 2.0);   a[3 + 2 * n] = dd_complex(-3.0,-3.0);  a[3 + 3 * n] = dd_complex(0.0, 4.0);

    printf("# Ex. 6.5 p. 116, Collection of Matrices for Testing Computational Algorithms, Robert T. Gregory, David L. Karney\n");
    printf("# octave check\n");
    printf("split_long_rows(0)\n");
    printf("a ="); printmat(n, n, a, n); printf("\n");
    Cgees("V", "S", cselect, n, a, n, sdim, w, vs, n, work, lwork, rwork, bwork, info);
    printf("w ="); printvec(w, n); printf("\n");
    printf("vs ="); printmat(n, n, vs, n); printf("\n");
    printf("t ="); printmat(n, n, a, n); printf("\n");
    printf("vs*t*vs'\n");
    printf("eig(a)\n");

    delete[] rwork;
    delete[] work;
    delete[] vs;
    delete[] w;
    delete[] a;
}

\end{lstlisting}
One can compile this source code by:
\begin{verbatim}
$ g++ -O2 -I/home/docker/MPLAPACK/include -I/home/docker/MPLAPACK/include/mplapack \
Cgees_test_dd.cpp -Wl,--rpath=/home/docker/MPLAPACK/lib -L/home/docker/MPLAPACK/lib \
-lmplapack_dd -lmpblas_dd -lqd
\end{verbatim}
The output of the executable is the following:
{\fontsize{8.5pt}{0.4cm}\selectfont
\begin{verbatim}
# Ex. 6.5 p. 116, Collection of Matrices for Testing Computational Algorithms, Robert T. Gregory, David L. Karney
# octave check
split_long_rows(0)
a =[ [ +5.0000000000000000e+00+9.0000000000000000e+00i, +5.0000000000000000e+00+5.0000000000000000e+00i,
-6.0000000000000000e+00-6.0000000000000000e+00i, -7.0000000000000000e+00-7.0000000000000000e+00i]; 
[ +3.0000000000000000e+00+3.0000000000000000e+00i, +6.0000000000000000e+00+1.0000000000000000e+01i,
-5.0000000000000000e+00-5.0000000000000000e+00i, -6.0000000000000000e+00-6.0000000000000000e+00i];
[ +2.0000000000000000e+00+2.0000000000000000e+00i, +3.0000000000000000e+00+3.0000000000000000e+00i,
-1.0000000000000000e+00+3.0000000000000000e+00i, -5.0000000000000000e+00-5.0000000000000000e+00i]; 
[ +1.0000000000000000e+00+1.0000000000000000e+00i, +2.0000000000000000e+00+2.0000000000000000e+00i,
-3.0000000000000000e+00-3.0000000000000000e+00i, +0.0000000000000000e+00+4.0000000000000000e+00i] ]
w =[ +2.0000000000000000e+00+6.0000000000000000e+00i, +4.0000000000000000e+00+8.0000000000000000e+00i,
+3.0000000000000000e+00+7.0000000000000000e+00i, +1.0000000000000000e+00+5.0000000000000000e+00i]
vs =[ [ +3.7428970594742688e-01-5.2577170701090183e-02i, -1.1134324587883832e-01+4.9471763536387269e-01i,
+3.8355427599552469e-01-6.7296813993349557e-01i, +0.0000000000000000e+00+0.0000000000000000e+00i];
[ +7.4857941189485376e-01-1.0515434140218037e-01i, +3.7114415292946107e-02-1.6490587845462423e-01i, 
-1.2785142533184156e-01+2.2432271331116519e-01i, +4.2849243249895099e-01-3.8694646738853330e-01i];
[ +3.7428970594742688e-01-5.2577170701090183e-02i, -1.1134324587883832e-01+4.9471763536387269e-01i, 
-2.5570285066368313e-01+4.4864542662233038e-01i, -4.2849243249895099e-01+3.8694646738853330e-01i];
[ +3.7428970594742688e-01-5.2577170701090183e-02i, +1.4845766117178443e-01-6.5962351381849692e-01i, 
+1.2785142533184156e-01-2.2432271331116519e-01i, -4.2849243249895099e-01+3.8694646738853330e-01i] ]
t =[ [ +2.0000000000000000e+00+6.0000000000000000e+00i, -4.6106496752391337e+00+2.0837249271958309e+00i,
+4.9279443181496018e+00-6.3330804742348119e-01i, +1.9999717105990234e+01+3.8559640218114672e+00i];
[ +0.0000000000000000e+00+0.0000000000000000e+00i, +4.0000000000000000e+00+8.0000000000000000e+00i,
-2.8979419726911526e-01-5.4495110256015637e-01i, -1.0301880205614183e+00-5.9845553818543799e+00i];
[ +0.0000000000000000e+00+0.0000000000000000e+00i, +0.0000000000000000e+00+0.0000000000000000e+00i, 
+3.0000000000000000e+00+7.0000000000000000e+00i, +2.2784477280226518e+00+4.5175962580412624e+00i];
[ +0.0000000000000000e+00+0.0000000000000000e+00i, +0.0000000000000000e+00+0.0000000000000000e+00i, 
+0.0000000000000000e+00+0.0000000000000000e+00i, +1.0000000000000000e+00+5.0000000000000000e+00i] ]
vs*t*vs'
eig(a)
\end{verbatim}
}
We see that we calculated all the eigenvalues correctly. Moreover, you can check the Schur matrix by octave.
{\footnotesize
\begin{verbatim}
octave: X11 DISPLAY environment variable not set
octave: disabling GUI features
a =

    5 +  9i    5 +  5i   -6 -  6i   -7 -  7i
    3 +  3i    6 + 10i   -5 -  5i   -6 -  6i
    2 +  2i    3 +  3i   -1 +  3i   -5 -  5i
    1 +  1i    2 +  2i   -3 -  3i    0 +  4i

w =

   2 + 6i   4 + 8i   3 + 7i   1 + 5i

vs =

   0.37429 - 0.05258i  -0.11134 + 0.49472i   0.38355 - 0.67297i   0.00000 + 0.00000i
   0.74858 - 0.10515i   0.03711 - 0.16491i  -0.12785 + 0.22432i   0.42849 - 0.38695i
   0.37429 - 0.05258i  -0.11134 + 0.49472i  -0.25570 + 0.44865i  -0.42849 + 0.38695i
   0.37429 - 0.05258i   0.14846 - 0.65962i   0.12785 - 0.22432i  -0.42849 + 0.38695i

t =

    2.00000 +  6.00000i   -4.61065 +  2.08372i    4.92794 -  0.63331i   19.99972 +  3.85596i
    0.00000 +  0.00000i    4.00000 +  8.00000i   -0.28979 -  0.54495i   -1.03019 -  5.98456i
    0.00000 +  0.00000i    0.00000 +  0.00000i    3.00000 +  7.00000i    2.27845 +  4.51760i
    0.00000 +  0.00000i    0.00000 +  0.00000i    0.00000 +  0.00000i    1.00000 +  5.00000i

ans =

    5.00000 +  9.00000i    5.00000 +  5.00000i   -6.00000 -  6.00000i   -7.00000 -  7.00000i
    3.00000 +  3.00000i    6.00000 + 10.00000i   -5.00000 -  5.00000i   -6.00000 -  6.00000i
    2.00000 +  2.00000i    3.00000 +  3.00000i   -1.00000 +  3.00000i   -5.00000 -  5.00000i
    1.00000 +  1.00000i    2.00000 +  2.00000i   -3.00000 -  3.00000i   -0.00000 +  4.00000i

ans =

   2.0000 + 6.0000i
   4.0000 + 8.0000i
   3.0000 + 7.0000i
   1.0000 + 5.0000i
\end{verbatim} }
We can see that we correctly calculated all the eigenvalues, the Schur form, and the Schur vectors.

\subsubsection{Eigenvalues and eigenvectors of non-symmetric matrix ({\tt Rgeev})}  

Another example shows how to solve real non-symmetric eigenvalue problems with eigenvectors. {\tt Rgeev} computes the eigenvalues and, optionally, the left and/or right eigenvectors  for an $N$ by $N$ real nonsymmetric matrix $A$.
For example, let $A$ be four times four matrices:
\[
A=\left[\begin{array}{rrrr}4 & -5 & 0 & 3 \\ 0 & 4 & -3 & -5 \\ 5 & -3 & 4 & 0 \\ 3 & 0 & 5 & 4\end{array}\right]
\]
Then, the eigenvalues are $\lambda_1= 12$, $\lambda_2= 1+5i$, $\lambda_3 = 1-5i$ and $\lambda_4=2$ and the right eigenvectors are
\[
{x}_{1}=\left[\begin{array}{r}1 \\ -1 \\ 1 \\ 1\end{array}\right], \quad {x}_{2}=\left[\begin{array}{c}1 \\ -i \\ -i \\ -1\end{array}\right], \quad {x}_{3}=\left[\begin{array}{c}1 \\ i \\ i \\ -1\end{array}\right], \quad {x}_{4}=\left[\begin{array}{c}1 \\ 1 \\ -1 \\ 1\end{array}\right],
\]
And the left eigenvectors are
\[
{y}_{1}=[1,-1,1,1], \quad {y}_{2}=[1, i, i,-1], \quad {y}_{3}=[1,-i,-i,-1], \quad {y}_{4}=[1,1,-1,1].
\]
Thus, $Ax_1 = 12 x_1$, $Ax_2 = (1+5i) x_2$, $Ax_3 = (1-5i) x_3$ and $Ax_4 = 2 x_4$,
and $y_1 A = 12 y_1$, and $y_2 A = (1+5i) y_2$, $y_3 A = (1-5i) y_3$, $y_4 A = 2 y_4$.

Let us solve this eigenvalue problem with left and right eigenvectors for $A$ using MPLAPACK $\tt qd\_real$.
The prototype definition of {\tt Rgeev} is the following:
\begin{verbatim}
void Rgeev(const char *jobvl, const char *jobvr, mplapackint const n, qd_real *a,
mplapackint const lda, qd_real *wr, qd_real *wi, qd_real *vl, mplapackint const ldvl, 
qd_real *vr, mplapackint const ldvr, qd_real *work, mplapackint const lwork, 
mplapackint &info);
\end{verbatim}
The corresponding LAPACK routine is {\tt DGEEV}. We show the 14 lines of {\tt DGEEV} are following. 
\begin{verbatim}
      SUBROUTINE dgeev( JOBVL, JOBVR, N, A, LDA, WR, WI, VL, LDVL, VR,
      $                  LDVR, WORK, LWORK, INFO )
       implicit none
 *
 *  -- LAPACK driver routine --
 *  -- LAPACK is a software package provided by Univ. of Tennessee,    --
 *  -- Univ. of California Berkeley, Univ. of Colorado Denver and NAG Ltd..--
 *
 *     .. Scalar Arguments ..
       CHARACTER          JOBVL, JOBVR
       INTEGER            INFO, LDA, LDVL, LDVR, LWORK, N
 *     ..
 *     .. Array Arguments ..
       DOUBLE PRECISION   A( LDA, * ), VL( LDVL, * ), VR( LDVR, * ),
      $                   wi( * ), work( * ), wr( * )
\end{verbatim}
A sample program is following:
\begin{lstlisting}
//public domain
#include <iostream>
#include <string>
#include <sstream>
#include <cstring>
#include <algorithm>

#include <mpblas_qd.h>
#include <mplapack_qd.h>

#define QD_PRECISION_SHORT 16

inline void printnum(qd_real rtmp) {
    std::cout.precision(QD_PRECISION_SHORT);
    if (rtmp >= 0.0) {
        std::cout << "+" << rtmp;
    } else {
        std::cout << rtmp;
    }
    return;
}

//Matlab/Octave format
void printvec(qd_real *a, int len) {
    qd_real tmp;
    printf("[ ");
    for (int i = 0; i < len; i++) {
        tmp = a[i];
        printnum(tmp);
        if (i < len - 1)
            printf(", ");
    }
    printf("]");
}

void printmat(int n, int m, qd_real * a, int lda)
{
    qd_real mtmp;
    printf("[ ");
    for (int i = 0; i < n; i++) {
        printf("[ ");
        for (int j = 0; j < m; j++) {
            mtmp = a[i + j * lda];
            printnum(mtmp);
            if (j < m - 1)
                printf(", ");
        }
        if (i < n - 1)
            printf("]; ");
        else
            printf("] ");
    }
    printf("]");
}
bool rselect(qd_real ar, qd_real ai) {
    // sorting rule for eigenvalues.
    return false;
}

int main() {
    mplapackint n = 4;
    qd_real *a = new qd_real[n * n];
    qd_real *vl = new qd_real[n * n];
    qd_real *vr = new qd_real[n * n];
    mplapackint lwork = 4 * n;
    qd_real *wr = new qd_real[n];
    qd_real *wi = new qd_real[n];
    qd_real *work = new qd_real[lwork];
    mplapackint info;
    // setting A matrix
    a[0 + 0 * n] = 4.0; a[0 + 1 * n] = -5.0; a[0 + 2 * n] =  0.0;  a[0 + 3 * n] =  3.0;
    a[1 + 0 * n] = 0.0; a[1 + 1 * n] =  4.0; a[1 + 2 * n] = -3.0;  a[1 + 3 * n] = -5.0;
    a[2 + 0 * n] = 5.0; a[2 + 1 * n] = -3.0; a[2 + 2 * n] =  4.0;  a[2 + 3 * n] =  0.0;
    a[3 + 0 * n] = 3.0; a[3 + 1 * n] =  0.0; a[3 + 2 * n] =  5.0;  a[3 + 3 * n] =  4.0;

    printf("# octave check\n");
    printf("split_long_rows(0)\n");
    printf("a =");
    printmat(n, n, a, n);
    printf("\n");
    Rgeev("V", "V", n, a, n, wr, wi, vl, n, vr, n, work, lwork, info);
    printf("# right vectors\n");
    for (int j = 1; j <= n; j = j + 1) {
        if (abs(wi[j - 1]) < 1e-15) {
            printf("vr_%d =[ ", j);
            for (int i = 1; i <= n - 1; i = i + 1) {
                printnum(vr[(i - 1) + (j - 1) * n]); printf(", ");
            }
            printnum(vr[(n - 1) + (j - 1) * n]); printf("];\n");
        } else {
            printf("vr_%d =[ ", j);
            for (int i = 1; i <= n - 1; i = i + 1) {
                printnum(vr[(i - 1) + (j - 1) * n]); printnum(-vr[(i - 1) + j * n]); printf("i, ");
            }
            printnum(vr[(n - 1) + (j - 1) * n]); printnum(-vr[(n - 1) + j * n]); printf("i ];\n");
            printf("vr_%d =[ ", j + 1);
            for (int i = 1; i <= n - 1; i = i + 1) {
                printnum(vr[(i - 1) + (j - 1) * n]); printnum(vr[(i - 1) + j * n]); printf("i, ");
            }
            printnum(vr[(n - 1) + (j - 1) * n]); printnum(vr[(n - 1) + j * n]); printf("i ];\n");
            j++;
        }
    }
    printf("# left vectors\n");
    for (int j = 1; j <= n; j = j + 1) {
        if (abs(wi[j - 1]) < 1e-15) {
            printf("vl_%d =[ ", j);
            for (int i = 1; i <= n - 1; i = i + 1) {
                printnum(vl[(i - 1) + (j - 1) * n]); printf(", ");
            }
            printnum(vl[(n - 1) + (j - 1) * n]); printf("];\n");
        } else {
            printf("vl_%d =[ ", j);
            for (int i = 1; i <= n - 1; i = i + 1) {
                printnum(vl[(i - 1) + (j - 1) * n]); printnum(-vl[(i - 1) + j * n]); printf("i, ");
            }
            printnum(vl[(n - 1) + (j - 1) * n]); printnum(-vl[(n - 1) + j * n]); printf("i ];\n");
            printf("vl_%d =[ ", j + 1);
            for (int i = 1; i <= n - 1; i = i + 1) {
                printnum(vl[(i - 1) + (j - 1) * n]); printnum(vl[(i - 1) + j * n]); printf("i, ");
            }
            printnum(vl[(n - 1) + (j - 1) * n]); printnum(vl[(n - 1) + j * n]); printf("i ];\n");
            j++;
        }
    }
    for (int i = 1; i <= n; i = i + 1) {
        printf("w_%d = ", (int)i); printnum(wr[i - 1]); printf(" "); printnum(wi[i - 1]); printf("i\n");
    }
    for (int i = 1; i <= n; i = i + 1) {
        printf("disp (\"a * vr_%d\")\n", i);
        printf("a * vr_%d'\n", i);
        printf("disp (\"w_%d * vr_%d\")\n", i, i);
        printf("w_%d * vr_%d'\n", i, i);
        printf("disp (\"vr_%d\")\n", i);
        printf("vr_%d'\n", i);
    }

    for (int i = 1; i <= n; i = i + 1) {
        printf("disp (\"vl_%d * a \")\n", i);
        printf("vl_%d * a\n", i);
        printf("disp (\"w_%d * vl_%d \")\n", i, i);
        printf("w_%d * vl_%d\n", i, i);
        printf("disp (\"vl_%d\")\n", i);
        printf("vl_%d\n", i);
    }
    delete[] work;
    delete[] wr;
    delete[] wi;
    delete[] vr;
    delete[] vl;
    delete[] a;
}

\end{lstlisting}
You can find this file as \\ {\tt /home/docker/mplapack/examples/mplapack/04\_NonsymmetricEigenproblems/Rgeev\_test\_qd.cpp },
Furthermore, compile this source code by:
\begin{verbatim}
$ g++ -O2 -I/home/docker/MPLAPACK/include -I/home/docker/MPLAPACK/include/mplapack \
Rgeev_test_qd.cpp -Wl,--rpath=/home/docker/MPLAPACK/lib -L/home/docker/MPLAPACK/lib \
-lmplapack_qd -lmpblas_qd -lqd
\end{verbatim}
The output of the executable is the following:
{\footnotesize
\begin{verbatim}
$ ./a.out
# octave check
split_long_rows(0)
a =[ [ +4.0000000000000000e+00, -5.0000000000000000e+00, +0.0000000000000000e+00, +3.0000000000000000e+00]; 
[ +0.0000000000000000e+00, +4.0000000000000000e+00, -3.0000000000000000e+00, -5.0000000000000000e+00];
[ +5.0000000000000000e+00, -3.0000000000000000e+00, +4.0000000000000000e+00, +0.0000000000000000e+00];
[ +3.0000000000000000e+00, +0.0000000000000000e+00, +5.0000000000000000e+00, +4.0000000000000000e+00] ]
# right vectors
vr_1 =[ -5.0000000000000000e-01, +5.0000000000000000e-01, -5.0000000000000000e-01, -5.0000000000000000e-01];
vr_2 =[ +2.8486703237279396e-65-5.0000000000000000e-01i, +5.0000000000000000e-01+0.0000000000000000e+00i,
+5.0000000000000000e-01+0.0000000000000000e+00i, -9.0207893584718088e-65+5.0000000000000000e-01i ];
vr_3 =[ +2.8486703237279396e-65+5.0000000000000000e-01i, +5.0000000000000000e-01+0.0000000000000000e+00i,
+5.0000000000000000e-01+0.0000000000000000e+00i, -9.0207893584718088e-65-5.0000000000000000e-01i ];
vr_4 =[ +5.0000000000000000e-01, +5.0000000000000000e-01, -5.0000000000000000e-01, +5.0000000000000000e-01];
# left vectors
vl_1 =[ -5.0000000000000000e-01, +5.0000000000000000e-01, -5.0000000000000000e-01, -5.0000000000000000e-01];
vl_2 =[ -4.7477838728798994e-65-5.0000000000000000e-01i, +5.0000000000000000e-01+0.0000000000000000e+00i,
+5.0000000000000000e-01-1.7092021942367638e-64i, -3.4184043884735275e-64+5.0000000000000000e-01i ];
vl_3 =[ -4.7477838728798994e-65+5.0000000000000000e-01i, +5.0000000000000000e-01+0.0000000000000000e+00i,
+5.0000000000000000e-01+1.7092021942367638e-64i, -3.4184043884735275e-64-5.0000000000000000e-01i ];
vl_4 =[ +5.0000000000000000e-01, +5.0000000000000000e-01, -5.0000000000000000e-01, +5.0000000000000000e-01];
w_1 = +1.2000000000000000e+01 +0.0000000000000000e+00i
w_2 = +1.0000000000000000e+00 +5.0000000000000000e+00i
w_3 = +1.0000000000000000e+00 -5.0000000000000000e+00i
w_4 = +2.0000000000000000e+00 +0.0000000000000000e+00i
...
\end{verbatim}
}
This can be confirmed by passing the output of the terminal to Octave, as shown below.
{\footnotesize
\begin{verbatim}
$ ./a.out | octave
a =

   4  -5   0   3
   0   4  -3  -5
   5  -3   4   0
   3   0   5   4

w_1 =  12
w_2 =  1 + 5i
w_3 =  1 - 5i
w_4 =  2
a * vr_1
ans =

  -6
   6
  -6
  -6

w_1 * vr_1
ans =

  -6
   6
  -6
  -6

vr_1
ans =

  -0.50000
   0.50000
  -0.50000
  -0.50000

a * vr_2
ans =

  -2.50000 + 0.50000i
   0.50000 + 2.50000i
   0.50000 + 2.50000i
   2.50000 - 0.50000i

w_2 * vr_2
ans =

  -2.50000 + 0.50000i
   0.50000 + 2.50000i
   0.50000 + 2.50000i
   2.50000 - 0.50000i

vr_2
ans =

   0.00000 + 0.50000i
   0.50000 - 0.00000i
   0.50000 - 0.00000i
  -0.00000 - 0.50000i

...
\end{verbatim}
}
We only show the first part of the output since the whole output is lengthy. 
{\tt w\_1}, {\tt w\_2}, {\tt w\_3} and {\tt w\_4} are the eigenvalues. {\tt vr\_1} and {\tt vr\_2} are right eigenvectors. You will see that you obtained the correct eigenvalues and eigenvectors.

\subsubsection{Eigenvalues and eigenvectors of complex non-symmetric matrix ({\tt Cgeev})}  

Another example shows how to solve real non-symmetric eigenvalue problems with eigenvectors. {\tt Cgeev} computes the eigenvalues, and, optionally, the left and/or right eigenvectors  for an $N$ by $N$ complex nonsymmetric matrix $A$.
For example, let $A$ be four times four matrices:
The following example shows how to solve a complex non-symmetric eigenvalue problem. For example, let $A$ be four times four matrices:
\[
A = \left[\begin{array}{llll}
5+9 i & 5+5 i & -6-6 i & -7-7 i \\
3+3 i & 6+10 i & -5-5 i & -6-6 i \\
2+2 i & 3+3 i & -1+3 i & -5-5 i \\
1+1 i & 2+2 i & -3-3 i & 4 i
\end{array}\right]
\]

with eigenvalues $\lambda_1 = 1+5i$, $\lambda_2 = 2+6i$, $\lambda_3 = 3+7i$, $\lambda_4=4+8i$
and the right eigenvectors are
\[
x_1=\left[\begin{array}{l}
2 \\
1 \\
1 \\
1
\end{array}\right], \quad x_2=\left[\begin{array}{l}
1 \\
2 \\
1 \\
1
\end{array}\right], \quad x_3=\left[\begin{array}{c}
-1 \\
-1 \\
0 \\
-1
\end{array}\right], \quad x_4=\left[\begin{array}{c}
-1 \\
-1 \\
-1 \\
0
\end{array}\right]
\]~\cite{Gregory1969ACO}.

Let us solve this eigenvalue problem with left and right eigenvectors for $A$ using MPLAPACK $\tt qd\_real$.
The prototype definition of {\tt Cgeev} is the following:
\begin{verbatim}
void Cgeev(const char *jobvl, const char *jobvr, mplapackint const n, qd_complex *a, 
mplapackint const lda, qd_complex *w, qd_complex *vl, mplapackint const ldvl, 
qd_complex *vr, mplapackint const ldvr, qd_complex *work, mplapackint const lwork, 
qd_real *rwork, mplapackint &info);
\end{verbatim}
The corresponding LAPACK routine is {\tt ZGEEV}. We show the 14 lines of {\tt ZGEEV} are following. 
\begin{verbatim}
*       SUBROUTINE ZGGEV( JOBVL, JOBVR, N, A, LDA, B, LDB, ALPHA, BETA,
*                         VL, LDVL, VR, LDVR, WORK, LWORK, RWORK, INFO )
*
*       .. Scalar Arguments ..
*       CHARACTER          JOBVL, JOBVR
*       INTEGER            INFO, LDA, LDB, LDVL, LDVR, LWORK, N
*       ..
*       .. Array Arguments ..
*       DOUBLE PRECISION   RWORK( * )
*       COMPLEX*16         A( LDA, * ), ALPHA( * ), B( LDB, * ),
*      $                   BETA( * ), VL( LDVL, * ), VR( LDVR, * ),
*      $                   WORK( * )

\end{verbatim}
A sample program is following:
\begin{lstlisting}
//public domain
#include <iostream>
#include <string>
#include <sstream>
#include <cstring>
#include <algorithm>

#include <mpblas_qd.h>
#include <mplapack_qd.h>

#define QD_PRECISION_SHORT 16

inline void printnum(qd_real rtmp) {
    std::cout.precision(QD_PRECISION_SHORT);
    if (rtmp >= 0.0) {
        std::cout << "+" << rtmp;
    } else {
        std::cout << rtmp;
    }
    return;
}

inline void printnum(qd_complex rtmp) {
    std::cout.precision(QD_PRECISION_SHORT);
    if (rtmp.real() >= 0.0) {
        std::cout << "+" << rtmp.real();
    } else {
        std::cout << rtmp.real();
    }
    if (rtmp.imag() >= 0.0) {
        std::cout << "+" << rtmp.imag() << "i";
    } else {
        std::cout << rtmp.imag() << "i";
    }
    return;
}

//Matlab/Octave format
template <class X> void printvec(X *a, int len) {
    X tmp;
    printf("[ ");
    for (int i = 0; i < len; i++) {
        tmp = a[i];
        printnum(tmp);
        if (i < len - 1)
            printf(", ");
    }
    printf("]");
}

template <class X> void printmat(int n, int m, X *a, int lda)
{
    X mtmp;

    printf("[ ");
    for (int i = 0; i < n; i++) {
        printf("[ ");
        for (int j = 0; j < m; j++) {
            mtmp = a[i + j * lda];
            printnum(mtmp);
            if (j < m - 1)
                printf(", ");
        }
        if (i < n - 1)
            printf("]; ");
        else
            printf("] ");
    }
    printf("]");
}
bool rselect(qd_real ar, qd_real ai) {
    // sorting rule for eigenvalues.
    return false;
}

int main() {
    mplapackint n = 4;
    qd_complex *a = new qd_complex[n * n];
    qd_complex *w = new qd_complex[n];
    qd_complex *vl = new qd_complex[n * n];
    qd_complex *vr = new qd_complex[n * n];
    mplapackint lwork = 4 * n;
    qd_complex *work = new qd_complex[lwork];
    qd_real *rwork = new qd_real[lwork];
    mplapackint info;
    // setting A matrix
    a[0 + 0 * n] = qd_complex(5.0, 9.0); a[0 + 1 * n] = qd_complex(5.0, 5.0);   a[0 + 2 * n] = qd_complex(-6.0, -6.0); a[0 + 3 * n] = qd_complex(-7.0, -7.0);
    a[1 + 0 * n] = qd_complex(3.0, 3.0); a[1 + 1 * n] = qd_complex(6.0, 10.0);  a[1 + 2 * n] = qd_complex(-5.0, -5.0); a[1 + 3 * n] = qd_complex(-6.0, -6.0);
    a[2 + 0 * n] = qd_complex(2.0, 2.0); a[2 + 1 * n] = qd_complex(3.0, 3.0);   a[2 + 2 * n] = qd_complex(-1.0,  3.0); a[2 + 3 * n] = qd_complex(-5.0, -5.0);
    a[3 + 0 * n] = qd_complex(1.0, 1.0); a[3 + 1 * n] = qd_complex(2.0, 2.0);   a[3 + 2 * n] = qd_complex(-3.0, -3.0); a[3 + 3 * n] = qd_complex(0.0, 4.0);

    printf("# Ex. 6.5 p. 116, Collection of Matrices for Testing Computational Algorithms, Robert T. Gregory, David L. Karney\n");
    printf("# octave check\n");
    printf("split_long_rows(0)\n");
    printf("a ="); printmat(n, n, a, n); printf("\n");
    Cgeev("V", "V", n, a, n, w, vl, n, vr, n, work, lwork, rwork, info);
    printf("lambda ="); printvec(w,n); printf("\n");
    printf("vr ="); printmat(n,n,vr,n); printf("\n");

    delete[] rwork;
    delete[] work;
    delete[] vr;
    delete[] vl;
    delete[] w;
    delete[] a;
}

\end{lstlisting}
You can find this file as \\ {\tt /home/docker/mplapack/examples/mplapack/04\_NonsymmetricEigenproblems/Cgeev\_test\_qd.cpp },
and compile this source code by:
\begin{verbatim}
$ g++ -O2 -I/home/docker/MPLAPACK/include -I/home/docker/MPLAPACK/include/mplapack \
Cgeev_test_qd.cpp -Wl,--rpath=/home/docker/MPLAPACK/lib -L/home/docker/MPLAPACK/lib \
-lmplapack_qd -lmpblas_qd -lqd
\end{verbatim}
The output of the executable is the following:
{\footnotesize
\begin{verbatim}
# Ex. 6.5 p. 116, Collection of Matrices for Testing Computational Algorithms, Robert T. Gregory, David L. Karney
# octave check
split_long_rows(0)
a =[ [ +5.0000000000000000e+00+9.0000000000000000e+00i, +5.0000000000000000e+00+5.0000000000000000e+00i,
-6.0000000000000000e+00-6.0000000000000000e+00i, -7.0000000000000000e+00-7.0000000000000000e+00i];
[+3.0000000000000000e+00+3.0000000000000000e+00i, +6.0000000000000000e+00+1.0000000000000000e+01i,
-5.0000000000000000e+00-5.0000000000000000e+00i, -6.0000000000000000e+00-6.0000000000000000e+00i];
[ +2.0000000000000000e+00+2.0000000000000000e+00i, +3.0000000000000000e+00+3.0000000000000000e+00i,
-1.0000000000000000e+00+3.0000000000000000e+00i, -5.0000000000000000e+00-5.0000000000000000e+00i];
[ +1.0000000000000000e+00+1.0000000000000000e+00i, +2.0000000000000000e+00+2.0000000000000000e+00i,
-3.0000000000000000e+00-3.0000000000000000e+00i, +0.0000000000000000e+00+4.0000000000000000e+00i] ]
lambda =[ +2.0000000000000000e+00+6.0000000000000000e+00i, +4.0000000000000000e+00+8.0000000000000000e+00i,
+3.0000000000000000e+00+7.0000000000000000e+00i, +1.0000000000000000e+00+5.0000000000000000e+00i]
vr =[ [ +3.7796447300922723e-01-1.5808636614229790e-64i, +5.7735026918962576e-01+0.0000000000000000e+00i,
+5.7735026918962576e-01+1.6142465167791658e-64i, +7.5592894601845445e-01+0.0000000000000000e+00i];
[ +7.5592894601845445e-01+0.0000000000000000e+00i, +5.7735026918962576e-01+1.2462932666309736e-65i,
+5.7735026918962576e-01+0.0000000000000000e+00i, +3.7796447300922723e-01+6.7655920188538566e-65i];
[ +3.7796447300922723e-01-1.2425840604802862e-66i, +5.7735026918962576e-01+1.0949576556829268e-64i,
-2.1515093228017323e-65-6.2460902724518248e-64i, +3.7796447300922723e-01-1.1275986698089761e-65i]; 
[ +3.7796447300922723e-01-5.0333927464828308e-65i, -4.9136820461357707e-65-1.1498598272944756e-64i,
+5.7735026918962576e-01+5.4614351362721592e-64i, +3.7796447300922723e-01+5.6379933490448805e-65i] ]
\end{verbatim}
}
This can be confirmed by passing the output of the terminal to Octave, as shown below.
{\footnotesize
\begin{verbatim}
$ ./a.out | octave
a =

    5 +  9i    5 +  5i   -6 -  6i   -7 -  7i
    3 +  3i    6 + 10i   -5 -  5i   -6 -  6i
    2 +  2i    3 +  3i   -1 +  3i   -5 -  5i
    1 +  1i    2 +  2i   -3 -  3i    0 +  4i

lambda =

   2 + 6i   4 + 8i   3 + 7i   1 + 5i

vr =

   0.37796 - 0.00000i   0.57735 + 0.00000i   0.57735 + 0.00000i   0.75593 + 0.00000i
   0.75593 + 0.00000i   0.57735 + 0.00000i   0.57735 + 0.00000i   0.37796 + 0.00000i
   0.37796 - 0.00000i   0.57735 + 0.00000i  -0.00000 - 0.00000i   0.37796 - 0.00000i
   0.37796 - 0.00000i  -0.00000 - 0.00000i   0.57735 + 0.00000i   0.37796 + 0.00000i

\end{verbatim}
}
{\tt lambda} is a list of eigenvalues, and {\tt vr} is a list of the corresponding eigenvectors. Aside from {\tt vr} is normalized
to unity, you will see you obtained correct eigenvalues and eigenvectors.

\subsubsection{Singular value decomposition of a real matrix ({\tt Rgesvd})}

We show an example of how to solve a singular value problem~\cite{wikipedia_svd}.
Let $A$ is $4 \times 5$ as follows:
\[
A=\left[\begin{array}{lllll}1 & 0 & 0 & 0 & 2 \\ 0 & 0 & 3 & 0 & 0 \\ 0 & 0 & 0 & 0 & 0 \\ 0 & 2 & 0 & 0 & 0\end{array}\right]
\]
Then, a singular value decomposition of this matrix is given by $A= U\Sigma V^t$ as follows.
\[
{U}=\left[\begin{array}{cccc}0 & -1 & 0 & 0 \\ -1 & 0 & 0 & 0 \\ 0 & 0 & 0 & -1 \\ 0 & 0 & -1 & 0\end{array}\right],
\]
\[
\Sigma=\left[\begin{array}{ccccc}3 & 0 & 0 & 0 & 0 \\ 0 & \sqrt{5} & 0 & 0 & 0 \\ 0 & 0 & 2 & 0 & 0 \\ 0 & 0 & 0 & 0 & 0\end{array}\right],
\]
and
\[
V^t=\left[\begin{array}{ccccc}0 & 0 & -1 & 0 & 0 \\ -\sqrt{0.2} & 0 & 0 & 0 & -\sqrt{0.8} \\ 0 & -1 & 0 & 0 & 0 \\ 0 & 0 & 0 & 1 & 0 \\ -\sqrt{0.8} & 0 & 0 & 0 & \sqrt{0.2}\end{array}\right].
\]
Let us solve the singular value problem for matrix $A$ using MPLAPACK $\tt dd\_real$.
The prototype definition of {\tt Rgesvd} is following:
\begin{verbatim}
void Rgesvd(const char *jobu, const char *jobvt, mplapackint const m, 
mplapackint const n, dd_real *a, mplapackint const lda, dd_real *s, dd_real *u, 
mplapackint const ldu, dd_real *vt, mplapackint const ldvt, dd_real *work, 
mplapackint const lwork, mplapackint &info);
\end{verbatim}
The corresponding LAPACK routine is {\tt DGESVD}. We show some lines of {\tt DGESVD} are following. 
\begin{verbatim}
       SUBROUTINE dgesvd( JOBU, JOBVT, M, N, A, LDA, S, U, LDU,
      $                   VT, LDVT, WORK, LWORK, INFO )
 *
 *  -- LAPACK driver routine --
 *  -- LAPACK is a software package provided by Univ. of Tennessee,    --
 *  -- Univ. of California Berkeley, Univ. of Colorado Denver and NAG Ltd..--
 *
 *     .. Scalar Arguments ..
       CHARACTER          JOBU, JOBVT
       INTEGER            INFO, LDA, LDU, LDVT, LWORK, M, N
 *     ..
 *     .. Array Arguments ..
       DOUBLE PRECISION   A( LDA, * ), S( * ), U( LDU, * ),
      $                   vt( ldvt, * ), work( * )
 *     ..
\end{verbatim}
A sample program for {\tt dd\_real} is following:
\begin{lstlisting}
//public domain
#include <iostream>
#include <string>
#include <sstream>
#include <cstring>
#include <algorithm>

#include <mpblas_dd.h>
#include <mplapack_dd.h>

#define DD_PRECISION_SHORT 16

inline void printnum(dd_real rtmp) {
    std::cout.precision(DD_PRECISION_SHORT);
    if (rtmp >= 0.0) {
        std::cout << "+" << rtmp;
    } else {
        std::cout << rtmp;
    }
    return;
}

//Matlab/Octave format
void printvec(dd_real *a, int len) {
    dd_real tmp;
    printf("[ ");
    for (int i = 0; i < len; i++) {
        tmp = a[i];
        printnum(tmp);
        if (i < len - 1)
            printf(", ");
    }
    printf("]");
}
void printmat(int n, int m, dd_real * a, int lda)
{
    dd_real mtmp;
    printf("[ ");
    for (int i = 0; i < n; i++) {
        printf("[ ");
        for (int j = 0; j < m; j++) {
            mtmp = a[i + j * lda];
            printnum(mtmp);
            if (j < m - 1)
                printf(", ");
        }
        if (i < n - 1)
            printf("]; ");
        else
            printf("] ");
    }
    printf("]");
}
int main() {
    mplapackint n = 5;
    mplapackint m = 4;

    dd_real *a = new dd_real[m * n];
    dd_real *s = new dd_real[std::min(m, n)];
    dd_real *u = new dd_real[m * m];
    dd_real *vt = new dd_real[n * n];
    mplapackint lwork = std::max({(mplapackint)1, 3 * std::min(m, n) + std::max(m, n), 5 * std::min(m, n)});
    dd_real *work = new dd_real[lwork];
    mplapackint info;

    // setting A matrix
    a[0 + 0 * m] = 1.0; a[0 + 1 * m] = 0.0; a[0 + 2 * m] = 0.0;  a[0 + 3 * m] = 0.0;  a[0 + 4 * m] = 2.0;
    a[1 + 0 * m] = 0.0; a[1 + 1 * m] = 0.0; a[1 + 2 * m] = 3.0;  a[1 + 3 * m] = 0.0;  a[1 + 4 * m] = 0.0;
    a[2 + 0 * m] = 0.0; a[2 + 1 * m] = 0.0; a[2 + 2 * m] = 0.0;  a[2 + 3 * m] = 0.0;  a[2 + 4 * m] = 0.0;
    a[3 + 0 * m] = 0.0; a[3 + 1 * m] = 2.0; a[3 + 2 * m] = 0.0;  a[3 + 3 * m] = 0.0;  a[3 + 4 * m] = 0.0;

    printf("# octave check\n");
    printf("a ="); printmat(m, n, a, m); printf("\n");
    Rgesvd("A", "A", m, n, a, m, s, u, m, vt, n, work, lwork, info);
    printf("s="); printvec(s, std::min(m, n)); printf("\n");
    if (m < n)
        printf("padding=zeros(%d, %d-%d)\n", (int)m, (int)n, (int)m);
    if (n < m)
        printf("padding=zeros(%d-%d,%d)\n", (int)m, (int)n, (int)n);
    printf("u ="); printmat(m, m, u, m); printf("\n");
    printf("vt ="); printmat(n, n, vt, n); printf("\n");
    printf("svd(a)\n");
    if (m < n)
        printf("sigma=[diag(s) padding] \n");
    if (n < m)
        printf("sigma=[diag(s); padding] \n");
    if (n == m)
        printf("sigma=[diag(s)] \n");
    printf("sigma \n");
    printf("u * sigma  * vt\n");
    delete[] work;
    delete[] vt;
    delete[] u;
    delete[] s;
    delete[] a;
}
\end{lstlisting}
You can find corresponding example files for various mutiple-precision verions as {\tt Rgesvd\_test\_\_Float128.cpp},
{\tt Rgesvd\_test\_dd.cpp}, {\tt Rgesvd\_test\_gmp.cpp}, {\tt Rgesvd\_test\_qd.cpp},
{\tt Rgesvd\_test\_\_Float64x.cpp}, \\
{\tt Rgesvd\_test\_double.cpp},  and {\tt Rgesvd\_test\_mpfr.cpp}\\
at {\tt /home/docker/mplapack/examples/mplapack/05\_SingularValueDecomposition/}.
In {\tt dd\_real} case, you can compile {\tt dd\_real} version by:
\begin{verbatim}
$ g++ -O2 -I/home/docker/MPLAPACK/include -I/home/docker/MPLAPACK/include/mplapack \
Rgesvd_test_dd.cpp -Wl,--rpath=/home/docker/MPLAPACK/lib -L/home/docker/MPLAPACK/lib \
-lmplapack_dd -lmpblas_dd -lqd
\end{verbatim}
The output of the executable is the following:
{\footnotesize
\begin{verbatim}
$ ./a.out
# octave check
a =[ [ +1.0000000000000000e+00, +0.0000000000000000e+00, +0.0000000000000000e+00, +0.0000000000000000e+00, 
+2.0000000000000000e+00]; [ +0.0000000000000000e+00, +0.0000000000000000e+00, +3.0000000000000000e+00,
+0.0000000000000000e+00, +0.0000000000000000e+00]; [ +0.0000000000000000e+00, +0.0000000000000000e+00, 
+0.0000000000000000e+00, +0.0000000000000000e+00, +0.0000000000000000e+00]; [ +0.0000000000000000e+00, 
+2.0000000000000000e+00, +0.0000000000000000e+00, +0.0000000000000000e+00, +0.0000000000000000e+00] ]
s=[ +3.0000000000000000e+00, +2.2360679774997897e+00, +2.0000000000000000e+00, +0.0000000000000000e+00]
padding=zeros(4, 5-4)
u =[ [ +0.0000000000000000e+00, +1.0000000000000000e+00, +0.0000000000000000e+00, +0.0000000000000000e+00];
[ +1.0000000000000000e+00, +0.0000000000000000e+00, +0.0000000000000000e+00, +0.0000000000000000e+00];
[ +0.0000000000000000e+00, +0.0000000000000000e+00, +0.0000000000000000e+00, -1.0000000000000000e+00];
[ +0.0000000000000000e+00, +0.0000000000000000e+00, +1.0000000000000000e+00, +0.0000000000000000e+00] ]
vt =[ [ +0.0000000000000000e+00, +0.0000000000000000e+00, +1.0000000000000000e+00, +0.0000000000000000e+00, 
+0.0000000000000000e+00]; [ +4.4721359549995794e-01, +0.0000000000000000e+00, +0.0000000000000000e+00, 
+0.0000000000000000e+00, +8.9442719099991588e-01]; [ +0.0000000000000000e+00, +1.0000000000000000e+00, 
+0.0000000000000000e+00, +0.0000000000000000e+00, +0.0000000000000000e+00]; [ +0.0000000000000000e+00, 
+0.0000000000000000e+00, +0.0000000000000000e+00, +1.0000000000000000e+00, +0.0000000000000000e+00];
[ -8.9442719099991588e-01, 
+0.0000000000000000e+00, +0.0000000000000000e+00, +0.0000000000000000e+00, +4.4721359549995794e-01] ]
svd(a)
sigma=[diag(s) padding]
sigma
u * sigma  * vt
\end{verbatim}
}
To better readability, you can pass the output to the octave. The output is the following:
{\footnotesize
\begin{verbatim}
$ ./a.out | octave
octave: X11 DISPLAY environment variable not set
octave: disabling GUI features
a =

   1   0   0   0   2
   0   0   3   0   0
   0   0   0   0   0
   0   2   0   0   0

s =

   3.00000   2.23607   2.00000   0.00000

padding =

   0
   0
   0
   0

u =

   0   1   0   0
   1   0   0   0
   0   0   0  -1
   0   0   1   0

vt =

   0.00000   0.00000   1.00000   0.00000   0.00000
   0.44721   0.00000   0.00000   0.00000   0.89443
   0.00000   1.00000   0.00000   0.00000   0.00000
   0.00000   0.00000   0.00000   1.00000   0.00000
  -0.89443   0.00000   0.00000   0.00000   0.44721

ans =

   3.00000
   2.23607
   2.00000
   0.00000

sigma =

   3.00000   0.00000   0.00000   0.00000   0.00000
   0.00000   2.23607   0.00000   0.00000   0.00000
   0.00000   0.00000   2.00000   0.00000   0.00000
   0.00000   0.00000   0.00000   0.00000   0.00000

sigma =

   3.00000   0.00000   0.00000   0.00000   0.00000
   0.00000   2.23607   0.00000   0.00000   0.00000
   0.00000   0.00000   2.00000   0.00000   0.00000
   0.00000   0.00000   0.00000   0.00000   0.00000

ans =

   1   0   0   0   2
   0   0   3   0   0
   0   0   0   0   0
   0   2   0   0   0
\end{verbatim}
}
You will see that {\tt Rgesvd}, the singular value decomposition solver solved the problem correctly.

\subsubsection{Singular value decomposition of a complex matrix ({\tt Cgesvd})}

We show an example of how to solve a singular value problem.
Let $A$ is $4 \times 4$ as follows:
\[
A=\left[\begin{array}{cccc} 
0.9-i    & 20-2.25i  & 1.75-0.5i  & 0.5i     \\ 
8 -2.25i & -0.25     & 1.25-0.25i & -3.75    \\
-1.75    & -80+1.25i & 1.5        & 30+2.25i \\ 
3+0.25i  & 1.75      & 2.25i      & -0.25-80.0i \\
\end{array}\right]
\]
Then, a singular value decomposition of this matrix is given by $A= U\Sigma V^t$ as follows.
\[
{U}=\left[\begin{array}{cccc}-0.091031 + 0.104196i &  0.060624 + 0.210082i & -0.099978 + 0.156181i &  0.873552 - 0.368191i \\
  -0.019857 + 0.021261i & -0.009166 - 0.038982i & -0.947158 + 0.255136i & -0.176226 - 0.065515i \\
   0.572129 - 0.526799i & -0.102805 - 0.570697i & -0.027491 + 0.035503i &  0.228321 - 0.068858i  \\
  -0.472776 - 0.389416i &  0.745095 - 0.243220i &  0.001372 + 0.037303i  & 0.034920 + 0.088893i \\
\end{array}\right],
\]
\[
\Sigma=\left[\begin{array}{cccc}97.5165 & 0 & 0 & 0\\ 0 & 68.1063 & 0 & 0 \\ 0 & 0 & 8.4782 & 0 \\ 0 & 0 & 0 & 2.1504 \end{array}\right],
\]
and
\[
V^t=\left[\begin{array}{cccc}
-0.02984 + 0.00000i & -0.50562 - 0.43706i & -0.00266 - 0.00443i &   0.48583 + 0.56247i \\ 
0.03250 + 0.00000i  & 0.14032 - 0.72984i & -0.01031 + 0.03209i &  0.22087 - 0.62982i \\ 
-0.98322 + 0.00000i &  0.01556 - 0.01112i & -0.17198 - 0.04195i & -0.01174 - 0.03782i \\ 
-0.17699 + 0.00000i  & 0.02458 + 0.00145i &  0.95397 + 0.23966i &  0.02386 - 0.00037i \\ 
\end{array}\right].
\]
Let us solve the singular value problem for matrix $A$ using MPLAPACK $\tt dd\_real$.
The prototype definition of {\tt Cgesvd} is the following:
\begin{verbatim}
void Cgesvd(const char *jobu, const char *jobvt, mplapackint const m, 
mplapackint const n, dd_complex *a, mplapackint const lda, dd_real *s, 
dd_complex *u, mplapackint const ldu, dd_complex *vt, mplapackint const ldvt, 
dd_complex *work, mplapackint const lwork, dd_real *rwork, mplapackint &info);
\end{verbatim}
The corresponding LAPACK routine is {\tt ZGESVD}. We show some lines of {\tt ZGESVD} are following. 
\begin{verbatim}
*       SUBROUTINE ZGESVD( JOBU, JOBVT, M, N, A, LDA, S, U, LDU, VT, LDVT,
*                          WORK, LWORK, RWORK, INFO )
*
*       .. Scalar Arguments ..
*       CHARACTER          JOBU, JOBVT
*       INTEGER            INFO, LDA, LDU, LDVT, LWORK, M, N
*       ..
*       .. Array Arguments ..
*       DOUBLE PRECISION   RWORK( * ), S( * )
*       COMPLEX*16         A( LDA, * ), U( LDU, * ), VT( LDVT, * ),
*      $                   WORK( * )

\end{verbatim}
A sample program for {\tt dd\_real} is following:
\begin{lstlisting}
//public domain
#include <iostream>
#include <string>
#include <sstream>
#include <cstring>
#include <algorithm>

#include <mpblas_dd.h>
#include <mplapack_dd.h>

#define DD_PRECISION_SHORT 16

inline void printnum(dd_real rtmp) {
    std::cout.precision(DD_PRECISION_SHORT);
    if (rtmp >= 0.0) {
        std::cout << "+" << rtmp;
    } else {
        std::cout << rtmp;
    }
    return;
}

inline void printnum(dd_complex rtmp) {
    std::cout.precision(DD_PRECISION_SHORT);
    if (rtmp.real() >= 0.0) {
        std::cout << "+" << rtmp.real();
    } else {
        std::cout << rtmp.real();
    }
    if (rtmp.imag() >= 0.0) {
        std::cout << "+" << rtmp.imag() << "i";
    } else {
        std::cout << rtmp.imag() << "i";
    }
    return;
}

//Matlab/Octave format
template <class X> void printvec(X *a, int len) {
    X tmp;
    printf("[ ");
    for (int i = 0; i < len; i++) {
        tmp = a[i];
        printnum(tmp);
        if (i < len - 1)
            printf(", ");
    }
    printf("]");
}

template <class X> void printmat(int n, int m, X *a, int lda)
{
    X mtmp;

    printf("[ ");
    for (int i = 0; i < n; i++) {
        printf("[ ");
        for (int j = 0; j < m; j++) {
            mtmp = a[i + j * lda];
            printnum(mtmp);
            if (j < m - 1)
                printf(", ");
        }
        if (i < n - 1)
            printf("]; ");
        else
            printf("] ");
    }
    printf("]");
}
int main() {
    mplapackint n = 4;
    mplapackint m = 4;

    dd_complex *a = new dd_complex[m * n];
    dd_real *s = new dd_real[std::min(m, n)];
    dd_complex *u = new dd_complex[m * m];
    dd_complex *vt = new dd_complex[n * n];
    mplapackint lwork = std::max((mplapackint)1, 2 * std::min(m, n) + std::max(m, n));
    dd_complex *work = new dd_complex[lwork];
    dd_real *rwork = new dd_real[5 * std::min(m, n)];
    mplapackint info;

    // setting A matrix
    a[0 + 0 * n] = dd_complex(0.9, -1.0); a[0 + 1 * n] = dd_complex(20.0, -2.25);  a[0 + 2 * n] = dd_complex(1.75, -0.5);  a[0 + 3 * n] = dd_complex(0.0, 0.5);
    a[1 + 0 * n] = dd_complex(8.0,-2.25); a[1 + 1 * n] = dd_complex(-0.25, 0.0);   a[1 + 2 * n] = dd_complex(1.25, -0.25); a[1 + 3 * n] = dd_complex(-3.75, 0.0);
    a[2 + 0 * n] = dd_complex(-1.75,0.0); a[2 + 1 * n] = dd_complex(-80.0,  1.25); a[2 + 2 * n] = dd_complex(1.5, 0.0);    a[2 + 3 * n] = dd_complex(30.0, 2.25);
    a[3 + 0 * n] = dd_complex(3.0, 0.25); a[3 + 1 * n] = dd_complex(1.75, 0.0);    a[3 + 2 * n] = dd_complex(0.0, 2.25);   a[3 + 3 * n] = dd_complex(-0.25, -80.0);

    printf("# octave check\n");
    printf("split_long_rows(0)\n");
    printf("a ="); printmat(m, n, a, m); printf("\n");
    Cgesvd("A", "A", m, n, a, m, s, u, m, vt, n, work, lwork, rwork, info);
    printf("s="); printvec(s, std::min(m, n)); printf("\n");
    if (m < n)
        printf("padding=zeros(%d, %d-%d)\n", (int)m, (int)n, (int)m);
    if (n < m)
        printf("padding=zeros(%d-%d,%d)\n", (int)m, (int)n, (int)n);
    printf("u ="); printmat(m, m, u, m); printf("\n");
    printf("vt ="); printmat(n, n, vt, n); printf("\n");
    printf("svd(a)\n");
    if (m < n)
        printf("sigma=[diag(s) padding] \n");
    if (n < m)
        printf("sigma=[diag(s); padding] \n");
    if (n == m)
        printf("sigma=[diag(s)] \n");
    printf("sigma \n");
    printf("u * sigma  * vt\n");
    delete[] rwork;
    delete[] work;
    delete[] vt;
    delete[] u;
    delete[] s;
    delete[] a;
}
\end{lstlisting}
You can find corresponding example files for various mutiple-precision verions as {\tt Cgesvd\_test\_\_Float128.cpp},
{\tt Cgesvd\_test\_dd.cpp}, {\tt Cgesvd\_test\_gmp.cpp}, {\tt Cgesvd\_test\_qd.cpp},
{\tt Cgesvd\_test\_\_Float64x.cpp}, \\
{\tt Cgesvd\_test\_double.cpp},  and {\tt Cgesvd\_test\_mpfr.cpp}\\
at {\tt /home/docker/mplapack/examples/mplapack/05\_SingularValueDecomposition/}.
In {\tt dd\_real} case, you can compile {\tt dd\_real} version by:
\begin{verbatim}
$ g++ -O2 -I/home/docker/MPLAPACK/include -I/home/docker/MPLAPACK/include/mplapack \
Cgesvd_test_dd.cpp -Wl,--rpath=/home/docker/MPLAPACK/lib -L/home/docker/MPLAPACK/lib \
-lmplapack_dd -lmpblas_dd -lqd
\end{verbatim}
The output of the executable is the following:
{\fontsize{7pt}{0.4cm}\selectfont
\begin{verbatim}
$ ./a.out
# octave check
split_long_rows(0)
a =[ [ +9.0000000000000002220446049250313080847263336181640625000000000000e-01-1.0000000000000000000000000000000000000000000000000000000000000000e+00i,
+2.0000000000000000000000000000000000000000000000000000000000000000e+01-2.2500000000000000000000000000000000000000000000000000000000000000e+00i,
+1.7500000000000000000000000000000000000000000000000000000000000000e+00-5.0000000000000000000000000000000000000000000000000000000000000000e-01i,
+0.0000000000000000000000000000000000000000000000000000000000000000e+00+5.0000000000000000000000000000000000000000000000000000000000000000e-01i];
[ +8.0000000000000000000000000000000000000000000000000000000000000000e+00-2.2500000000000000000000000000000000000000000000000000000000000000e+00i,
-2.5000000000000000000000000000000000000000000000000000000000000000e-01+0.0000000000000000000000000000000000000000000000000000000000000000e+00i,
+1.2500000000000000000000000000000000000000000000000000000000000000e+00-2.5000000000000000000000000000000000000000000000000000000000000000e-01i,
-3.7500000000000000000000000000000000000000000000000000000000000000e+00+0.0000000000000000000000000000000000000000000000000000000000000000e+00i];
[ -1.7500000000000000000000000000000000000000000000000000000000000000e+00+0.0000000000000000000000000000000000000000000000000000000000000000e+00i,
-8.0000000000000000000000000000000000000000000000000000000000000000e+01+1.2500000000000000000000000000000000000000000000000000000000000000e+00i,
+1.5000000000000000000000000000000000000000000000000000000000000000e+00+0.0000000000000000000000000000000000000000000000000000000000000000e+00i,
+3.0000000000000000000000000000000000000000000000000000000000000000e+01+2.2500000000000000000000000000000000000000000000000000000000000000e+00i];
[ +3.0000000000000000000000000000000000000000000000000000000000000000e+00+2.5000000000000000000000000000000000000000000000000000000000000000e-01i,
+1.7500000000000000000000000000000000000000000000000000000000000000e+00+0.0000000000000000000000000000000000000000000000000000000000000000e+00i,
+0.0000000000000000000000000000000000000000000000000000000000000000e+00+2.2500000000000000000000000000000000000000000000000000000000000000e+00i, 
-2.5000000000000000000000000000000000000000000000000000000000000000e-01-8.0000000000000000000000000000000000000000000000000000000000000000e+01i] ]
s=[ +9.7516494435839724342537934011744964189236290378865784643347960368e+01, +6.8106277354635765125630927080704091707844321939430382469272564138e+01,
+8.4781583569581929193257165114938017510166457266377499488969146190e+00, +2.1503787370153202482459981470702526329803293729092259494247941795e+00]
u =[ [ -9.1031344230065289625000603692895258254109199563673856613631987783e-02+1.0419550007707547081892003210016871725842634797582791410249228936e-01i,
+6.0624015933791684880101970656199134272860397091935547317571739868e-02+2.1008219841477778090011021923228197915008235604381081943292731807e-01i,
-9.9978469925849530955490215397374086462581292970446108678666366672e-02+1.5618105933067197645819512610919051343301177207836333131951543727e-01i,
+8.7355245250617254255021322838492243550830273310262636819188476464e-01-3.6819109944526785319692942345862467313117969007310991253602368805e-01i];
[ -1.9857057918844668508194923848264672674827541149575776877467762232e-02+2.1261406272041457850596651639122912053764705903056249828047829994e-02i,
-9.1661574892754928734045271369121793400923378513876808480540487939e-03-3.8982298938279813017463111582481151532286050557011081307982359715e-02i,
-9.4715781555964404355021035921467887158437034362335955910113579662e-01+2.5513564301397109620885117891155720411045767100304996091477259914e-01i,
-1.7622605699795972492191618082958299743466785925908687025400420878e-01-6.5515377611625650642778502070271900697881487135131290739710052841e-02i];
[+5.7212865453775911876209433088888378678015704682520653614966501059e-01-5.2679881119508889212570067087606726528213619071689150491978561913e-01i,
-1.0280453437606294917501451825749672199318907430352403992267112353e-01-5.7069659877524894683413175387466614439142123832093715941506060498e-01i,
-2.7491348747212516867749761285866090287970541616576302889303235643e-02+3.5503263645948459662712416388082136298992694820774724140988646759e-02i,
+2.2832147261394758144289270273679539897363368839194944166497880608e-01-6.8858435912252448915976238595384806375580376776718454679051942621e-02i];
[ -4.7277573051874758750884260023823098397257693984940907812279232275e-01-3.8941644263977541772805124956744968110703693808087656417153595879e-01i,
+7.4509533337138732928486540830547081945080989089341576817124012507e-01-2.4322032105460738462953739377692881294892930945862711758541460737e-01i,
+1.3722197565686791725123297720384344172881445692984491516542700100e-03+3.7303074581742350398412376125544143984966926055469528482835254700e-02i,
+3.4919914776892904123087051825643190789294671627841834161144819628e-02+8.8892967326921520048953580964935184955357406567111076952094885644e-02i] ]
vt =[ [ -2.9838283275437904677981078252061374697499430617316993098822847353e-02+0.0000000000000000000000000000000000000000000000000000000000000000e+00i,
-5.0561959990428189803987557498007050028508558598803909581266186879e-01-4.3706492603990562617863033812586628591798259998022451509139092210e-01i,
-2.6614296590750835969117035558529485216857552931590475544750949781e-03-4.4298774864078094877565754305104919310123158810553060779232926373e-03i,
+4.8583199843346362348703182426050056752793902330663066922401758006e-01+5.6247071901149899446413169974532320544220661587531798620308731958e-01i];
[ +3.2496992488762597907964387515542319276603362918117020319359346521e-02+0.0000000000000000000000000000000000000000000000000000000000000000e+00i,
+1.4032474124032111114640285436525182953802654632942948451711455109e-01-7.2983568501622474240687954437213659452798801125293414002260123392e-01i,
-1.0309089626661974290949324966638353418737248637627508120396185556e-02+3.2090624256563363535957212538268213661898668484938870563587187365e-02i,
+2.2086896019824552483954657586346671754727421017002976965842749182e-01-6.2982029363839933799110165363797820437639004016585224523835087573e-01i];
[ -9.8322380454928627593357750706161049590724043016230250792627017777e-01+0.0000000000000000000000000000000000000000000000000000000000000000e+00i,
+1.5557746179521022745472470501153401319301822933632620168302167001e-02-1.1119064935323447671186965031169351281860527363722541382850349940e-02i,
-1.7198182390092257164272666175190298478482096240869947698066647614e-01-4.1946034125215249890874708936693176256949905794670062763938078790e-02i,
-1.1737785317001863645752020621778756342758817320563642174518560436e-02-3.7819042308536459610100091608526996733397032526690882424269244982e-02i]; 
[ -1.7698749249025177994139125324872686755256471092406618305781220382e-01+0.0000000000000000000000000000000000000000000000000000000000000000e+00i,
+2.4579174853446943521565353394166564952277757025400163843805728908e-02+1.4483036290672481703529854195308447371511910768799389582350077765e-03i,
+9.5397148644507469609163061186547701562956565817170822244991006671e-01+2.3966308226794491229599443416109686062123262637446982923420847550e-01i,
+2.3855098594784970246426131859255500566423607296986688880428268896e-02-3.7202254072076200340788215932734883131231597648245161542298777760e-04i] ]
svd(a)
sigma=[diag(s)]
sigma
u * sigma  * vt
\end{verbatim}
}
To better readability, you can pass the output to the octave. The output is the following:
{\footnotesize
\begin{verbatim}
$ ./a.out | octave
octave: X11 DISPLAY environment variable not set
octave: disabling GUI features
a =

    0.90000 -  1.00000i   20.00000 -  2.25000i    1.75000 -  0.50000i    0.00000 +  0.50000i
    8.00000 -  2.25000i   -0.25000 +  0.00000i    1.25000 -  0.25000i   -3.75000 +  0.00000i
   -1.75000 +  0.00000i  -80.00000 +  1.25000i    1.50000 +  0.00000i   30.00000 +  2.25000i
    3.00000 +  0.25000i    1.75000 +  0.00000i    0.00000 +  2.25000i   -0.25000 - 80.00000i

s =

   97.5165   68.1063    8.4782    2.1504

u =

  -0.091031 + 0.104196i   0.060624 + 0.210082i  -0.099978 + 0.156181i   0.873552 - 0.368191i
  -0.019857 + 0.021261i  -0.009166 - 0.038982i  -0.947158 + 0.255136i  -0.176226 - 0.065515i
   0.572129 - 0.526799i  -0.102805 - 0.570697i  -0.027491 + 0.035503i   0.228321 - 0.068858i
  -0.472776 - 0.389416i   0.745095 - 0.243220i   0.001372 + 0.037303i   0.034920 + 0.088893i

vt =

  -0.02984 + 0.00000i  -0.50562 - 0.43706i  -0.00266 - 0.00443i   0.48583 + 0.56247i
   0.03250 + 0.00000i   0.14032 - 0.72984i  -0.01031 + 0.03209i   0.22087 - 0.62982i
  -0.98322 + 0.00000i   0.01556 - 0.01112i  -0.17198 - 0.04195i  -0.01174 - 0.03782i
  -0.17699 + 0.00000i   0.02458 + 0.00145i   0.95397 + 0.23966i   0.02386 - 0.00037i

ans =

   97.5165
   68.1063
    8.4782
    2.1504

sigma =

   97.51649    0.00000    0.00000    0.00000
    0.00000   68.10628    0.00000    0.00000
    0.00000    0.00000    8.47816    0.00000
    0.00000    0.00000    0.00000    2.15038

sigma =

   97.51649    0.00000    0.00000    0.00000
    0.00000   68.10628    0.00000    0.00000
    0.00000    0.00000    8.47816    0.00000
    0.00000    0.00000    0.00000    2.15038

ans =

   9.0000e-01 - 1.0000e+00i   2.0000e+01 - 2.2500e+00i   1.7500e+00 - 5.0000e-01i  -3.0693e-16 + 5.0000e-01i
   8.0000e+00 - 2.2500e+00i  -2.5000e-01 - 8.9405e-17i   1.2500e+00 - 2.5000e-01i  -3.7500e+00 + 2.1264e-16i
  -1.7500e+00 + 2.1460e-17i  -8.0000e+01 + 1.2500e+00i   1.5000e+00 + 2.9835e-17i   3.0000e+01 + 2.2500e+00i
   3.0000e+00 + 2.5000e-01i   1.7500e+00 - 2.3781e-15i  -2.0181e-17 + 2.2500e+00i  -2.5000e-01 - 8.0000e+01i

\end{verbatim}
}
You will see that {\tt Cgesvd}, the singular value decomposition solver solved the problem correctly.

\subsection{Examples provided for MPLAPACK}
There are more examples in {\tt /home/docker/mplapack/examples/mplapack/} as follows:
\begin{itemize}
\item {\tt 00\_LinearEquations/Cgetri\_test\_*.cpp} ... samples of inversion of complex matrices
\item {\tt 00\_LinearEquations/Rgesv\_test\_*.cpp} ... samples of solving linear equation of real matrices
\item {\tt 00\_LinearEquations/Rgetri\_Hilbert\_*.cpp} ... samples of inversion of Hilbert matrices of various orders.
\item {\tt 00\_LinearEquations/Rgetri\_test*.cpp} ... samples of inversion of real matrices.
\item {\tt 03\_SymmetricEigenproblems/Cheev\_test\_*.cpp} ... samples of diagonalization of Hermitian matrices.
\item {\tt 03\_SymmetricEigenproblems/Rsyev\_Frank\_*.cpp} ... samples of diagonalization of real Frank matrices.
\item {\tt 03\_SymmetricEigenproblems/Rsyev\_test\_*.cpp} ... samples of diagonalization of real matrices.
\item {\tt 03\_SymmetricEigenproblems/Rsyevd\_DingDong\_*.cpp} ... samples of diagonalization of Ding Dong matrices.
\item {\tt 03\_SymmetricEigenproblems/Rsyevd\_Frank\_*.cpp} ... samples of diagonalization of real Frank matrices using divide-and-conquer driver.
\item {\tt 03\_SymmetricEigenproblems/Rsyevr\_Frank\_*.cpp}  ... samples of diagonalization of real Frank matrices using relatively robust representation driver.
\item {\tt 04\_NonsymmetricEigenproblems/Cgees\_test*\_*.cpp} ... samples of solving non-symmetric complex eigenvalues problem (Schur form)

\item {\tt 04\_NonsymmetricEigenproblems/Cgeev\_NPR\_*.cpp} ... samples of solving non-symmetric complex eigenvalues problem with eigenvectors; GMP version is not available because no high precision cosine is in GMP.

\item {\tt 04\_NonsymmetricEigenproblems/Cgeev\_test*\_*.cpp} ... samples of solving non-symmetric complex eigenvalues problem with eigenvectors.

\item {\tt 04\_NonsymmetricEigenproblems/Rgees\_Grcar\_*.cpp} ... samples of solving non-symmetric eigenvalues problem using Grcar matrix (Schur form)
\item {\tt 04\_NonsymmetricEigenproblems/Rgees\_readfromfile\_*.cpp} ... samples of solving non-symmetric eigenvalues problem using matrix from a file (Schur form)
\item {\tt 04\_NonsymmetricEigenproblems/Rgees\_test\_*.cpp} ... samples of solving simple non-symmetric eigenvalues problem (Schur form)
\item {\tt 04\_NonsymmetricEigenproblems/Rgeev\_Frank\_*.cpp} ... samples of solving non-symmetric eigenvalues problem for Frank matrix
\item {\tt 04\_NonsymmetricEigenproblems/Rgeev\_random\_highcond\_*.cpp}  ... samples of solving non-symmetric eigenvalues problem for high conditioned random matrix
\item {\tt 04\_NonsymmetricEigenproblems/Rgeev\_readfromfile\_*.cpp} ... samples of solving non-symmetric eigenvalues problem using matrix from a file
\item {\tt 04\_NonsymmetricEigenproblems/Rgeev\_test\_*.cpp} .. samples of solving simple non-symmetric eigenvalues problem
\item {\tt 05\_SingularValueDecomposition/Cgesvd\_test\_*.cpp} ... samples of simple singular value decomposition problems
\item {\tt 05\_SingularValueDecomposition/Rgesvd\_random\_highcond\_*.cpp} ...samples of solving singular value decomposition problem for matrices with high condition numbers
\item {\tt 05\_SingularValueDecomposition/Rgesvd\_readfromfile\_*.cpp} ... samples of singular value decomposition problems using matrix from a file
\item {\tt 05\_SingularValueDecomposition/Rgesvd\_test\_*.cpp} ... samples of simple singular value decomposition problems
\end{itemize}

\subsection{Accuracy of each precision, Rlamch values, and specifying precision at runtime}
For MPLAPACK, {\tt Rlamch} routines are vital since they return precision-specific constants. These values control convergence behaviors such as eigenvalue and singular value problems. Thus, we cannot say the binary128 version of MPLAPACK routines runs output if we use the same {\tt Rlamch} values for binary64 and vice versa.
In table \ref{rlamchmpfr}, we show {\tt Rlamch} values for {\tt MPFR}. The default accuracy for {\tt MPFR} is approximately 154 decimal digits ($154.127 =512 \log_{10}2$).
In table \ref{rlamchmpfr_65536}, we show {\tt Rlamch} values for {\tt MPFR} with {\tt MPLAPACK\_MPFR\_PRECISION=65536}. This accuracy is approximately 19728 decimal digits ($19728 = 65536 \log_{10}2)$.

In table \ref{rlamchgmp}, we show {\tt Rlamch} values for {\tt GMP}. The default accuracy for {\tt MPFR} is approximately 154 decimal digits ($154.127 =512 \log_{10}2$).
In table \ref{rlamchgmp1024}, we show Rlamch values with for {\tt GMP} with {\tt MPLAPACK\_GMP\_PRECISION=1024}. This accuracy is approximately 308 decimal digits ($308.25 = 1024 \log_{10}2)$.
In table \ref{rlamch_float128}, we show Rlamch values for {\tt \_Float128}.
In table \ref{rlamch_float64x}, we show Rlamch values for {\tt \_Float64x}.
In table \ref{rlamch_double}, we show Rlamch values for {\tt double}.
In table \ref{rlamch_dd}, we show Rlamch values for {\tt dd\_real}. We intentionally reduce the overflow values
so that {\tt sqrt} does not fail.
In table \ref{rlamch_qd}, we show Rlamch values for {\tt qd\_real}, respectively.

We can change the default precision at runtime for {\tt GMP} and {\tt MPFR} by setting the environment variable,
{\tt MPLAPACK\_GMP\_PRECISION} and {\tt MPLAPACK\_MPFR\_PRECISION}, respectively.

Let us show an example by inverting the Hilerbt matrix.
{\tt Rgetri\_Hilbert\_mpfr} inverts Hilbert matrix of order $n$, where $n=1, 2, \cdots $ via LU decomposition. 
${\tt a}$ is the Hilbert matrix and {\tt ainv} is the inverted matrix, and {\tt InfnormL} is the left residuals. 

The output is the following:
\begin{verbatim}
$ cd /home/docker/mplapack/examples/mplapack/00_LinearEquations
$ g++ -O2 -I/home/docker/MPLAPACK/include -I/home/docker/MPLAPACK/include/mplapack \
Rgetri_Hilbert_mpfr.cpp -Wl,--rpath=/home/docker/MPLAPACK/lib \ 
-L/home/docker/MPLAPACK/lib -lmplapack_mpfr -lmpblas_mpfr -lmpfr -lmpc -lgmp
$ ./a.out | grep InfnromL 
InfnormL:(ainv * a - I)=+0.0000000000000000e+00
InfnormL:(ainv * a - I)=+0.0000000000000000e+00
InfnormL:(ainv * a - I)=+4.7733380679681323e-153
InfnormL:(ainv * a - I)=+7.6373409087490117e-152
InfnormL:(ainv * a - I)=+2.4439490907996837e-150
InfnormL:(ainv * a - I)=+1.1730955635838482e-148
InfnormL:(ainv * a - I)=+2.1897783853565166e-147
InfnormL:(ainv * a - I)=+6.0062492855493028e-146
InfnormL:(ainv * a - I)=+2.7228330094490173e-144
InfnormL:(ainv * a - I)=+8.2005323578699814e-143
...
$ MPLAPACK_MPFR_PRECISION=65536 ./a.out | grep InfnromL 
InfnormL:(ainv * a - I)=+0.0000000000000000e+00
InfnormL:(ainv * a - I)=+0.0000000000000000e+00
InfnormL:(ainv * a - I)=+3.9929525776415436e-19728
InfnormL:(ainv * a - I)=+5.1109792993811758e-19726
InfnormL:(ainv * a - I)=+3.2710267516039525e-19724
InfnormL:(ainv * a - I)=+5.2336428025663240e-19723
InfnormL:(ainv * a - I)=+8.3738284841061184e-19722
InfnormL:(ainv * a - I)=+5.0242970904636710e-19720
InfnormL:(ainv * a - I)=+1.0718500459655832e-19718
...
\end{verbatim}
In the first case, we calculate in 512 bit accuracy or 154 decimal digit accuracy ($154.127 =512 \log_{10}2$), and the next case we calculate in 65536 bit accuracy or 19728 decimal digits accuracy ($19728 = 65536 \log_{10}2$). The results show we can specify how many bits are for multiple-precision calculations at the runtime.

\begin{table}
\caption{Rlamch values for {\tt MPFR} (default setting) }\label{rlamchmpfr}
\begin{center}
\scriptsize
{
\begin{verbatim}
Rlamch E: Epsilon                      +7.4583407312002067432909653154629338373764715346004068942715183332e-155
Rlamch S: Safe minimum                 +9.5302596195518042928646791269306980816607845022369591518200950104e-323228497
Rlamch B: Base                         +2.0000000000000000000000000000000000000000000000000000000000000000e+00
Rlamch P: Precision                    +1.4916681462400413486581930630925867674752943069200813788543036666e-154
Rlamch N: Number of digits in mantissa +5.1200000000000000000000000000000000000000000000000000000000000000e+02
Rlamch R: Rounding mode                +1.0000000000000000000000000000000000000000000000000000000000000000e+00
Rlamch M: Minimum exponent:            -1.0737418230000000000000000000000000000000000000000000000000000000e+09
Rlamch U: Underflow threshold          +9.5302596195518042928646791269306980816607845022369591518200950104e-323228497
Rlamch L: Largest exponent             +1.0737418220000000000000000000000000000000000000000000000000000000e+09
Rlamch O: Overflow threshold           +2.0985787164673876924043581168838390706380979654733526277866462257e+323228496
Rlamch -: Reciprocal of safe minimum   +1.0492893582336938462021790584419195353190489827366763138933231129e+323228496
\end{verbatim}
}
\end{center}
\end{table}

\begin{table}
\caption{Rlamch values for {\tt MPFR} ({\tt MPLAPACK\_MPFR\_PRECISION=65536)} }\label{rlamchmpfr_65536}
\begin{center}
\scriptsize
{
\begin{verbatim}
Rlamch E: Epsilon                      +4.9911907220519294656590574792132451973746770423207674161425040336e-19729
Rlamch S: Safe minimum                 +9.5302596195518042928646791269306980816607845022369591518200950104e-323228497
Rlamch B: Base                         +2.0000000000000000000000000000000000000000000000000000000000000000e+00
Rlamch P: Precision                    +9.9823814441038589313181149584264903947493540846415348322850080672e-19729
Rlamch N: Number of digits in mantissa +6.5536000000000000000000000000000000000000000000000000000000000000e+04
Rlamch R: Rounding mode                +1.0000000000000000000000000000000000000000000000000000000000000000e+00
Rlamch M: Minimum exponent:            -1.0737418230000000000000000000000000000000000000000000000000000000e+09
Rlamch U: Underflow threshold          +9.5302596195518042928646791269306980816607845022369591518200950104e-323228497
Rlamch L: Largest exponent             +1.0737418220000000000000000000000000000000000000000000000000000000e+09
Rlamch O: Overflow threshold           +2.0985787164673876924043581168838390706380979654733526277866462257e+323228496
Rlamch -: Reciprocal of safe minimum   +1.0492893582336938462021790584419195353190489827366763138933231129e+323228496
\end{verbatim}
}
\end{center}
\end{table}

\begin{table}
\caption{Rlamch values for {\tt GMP} (default setting) on Linux and macOS}\label{rlamchgmp}
\begin{center}
\scriptsize
{
\begin{verbatim}
Rlamch E: Epsilon                      +7.4583407312002067432909653154629338373764715346004068942715183332e-155
Rlamch S: Safe minimum                 +1.7019382623481672278259575819240965611355119939659392498165297957e-1388255822130839283
Rlamch B: Base                         +2.0000000000000000000000000000000000000000000000000000000000000000e+00
Rlamch P: Precision                    +1.4916681462400413486581930630925867674752943069200813788543036666e-154
Rlamch N: Number of digits in mantissa +5.1200000000000000000000000000000000000000000000000000000000000000e+02
Rlamch R: Rounding mode                +1.0000000000000000000000000000000000000000000000000000000000000000e+00
Rlamch M: Minimum exponent:            -4.6116860184273879030000000000000000000000000000000000000000000000e+18
Rlamch U: Underflow threshold          +1.7019382623481672278259575819240965611355119939659392498165297957e-1388255822130839283
Rlamch L: Largest exponent             +0.0000000000000000000000000000000000000000000000000000000000000000e+00
Rlamch O: Overflow threshold           +5.8756537891115875909369119988784425899385163927454983083337796065e+1388255822130839282
Rlamch -: Reciprocal of safe minimum   +5.8756537891115875909369119988784425899385163927454983083337796065e+1388255822130839282
\end{verbatim}
}
\end{center}
\end{table}

\begin{table}
\caption{Rlamch values for {\tt GMP} (default setting) on Windows (ming64) }\label{rlamchgmp}
\begin{center}
\scriptsize
{
\begin{verbatim}
Rlamch E: Epsilon                      +7.4583407312002067432909653154629338373764715346004068942715183332e-155
Rlamch S: Safe minimum                 +4.7651298097759021464323395634653490408303922511184795759100475052e-323228497
Rlamch B: Base                         +2.0000000000000000000000000000000000000000000000000000000000000000e+00
Rlamch P: Precision                    +1.4916681462400413486581930630925867674752943069200813788543036666e-154
Rlamch N: Number of digits in mantissa +5.1200000000000000000000000000000000000000000000000000000000000000e+02
Rlamch R: Rounding mode                +1.0000000000000000000000000000000000000000000000000000000000000000e+00
Rlamch M: Minimum exponent:            -1.0737418230000000000000000000000000000000000000000000000000000000e+09
Rlamch U: Underflow threshold          +4.7651298097759021464323395634653490408303922511184795759100475052e-323228497
Rlamch L: Largest exponent             +0.0000000000000000000000000000000000000000000000000000000000000000e+00
Rlamch O: Overflow threshold           +2.0985787164673876924043581168838390706380979654733526277866462257e+323228496
Rlamch -: Reciprocal of safe minimum   +2.0985787164673876924043581168838390706380979654733526277866462257e+323228496
\end{verbatim}
}
\end{center}
\end{table}

\begin{table}
\caption{Rlamch values for {\tt GMP} with {\tt MPLAPACK\_GMP\_PRECISION=1024} on Linux and macOS} \label{rlamchgmp1024}
\begin{center}
\scriptsize
{
\begin{verbatim}
Rlamch E: Epsilon                      +5.5626846462680034577255817933310101605480399511558295763833185422e-309
Rlamch S: Safe minimum                 +1.7019382623481672278259575819240965611355119939659392498165297957e-1388255822130839283
Rlamch B: Base                         +2.0000000000000000000000000000000000000000000000000000000000000000e+00
Rlamch P: Precision                    +1.1125369292536006915451163586662020321096079902311659152766637084e-308
Rlamch N: Number of digits in mantissa +1.0240000000000000000000000000000000000000000000000000000000000000e+03
Rlamch R: Rounding mode                +1.0000000000000000000000000000000000000000000000000000000000000000e+00
Rlamch M: Minimum exponent:            -4.6116860184273879030000000000000000000000000000000000000000000000e+18
Rlamch U: Underflow threshold          +1.7019382623481672278259575819240965611355119939659392498165297957e-1388255822130839283
Rlamch L: Largest exponent             +0.0000000000000000000000000000000000000000000000000000000000000000e+00
Rlamch O: Overflow threshold           +5.8756537891115875909369119988784425899385163927454983083337796065e+1388255822130839282
Rlamch -: Reciprocal of safe minimum   +5.8756537891115875909369119988784425899385163927454983083337796065e+1388255822130839282
Rlamch Z: dummy (error)                +0.0000000000000000000000000000000000000000000000000000000000000000e+00
\end{verbatim}
}
\end{center}
\end{table}

\begin{table}
\caption{Rlamch values for {\tt GMP} with {\tt MPLAPACK\_GMP\_PRECISION=1024} on Windows (mingw64)} \label{rlamchgmp1024}
\begin{center}
\scriptsize
{
\begin{verbatim}
Rlamch E: Epsilon                      +5.5626846462680034577255817933310101605480399511558295763833185422e-309
Rlamch S: Safe minimum                 +4.7651298097759021464323395634653490408303922511184795759100475052e-323228497
Rlamch B: Base                         +2.0000000000000000000000000000000000000000000000000000000000000000e+00
Rlamch P: Precision                    +1.1125369292536006915451163586662020321096079902311659152766637084e-308
Rlamch N: Number of digits in mantissa +1.0240000000000000000000000000000000000000000000000000000000000000e+03
Rlamch R: Rounding mode                +1.0000000000000000000000000000000000000000000000000000000000000000e+00
Rlamch M: Minimum exponent:            -1.0737418230000000000000000000000000000000000000000000000000000000e+09
Rlamch U: Underflow threshold          +4.7651298097759021464323395634653490408303922511184795759100475052e-323228497
Rlamch L: Largest exponent             +0.0000000000000000000000000000000000000000000000000000000000000000e+00
Rlamch O: Overflow threshold           +2.0985787164673876924043581168838390706380979654733526277866462257e+323228496
Rlamch -: Reciprocal of safe minimum   +2.0985787164673876924043581168838390706380979654733526277866462257e+323228496

\end{verbatim}
}
\end{center}
\end{table}

\begin{table}
\caption{Rlamch values for {\tt \_Float128}}\label{rlamch_float128}
\begin{center}
\begin{verbatim}
Rlamch E: Epsilon                      +1.9259299443872358530559779425849273185381e-34
Rlamch S: Safe minimum                 +3.3621031431120935062626778173217526025981e-4932
Rlamch B: Base                         +2.0000000000000000000000000000000000000000e+00
Rlamch P: Precision                    +3.8518598887744717061119558851698546370762e-34
Rlamch N: Number of digits in mantissa +1.1300000000000000000000000000000000000000e+02
Rlamch R: Rounding mode                +1.0000000000000000000000000000000000000000e+00
Rlamch M: Minimum exponent:            -1.6381000000000000000000000000000000000000e+04
Rlamch U: Underflow threshold          +3.3621031431120935062626778173217526025981e-4932
Rlamch L: Largest exponent             +1.6384000000000000000000000000000000000000e+04
Rlamch O: Overflow threshold           +1.1897314953572317650857593266280070161965e+4932
Rlamch -: Reciprocal of safe minimum   +2.9743287383930794127143983165700178269086e+4931

\end{verbatim}
\end{center}
\end{table}

\begin{table}
\caption{Rlamch values for {\tt \_Float64x}}\label{rlamch_float64x}
\begin{center}
\begin{verbatim}
Rlamch E: Epsilon                      +5.421010862427522170037e-20
Rlamch S: Safe minimum                 +3.362103143112093506263e-4932
Rlamch B: Base                         +2.000000000000000000000e+00
Rlamch P: Precision                    +1.084202172485504434007e-19
Rlamch N: Number of digits in mantissa +6.400000000000000000000e+01
Rlamch R: Rounding mode                +1.000000000000000000000e+00
Rlamch M: Minimum exponent:            -1.638100000000000000000e+04
Rlamch U: Underflow threshold          +3.362103143112093506263e-4932
Rlamch L: Largest exponent             +1.638400000000000000000e+04
Rlamch O: Overflow threshold           +1.189731495357231765021e+4932
Rlamch -: Reciprocal of safe minimum   +2.974328738393079412714e+4931
\end{verbatim}
\end{center}
\end{table}

\begin{table}
\caption{Rlamch values for {\tt double}}\label{rlamch_double}
\begin{center}
\begin{verbatim}
Rlamch E: Epsilon                      +1.1102230246251565e-16
Rlamch S: Safe minimum                 +2.2250738585072014e-308
Rlamch B: Base                         +2.0000000000000000e+00
Rlamch P: Precision                    +2.2204460492503131e-16
Rlamch N: Number of digits in mantissa +5.3000000000000000e+01
Rlamch R: Rounding mode                +1.0000000000000000e+00
Rlamch M: Minimum exponent:            -1.0210000000000000e+03
Rlamch U: Underflow threshold          +2.2250738585072014e-308
Rlamch L: Largest exponent             +1.0240000000000000e+03
Rlamch O: Overflow threshold           +1.7976931348623157e+308
Rlamch -: Reciprocal of safe minimum   +4.4942328371557898e+307
\end{verbatim}
\end{center}
\end{table}

\begin{table}
\caption{Rlamch values for {\tt dd\_real}}\label{rlamch_dd}
\begin{center}
\begin{verbatim}
Rlamch E: Epsilon                      +4.93038065763131995214781514484568e-32
Rlamch S: Safe minimum                 +2.00416836000897277799610805134985e-292
Rlamch B: Base                         +2.00000000000000000000000000000000e+00
Rlamch P: Precision                    +9.86076131526263990429563028969136e-32
Rlamch N: Number of digits in mantissa +1.06000000000000000000000000000000e+02
Rlamch R: Rounding mode                +1.00000000000000000000000000000000e+00
Rlamch M: Minimum exponent:            -9.68000000000000000000000000000001e+02
Rlamch U: Underflow threshold          +2.00416836000897277799610805134985e-292
Rlamch L: Largest exponent             +1.02400000000000000000000000000000e+03
Rlamch O: Overflow threshold           +1.79769313486231580793728971405328e+308
Rlamch -: Reciprocal of safe minimum   +4.98960077383679952914093178259285e+291

\end{verbatim}
\end{center}
\end{table}

\begin{table}
\caption{Rlamch values for {\tt qd\_real}}\label{rlamch_qd}
\begin{center}
\scriptsize
{
\begin{verbatim}
Rlamch E: Epsilon                      +1.2154326714572500565324311366323150942261000827598106963711353150e-63
Rlamch S: Safe minimum                 +1.6259745436952323153369025467109371916500580246095533056887340823e-260
Rlamch B: Base                         +2.0000000000000000000000000000000000000000000000000000000000000000e+00
Rlamch P: Precision                    +2.4308653429145001130648622732646301884522001655196213927422706300e-63
Rlamch N: Number of digits in mantissa +2.1200000000000000000000000000000000000000000000000000000000000000e+02
Rlamch R: Rounding mode                +1.0000000000000000000000000000000000000000000000000000000000000000e+00
Rlamch M: Minimum exponent:            -8.6200000000000000000000000000000000000000000000000000000000000000e+02
Rlamch U: Underflow threshold          +1.6259745436952323153369025467109371916500580246095533056887340823e-260
Rlamch L: Largest exponent             +1.0240000000000000000000000000000000000000000000000000000000000000e+03
Rlamch O: Overflow threshold           +1.7976931348623158079372897140530286112296785259868571699620068630e+308
Rlamch -: Reciprocal of safe minimum   +6.1501577861568104283923723841611832207865934590357532972465351809e+259
\end{verbatim}
}
\end{center}
\end{table}

\subsection{A basic strategy to use MPBLAS and MPLAPACK}

Unfortunately, We do not provide concise manuals for MPBLAS and MPLAPACK since the API conversion from BLAS and LAPACK is trivial, and BLAS and LAPACK are very well documented. If you want to use some specific routines,
first, consult LAPACK users' guide~\cite{laug}, or \url{https://www.netlib.org/lapack/lug/}.

Following is the basic strategy for using MPLAPACK routines.
\begin{enumerate}
\item See \url{https://www.netlib.org/lapack/lug/node25.html} to find the name of the driver routine.
\item For example, If you want to compute the generalized singular values and vectors of a matrix $A$, you will find the routine name as {\tt DGGSVD} in \url{https://www.netlib.org/lapack/lug/node36.html}.
\item Find the definition part of {\tt DGGSVD} by download LAPACK and extract to find {\tt dggsvd.f} or you can find on the internet {\tt https://www.netlib.org/lapack/explore-html/dd/db4/dggsvd\_8f.html}.
\item Look for {\tt Rggsvd} prototype definition of MPLAPACK and change the variables appropriately to fit MPLAPACK.
\end{enumerate}

\subsection{How to link MPBLAS and MPLAPACK statically}
In this subsection, we show how to link MPBLAS and MPLAPACK statically.
For MPBLAS, we need to specify the full path of the library name and enclose it with {\tt -Wl,--whole-archive} and {\tt --no-whole-archive}. Note that {\tt -Wl,--whole-archive} is necessary as we call some initialization function at runtime; otherwise, MPBLAS will not work correctly.
For example, when one builds {\tt Rgemm\_mpfr} statically, one should type the following: 
\begin{verbatim}
$ cd /home/docker/MPLAPACK/share/examples/mpblas
$ g++ -fopenmp -O2 -I/home/docker/MPLAPACK/include \
-I/home/docker/MPLAPACK/include/mplapack Rgemm_mpfr.cpp \
-Wl,--whole-archive,/home/docker/MPLAPACK/lib/libmpblas_mpfr.a,--no-whole-archive \
-L/home/docker/MPLAPACK/lib -lmpfr -lmpc -lgmp
$ ldd ./a.out
        linux-vdso.so.1 (0x00007fff1dd12000)
        libmpfr.so.6 => /usr/lib/x86_64-linux-gnu/libmpfr.so.6 (0x00007fc36388a000)
        libmpc.so.3 => /usr/lib/x86_64-linux-gnu/libmpc.so.3 (0x00007fc363672000)
        libstdc++.so.6 => /usr/lib/x86_64-linux-gnu/libstdc++.so.6 (0x00007fc363458000)
        libgcc_s.so.1 => /usr/lib/x86_64-linux-gnu/libgcc_s.so.1 (0x00007fc36343d000)
        libpthread.so.0 => /usr/lib/x86_64-linux-gnu/libpthread.so.0 (0x00007fc36341a000)
        libc.so.6 => /usr/lib/x86_64-linux-gnu/libc.so.6 (0x00007fc363228000)
        libgmp.so.10 => /usr/lib/x86_64-linux-gnu/libgmp.so.10 (0x00007fc3631a2000)
        /lib64/ld-linux-x86-64.so.2 (0x00007fc3639d4000)
        libm.so.6 => /usr/lib/x86_64-linux-gnu/libm.so.6 (0x00007fc363053000)
$ ./a.out
...
\end{verbatim}
For MPLAPACK, we need to specify the full path of library name, but enclosed with {\tt -Wl,--whole-archive} and {\tt --no-whole-archive} is {\it not necessary}. For example,
To link MPBLAS and MLAPACK statically, we compile as follows:
\begin{verbatim}
$ cd /home/docker/MPLAPACK/share/examples/mplapack/00_LinearEquations
$ g++ -O2 -I/home/docker/MPLAPACK/include -I/home/docker/MPLAPACK/include/mplapack \
Rgetri_Hilbert_mpfr.cpp \
/home/docker/MPLAPACK/lib/libmplapack_mpfr.a \
-Wl,--whole-archive,/home/docker/MPLAPACK/lib/libmpblas_mpfr.a,--no-whole-archive \
-L/home/docker/MPLAPACK/lib -lmpfr -lmpc -lgmp
$ ldd ./a.out
        linux-vdso.so.1 (0x00007ffd64db1000)
        libmpfr.so.6 => /usr/lib/x86_64-linux-gnu/libmpfr.so.6 (0x00007fe59c609000)
        libmpc.so.3 => /usr/lib/x86_64-linux-gnu/libmpc.so.3 (0x00007fe59c3f1000)
        libstdc++.so.6 => /usr/lib/x86_64-linux-gnu/libstdc++.so.6 (0x00007fe59c1d7000)
        libgcc_s.so.1 => /usr/lib/x86_64-linux-gnu/libgcc_s.so.1 (0x00007fe59c1bc000)
        libc.so.6 => /usr/lib/x86_64-linux-gnu/libc.so.6 (0x00007fe59bfca000)
        libgmp.so.10 => /usr/lib/x86_64-linux-gnu/libgmp.so.10 (0x00007fe59bf46000)
        /lib64/ld-linux-x86-64.so.2 (0x00007fe59c75f000)
        libm.so.6 => /usr/lib/x86_64-linux-gnu/libm.so.6 (0x00007fe59bdf5000)
$ ./a.out | grep InfnromL
...
\end{verbatim}

\subsection{How to use MPBLAS and MPLAPACK on MacOS}
This section explains how to configure and build MPLAPACK on macOS using MacPorts.
We assume the home directory is {\tt /Volumes/Users/maho}.
You may build MPLAPACK on macOS as follows:
{\footnotesize
\begin{verbatim}
$ sudo port install gcc10 gsed
$ rm -rf /Volumes/Users/maho/tmp /Volumes/Users/maho/MPLAPACK
$ mkdir /Volumes/Users/maho/tmp
$ cd /Volumes/Users/maho/tmp
$ wget https://github.com/nakatamaho/mplapack/releases/download/v2.0.1/mplapack-2.0.1.tar.xz
$ tar xvfz mplapack-2.0.1.tar.xz
$ cd mplapack-2.0.1
$ CXX="g++-mp-10" ; export CXX
$ CC="gcc-mp-10" ; export CC
$ FC="gfortran-mp-10"; export FC
$ ./configure --prefix=$HOME/MPLAPACK --enable-gmp=yes --enable-mpfr=yes \
--enable-_Float128=yes --enable-qd=yes --enable-dd=yes --enable-double=yes \
--enable-test=yes --enable-_Float64x=yes 
$ make -j4
$ make install
$ cd 
$ rm -rf /Volumes/Users/maho/tmp
\end{verbatim}
}

You can test the installation, and how libraries are linked can be checked by {\tt otool -L}

\begin{verbatim}
$ cd /Volumes/Users/maho/MPLAPACK/share/examples/mpblas
$ g++-mp-10 -c -O2 -fopenmp -I/Volumes/Users/maho/MPLAPACK/include \
-I/Volumes/Users/maho/MPLAPACK/include/mplapack Rgemm_mpfr.cpp
$ g++-mp-10 -o Rgemm_mpfr Rgemm_mpfr.o -L/Volumes/Users/maho/MPLAPACK/lib \
-lmpblas_mpfr -lmpfr -lmpc -lgmp
$ ./Rgemm_mpfr
...
\end{verbatim}
{\scriptsize
\begin{verbatim}
$ otool -L Rgemm_mpfr
Rgemm_mpfr:
        /Volumes/Users/maho/MPLAPACK/lib/libmpblas_mpfr.0.dylib (compatibility version 1.0.0, current version 1.0.0)
        /Volumes/Users/maho/MPLAPACK/lib/libmpfr.6.dylib (compatibility version 8.0.0, current version 8.0.0)
        /Volumes/Users/maho/MPLAPACK/lib/libmpc.3.dylib (compatibility version 6.0.0, current version 6.1.0)
        /Volumes/Users/maho/MPLAPACK/lib/libgmp.10.dylib (compatibility version 15.0.0, current version 15.1.0)
        @rpath/libstdc++.6.dylib (compatibility version 7.0.0, current version 7.30.0)
        @rpath/libgcc_s.1.1.dylib (compatibility version 1.0.0, current version 1.1.0)
        /usr/lib/libSystem.B.dylib (compatibility version 1.0.0, current version 1292.100.5)
\end{verbatim}
}
We don't have to pass {\tt -Wl,--rpath=/Volumes/Users/maho/docker/MPLAPACK/lib} to linker. However,
it is hard coded, and you cannot move the top directory of MPLAPACK to another place.

\subsubsection{Static linking}
We do not support making a static binary for macOS since it is complicated to make a static binary on macOS. 

\subsection{How to use MPBLAS and MPLAPACK on Windows(mingw64)}
\subsubsection{Dynamic linking}
By default, the mingw64 system links dynamically to make an executable.
For example, to make {\tt Rgemm\_mpfr.exe}, type as follows:
\begin{verbatim}
$ cd /home/docker/MPLAPACK_MINGW/share/examples/mpblas/
$ x86_64-w64-mingw32-g++ -c -O2 -fopenmp -I/home/docker/MPLAPACK_MINGW/include \
-I/home/docker/MPLAPACK_MINGW/include/mplapack Rgemm_mpfr.cpp
$ x86_64-w64-mingw32-g++ -o Rgemm_mpfr.exe Rgemm_mpfr.o -L/home/docker/MPLAPACK_MINGW/lib \
-lmpblas_mpfr -lmpfr -lmpc -lgmp
$ wine64 ./Rgemm_mpfr.exe
...
\end{verbatim}

For example, to make {\tt Rgess\_test\_gmp.exe}, type as follows:
\begin{verbatim}
$ cd /home/docker/MPLAPACK_MINGW/share/examples/mplapack/04_NonsymmetricEigenproblems
$ x86_64-w64-mingw32-g++ -c -O2 -fopenmp -I/home/docker/MPLAPACK_MINGW/include \
 -I/home/docker/MPLAPACK_MINGW/include/mplapack Rgees_test_gmp.cpp
$ x86_64-w64-mingw32-g++ -O2 -fopenmp -o Rgees_test_gmp.exe \
Rgees_test_gmp.o -L/home/docker/MPLAPACK_MINGW/lib -lmplapack_gmp -lmpblas_gmp -lgmp
$ wine64 ./Rgees_test_gmp.exe
...
\end{verbatim}

By default, the mingw64 system links dynamically to make an executable.
\begin{verbatim}
$ cd /home/docker/MPLAPACK_MINGW/share/examples/mpblas/
$ x86_64-w64-mingw32-g++ -c -O2 -fopenmp -I/home/docker/MPLAPACK_MINGW/include \
-I/home/docker/MPLAPACK_MINGW/include/mplapack Rgemm_mpfr.cpp
$ x86_64-w64-mingw32-g++ -o Rgemm_mpfr.exe Rgemm_mpfr.o -L/home/docker/MPLAPACK_MINGW/lib \
-lmpblas_mpfr -lmpfr -lmpc -lgmp
$ x86_64-w64-mingw32-objdump -p Rgemm_mpfr.exe | grep 'DLL Name'
        DLL Name: libmpfr-6.dll
        DLL Name: libmpblas_mpfr-0.dll
        DLL Name: KERNEL32.dll
        DLL Name: msvcrt.dll
        DLL Name: libgcc_s_seh-1.dll
        DLL Name: libstdc++-6.dll
\end{verbatim}
To run {\tt Rgemm\_mpfr.exe} on other computers, we must first install mingw64 runtime, then, copy DLLs in {\tt /home/docker/MPLAPACK\_MINGW/bin}
to somewhere under {\tt C:\textbackslash{}Program File}, like \\
{\tt C:\textbackslash{}Program File\textbackslash{}MPLAPACK} or current directory.

\subsubsection{Static link for mingw64 (Windows)}
To link mpblas statically on the mingw64 system, we should add\\
``{\tt -Wl,--whole-archive,/home/docker/MPLAPACK\_MINGW/lib/libmpblas\_mpfr.a,--no-whole-archive}"\\
when linking. This is required because MPBLAS initializes some values at the runtime. If you do not specify
them, simply link like ``{\tt /home/docker/MPLAPACK\_MINGW/lib/libmpblas\_mpfr.a}", MPBLAS will not work properly.
The following example shows how we link Rgemm\_mpfr.exe against {\tt libmpblas\_mpfr.a}, which is a static library: 
\begin{verbatim}
$ cd /home/docker/MPLAPACK_MINGW/share/examples/mpblas/
$ x86_64-w64-mingw32-g++ -c -O2 -fopenmp -I/home/docker/MPLAPACK_MINGW/include \
-I/home/docker/MPLAPACK_MINGW/include/mplapack Rgemm_mpfr.cpp
$ x86_64-w64-mingw32-g++ -o Rgemm_mpfr.exe Rgemm_mpfr.o -L/home/docker/MPLAPACK_MINGW/lib \
-static-libgcc -static-libstdc++ \
-Wl,--whole-archive,/home/docker/MPLAPACK_MINGW/lib/libmpblas_mpfr.a,--no-whole-archive \
-lmpfr -lmpc -lgmp
\end{verbatim}
However, the resultant file {\tt Rgemm\_mpfr.exe} is not a completely static executable.
\begin{verbatim}
$ x86_64-w64-mingw32-objdump -p Rgemm_mpfr.exe | grep 'DLL Name'
        DLL Name: libmpc-3.dll
        DLL Name: libmpfr-6.dll
        DLL Name: KERNEL32.dll
        DLL Name: msvcrt.dll
\end{verbatim}
Moreover, this executable implicitly depends on \\
{\tt /usr/lib/gcc/x86\_64-w64-mingw32/9.3-win32/libgcc\_s\_seh-1.dll}.
Therefore, to run {\tt Rgemm\_mpfr.exe} on other computers, we must copy the contents of {\tt /home/docker/MPLAPACK\_MINGW/bin} and \\
 {\tt /usr/lib/gcc/x86\_64-w64-mingw32/9.3-win32/libgcc\_s\_seh-1.dll}
to current directory or somewhere under {\tt C:\textbackslash{}Program File}, like {\tt C:\textbackslash{}Program File\textbackslash{}MPLAPACK}.
Currently, it is not possible to build complete static executables using MPLAPACK. However, we are planning to handle this issue in the next version.

\section{Quality assurance of MPBLAS and MPLAPACK}
\label{sec:howtotest}
\subsection{Testing MPBLAS routines}
We performed quality assurance of MPBLAS as follows. 
\begin{enumerate}
    \item We believe there are no bugs in the reference BLAS~\cite{10.1145/567806.567807}.
    \item We input random values of binary64 for matrices, vectors, and scalars to BLAS and MPFR version of MPBLAS for various sizes of matrices, vectors, and leading dimensions (It means MPBLAS and BLAS may fail with error codes) to a routine.
    \item If the difference between the two results is within tolerance, we regard the quality as assured for the MPFR version of the MPBLAS routine.
    \item We also check the error codes by {\tt xerbla} and {\tt Mxerbla}, and invariably these error codes are the same for the routine.
    \item Do the same checks for all routines in Table~\ref{mpblasroutines}.
    \item If there are no errors, we regard the MPFR version of {\tt MPBLAS} as bug-free. 
    \item Then, we input random values of a multiple precision type (e.g., {\tt \_Float128}) for matrices, vectors, and scalars to this version of MPBLAS and MPFR version MPBLAS for various sizes of the matrix, vector, and leading dimensions.
    \item If the difference between the two results is within tolerance, we regard the quality as assured for the multiple-precision type version of MPBLAS.
\end{enumerate}
Since MPBLAS routines only consist of algebraic operations, this method ensures that all routines are correctly implemented. 
Test programs and input files are installed at \\
{\tt /home/docker/MPLAPACK/lib/*/mplapack/test/compare/}.\\
For example, if you check the MPFR version of {\tt Rgemm},
\begin{verbatim}
$ cd /home/docker/MPLAPACK/lib/x86_64-pc-linux-gnu/mplapack/test/compare/mpfr
$ ./Rgemm.test.mpfr
*** Testing Rgemm start ***
maxerror: +0.0000000000000000e+00
maxerror: +8.8494802737544305e-17
maxerror: +5.2155910699049989e-16
maxerror: +5.2155910699049989e-16
...
maxerror: +5.7344511452715043e-16
*** Testing Rgemm successful ***
\end{verbatim}
. There are 256 error check programs for each precision.

\subsection{Testing MPLAPACK routines and driver routines}
We converted LAPACK's original test programs to C++ and extended them to each multi-precision version. As a result, there are two tests: testing for linear equation solver and eigenvalue problems, singluarvarlue decomposition, and least square fitting. The former is in the ``lin" directory, and the latter is in the ``eig" directory.

In MPLAPACK, all testing programs are installed at {\tt /home/docker/MPLAPACK/lib/*/test/lin/} and \\
 {\tt /home/docker/MPLAPACK/lib/*/test/eig/}. To test all MPLAPACK routines and drivers on Linux, you type the following:
\begin{verbatim}
$ cd /home/docker/MPLAPACK/lib/x86_64-pc-linux-gnu/mplapack/test/lin
$ bash test_lin_all.sh
$ less log.*
$ cd /home/docker/MPLAPACK/lib/x86_64-pc-linux-gnu/mplapack/test/eig
$ bash test_eig_all.sh
$ less log.*
\end{verbatim}
On windows (mingw64, using wine64):
\begin{verbatim}
$ cd /home/docker/MPLAPACK_MINGW/lib/x86_64-w64-mingw32/mplapack/test/lin
$ bash test_lin_all_mingw.sh
$ less log.*
$ cd /home/docker/MPLAPACK_MINGW/lib/x86_64-w64-mingw32/mplapack/test/eig
$ bash test_eig_all_migw.sh
$ less log.*
\end{verbatim}
On macOS (x86\_64):
\begin{verbatim}
$ cd /Volumes/Users/maho/MPLAPACK/lib/x86_64-apple-darwin20.6.0/mplapack/test/lin
$ bash test_lin_all.sh
$ less log.*
$ cd /Volumes/Users/maho/MPLAPACK/lib/x86_64-apple-darwin20.6.0/mplapack/test/eig
$ bash test_eig_all.sh
$ less log.*
\end{verbatim}

It may take a day to run through all tests. The MPLAPACK libraries are reliable when there are not so many errors.
You can find complete log files at \url{https://github.com/nakatamaho/mplapack/tree/master/mplapack/test/lin/results} \\
and \url{https://github.com/nakatamaho/mplapack/tree/master/mplapack/test/eig/results} for several platforms.

Note that we only tested for default precision (512bits) for {\tt GMP} and {\tt MPFR}. For other precisions, users should check by themselves.

\subsubsection{Background}
Until version 0.9.4, we did the same method for quality assurance for MPBLAS and MPLAPACK. From version 2.0.1, we ported {\tt lapack-3.9.1/TESTING/LIN} and {\tt lapack-3.9.1/TESTING/EIG} for all precisions using FABLE. The original test programs
use the {\tt FORMAT} statement extensively. Interestingly, FABLE and FABLE's Fortran EMulator library translate these {\tt FORMAT} statements almost perfectly except for printing out multiple-precision floating-point numbers. Thus our modifications to the testing routines were minimal.
Then we extended the program to support multiple precision arithmetic so that our libraries are accurate in more than 16 decimal significant digits.

\subsubsection{Details of testing real linear MPLAPACK routines and drivers}
First, we review how to test LAPACK linear routines:
\begin{verbatim}
$ cd /home/docker/mplapack/external/lapack/work/internal/lapack-3.9.1/TESTING/LIN/
$ ./xlintstd < ../dtest.in
 Tests of the DOUBLE PRECISION LAPACK routines
 LAPACK VERSION 3.9.1

 The following parameter values will be used:
    M   :       0     1     2     3     5    10    50
    N   :       0     1     2     3     5    10    50
    NRHS:       1     2    15
    NB  :       1     3     3     3    20
    NX  :       1     0     5     9     1
    RANK:      30    50    90

 Routines pass computational tests if test ratio is less than   30.00

 Relative machine underflow is taken to be    0.222507-307
 Relative machine overflow  is taken to be    0.179769+309
 Relative machine precision is taken to be    0.111022D-15


 DGE routines passed the tests of the error exits

 All tests for DGE routines passed the threshold (   3653 tests run)

 DGE drivers passed the tests of the error exits

 All tests for DGE drivers  passed the threshold (   5748 tests run)

 DGB routines passed the tests of the error exits
...
\end{verbatim}
We omit the details of how we test the linear routines, but {\tt ./xlintstd} tests all the linear routines used in LAPACK.
Details can be found in the literature~\cite{lawn41}.

Corresponding ported testing programs for linear routines \\
{\tt /home/docker/MPLAPACK/lib/*/mplapack/test/lin} are following:
\begin{itemize}
    \item {\tt xlintstR\_mpfr}
    \item {\tt xlintstR\_double}
    \item {\tt xlintstR\_\_Float64x}
    \item {\tt xlintstR\_\_Float128}
    \item {\tt xlintstR\_dd}
    \item {\tt xlintstR\_gmp}
    \item {\tt xlintstR\_qd}
   \item {\tt xlintstrfR\_\_Float128}
   \item {\tt xlintstrfR\_\_Float64x}
   \item {\tt xlintstrfR\_dd}
   \item {\tt xlintstrfR\_double}
   \item {\tt xlintstrfR\_gmp}
   \item {\tt xlintstrfR\_mpfr}
   \item {\tt xlintstrfR\_qd}
   \end{itemize}
where we add suffix ``R" + ``mplib" for {\tt xlintst}.
Test input files are {\tt Rtest.in} (usual matrix) and {\tt Rtest\_rfp.in} (matrix in recutangular full packed format).
Input files can be used for all precisions.

We tested real MPLAPACK routines for MPFR as follows:
\begin{verbatim}
$ cd ~/MPLAPACK/lib/x86_64-pc-linux-gnu/mplapack/test/lin/
$ ./xlintstR_mpfr < Rtest.in
 Tests of the Multiple precision version of LAPACK MPLAPACK VERSION 2.0.1
 Based on the original LAPACK VERSION 3.9.1

The following parameter values will be used:
        M  :     0     1     2     3     5    10    50
        N  :     0     1     2     3     5    10    50
     NRHS  :     1     2    15
       NB  :     1     3     3     3    20
       NX  :     1     0     5     9     1
     RANK  :    30    50    90

 Routines pass computational tests if test ratio is less than   30.00

 Relative machine underflow is taken to be : +9.5302596195518043e-323228497
 Relative machine overflow  is taken to be : +2.0985787164673877e+323228496
 Relative machine precision is taken to be : +7.4583407312002067e-155
 RGE routines passed the tests of the error exits

 All tests for RGE routines passed the threshold (   3653 tests run)
 RGE drivers passed the tests of the error exits

 All tests for RGE drivers  passed the threshold (   5748 tests run)
 RGB routines passed the tests of the error exits

 All tests for RGB routines passed the threshold (  28938 tests run)
 RGB drivers passed the tests of the error exits

 All tests for RGB drivers  passed the threshold (  36567 tests run)
 RGT routines passed the tests of the error exits

 All tests for RGT routines passed the threshold (   2694 tests run)
 RGT drivers passed the tests of the error exits

 All tests for RGT drivers  passed the threshold (   2033 tests run)
 RPO routines passed the tests of the error exits

 All tests for RPO routines passed the threshold (   1628 tests run)
 RPO drivers passed the tests of the error exits

 All tests for RPO drivers  passed the threshold (   1910 tests run)
 RPS routines passed the tests of the error exits

 All tests for RPS routines passed the threshold (    150 tests run)
 RPP routines passed the tests of the error exits
...
\end{verbatim}
The file {\tt Rtest.in} is same as {\tt dtest.in} of LAPACK, except for the first comment line. 
For other precisions, we similarly performed tests. We verified that all tests passed for all precisions.

The RFP (Rectangular Full Packed) version can be tested, for example, the dd version, as follows:
\begin{verbatim}
$ ./xlintstrfR_dd < Rtest_rfp.in
 Tests of the Multiple precision version of LAPACK MPLAPACK VERSION 2.0.1
 Based on the original LAPACK VERSION 3.9.1

The following parameter values will be used:
        N  :     0     1     2     3     5     6    10    11    50
     NRHS  :     1     2    15
        M  :     1     2     3     4     5     6     7     8     9
 Routines pass computational tests if test ratio is less than   30.00

 Relative machine underflow is taken to be : 2.0041683600089728e-292
 Relative machine overflow  is taken to be : 1.7976931348623158e+308
 Relative machine precision is taken to be : 4.9303806576313200e-32
 MULTIPLE PRECISION RFP routines passed the tests of the error exits

 All tests for RPF drivers  passed the threshold (   2304 tests run)
 All tests for Rlansf auxiliary routine passed the threshold (   384 tests run)
 All tests for the RFP conversion routines passed (    72 tests run)
 All tests for Rtfsm auxiliary routine passed the threshold (  7776 tests run)
 All tests for Rsfrk auxiliary routine passed the threshold (  2592 tests run)

 End of tests
 Total time used =  15 seconds
\end{verbatim}

Since we use random number with no fixed random seed, we may obtain failed results like as follows:
\begin{verbatim}
$ ./xlintstR__Float128 < Rtest.in
 Tests of the Multiple precision version of LAPACK MPLAPACK VERSION 2.0.1
 Based on the original LAPACK VERSION 3.9.1

The following parameter values will be used:
        M  :     0     1     2     3     5    10    50
        N  :     0     1     2     3     5    10    50
     NRHS  :     1     2    15
       NB  :     1     3     3     3    20
       NX  :     1     0     5     9     1
     RANK  :    30    50    90

 Routines pass computational tests if test ratio is less than   30.00

 Relative machine underflow is taken to be : +3.3621031431120935e-4932
 Relative machine overflow  is taken to be : +1.1897314953572318e+4932
 Relative machine precision is taken to be : +1.9259299443872359e-34
 RGE routines passed the tests of the error exits

 All tests for RGE routines passed the threshold (   3653 tests run)
 RGE drivers passed the tests of the error exits

 All tests for RGE drivers  passed the threshold (   5748 tests run)
 RGB routines passed the tests of the error exits

 All tests for RGB routines passed the threshold (  28938 tests run)
 RGB drivers passed the tests of the error exits

 RGB drivers:  General band matrices
 Matrix types:
    1. Random, CNDNUM = 2              5. Random, CNDNUM = sqrt(0.1/EPS)
    2. First column zero               6. Random, CNDNUM = 0.1/EPS
    3. Last column zero                7. Scaled near underflow
    4. Last n/2 columns zero           8. Scaled near overflow
 Test ratios:
    1: norm( L * U - A )  / ( N * norm(A) * EPS )
    2: norm( B - A * X )  / ( norm(A) * norm(X) * EPS )
    3: norm( X - XACT )   / ( norm(XACT) * CNDNUM * EPS )
    4: norm( X - XACT )   / ( norm(XACT) * (error bound) )
    5: (backward error)   / EPS
    6: RCOND * CNDNUM - 1.0
    7: abs( WORK(1) - RPVGRW ) / ( max( WORK(1), RPVGRW ) * EPS )
 Messages:
 Rgbsvx( ' ',' ',   50,   12,   12,...), type 6, test(5)=+4.5401871787002674e+01
 Rgbsvx( ' ',' ',   50,   12,   12,...), type 6, test(5)=+4.5401871787002674e+01
 Rgbsvx( ' ',' ',   50,   12,   12,...), EQUED=' ', type 6, test(5)=+8.2754819479199405e+01
 Rgbsvx( ' ',' ',   50,   12,   12,...), EQUED=' ', type 6, test(5)=+8.2754819479199405e+01
 Rgbsvx( ' ',' ',   50,   12,   12,...), EQUED=' ', type 6, test(5)=+4.5401871787002674e+01
 Rgbsvx( ' ',' ',   50,   12,   12,...), EQUED=' ', type 6, test(5)=+4.5401871787002674e+01
 RGB drivers:      6 out of  36567 tests failed to pass the threshold
 RGT routines passed the tests of the error exits
\end{verbatim}
However, this result is not fatal since errors are only slightly above the threshold.

\subsubsection{Details of testing complex linear MPLAPACK routines and drivers}

Corresponding ported testing programs for linear routines \\
{\tt /home/docker/MPLAPACK/lib/*/mplapack/test/lin} are following:
\begin{itemize}
    \item {\tt xlintstC\_mpfr}
    \item {\tt xlintstC\_double}
    \item {\tt xlintstC\_\_Float64x}
    \item {\tt xlintstC\_\_Float128}
    \item {\tt xlintstC\_dd}
    \item {\tt xlintstC\_gmp}
    \item {\tt xlintstC\_qd}
   \item {\tt xlintstrfC\_\_Float128}
   \item {\tt xlintstrfC\_\_Float64x}
   \item {\tt xlintstrfC\_dd}
   \item {\tt xlintstrfC\_double}
   \item {\tt xlintstrfC\_gmp}
   \item {\tt xlintstrfC\_mpfr}
   \item {\tt xlintstrfC\_qd}
   \end{itemize}
where we add suffix ``C" + ``mplib" for {\tt xlintst}.
Test input files are {\tt Ctest.in} (usual matrix) and {\tt Ctest\_rfp.in} (matrix in recutangular full packed format).
Input files can be used for all precisions.

We test complex MPLAPACK routines for the GMP version as follows:
\begin{verbatim}
$ cd ~/MPLAPACK/lib/x86_64-pc-linux-gnu/mplapack/test/lin/
$ ./xlintstC_gmp < Ctest.in
 Tests of the Multiple precision version of LAPACK MPLAPACK VERSION 2.0.1
 Based on the original LAPACK VERSION 3.9.1

The following parameter values will be used:
        M  :     0     1     2     3     5    10    50
        N  :     0     1     2     3     5    10    50
     NRHS  :     1     2    15
       NB  :     1     3     3     3    20
       NX  :     1     0     5     9     1
     RANK  :    30    50    90

 Routines pass computational tests if test ratio is less than   60.00

 Relative machine underflow is taken to be : +1.7019382623481672e-1388255822130839283
 Relative machine overflow  is taken to be : +5.8756537891115876e+1388255822130839282
 Relative machine precision is taken to be : +7.4583407312002067e-155
 CGE routines passed the tests of the error exits

 All tests for CGE routines passed the threshold (   3653 tests run)
 CGE drivers passed the tests of the error exits
...
 All tests for CXQ routines passed the threshold (   1482 tests run)
 CTQ routines passed the tests of the error exits

 All tests for CTQ routines passed the threshold (    510 tests run)
 CTS routines passed the tests of the error exits

 All tests for CTS routines passed the threshold (  10800 tests run)
 CHH routines passed the tests of the error exits

 All tests for CHH routines passed the threshold (  15900 tests run)

 End of tests
 Total time used =  2066 seconds
\end{verbatim}
We can test the RFP version, for example, the qd version, as follows:
\begin{verbatim}
$ ./xlintstrfC_qd < Ctest_rfp.in
 Tests of the Multiple precision version of LAPACK MPLAPACK VERSION 2.0.1
 Based on original LAPACK VERSION 3.9.1

The following parameter values will be used:
        N  :     0     1     2     3     5     6    10    11    50
     NRHS  :     1     2    15
        M  :     1     2     3     4     5     6     7     8     9
 Routines pass computational tests if test ratio is less than   30.00

 Relative machine underflow is taken to be : 1.6259745436952323e-260
 Relative machine overflow  is taken to be : 1.7976931348623158e+308
 Relative machine precision is taken to be : 1.2154326714572501e-63
 MULTIPLE PRECISION COMPLEX RFP routines passed the tests of the error exits

 All tests for ZPF drivers  passed the threshold (   2304 tests run)
 All tests for Clanhf auxiliary routine passed the threshold (   384 tests run)
 All tests for the RFP conversion routines passed (   72 tests run)
 All tests for Ctfsm auxiliary routine passed the threshold (  7776 tests run)
 All tests for Chfrk auxiliary routine passed the threshold (   2592 tests run)

 End of tests
 Total time used =  134 seconds
\end{verbatim}
\subsubsection{Details of testing other (EIG) MPLAPACK routines and drivers}
Next, we explain how to test MPLAPACK EIG.

There are 20 input files for testing eigenvalue problems, singular value problems, least square fitting problems,
and other related routines. For MPLAPCK, we use the following input files:
\begin{verbatim}
Cbak.in Cbal.in Cbal_double.in Cbb.in Cec.in Ced.in
Cgbak.in Cgbal.in Cgd.in Cgg.in Csb.in Csg.in Rbak.in 
Rbal.in Rbal_double.in Rbb.in Rec.in Red.in Rgbak.in 
Rgbal.in Rgd.in Rgg.in Rsb.in Rsg.in csd.in glm.in 
gqr.in gsv.in lse.in nep.in se2.in sep.in svd.in
\end{verbatim}
We modified the original input files; do not break the line for each matrix row so that the testing program can process the data correctly by C++.
The original input files for the balancing test depend on the precision. There are two
designed for double precision, {\tt Rbal\_double.in} and {\tt Cbal\_double.in}, which are identical to {\tt dbal.in} and {\tt zbal.in}
. The final test matrices in each input file of {\tt Rbal\_double.in} and {\tt Cbal\_double.in} are tricky. We omitted the final matrix for each input and make them {\tt Rbal.in} and {\tt Cbal.in}

Recall that we test LAPACK EIG as follows: 
\begin{verbatim}
$ cd /home/docker/mplapack/external/lapack/work/internal/lapack-3.9.1/TESTING/EIG/
$  ./xeigtstd < ../nep.in 
 Tests of the Nonsymmetric Eigenvalue Problem routines

 LAPACK VERSION 3.9.1

 The following parameter values will be used:
    M:         0     1     2     3     5    10    16
    N:         0     1     2     3     5    10    16
    NB:        1     3     3     3    20
    NBMIN:     2     2     2     2     2
    NX:        1     0     5     9     1
    INMIN:     11    12    11    15    11
    INWIN:      2     3     5     3     2
    INIBL:      0     5     7     3   200
    ISHFTS:      1     2     4     2     1
    IACC22:      0     1     2     0     1

 Relative machine underflow is taken to be    0.222507-307
 Relative machine overflow  is taken to be    0.179769+309
 Relative machine precision is taken to be    0.111022D-15

 Routines pass computational tests if test ratio is less than   20.00


 DHS routines passed the tests of the error exits ( 66 tests done)
...
\end{verbatim}

Corresponding ported testing programs for eigenvalues and sigular value problme solver
routines are in {\tt /home/docker/mplapack/mplapack/test/eig}, and we list them as follows:
\begin{itemize}
\item {\tt mpfr/xeigtstC\_mpfr}
\item {\tt mpfr/xeigtstR\_mpfr}
\item {\tt \_Float128/xeigtstC\_\_Float128}
\item {\tt \_Float128/xeigtstR\_\_Float128}
\item {\tt \_Float64x/xeigtstC\_\_Float64x}
\item {\tt \_Float64x/xeigtstR\_\_Float64x}
\item {\tt dd/xeigtstC\_dd}
\item {\tt dd/xeigtstR\_dd}
\item {\tt double/xeigtstC\_double}
\item {\tt double/xeigtstR\_double}
\item {\tt gmp/xeigtstC\_gmp}
\item {\tt gmp/xeigtstR\_gmp}
\item {\tt qd/xeigtstC\_qd}
\item {\tt qd/xeigtstR\_qd}
\end{itemize}
where we add suffix ``R" + ``mplib" or ``C" + ``mplib" for {\tt xeigtst}, respectively.

We test MPLAPACK routines for MPFR as follows:
\begin{verbatim}
$ cd /home/docker/mplapack/mplapack/test/eig/mpfr
$ ./xeigtstR_mpfr < ../svd.in
 Tests of the Singular Value Decomposition routines
 Tests of the Multiple precision version of LAPACK MPLAPACK VERSION 2.0.1
 Based on original LAPACK VERSION 3.9.1

The following parameter values will be used:
    M:         0     0     0     1     1     1     2     2     3     3
               3    10    10    16    16    30    30    40    40
    N:         0     1     3     0     1     2     0     1     0     1
               3    10    16    10    16    30    40    30    40
    NB:        1     3     3     3    20
    NBMIN:     2     2     2     2     2
    NX:        1     0     5     9     1
    NS:        2     0     2     2     2

 Relative machine underflow is taken to be+9.5302596195518043e-323228497
 Relative machine overflow  is taken to be+2.0985787164673877e+323228496
 Relative machine precision is taken to be+7.4583407312002067e-155

 Routines pass computational tests if test ratio is less than+5.0000000000000000e+01

 DBD routines passed the tests of the error exits ( 55 tests done)
 Rgesvd passed the tests of the error exits (  8 tests done)
 Rgesdd passed the tests of the error exits (  6 tests done)
 Rgejsv passed the tests of the error exits ( 11 tests done)
 Rgesvdx passed the tests of the error exits ( 12 tests done)
 Rgesvdq passed the tests of the error exits ( 11 tests done)


 SVD:  NB =   1, NBMIN =   2, NX =   1, NRHS =   2

 All tests for DBD routines passed the threshold (  10260 tests run)
....
\end{verbatim}
We performed tests for all the 20 testing files for all real precision versions. 
Almost all the tests passed for every input file.
In GMP version, {\tt gmp/xeigtstR\_gmp}, and the errors are related to precision loss in arithmetic; we should raise the error threshold by a factor of 10 or 100, and we do not consider such precision loesses are profound
since we can raise precision at runtime.
We do not perform test for {\tt csd.in}, because {\tt GMP} does not have trigonometric functions. MPLAPACK stops
when users try to use cosine sine decomposition.
We do not fix these failures and planning to drop the GMP version and integrate it into the {\tt MPFR} version in MPLAPACK version 3.0.0.  
Besides, some failures for {\tt svd.in} are using {\tt xeigtstR\_dd}, {\tt xeigstsC\_dd}, {\tt xeigtstC\_dd} and {\tt xeigstsC\_qd} are observed.
We do not know the root cause of the error; however, we suspect the errors might be related to underflow.

In summary, MPLAPACK routines for all real and complex precisions are reliable. 
\section{Fortran90 to C++ Conversion}
\label{sec:howtoconvert}
Until MPLAPACK 0.9.3, we used f2c to convert FORTRAN77 routines to C (until version 3.1.1, LAPACK only required the FORTRAN77 compiler). Unfortunately, the resultant source code was tough to read by a human; we replaced many tokens heavily using {\tt sed} and finally converted them to user-friendly, easy-to-read C++ programs by hand.

We simply use REAL, COMPLEX, INTEGER, LOGICAL, and BOOL to support all floating-point types. These types are abstract, and using typedef to specify the actual floating-point type (e.g., {\tt typedef mpf\_class REAL} for GMP. See {\tt mplapack/include/mplapack.h} for details).

The main differences between Fortran and C++ are: (i) array index starts from one, (ii) loop index may be incremented or decremented depending on its context, (iii) treatment of two-dimension arrays, (iv) when calling subroutines, FORTRAN always references the values, but we can choose call by value or call by reference in C++.  For (i), we use the same array index and decrement the index when accessing the array. (ii) we judge the direction of loop by hand, (iii) we expand two-dimension array using the leading dimension; we translate {\tt a(i,j)} to {\tt a[(i-1)  + (j-1) * lda]}, and (iv) we changed call by value when possible.

Since MPLAPACK 1.0.0, we used FABLE~\cite{fable} to convert all BLAS and LAPACK routines. In the original paper, they could translate {\tt dsyev.f} without modifications. However, the resultant source code also depends on Fortran Emulation Module (FEM), which is unnecessary for the MPBLAS and MPLAPACK parts, and the treatment of the array is a bit unnatural.  To make the resultant C++ code more natural and readable, we further patched {\tt cout.py} and passed through sed. Then, we modified the source codes by hand. For MPBLAS, conversion is almost automatic. For MPLAPACK, there are many hand-corrected minor syntax errors, interpretation of numbers, casts, and other corrections, but we can compile many without modifications. 

More important part is translation of {\tt TESTING/LIN} and {\tt TESTING/EIG} from Fortran90 to C++. Thanks to FEM, the written statement of Fortran can be used  directly without modifications. According to ChangeLog, it took approximately three weeks to just compile programs in \\
{\tt TESTING/LIN/xlint\{C,R\}\_mpfr} and {\tt TESTING/EIG/xeigtst\{C,R\}\_mpfr} for MPFR, although through check the results are required, and it took four months for real version and another four-month for complex version.

\section{Benchmarks}
\label{sec:benchmarks}
We show some benchmarks for MPBLAS and MPLAPACK. We do not perform any optimization for the reference implementations. Nevertheless, 
we provide a simple OpenMP accelerated version for MPBLAS and CUDA accelerated version of MPBLAS double-double routines ({\tt Rgemm} and {\tt Rsyrk}. They are shown in table~\ref{mpblasroutinesopt} and table~\ref{mpblasroutines_cuda_dd}.

Tables~\ref{benchmarkcondition} and~\ref{benchmarkoscondition} and, we show CPUs, OSes, machine names, Compilers, and their abbreviations Used to take the benchmarks. For A, D, E, and F, we took the benchmark inside the Docker, and for B, C and G, we ran on the native OS.

\begin{table}
{
\caption{CPU and memory configurations and their abbreviations}\label{benchmarkcondition}
\begin{center}
\begin{tabular}{c|c|c}
Abbrev.  &  CPU or GPU or Machine name        & Memory   \\ \hline
A  &  AMD Ryzen 3970X (1 CPU, 32 cores)       & DDR4 ECC 3200 MT/s \\
B  &  Intel Xeon E5 2623 (2 CPUs, 8 cores)    & DDR4 ECC 2133 MT/s  \\
C  &  Intel Core i5-8500B (1 CPU, 6 cores)    & DDR4 w/o ECC 2667 MT/s \\
D  &  same as A                               & same as A \\
E  &  NVIDIA A100 80GB PCIe                   & HBM2e 1935GB/s\\
F  &  NVIDIA V100 40GB PCIe                   & HBM2 900GB/s \\
G  &  Raspberry Pi4, ARM Cortex A72 (1 CPU, 4 cores) & LPDDR4-3200 SDRAM 4GB\\ \hline
\end{tabular}
\end{center}
}
\end{table}

\begin{table}
{
\caption{OSes, Compilers, and their abbreviations}\label{benchmarkoscondition}
\begin{center}
\begin{tabular}{c|c|c|c}
Abbrev.  & Docker filename                        &  OS            & Compiler/driver \\ \hline
 A  & \tt Dockerfile\_ubuntu20.04                 & Ubuntu 20.04  & GCC 9.3\\
 B  &  -                                          & Ubuntu 20.04  & GCC 9.3\\
 C  &  -                                          & macOS Big Sur & GCC 10.4\\
 D  & \tt Dockerfile\_ubuntu20.04\_inteloneapi    & Ubuntu 20.04  & Intel oneAPI 2021.6.0\\
 E  & \tt Dockerfile\_ubuntu20.04\_cuda           & Ubuntu 20.04  & Nvidia 11.7 \\
 F  &  -                                          & Ubuntu 20.04  & Nvidia 11.7 \\
 G  & \tt Dockerfile\_ubuntu20.04                 & Ubuntu 20.04  & GCC 9.3   \\  \hline
\end{tabular}
\end{center}
}
\end{table}

\subsection{{\tt Rgemm} benchmarks}
%%%%% AMD Ryzen 3970X %%%%
\subsubsection{{\tt Rgemm} on AMD Ryzen 3970X}
\begin{figure}
\caption{ {\tt Rgemm} performance on AMD Ryzen 3970X for {\tt \_Float128}, {\tt \_Float64x} and {\tt double-double} with/without simple OpenMP acceleration. }
\label{Rgemm1.A}
\begin{center}
\includegraphics{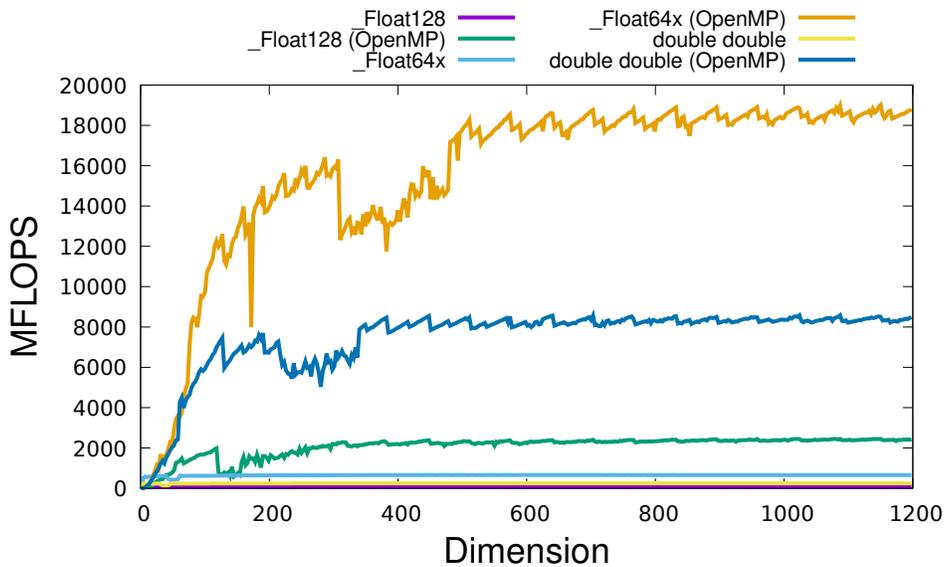}
\end{center}
\end{figure}

\begin{figure}
\caption{{\tt Rgemm} performance on AMD Ryzen 3970X for {\tt MPFR 512bit}, {\tt GMP 512bit} and {\tt quad-double} with/without simple OpenMP acceleration. }
\label{Rgemm2.A}
\begin{center}
\includegraphics{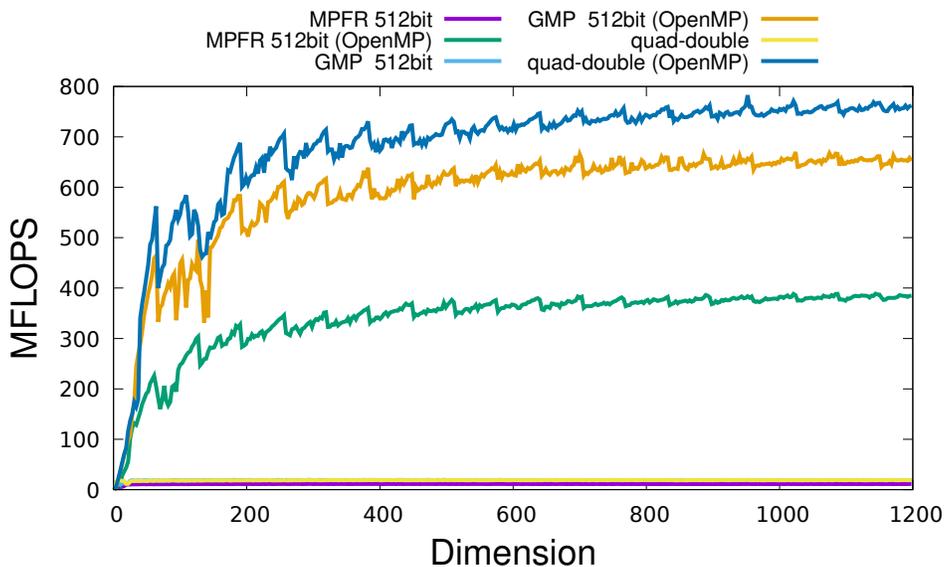}
\end{center}
\end{figure}

In Figure~\ref{Rgemm1.A}, we show the result of {\tt Rgemm} performance for {\tt \_Float128}, {\tt \_Float64x} and {\tt double-double}, and in Figure~\ref{Rgemm2.A} we show the result of {\tt Rgemm} performance for {\tt MPFR}, {\tt GMP} and {\tt quad-double} on AMD Ryzen 3970X.
The peak performances of the reference {\tt Rgemm}s of {\tt \_Float128}, {\tt \_Float64x} and {\tt double-double} are 64.1 MFlops, 658 MFlops, 245 MFlops, respectively.
The peak performances of simple OpenMP parallelized {\tt Rgemm}s of {\tt \_Float128}, {\tt \_Float64x} and {\tt double-double} are 2450 MFlops, 19020 MFlops, 8582 MFlops, respectively.
The peak performances of the reference {\tt Rgemm}s of {\tt MPFR 512bit}, {\tt GMP 512bit} and {\tt quad-double} are 11.5 MFlops, 19.5 MFlops, 19.8 MFlops, respectively.
The peak performances of simple OpenMP parallelized {\tt MPFR 512bit}, {\tt GMP 512bit} and {\tt quad-double} are 389 MFlops, 670 MFlops, 783 MFlops, respectively.\\

%%%%% Intel Xeon E5 2623 %%%%
\subsubsection{{\tt Rgemm} on Intel Xeon E5 2623}
\begin{figure}
\caption{ {\tt Rgemm} performance on Intel Xeon E5 2623 for {\tt \_Float128}, {\tt \_Float64x} and {\tt double-double} with/without simple OpenMP acceleration. }
\label{Rgemm1.B}
\begin{center}
\includegraphics{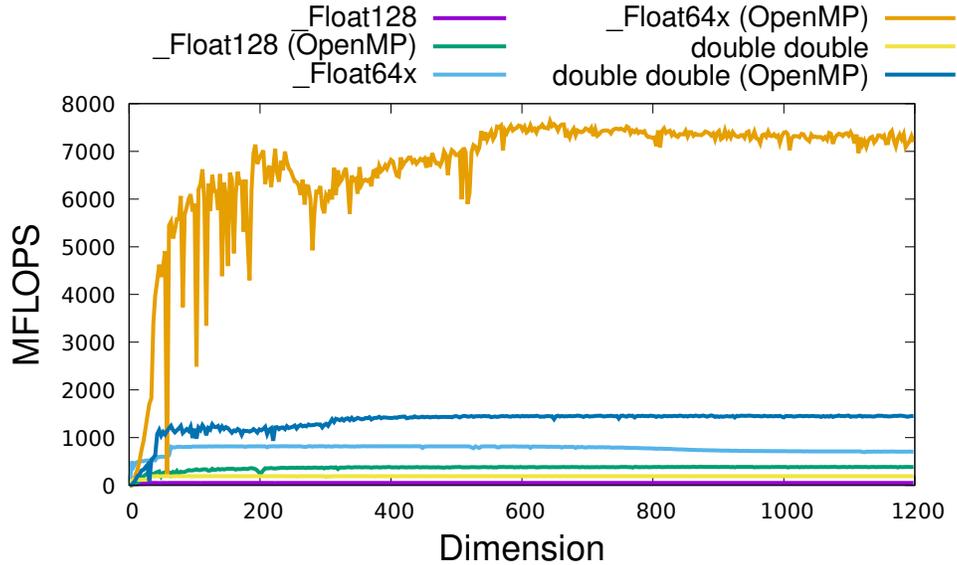}
\end{center}
\end{figure}

\begin{figure}
\caption{{\tt Rgemm} performance on Intel Xeon E5 2623 for {\tt MPFR 512bit}, {\tt GMP 512bit} and {\tt quad-double} with/without simple OpenMP acceleration. }
\label{Rgemm2.B}
\begin{center}
\includegraphics{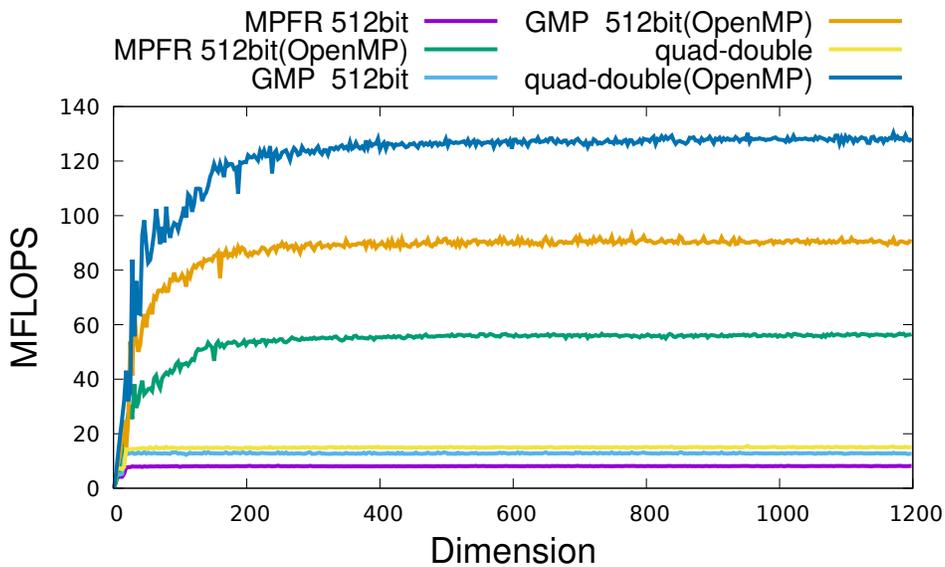}
\end{center}
\end{figure}

In Figure~\ref{Rgemm1.B}, we show the result of {\tt Rgemm} performance for {\tt \_Float128}, {\tt \_Float64x} and {\tt double-double}, and in Figure~\ref{Rgemm2.B} we show the result of {\tt Rgemm} performance for {\tt MPFR}, {\tt GMP} and {\tt quad-double} on Intel Xeon E5 2623.
The peak performances of the reference {\tt Rgemm}s of {\tt \_Float128}, {\tt \_Float64x} and {\tt double-double} are 51.0 MFlops, 820 MFlops, 195 MFlops, respectively.
The peak performances of simple OpenMP parallelized {\tt Rgemm}s of {\tt \_Float128}, {\tt \_Float64x} and {\tt double-double} are 389 MFlops, 7633 MFlops, 1464 MFlops, respectively.
The peak performances of the reference {\tt Rgemm}s of {\tt MPFR 512bit}, {\tt GMP 512bit} and {\tt quad-double} are 8.3 MFlops, 13.2 MFlops, 15.5 MFlops, respectively.
The peak performances of simple OpenMP parallelized {\tt MPFR 512bit}, {\tt GMP 512bit} and {\tt quad-double} are 56.7 MFlops, 92.9 MFlops, 130 MFlops, respectively.\\

%%%%% Intel Core i5-8500B %%%%
\subsubsection{{\tt Rgemm} on Intel Core i5-8500B}
\begin{figure}
\caption{ {\tt Rgemm} performance on Intel Core i5-8500B for {\tt \_Float128}, {\tt \_Float64x} and {\tt double-double} with/without simple OpenMP acceleration. }
\label{Rgemm1.C}
\begin{center}
\includegraphics{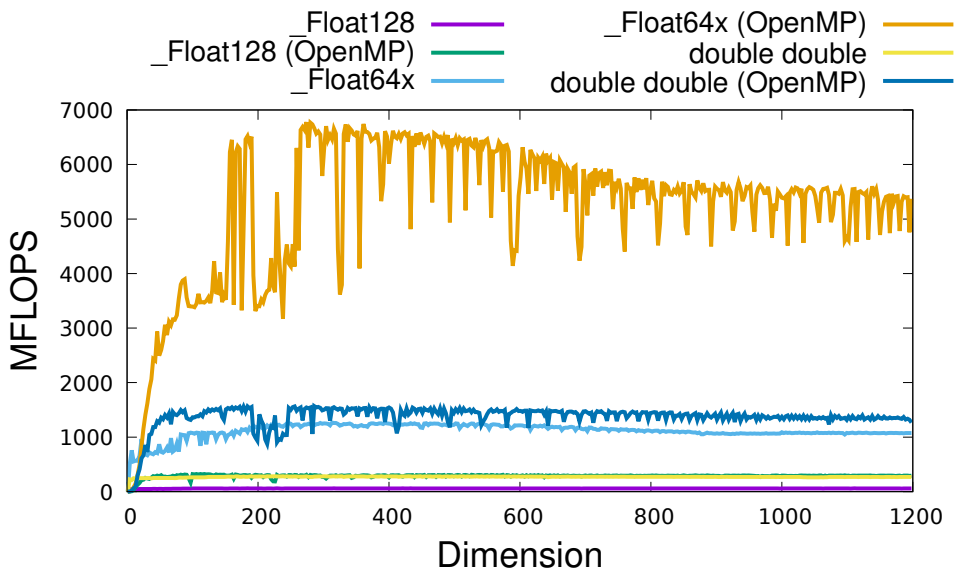}
\end{center}
\end{figure}

\begin{figure}
\caption{{\tt Rgemm} performance on Intel Core i5-8500B for {\tt MPFR 512bit}, {\tt GMP 512bit} and {\tt quad-double} with/without simple OpenMP acceleration. }
\label{Rgemm2.C}
\begin{center}
\includegraphics{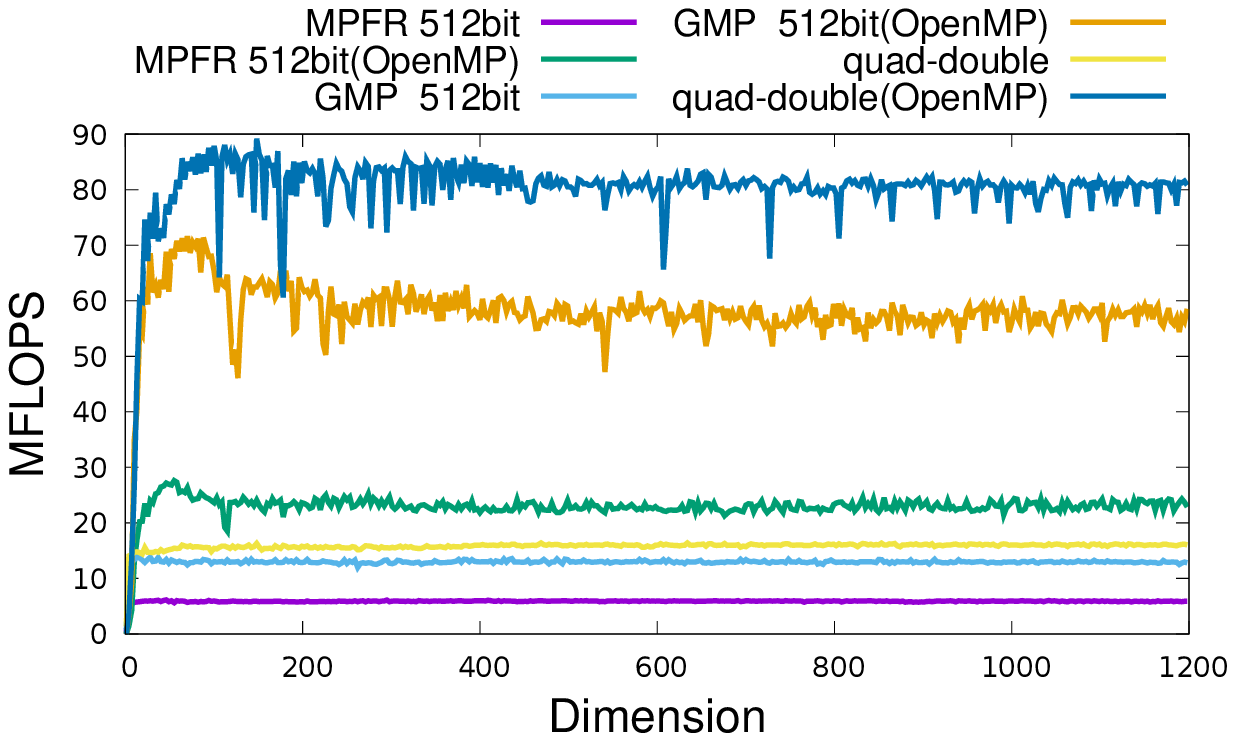}
\end{center}
\end{figure}

In Figure~\ref{Rgemm1.C}, we show the result of {\tt Rgemm} performance for {\tt \_Float128}, {\tt \_Float64x} and {\tt double-double}, and in Figure~\ref{Rgemm2.C} we show the result of {\tt Rgemm} performance for {\tt MPFR}, {\tt GMP} and {\tt quad-double} on Intel Core i5-8500B.
The peak performances of the reference {\tt Rgemm}s of {\tt \_Float128}, {\tt \_Float64x} and {\tt double-double} are 60.0 MFlops, 1268 MFlops, 282 MFlops, respectively.
The peak performances of simple OpenMP parallelized {\tt Rgemm}s of {\tt \_Float128}, {\tt \_Float64x} and {\tt double-double} are 324 MFlops, 6777 MFlops, 1568 MFlops, respectively.
The peak performances of the reference {\tt Rgemm}s of {\tt MPFR 512bit}, {\tt GMP 512bit} and {\tt quad-double} are 6.1 MFlops, 13.8 MFlops, 16.4 MFlops, respectively.
The peak performances of simple OpenMP parallelized {\tt MPFR 512bit}, {\tt GMP 512bit} and {\tt quad-double} are 27.6 MFlops, 71.6 MFlops, 89.2 MFlops, respectively.\\

%%%%% Raspberry Pi4 ARM Cortex A72 %%%%
\subsubsection{{\tt Rgemm} on Raspberry Pi4 ARM Cortex A72}
\begin{figure}
\caption{ {\tt Rgemm} performance on Raspberry Pi4 ARM Cortex A72 for {\tt \_Float128} and {\tt double-double} with/without simple OpenMP acceleration. }
\label{Rgemm1.G}
\begin{center}
\includegraphics{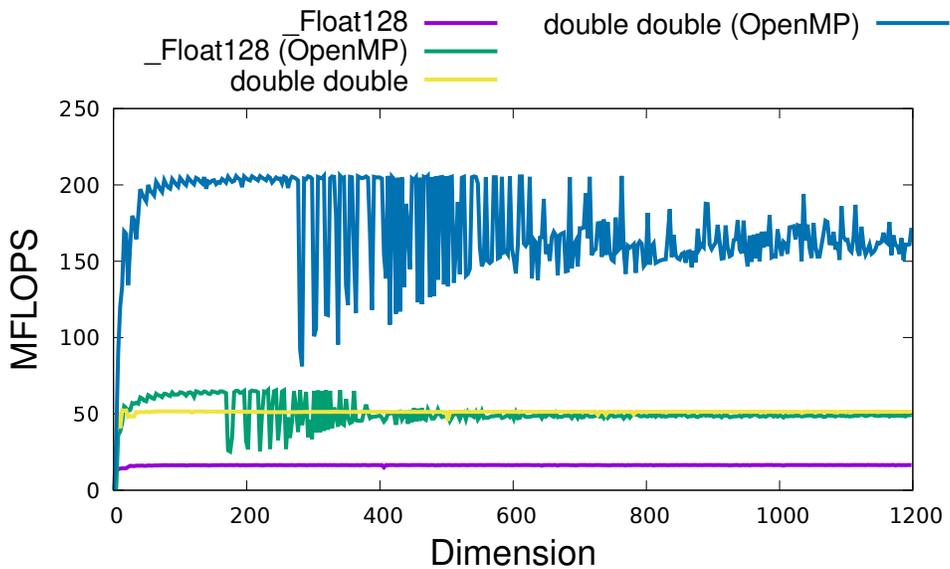}
\end{center}
\end{figure}

\begin{figure}
\caption{{\tt Rgemm} performance on Raspberry Pi4 ARM Cortex A72 for {\tt MPFR 512bit}, {\tt GMP 512bit} and {\tt quad-double} with/without simple OpenMP acceleration. }
\label{Rgemm2.G}
\begin{center}
\includegraphics{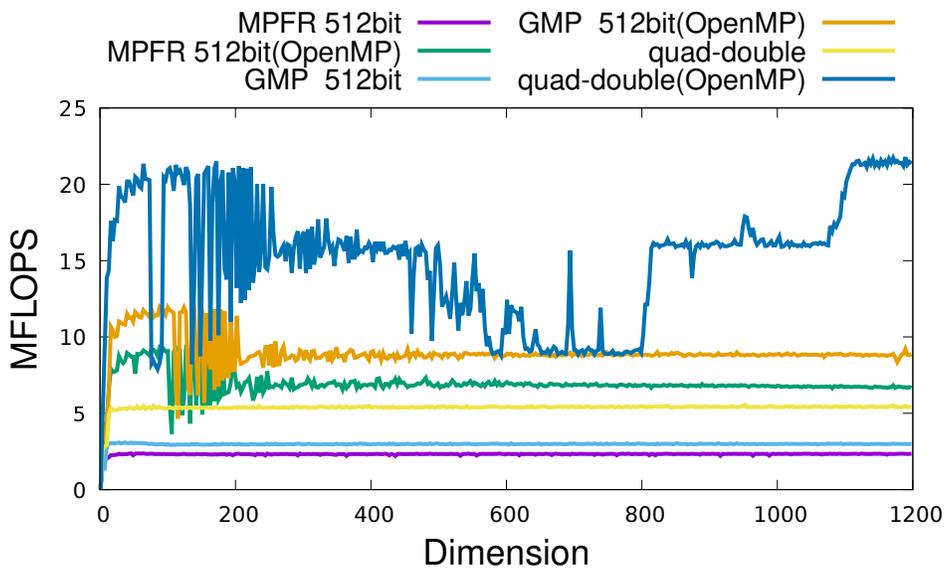}
\end{center}
\end{figure}

In Figure~\ref{Rgemm1.G}, we show the result of {\tt Rgemm} performance for {\tt \_Float128} and {\tt double-double}, and in Figure~\ref{Rgemm2.G} we show the result of {\tt Rgemm} performance for {\tt MPFR}, {\tt GMP} and {\tt quad-double} on Raspberry Pi4 ARM Cortex A72.
The peak performances of the reference {\tt Rgemm}s of {\tt \_Float128}, {\tt \_Float64x} and {\tt double-double} are 16.5 MFlops and 52.4 MFlops, respectively.
The peak performances of simple OpenMP parallelized {\tt Rgemm}s of {\tt \_Float128}, {\tt \_Float64x} and {\tt double-double} are 65.7 MFlops and 206 MFlops, respectively.
The peak performances of the reference {\tt Rgemm}s of {\tt MPFR 512bit}, {\tt GMP 512bit} and {\tt quad-double} are 2.4 MFlops, 3.1 MFlops, 5.6 MFlops, respectively.
The peak performances of simple OpenMP parallelized {\tt MPFR 512bit}, {\tt GMP 512bit} and {\tt quad-double} are 9.3 MFlops, 12.0 MFlops, 21.8 MFlops, respectively.\\

\subsubsection{{\tt Rgemm} on NVIDIA A100 80GB PCIe}

\begin{figure}
\caption{Rgemm {\tt double-double} performance on NVIDIA A100 80GB PCIe} \label{Rgemm4.E}
\begin{center}
\includegraphics{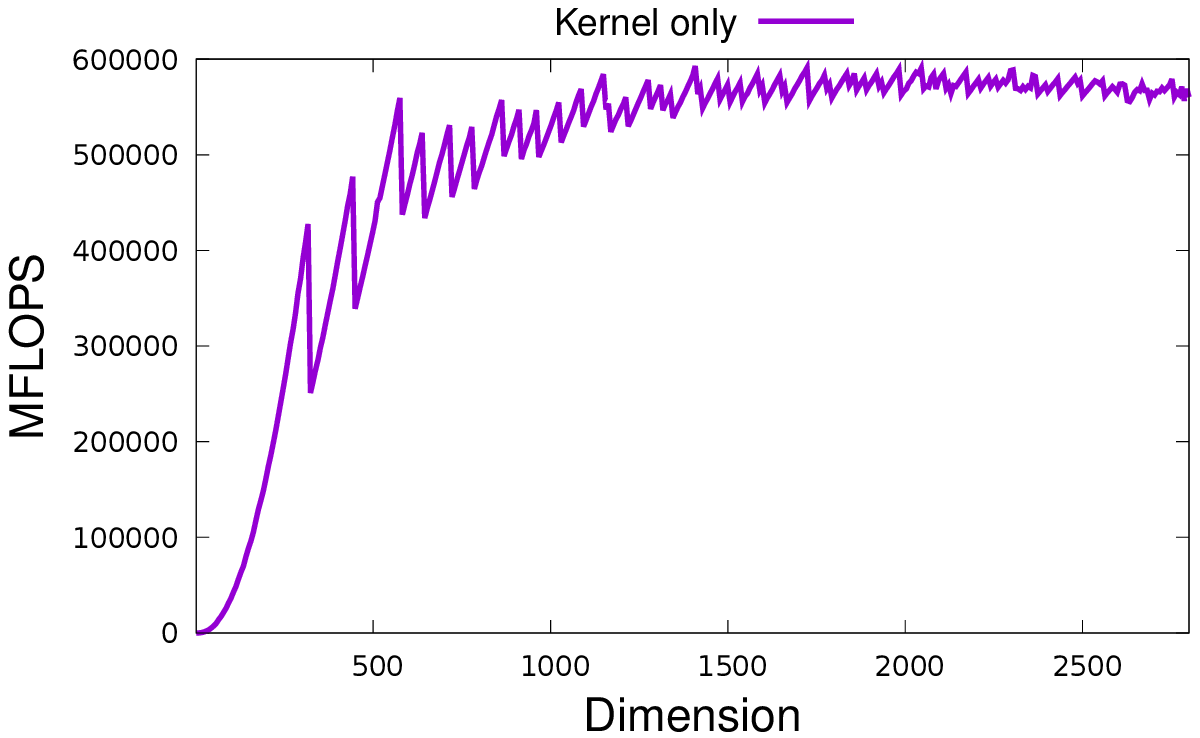}
\end{center}
\end{figure}

In Figure~\ref{Rgemm4.E}, we show the result of {\tt Rgemm} performance  {\tt double-double} on NVIDIA A100 80GB PCIe.
The peak performance of {\tt double-double} on NVIDIA A100 80GB PCIe without host-GPU transfer was 592GFlops. Unfortunately, our hardware configuration
does not equip with appropreate cooling system, we could not take a stable benchmark.
%%%%%%%

\subsubsection{{\tt Rgemm} on NVIDIA V100 40GB PCIe}
\begin{figure}
\caption{Rgemm {\tt double-double} performance on NVIDIA V100 40GB PCIe} \label{Rgemm4.F}
\begin{center}
\includegraphics{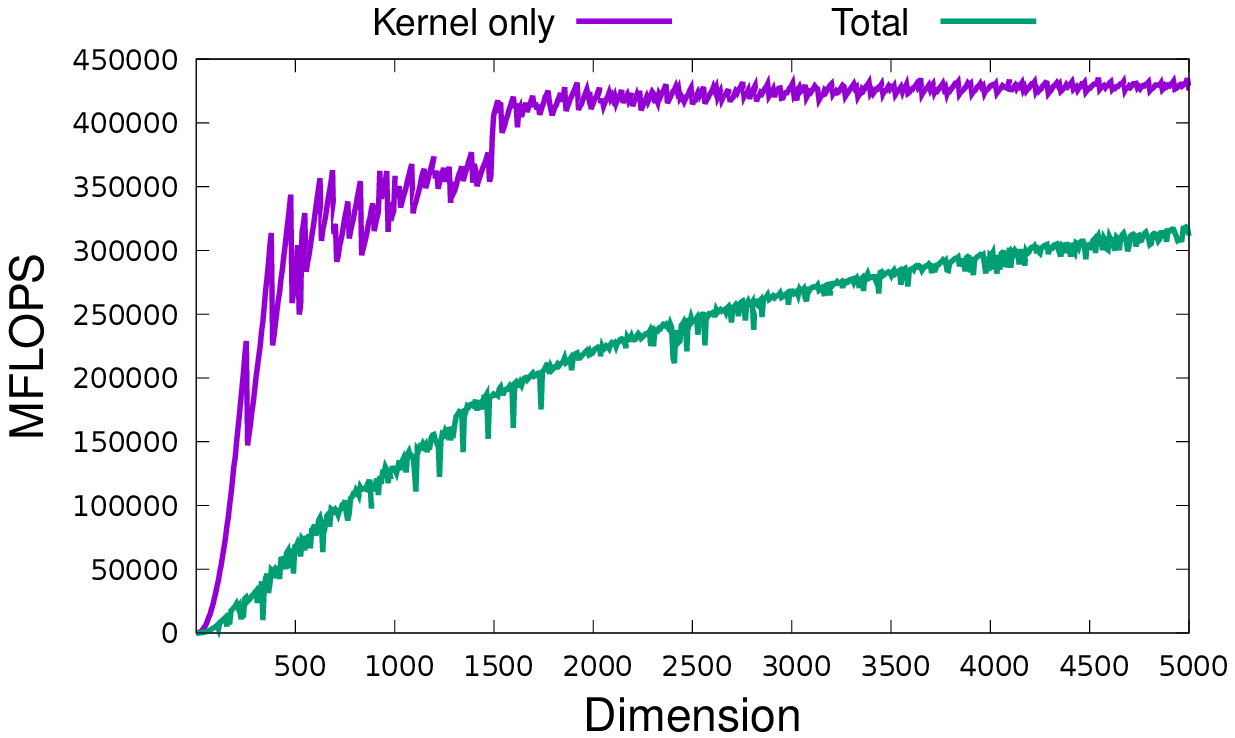}
\end{center}
\end{figure}

In Figure~\ref{Rgemm4.F}, we show the result of {\tt Rgemm} performance  {\tt double-double} on NVIDIA V100 40GB PCIe.
The peak performance of {\tt double-double} on NVIDIA V100 40GB PCIe without host-GPU transfer was 435GFlops,
and The peak performance of {\tt double-double} on NVIDIA V100 40GB PCIe without host-GPU transfer was 318GFlops.
%%%%%%%

\subsection{{\tt Rsyrk} benchmarks}
%%%%% AMD Ryzen 3970X %%%%
\subsubsection{{\tt Rsyrk} on AMD Ryzen 3970X}
\begin{figure}
\caption{ {\tt Rsyrk} performance on AMD Ryzen 3970X for {\tt \_Float128}, {\tt \_Float64x} and {\tt double-double} with/without simple OpenMP acceleration. }
\label{Rsyrk1.A}
\begin{center}
\includegraphics{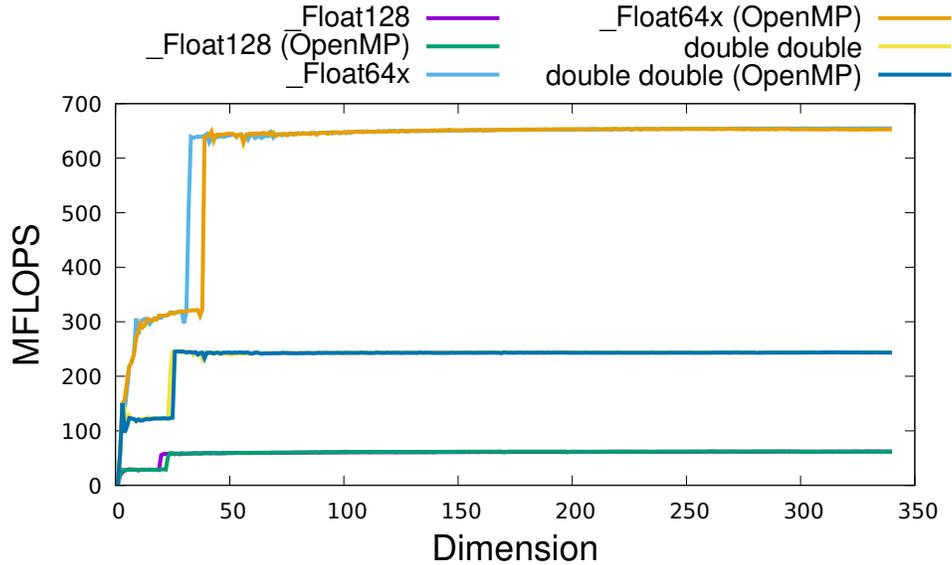}
\end{center}
\end{figure}

\begin{figure}
\caption{{\tt Rsyrk} performance on AMD Ryzen 3970X for {\tt MPFR 512bit}, {\tt GMP 512bit} and {\tt quad-double} with/without simple OpenMP acceleration. }
\label{Rsyrk2.A}
\begin{center}
\includegraphics{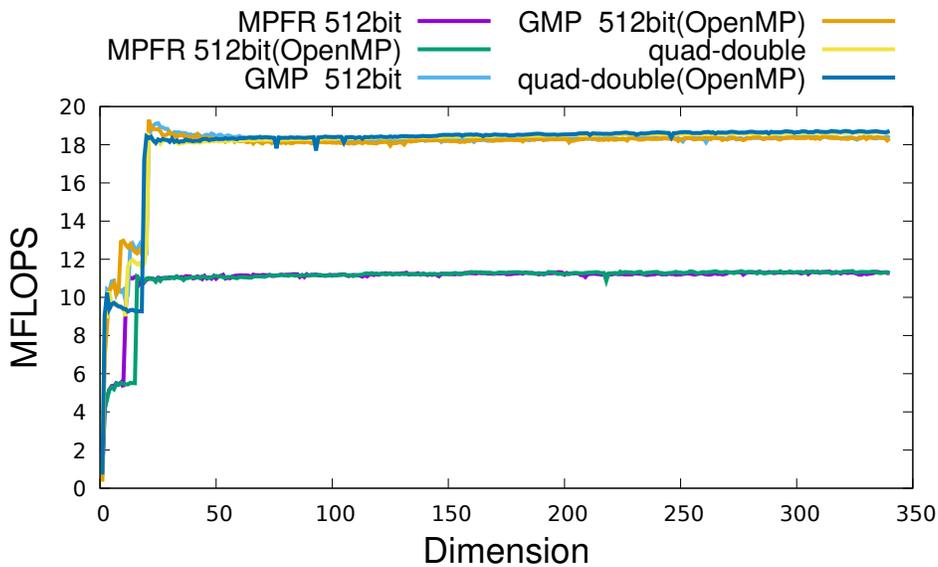}
\end{center}
\end{figure}

In Figure~\ref{Rsyrk1.A}, we show the result of {\tt Rsyrk} performance for {\tt \_Float128}, {\tt \_Float64x} and {\tt double-double}, and in Figure~\ref{Rsyrk2.A} we show the
 result of {\tt Rsyrk} performance for {\tt MPFR}, {\tt GMP} and {\tt quad-double} on AMD Ryzen 3970X.
The peak performances of the reference {\tt Rsyrk}s of {\tt \_Float128}, {\tt \_Float64x} and {\tt double-double} are 62.2 MFlops, 655 MFlops, 246 MFlops, respectively.
The peak performances of simple OpenMP parallelized {\tt Rsyrk}s of {\tt \_Float128}, {\tt \_Float64x} and {\tt double-double} are 62.3 MFlops, 654 MFlops, 245 MFlops, respectively.
The peak performances of the reference {\tt Rsyrk}s of {\tt MPFR 512bit}, {\tt GMP 512bit} and {\tt quad-double} are 11.4 MFlops, 19.1 MFlops, 18.6 MFlops, respectively.
The peak performances of simple OpenMP parallelized {\tt MPFR 512bit}, {\tt GMP 512bit} and {\tt quad-double} are 11.4 MFlops, 19.3 MFlops, 18.7 MFlops, respectively.\\

%%%%% Intel Xeon E5 2623 %%%%
\subsubsection{{\tt Rsyrk} on Intel Xeon E5 2623}
\begin{figure}
\caption{ {\tt Rsyrk} performance on Intel Xeon E5 2623 for {\tt \_Float128}, {\tt \_Float64x} and {\tt double-double} with/without simple OpenMP acceleration. }
\label{Rsyrk1.B}
\begin{center}
\includegraphics{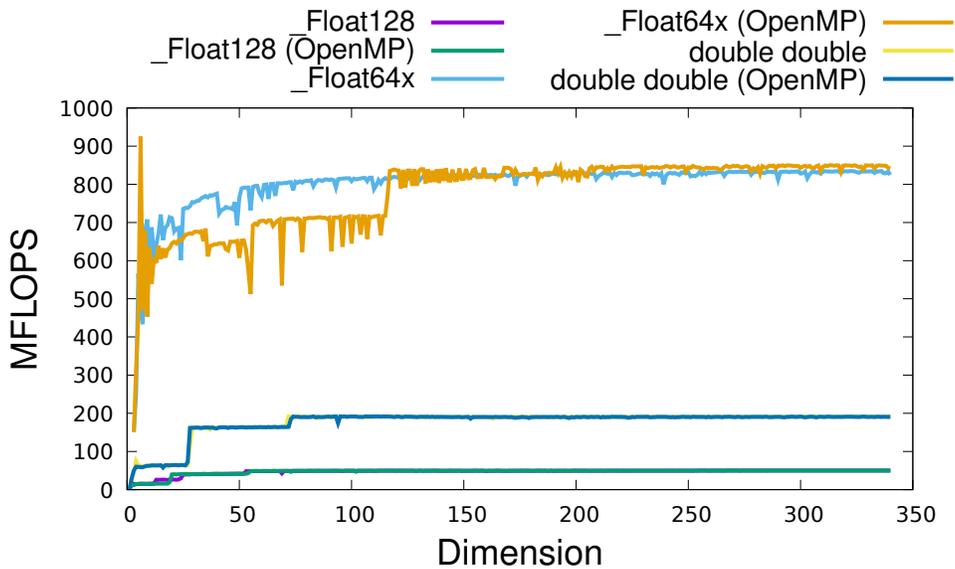}
\end{center}
\end{figure}

\begin{figure}
\caption{{\tt Rsyrk} performance on Intel Xeon E5 2623 for {\tt MPFR 512bit}, {\tt GMP 512bit} and {\tt quad-double} with/without simple OpenMP acceleration. }
\label{Rsyrk2.B}
\begin{center}
\includegraphics{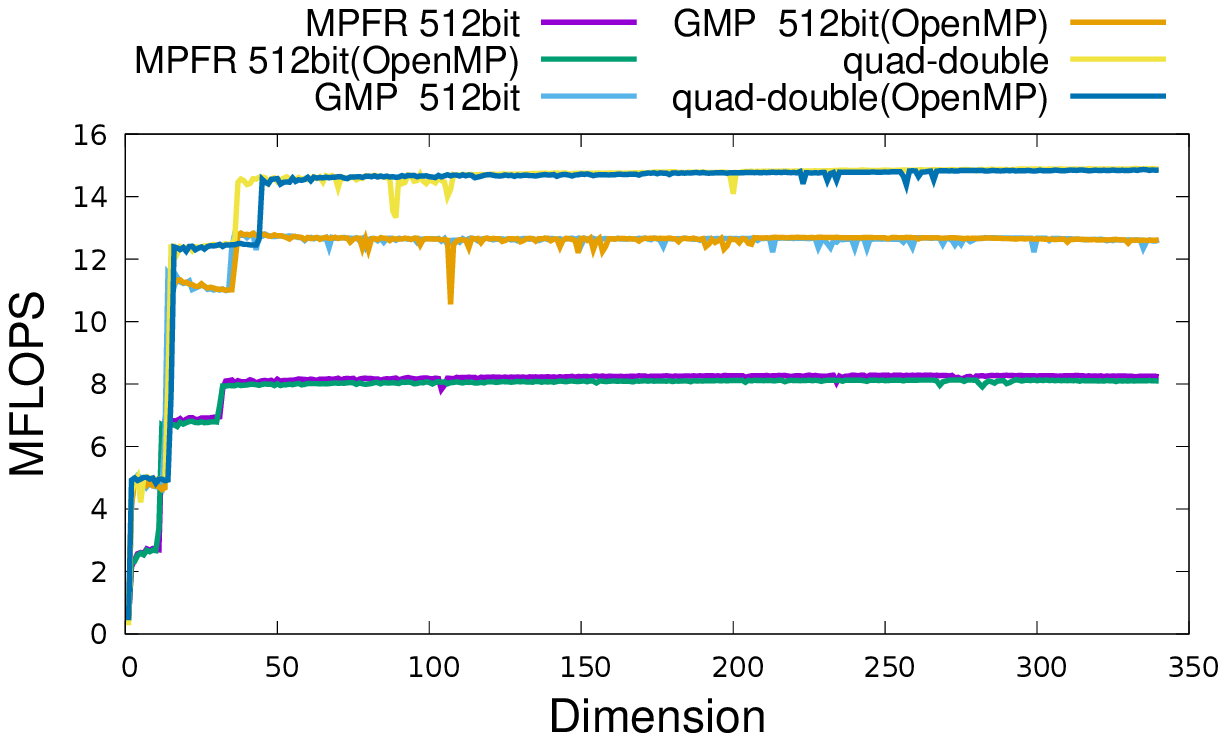}
\end{center}
\end{figure}

In Figure~\ref{Rsyrk1.B}, we show the result of {\tt Rsyrk} performance for {\tt \_Float128}, {\tt \_Float64x} and {\tt double-double}, and in Figure~\ref{Rsyrk2.B} we show the
 result of {\tt Rsyrk} performance for {\tt MPFR}, {\tt GMP} and {\tt quad-double} on Intel Xeon E5 2623.
The peak performances of the reference {\tt Rsyrk}s of {\tt \_Float128}, {\tt \_Float64x} and {\tt double-double} are 50.1 MFlops, 835 MFlops, 192 MFlops, respectively.
The peak performances of simple OpenMP parallelized {\tt Rsyrk}s of {\tt \_Float128}, {\tt \_Float64x} and {\tt double-double} are 50.2 MFlops, 925 MFlops, 192 MFlops, respectively.
The peak performances of the reference {\tt Rsyrk}s of {\tt MPFR 512bit}, {\tt GMP 512bit} and {\tt quad-double} are 8.3 MFlops, 12.8 MFlops, 14.9 MFlops, respectively.
The peak performances of simple OpenMP parallelized {\tt MPFR 512bit}, {\tt GMP 512bit} and {\tt quad-double} are 8.1 MFlops, 12.8 MFlops, 14.9 MFlops, respectively.\\

%%%%% Intel Core i5-8500B %%%%
\subsubsection{{\tt Rsyrk} on Intel Core i5-8500B}
\begin{figure}
\caption{ {\tt Rsyrk} performance on Intel Core i5-8500B for {\tt \_Float128}, {\tt \_Float64x} and {\tt double-double} with/without simple OpenMP acceleration. }
\label{Rsyrk1.C}
\begin{center}
\includegraphics{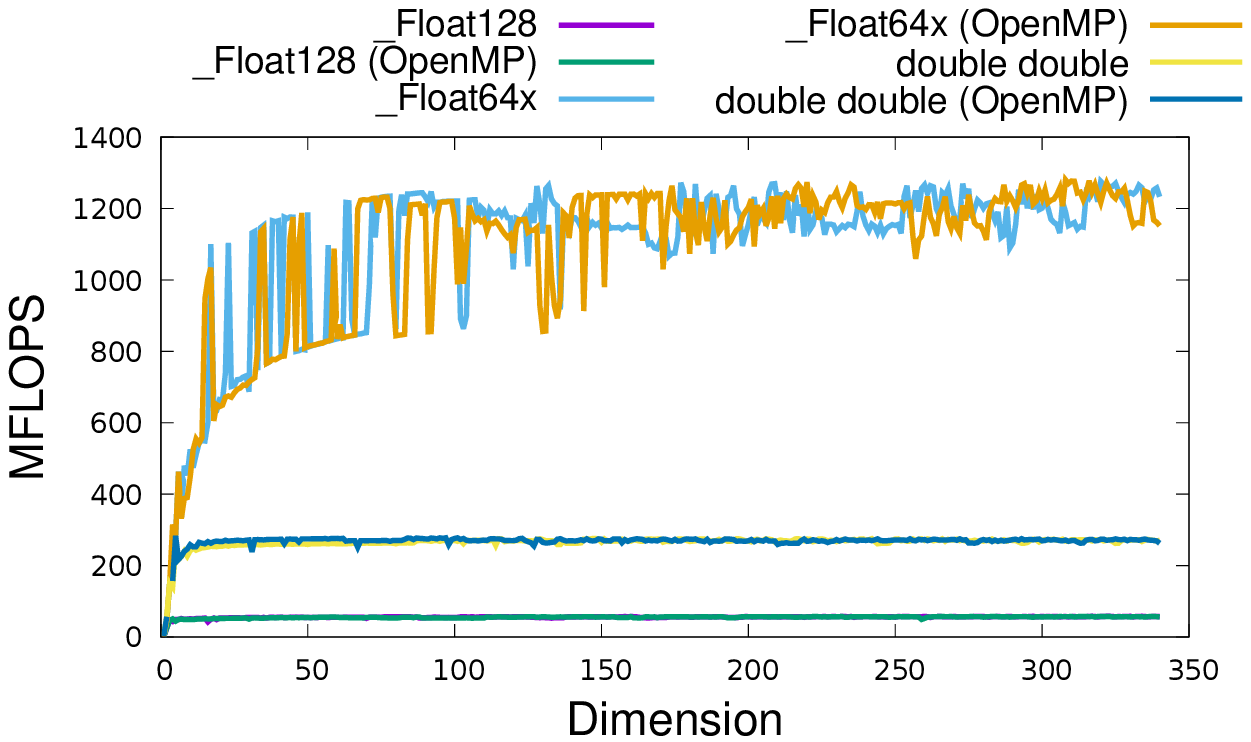}
\end{center}
\end{figure}

\begin{figure}
\caption{{\tt Rsyrk} performance on Intel Core i5-8500B for {\tt MPFR 512bit}, {\tt GMP 512bit} and {\tt quad-double} with/without simple OpenMP acceleration. }
\label{Rsyrk2.C}
\begin{center}
\includegraphics{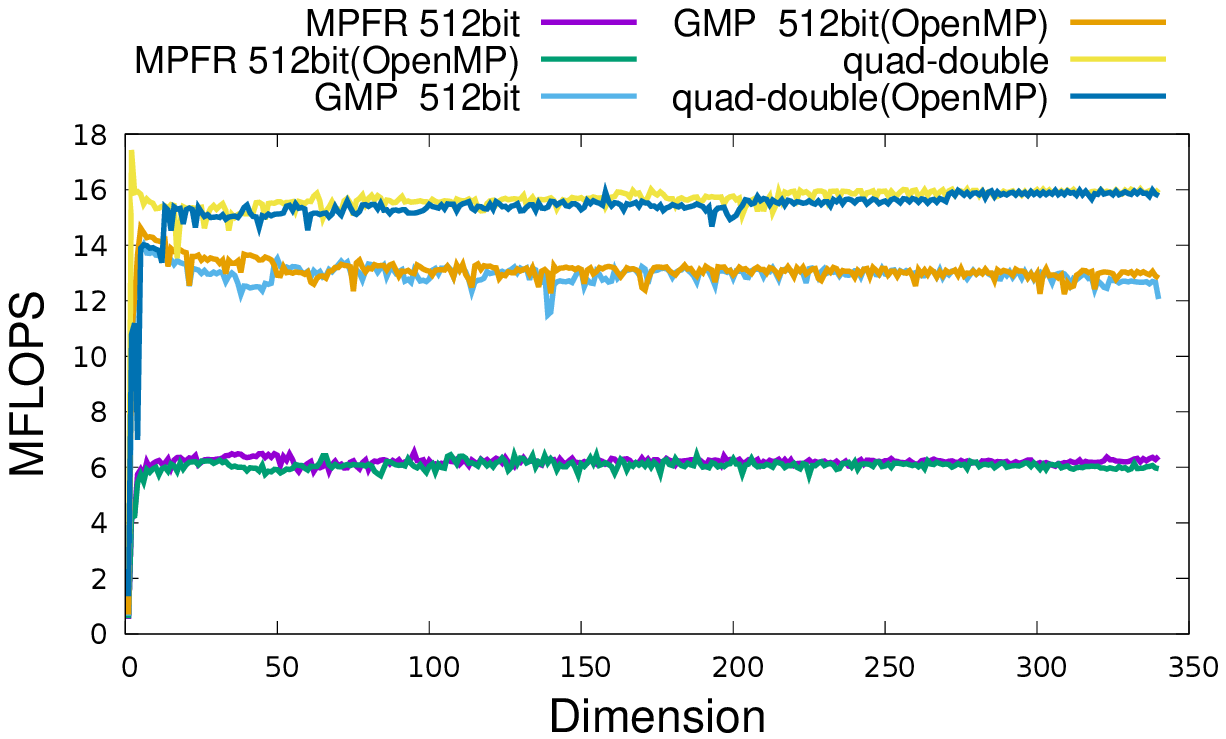}
\end{center}
\end{figure}

In Figure~\ref{Rsyrk1.C}, we show the result of {\tt Rsyrk} performance for {\tt \_Float128}, {\tt \_Float64x} and {\tt double-double}, and in Figure~\ref{Rsyrk2.C} we show the result of {\tt Rsyrk} performance for {\tt MPFR}, {\tt GMP} and {\tt quad-double} on Intel Core i5-8500B.
The peak performances of the reference {\tt Rsyrk}s of {\tt \_Float128}, {\tt \_Float64x} and {\tt double-double} are 57.8 MFlops, 1276 MFlops, 276 MFlops, respectively.
The peak performances of simple OpenMP parallelized {\tt Rsyrk}s of {\tt \_Float128}, {\tt \_Float64x} and {\tt double-double} are 57.4 MFlops, 1280 MFlops, 283 MFlops, respectively.
The peak performances of the reference {\tt Rsyrk}s of {\tt MPFR 512bit}, {\tt GMP 512bit} and {\tt quad-double} are 6.5 MFlops, 14.0 MFlops, 17.4 MFlops, respectively.
The peak performances of simple OpenMP parallelized {\tt MPFR 512bit}, {\tt GMP 512bit} and {\tt quad-double} are 6.5 MFlops, 14.6 MFlops, 16.0 MFlops, respectively.\\

%%%%% Raspberry Pi4 ARM Cortex A72 %%%%
\subsubsection{{\tt Rsyrk} on Raspberry Pi4 ARM Cortex A72}
\begin{figure}
\caption{ {\tt Rsyrk} performance on Raspberry Pi4 ARM Cortex A72 for {\tt \_Float128} and {\tt double-double} with/without simple OpenMP acceleration. }
\label{Rsyrk1.G}
\begin{center}
\includegraphics{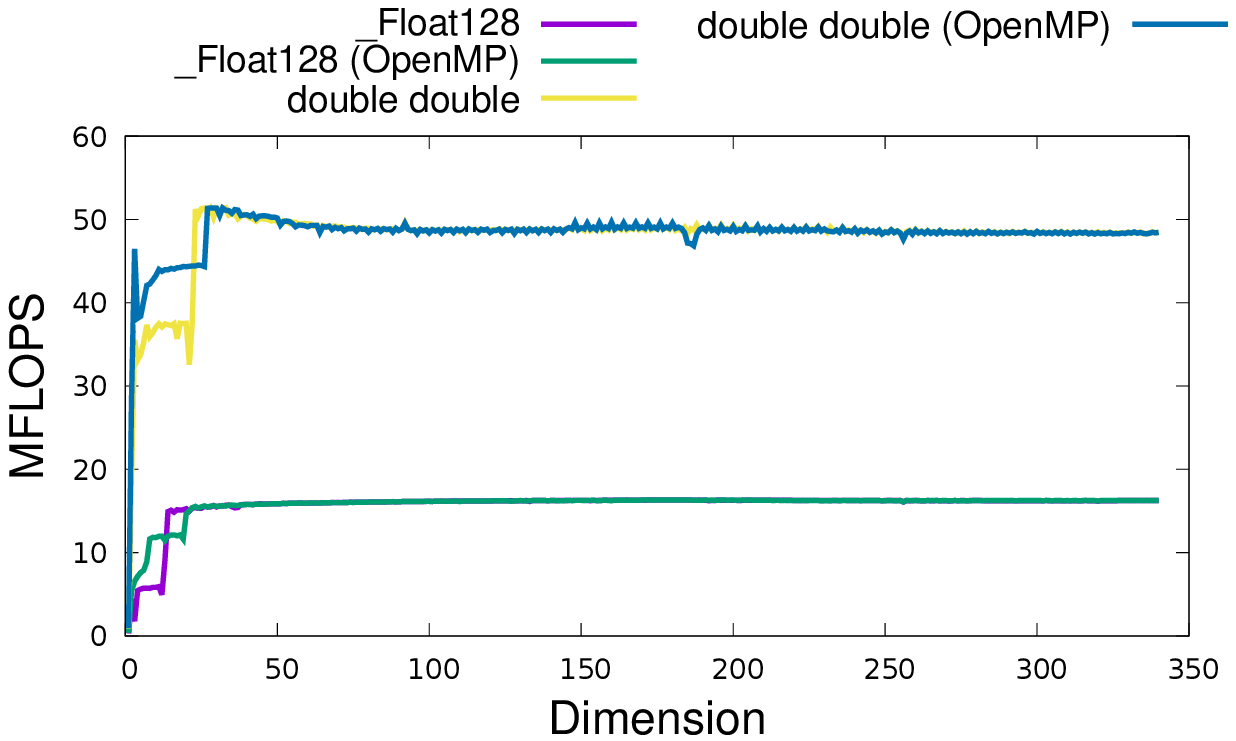}
\end{center}
\end{figure}

\begin{figure}
\caption{{\tt Rsyrk} performance on Raspberry Pi4 ARM Cortex A72 for {\tt MPFR 512bit}, {\tt GMP 512bit} and {\tt quad-double} with/without simple OpenMP acceleration. }
\label{Rsyrk2.G}
\begin{center}
\includegraphics{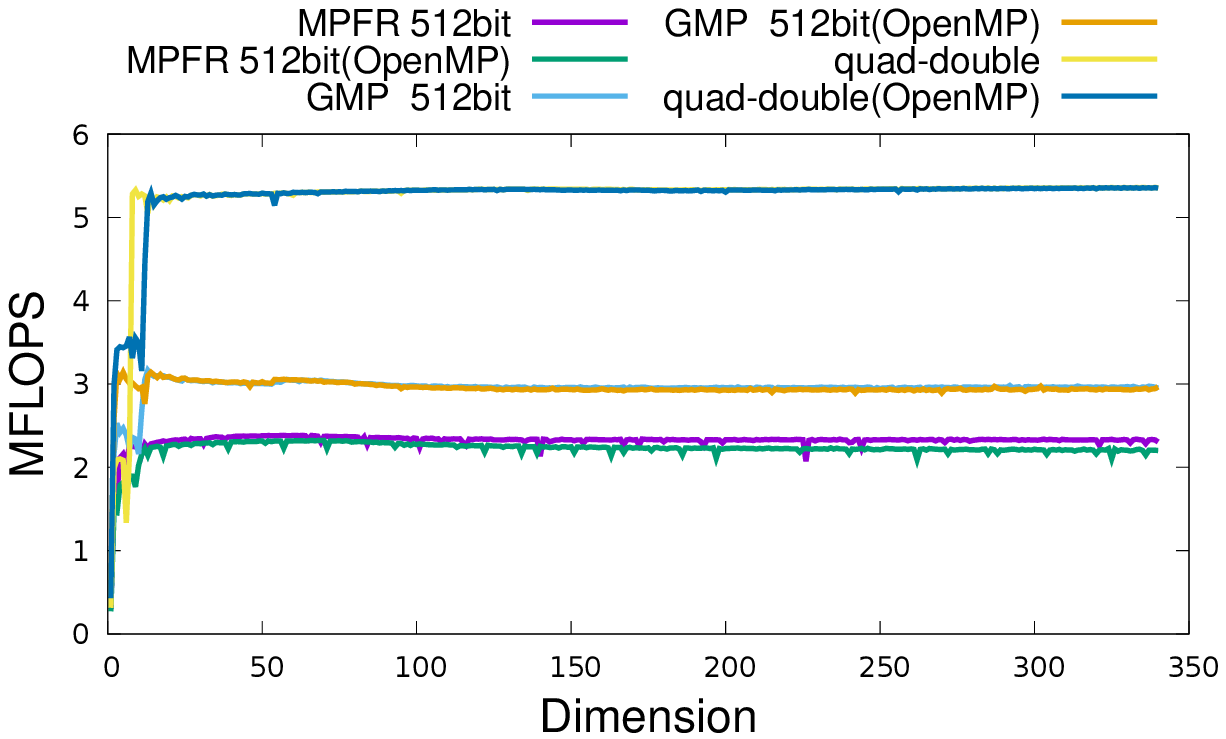}
\end{center}
\end{figure}

In Figure~\ref{Rsyrk1.G}, we show the result of {\tt Rsyrk} performance for {\tt \_Float128} and {\tt double-double}, and in Figure~\ref{Rsyrk2.G} we show the result of {\tt Rsyrk} performance for {\tt MPFR}, {\tt GMP} and {\tt quad-double} on Raspberry Pi4 ARM Cortex A72.
The peak performances of the reference {\tt Rsyrk}s of {\tt \_Float128} and {\tt double-double} are 16.3 MFlops and 51.4 MFlops, respectively.
The peak performances of simple OpenMP parallelized {\tt Rsyrk}s of {\tt \_Float128} and {\tt double-double} are 16.3 MFlops and 51.4 MFlops, respectively.
The peak performances of the reference {\tt Rsyrk}s of {\tt MPFR 512bit}, {\tt GMP 512bit} and {\tt quad-double} are 2.4 MFlops, 3.2 MFlops, 5.4 MFlops, respectively.
The peak performances of simple OpenMP parallelized {\tt MPFR 512bit}, {\tt GMP 512bit} and {\tt quad-double} are 2.3 MFlops, 3.1 MFlops, 5.4 MFlops, respectively.\\

\subsubsection{{\tt Rsyrk} on NVIDIA V100}

\begin{figure}
\caption{Rsyrk {\tt double-double} performance on NVIDIA V100 40GB PCIe} \label{Rsyrk4.F}
\begin{center}
\includegraphics{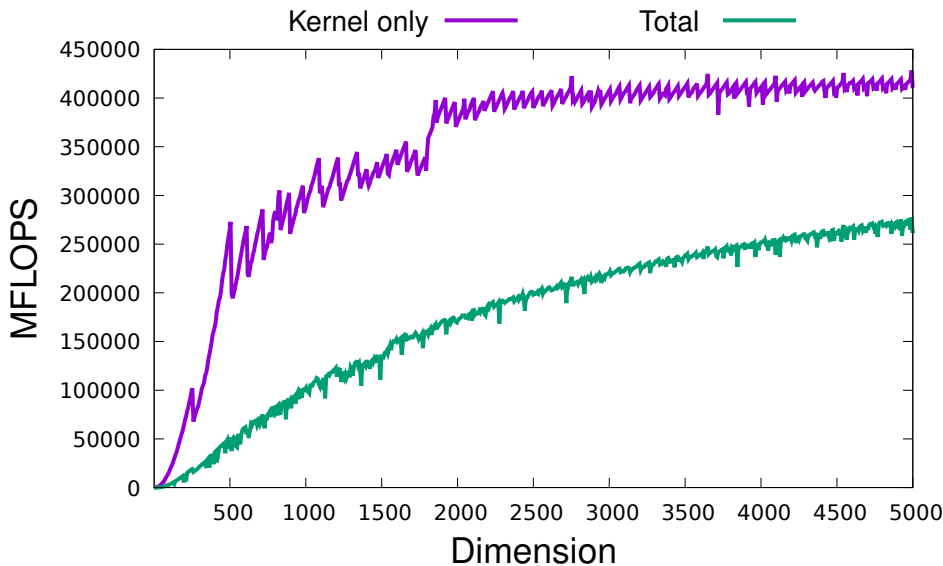}
\end{center}
\end{figure}

%%%%%%
In Figure~\ref{Rsyrk4.F}, we show the result of {\tt Rsyrk} performance  {\tt double-double} on NVIDIA V100 40GB PCIe.
The peak performance of {\tt double-double} on NVIDIA V100 40GB PCIe without host-GPU transfer was 428GFlops,
and The peak performance of {\tt double-double} on NVIDIA V100 40GB PCIe without host-GPU transfer was 277GFlops.
%%%%%%%

\subsection{{\tt Rgemv} benchmarks}
%%%%% AMD Ryzen 3970X %%%%
\subsubsection{{\tt Rgemv} on AMD Ryzen 3970X}
\begin{figure}
\caption{ {\tt Rgemv} performance on AMD Ryzen 3970X for {\tt \_Float128}, {\tt \_Float64x} and {\tt double-double} with/without simple OpenMP acceleration. }
\label{Rgemv1.A}
\begin{center}
\includegraphics{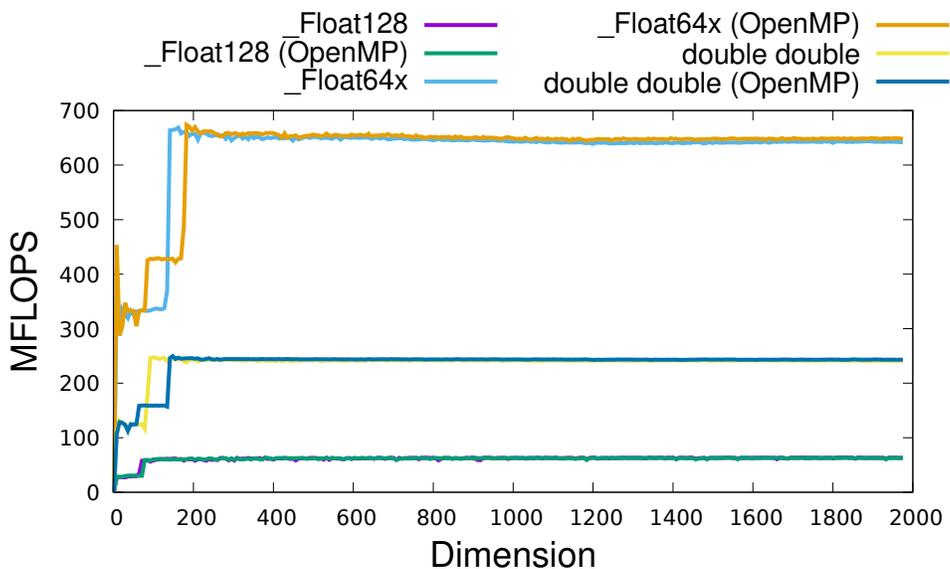}
\end{center}
\end{figure}

\begin{figure}
\caption{{\tt Rgemv} performance on AMD Ryzen 3970X for {\tt MPFR 512bit}, {\tt GMP 512bit} and {\tt quad-double} with/without simple OpenMP acceleration. }
\label{Rgemv2.A}
\begin{center}
\includegraphics{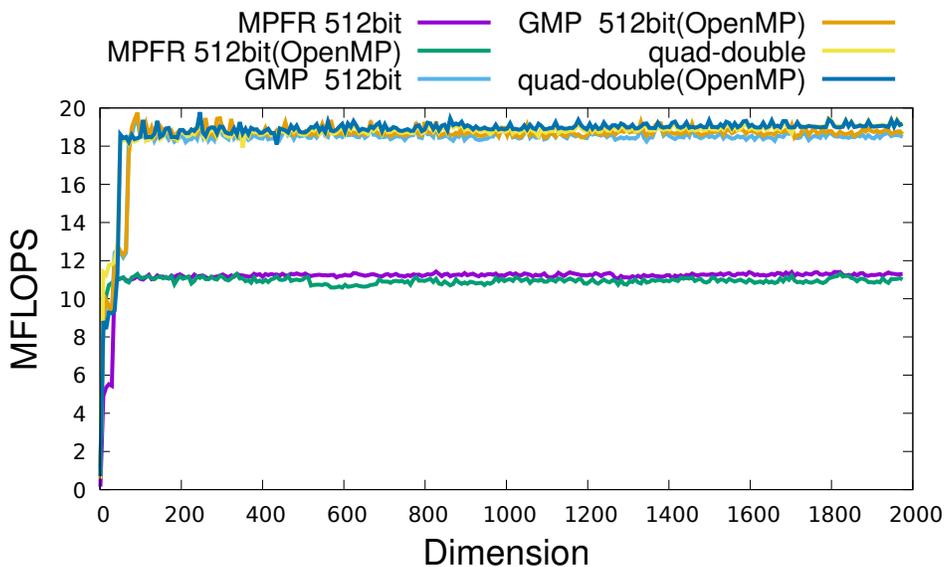}
\end{center}
\end{figure}

In Figure~\ref{Rgemv1.A}, we show the result of {\tt Rgemv} performance for {\tt \_Float128}, {\tt \_Float64x} and {\tt double-double}, and in Figure~\ref{Rgemv2.A} we show
 the result of {\tt Rgemv} performance for {\tt MPFR}, {\tt GMP} and {\tt quad-double} on AMD Ryzen 3970X.
The peak performances of the reference {\tt Rgemv}s of {\tt \_Float128}, {\tt \_Float64x} and {\tt double-double} are 63.6 MFlops, 669 MFlops, 247 MFlops, respectively.
The peak performances of simple OpenMP parallelized {\tt Rgemv}s of {\tt \_Float128}, {\tt \_Float64x} and {\tt double-double} are 63.3 MFlops, 674 MFlops, 250 MFlops, resp
ectively.
The peak performances of the reference {\tt Rgemv}s of {\tt MPFR 512bit}, {\tt GMP 512bit} and {\tt quad-double} are 11.4 MFlops, 19.4 MFlops, 19.6 MFlops, respectively.
The peak performances of simple OpenMP parallelized {\tt MPFR 512bit}, {\tt GMP 512bit} and {\tt quad-double} are 11.3 MFlops, 19.8 MFlops, 19.8 MFlops, respectively.\\

%%%%% Intel Xeon E5 2623 %%%%
\subsubsection{{\tt Rgemv} on Intel Xeon E5 2623}
\begin{figure}
\caption{ {\tt Rgemv} performance on Intel Xeon E5 2623 for {\tt \_Float128}, {\tt \_Float64x} and {\tt double-double} with/without simple OpenMP acceleration. }
\label{Rgemv1.B}
\begin{center}
\includegraphics{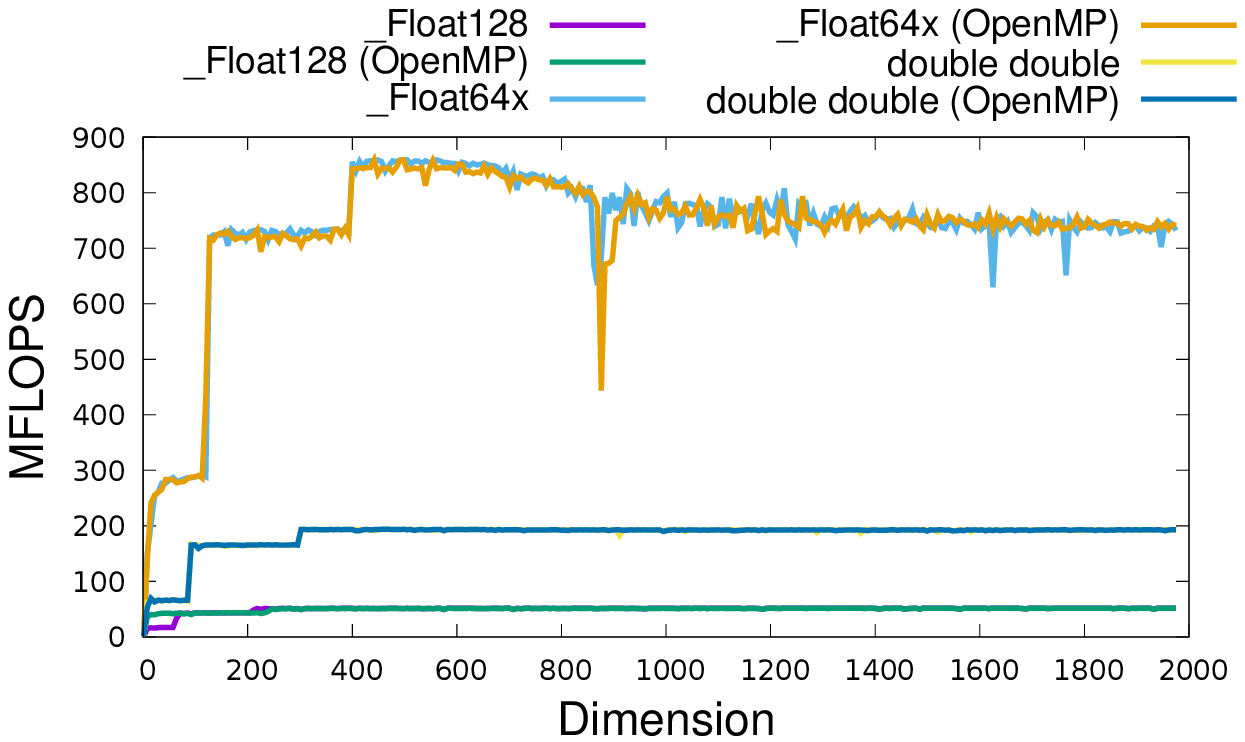}
\end{center}
\end{figure}

\begin{figure}
\caption{{\tt Rgemv} performance on Intel Xeon E5 2623 for {\tt MPFR 512bit}, {\tt GMP 512bit} and {\tt quad-double} with/without simple OpenMP acceleration. }
\label{Rgemv2.B}
\begin{center}
\includegraphics{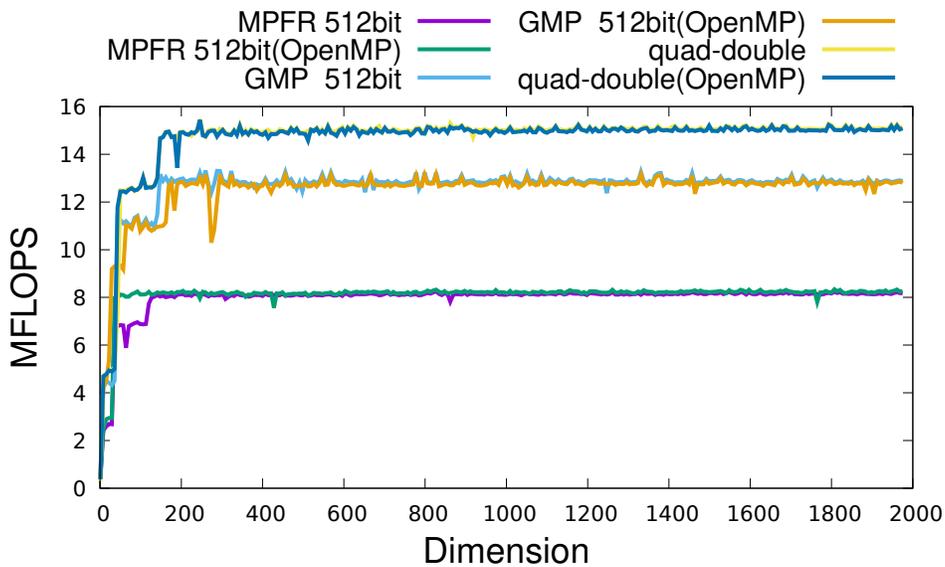}
\end{center}
\end{figure}

In Figure~\ref{Rgemv1.B}, we show the result of {\tt Rgemv} performance for {\tt \_Float128}, {\tt \_Float64x} and {\tt double-double}, and in Figure~\ref{Rgemv2.B} we show
 the result of {\tt Rgemv} performance for {\tt MPFR}, {\tt GMP} and {\tt quad-double} on Intel Xeon E5 2623.
The peak performances of the reference {\tt Rgemv}s of {\tt \_Float128}, {\tt \_Float64x} and {\tt double-double} are 52.0 MFlops, 859 MFlops, 194 MFlops, respectively.
The peak performances of simple OpenMP parallelized {\tt Rgemv}s of {\tt \_Float128}, {\tt \_Float64x} and {\tt double-double} are 51.8 MFlops, 859 MFlops, 194 MFlops, respectively.
The peak performances of the reference {\tt Rgemv}s of {\tt MPFR 512bit}, {\tt GMP 512bit} and {\tt quad-double} are 8.3 MFlops, 13.3 MFlops, 15.5 MFlops, respectively.
The peak performances of simple OpenMP parallelized {\tt MPFR 512bit}, {\tt GMP 512bit} and {\tt quad-double} are 8.3 MFlops, 13.2 MFlops, 15.4 MFlops, respectively.\\

%%%%% Intel Core i5-8500B %%%%
\subsubsection{{\tt Rgemv} on Intel Core i5-8500B}
\begin{figure}
\caption{ {\tt Rgemv} performance on Intel Core i5-8500B for {\tt \_Float128}, {\tt \_Float64x} and {\tt double-double} with/without simple OpenMP acceleration. }
\label{Rgemv1.C}
\begin{center}
\includegraphics{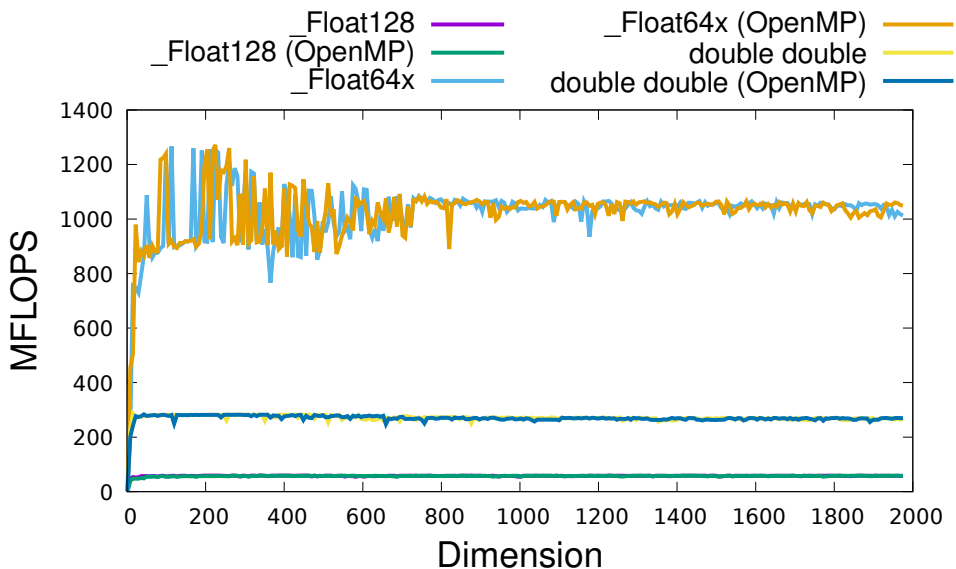}
\end{center}
\end{figure}

\begin{figure}
\caption{{\tt Rgemv} performance on Intel Core i5-8500B for {\tt MPFR 512bit}, {\tt GMP 512bit} and {\tt quad-double} with/without simple OpenMP acceleration. }
\label{Rgemv2.C}
\begin{center}
\includegraphics{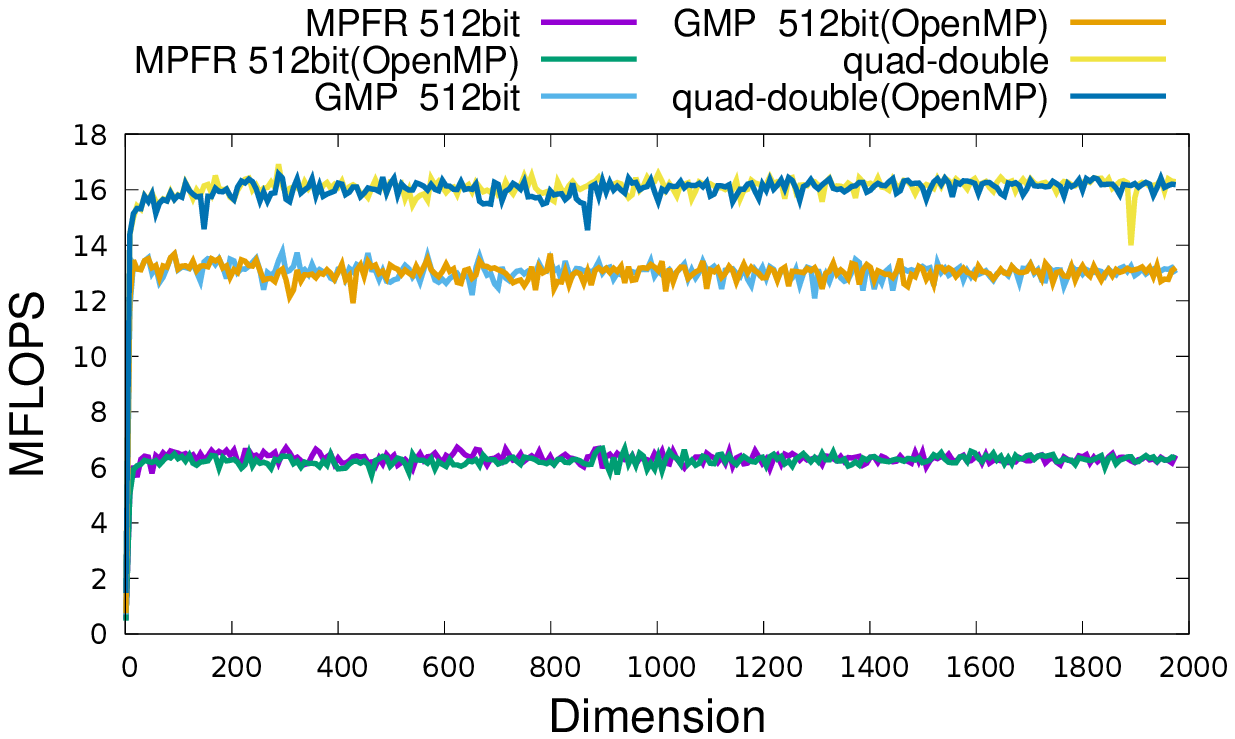}
\end{center}
\end{figure}

In Figure~\ref{Rgemv1.C}, we show the result of {\tt Rgemv} performance for {\tt \_Float128}, {\tt \_Float64x} and {\tt double-double}, and in Figure~\ref{Rgemv2.C} we show
 the result of {\tt Rgemv} performance for {\tt MPFR}, {\tt GMP} and {\tt quad-double} on Intel Core i5-8500B.
The peak performances of the reference {\tt Rgemv}s of {\tt \_Float128}, {\tt \_Float64x} and {\tt double-double} are 59.1 MFlops, 1266 MFlops, 288 MFlops, respectively.
The peak performances of simple OpenMP parallelized {\tt Rgemv}s of {\tt \_Float128}, {\tt \_Float64x} and {\tt double-double} are 59.0 MFlops, 1272 MFlops, 283 MFlops, respectively.
The peak performances of the reference {\tt Rgemv}s of {\tt MPFR 512bit}, {\tt GMP 512bit} and {\tt quad-double} are 6.7 MFlops, 13.8 MFlops, 16.9 MFlops, respectively.
The peak performances of simple OpenMP parallelized {\tt MPFR 512bit}, {\tt GMP 512bit} and {\tt quad-double} are 6.7 MFlops, 13.7 MFlops, 16.6 MFlops, respectively.\\

%%%%% Raspberry Pi4 ARM Cortex A72 %%%%
\subsubsection{{\tt Rgemv} on Raspberry Pi4 ARM Cortex A72}
\begin{figure}
\caption{ {\tt Rgemv} performance on Raspberry Pi4 ARM Cortex A72 for {\tt \_Float128} and {\tt double-double} with/without simple OpenMP acceleration. }
\label{Rgemv1.G}
\begin{center}
\includegraphics{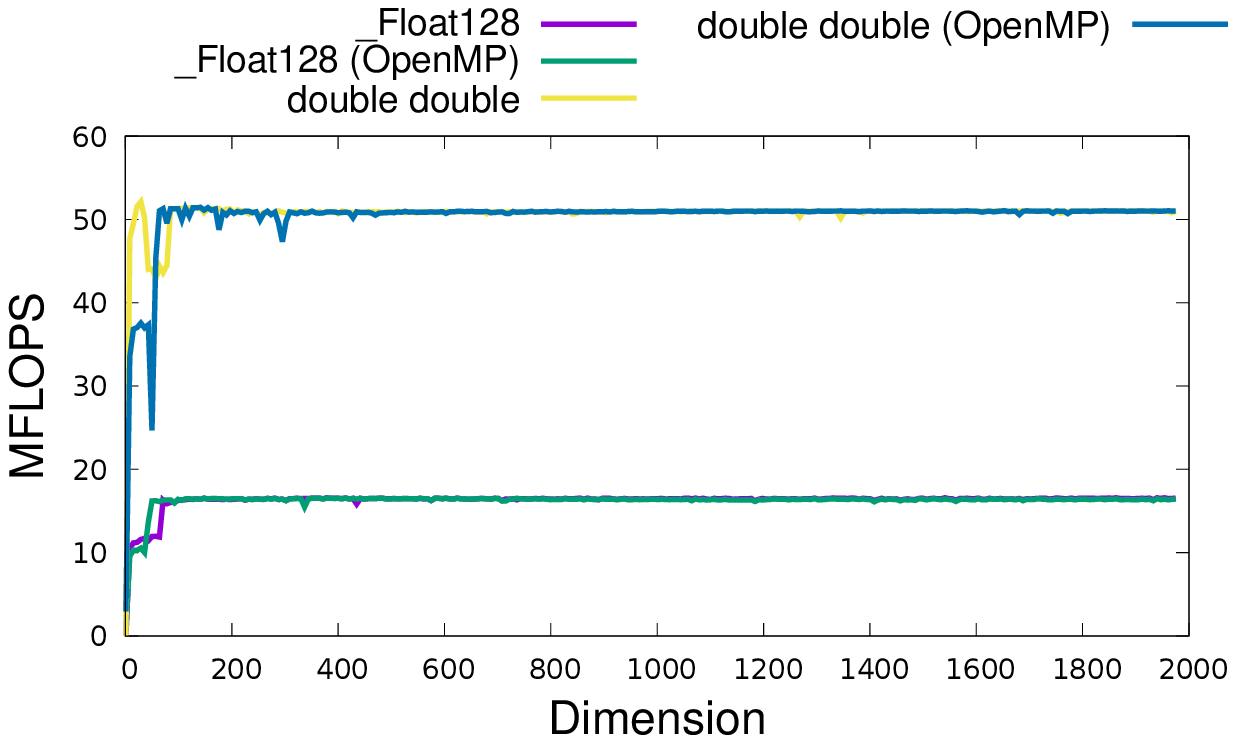}
\end{center}
\end{figure}

\begin{figure}
\caption{{\tt Rgemv} performance on Raspberry Pi4 ARM Cortex A72 for {\tt MPFR 512bit}, {\tt GMP 512bit} and {\tt quad-double} with/without simple OpenMP acceleration. }
\label{Rgemv2.G}
\begin{center}
\includegraphics{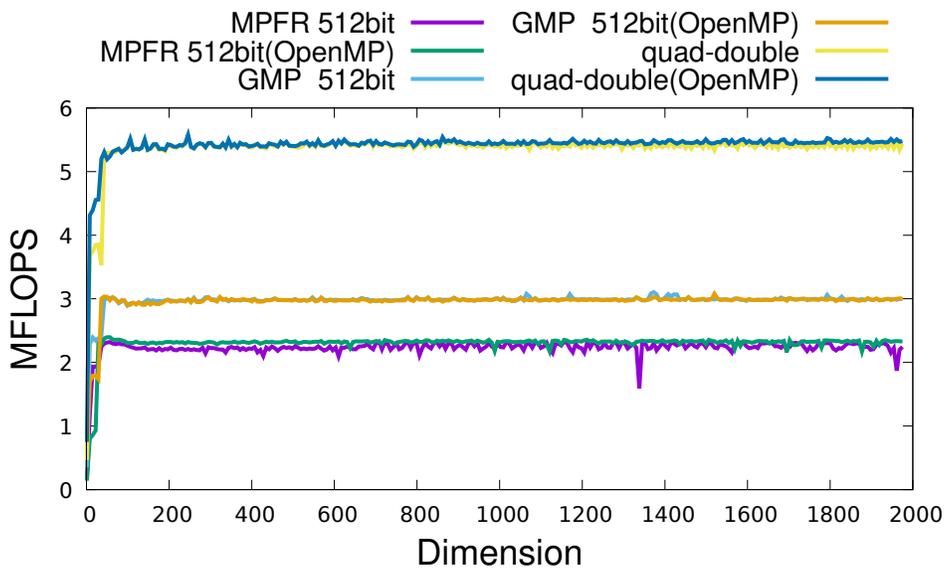}
\end{center}
\end{figure}

In Figure~\ref{Rgemv1.G}, we show the result of {\tt Rgemv} performance for {\tt \_Float128} and {\tt double-double}, and in Figure~\ref{Rgemv2.G} we show the result of {\tt Rgemv} performance for {\tt MPFR}, {\tt GMP} and {\tt quad-double} on Raspberry Pi4 ARM Cortex A72.
The peak performances of the reference {\tt Rgemv}s of {\tt \_Float128} and {\tt double-double} are 16.6 MFlops and 52.1 MFlops, respectively.
The peak performances of simple OpenMP parallelized {\tt Rgemv}s of {\tt \_Float128} and {\tt double-double} are 16.6 MFlops and 51.5 MFlops, respectively.
The peak performances of the reference {\tt Rgemv}s of {\tt MPFR 512bit}, {\tt GMP 512bit} and {\tt quad-double} are 2.3 MFlops, 3.1 MFlops, 5.6 MFlops, respectively.
The peak performances of simple OpenMP parallelized {\tt MPFR 512bit}, {\tt GMP 512bit} and {\tt quad-double} are 2.4 MFlops, 3.1 MFlops, 5.6 MFlops, respectively.\\

\subsection{{\tt Rgetrf} benchmarks}
%%%%% AMD Ryzen 3970X %%%%
\subsubsection{{\tt Rgetrf} on AMD Ryzen 3970X}
\begin{figure}
\caption{ {\tt Rgetrf} performance on AMD Ryzen 3970X for {\tt \_Float128}, {\tt \_Float64x} and {\tt double-double} with/without simple OpenMP acceleration. }
\label{Rgetrf1.A}
\begin{center}
\includegraphics{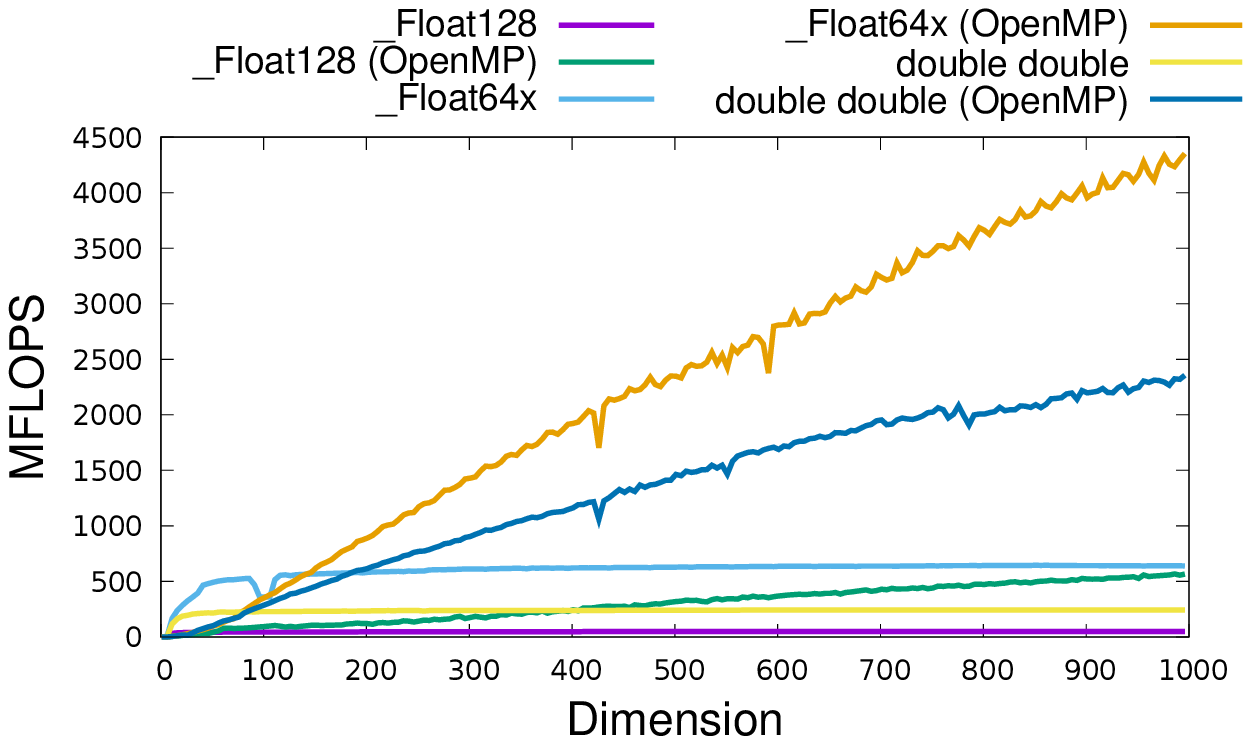}
\end{center}
\end{figure}

\begin{figure}
\caption{{\tt Rgetrf} performance on AMD Ryzen 3970X for {\tt MPFR 512bit}, {\tt GMP 512bit} and {\tt quad-double} with/without simple OpenMP acceleration. }
\label{Rgetrf2.A}
\begin{center}
\includegraphics{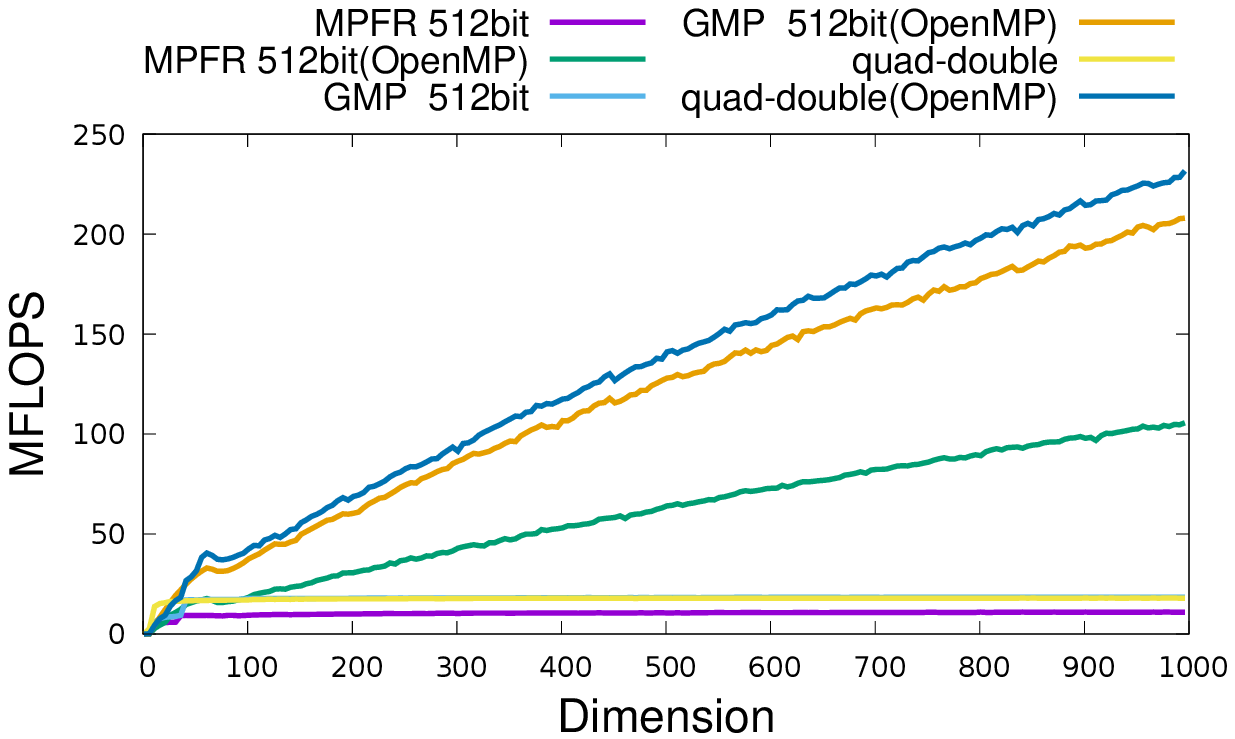}
\end{center}
\end{figure}

In Figure~\ref{Rgetrf1.A}, we show the result of {\tt Rgetrf} performance for {\tt \_Float128}, {\tt \_Float64x} and {\tt double-double}, and in Figure~\ref{Rgetrf2.A} we show the result of {\tt Rgetrf} performance for {\tt MPFR}, {\tt GMP} and {\tt quad-double} on AMD Ryzen 3970X.
The peak performances of the reference {\tt Rgetrf}s of {\tt \_Float128}, {\tt \_Float64x} and {\tt double-double} are 48.5 MFlops, 645 MFlops, 242 MFlops, respectively.
The peak performances of simple OpenMP parallelized {\tt Rgetrf}s of {\tt \_Float128}, {\tt \_Float64x} and {\tt double-double} are 567 MFlops, 4352 MFlops, 2355 MFlops, respectively.
The peak performances of the reference {\tt Rgetrf}s of {\tt MPFR 512bit}, {\tt GMP 512bit} and {\tt quad-double} are 11.0 MFlops, 18.4 MFlops, 17.9 MFlops, respectively.
The peak performances of simple OpenMP parallelized {\tt MPFR 512bit}, {\tt GMP 512bit} and {\tt quad-double} are 106 MFlops, 208 MFlops, 232 MFlops, respectively.\\

%%%%% Intel Xeon E5 2623 %%%%
\subsubsection{{\tt Rgetrf} on Intel Xeon E5 2623}
\begin{figure}
\caption{ {\tt Rgetrf} performance on Intel Xeon E5 2623 for {\tt \_Float128}, {\tt \_Float64x} and {\tt double-double} with/without simple OpenMP acceleration. }
\label{Rgetrf1.B}
\begin{center}
\includegraphics{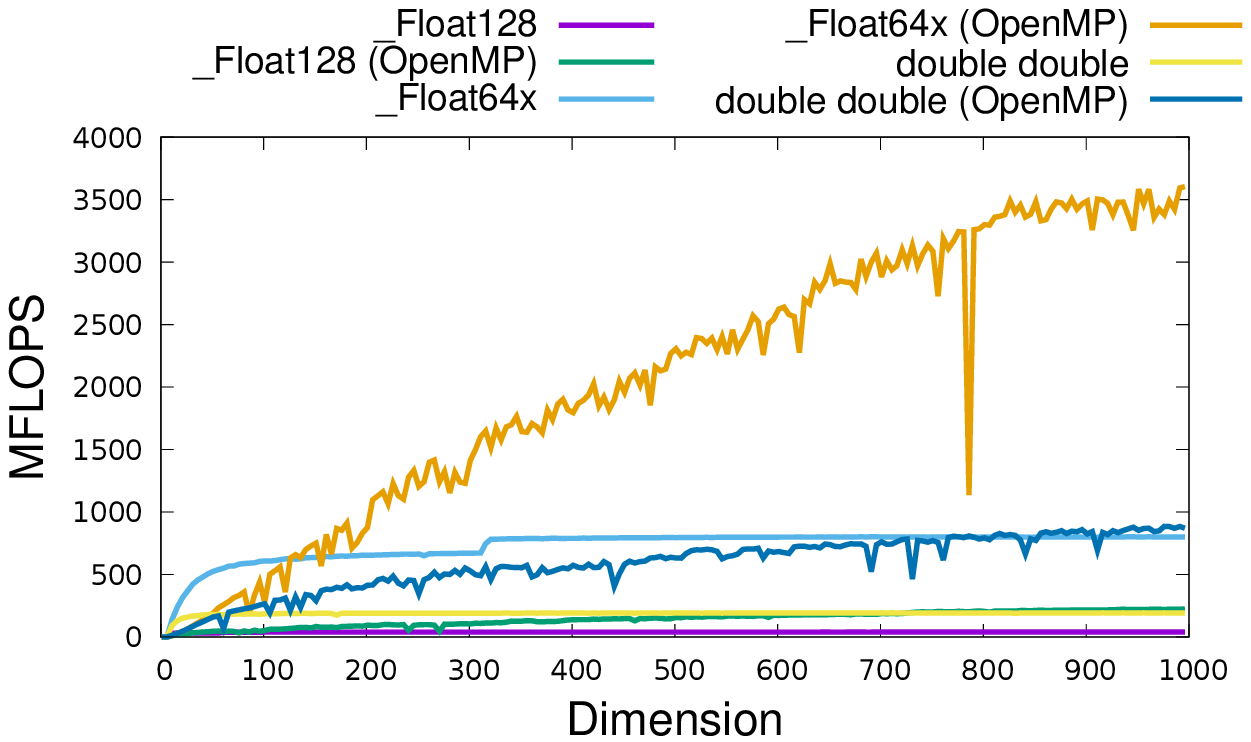}
\end{center}
\end{figure}

\begin{figure}
\caption{{\tt Rgetrf} performance on Intel Xeon E5 2623 for {\tt MPFR 512bit}, {\tt GMP 512bit} and {\tt quad-double} with/without simple OpenMP acceleration. }
\label{Rgetrf2.B}
\begin{center}
\includegraphics{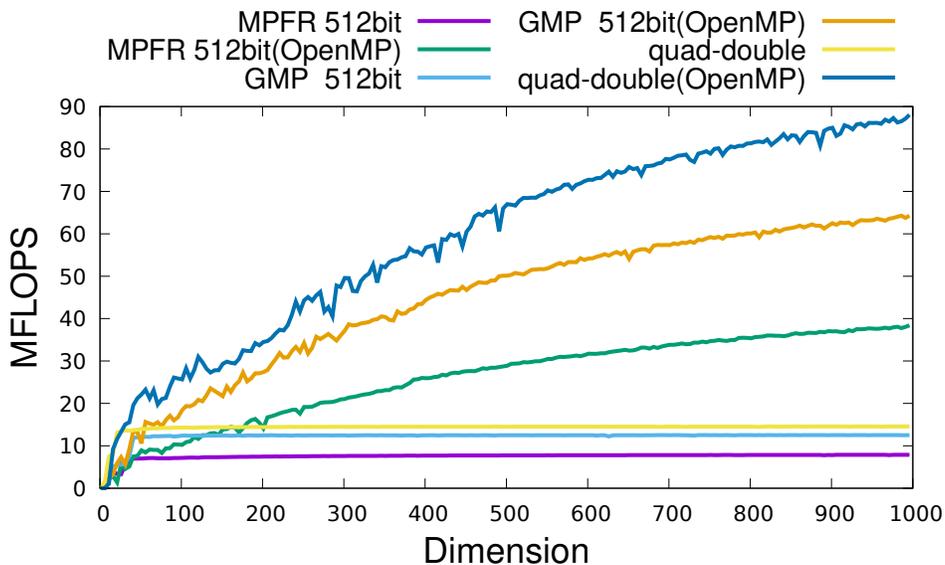}
\end{center}
\end{figure}

In Figure~\ref{Rgetrf1.B}, we show the result of {\tt Rgetrf} performance for {\tt \_Float128}, {\tt \_Float64x} and {\tt double-double}, and in Figure~\ref{Rgetrf2.B} we show the result of {\tt Rgetrf} performance for {\tt MPFR}, {\tt GMP} and {\tt quad-double} on Intel Xeon E5 2623.
The peak performances of the reference {\tt Rgetrf}s of {\tt \_Float128}, {\tt \_Float64x} and {\tt double-double} are 39.8 MFlops, 802 MFlops, 193 MFlops, respectively.
The peak performances of simple OpenMP parallelized {\tt Rgetrf}s of {\tt \_Float128}, {\tt \_Float64x} and {\tt double-double} are 224 MFlops, 3604 MFlops, 885 MFlops, respectively.
The peak performances of the reference {\tt Rgetrf}s of {\tt MPFR 512bit}, {\tt GMP 512bit} and {\tt quad-double} are 7.9 MFlops, 12.5 MFlops, 14.6 MFlops, respectively.
The peak performances of simple OpenMP parallelized {\tt MPFR 512bit}, {\tt GMP 512bit} and {\tt quad-double} are 38.4 MFlops, 64.3 MFlops, 88.1 MFlops, respectively.\\

%%%%% Intel Core i5-8500B %%%%
\subsubsection{{\tt Rgetrf} on Intel Core i5-8500B}
\begin{figure}
\caption{ {\tt Rgetrf} performance on Intel Core i5-8500B for {\tt \_Float128}, {\tt \_Float64x} and {\tt double-double} with/without simple OpenMP acceleration. }
\label{Rgetrf1.C}
\begin{center}
\includegraphics{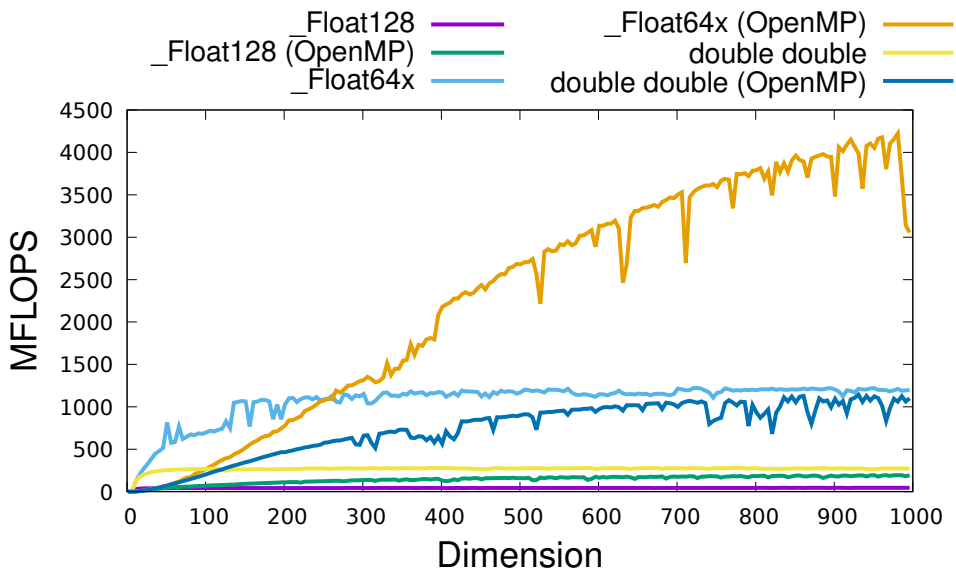}
\end{center}
\end{figure}

\begin{figure}
\caption{{\tt Rgetrf} performance on Intel Core i5-8500B for {\tt MPFR 512bit}, {\tt GMP 512bit} and {\tt quad-double} with/without simple OpenMP acceleration. }
\label{Rgetrf2.C}
\begin{center}
\includegraphics{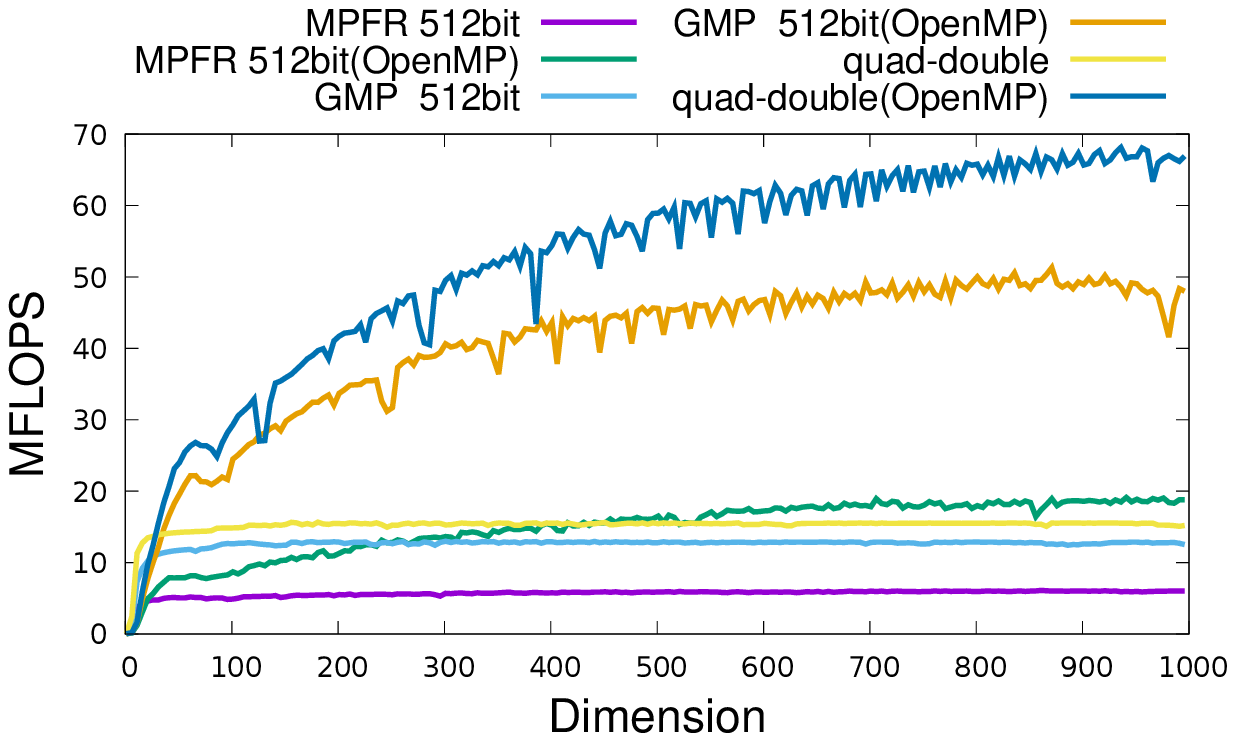}
\end{center}
\end{figure}

In Figure~\ref{Rgetrf1.C}, we show the result of {\tt Rgetrf} performance for {\tt \_Float128}, {\tt \_Float64x} and {\tt double-double}, and in Figure~\ref{Rgetrf2.C} we show the result of {\tt Rgetrf} performance for {\tt MPFR}, {\tt GMP} and {\tt quad-double} on Intel Core i5-8500B.
The peak performances of the reference {\tt Rgetrf}s of {\tt \_Float128}, {\tt \_Float64x} and {\tt double-double} are 44.8 MFlops, 1222 MFlops, 281 MFlops, respectively.
The peak performances of simple OpenMP parallelized {\tt Rgetrf}s of {\tt \_Float128}, {\tt \_Float64x} and {\tt double-double} are 195 MFlops, 4223 MFlops, 1142 MFlops, respectively.
The peak performances of the reference {\tt Rgetrf}s of {\tt MPFR 512bit}, {\tt GMP 512bit} and {\tt quad-double} are 6.1 MFlops, 13.0 MFlops, 15.7 MFlops, respectively.
The peak performances of simple OpenMP parallelized {\tt MPFR 512bit}, {\tt GMP 512bit} and {\tt quad-double} are 19.1 MFlops, 51.3 MFlops, 68.1 MFlops, respectively.\\
%%%%% Raspberry Pi4 ARM Cortex A72 %%%%
\subsubsection{{\tt Rgetrf} on Raspberry Pi4 ARM Cortex A72}
\begin{figure}
\caption{ {\tt Rgetrf} performance on Raspberry Pi4 ARM Cortex A72 for {\tt \_Float128} and {\tt double-double} with/without simple OpenMP acceleration. }
\label{Rgetrf1.G}
\begin{center}
\includegraphics{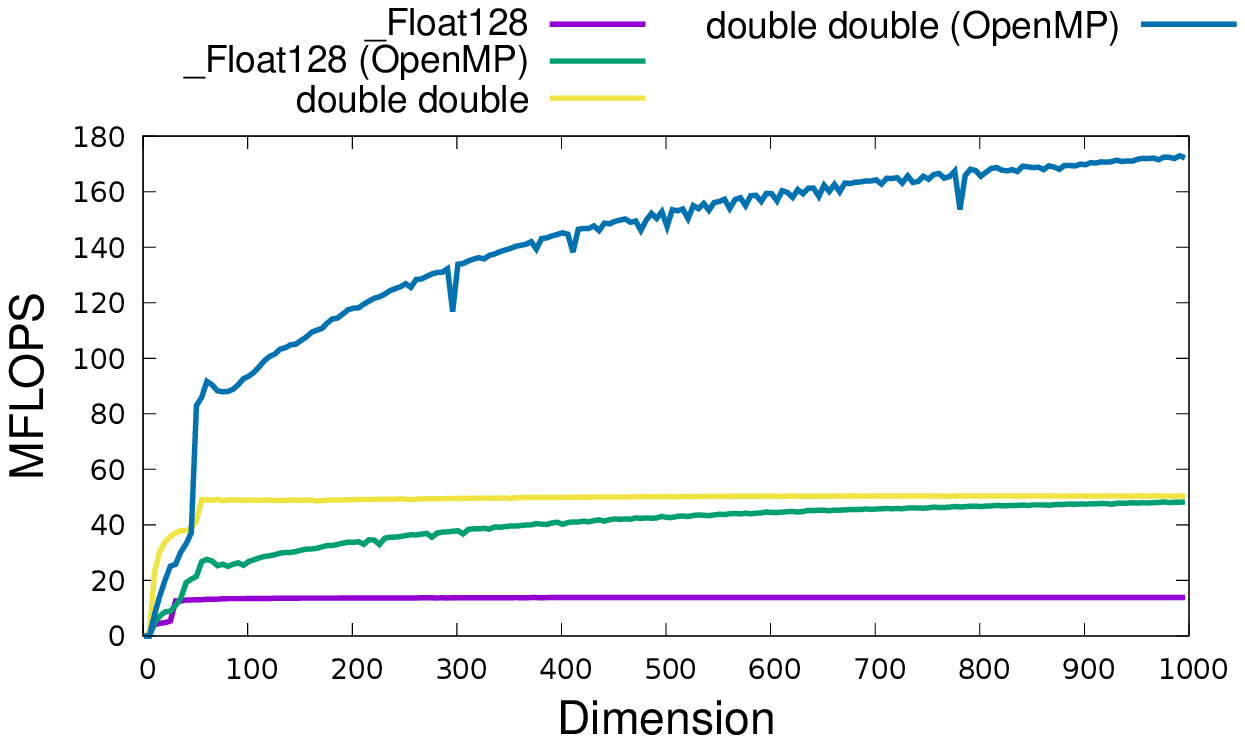}
\end{center}
\end{figure}

\begin{figure}
\caption{{\tt Rgetrf} performance on Raspberry Pi4 ARM Cortex A72 for {\tt MPFR 512bit}, {\tt GMP 512bit} and {\tt quad-double} with/without simple OpenMP acceleration. }
\label{Rgetrf2.G}
\begin{center}
\includegraphics{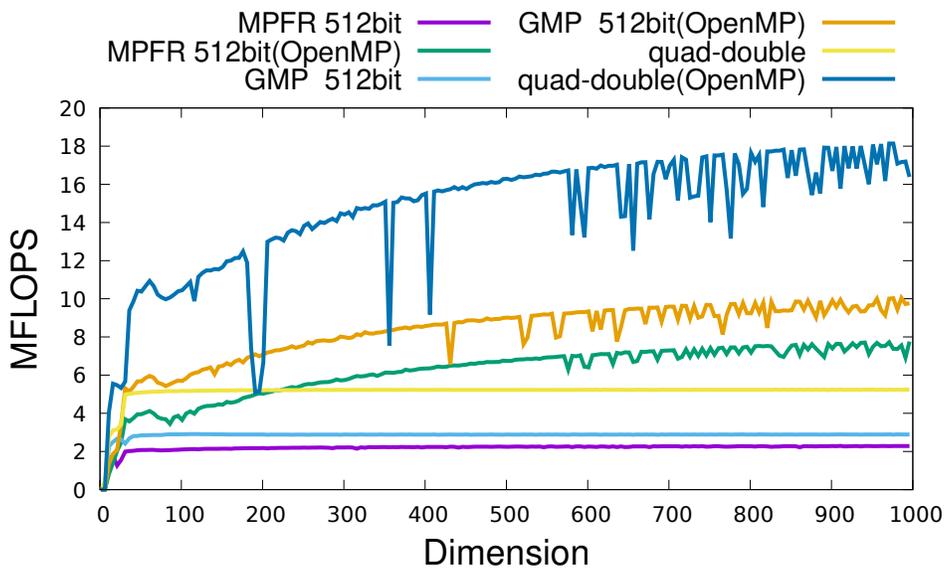}
\end{center}
\end{figure}

In Figure~\ref{Rgetrf1.G}, we show the result of {\tt Rgetrf} performance for {\tt \_Float128} and {\tt double-double}, and in Figure~\ref{Rgetrf2.G} we show the result of {\tt Rgetrf} performance for {\tt MPFR}, {\tt GMP} and {\tt quad-double} on Raspberry Pi4 ARM Cortex A72.
The peak performances of the reference {\tt Rgetrf}s of {\tt \_Float128} and {\tt double-double} are 13.9 MFlops and 50.5 MFlops, respectively.
The peak performances of simple OpenMP parallelized {\tt Rgetrf}s of {\tt \_Float128} and {\tt double-double} are 48.3 MFlops and 173 MFlops, respectively.
The peak performances of the reference {\tt Rgetrf}s of {\tt MPFR 512bit}, {\tt GMP 512bit} and {\tt quad-double} are 2.3 MFlops, 2.9 MFlops, 5.2 MFlops, respectively.
The peak performances of simple OpenMP parallelized {\tt MPFR 512bit}, {\tt GMP 512bit} and {\tt quad-double} are 7.8 MFlops, 10.0 MFlops, 18.1 MFlops, respectively.\\

\subsection{{\tt Rpotrf} benchmarks}
%%%%% AMD Ryzen 3970X %%%%
\subsubsection{{\tt Rpotrf} on AMD Ryzen 3970X}
\begin{figure}
\caption{ {\tt Rpotrf} performance on AMD Ryzen 3970X for {\tt \_Float128}, {\tt \_Float64x} and {\tt double-double} with/without simple OpenMP acceleration. }
\label{Rpotrf1.A}
\begin{center}
\includegraphics{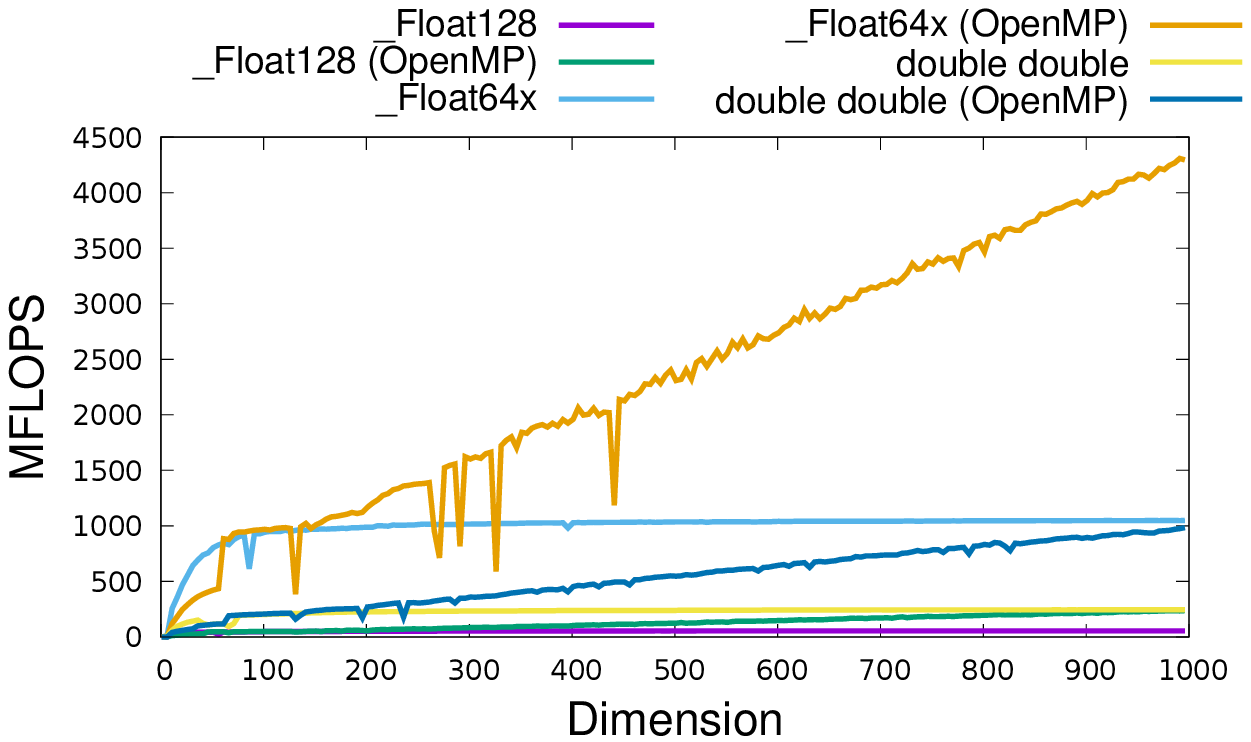}
\end{center}
\end{figure}

\begin{figure}
\caption{{\tt Rpotrf} performance on AMD Ryzen 3970X for {\tt MPFR 512bit}, {\tt GMP 512bit} and {\tt quad-double} with/without simple OpenMP acceleration. }
\label{Rpotrf2.A}
\begin{center}
\includegraphics{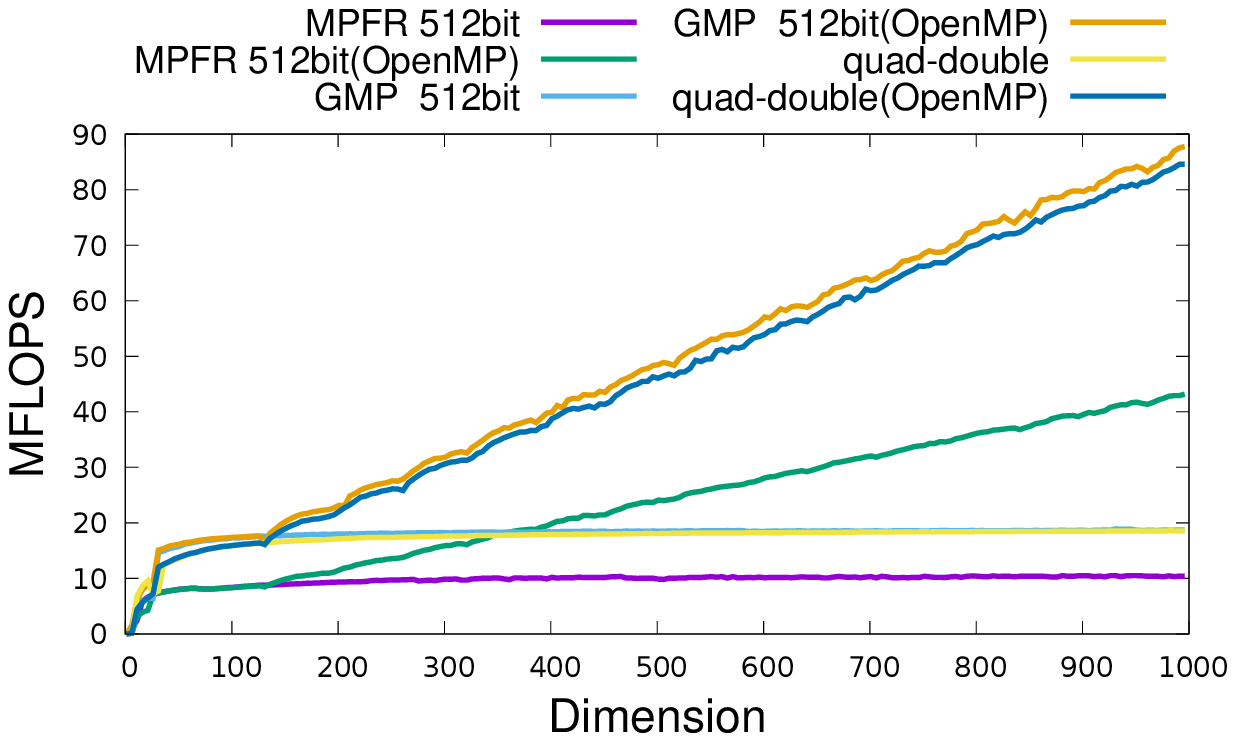}
\end{center}
\end{figure}

In Figure~\ref{Rpotrf1.A}, we show the result of {\tt Rpotrf} performance for {\tt \_Float128}, {\tt \_Float64x} and {\tt double-double}, and in Figure~\ref{Rpotrf2.A} we show the result of {\tt Rpotrf} performance for {\tt MPFR}, {\tt GMP} and {\tt quad-double} on AMD Ryzen 3970X.
The peak performances of the reference {\tt Rpotrf}s of {\tt \_Float128}, {\tt \_Float64x} and {\tt double-double} are 54.9 MFlops, 1049 MFlops, 246 MFlops, respectively.
The peak performances of simple OpenMP parallelized {\tt Rpotrf}s of {\tt \_Float128}, {\tt \_Float64x} and {\tt double-double} are 238 MFlops, 4307 MFlops, 985 MFlops, respectively.
The peak performances of the reference {\tt Rpotrf}s of {\tt MPFR 512bit}, {\tt GMP 512bit} and {\tt quad-double} are 10.5 MFlops, 18.9 MFlops, 18.6 MFlops, respectively.
The peak performances of simple OpenMP parallelized {\tt MPFR 512bit}, {\tt GMP 512bit} and {\tt quad-double} are 43.2 MFlops, 87.8 MFlops, 84.6 MFlops, respectively.\\
%%%%% Intel Xeon E5 2623 %%%%
\subsubsection{{\tt Rpotrf} on Intel Xeon E5 2623}
\begin{figure}
\caption{ {\tt Rpotrf} performance on Intel Xeon E5 2623 for {\tt \_Float128}, {\tt \_Float64x} and {\tt double-double} with/without simple OpenMP acceleration. }
\label{Rpotrf1.B}
\begin{center}
\includegraphics{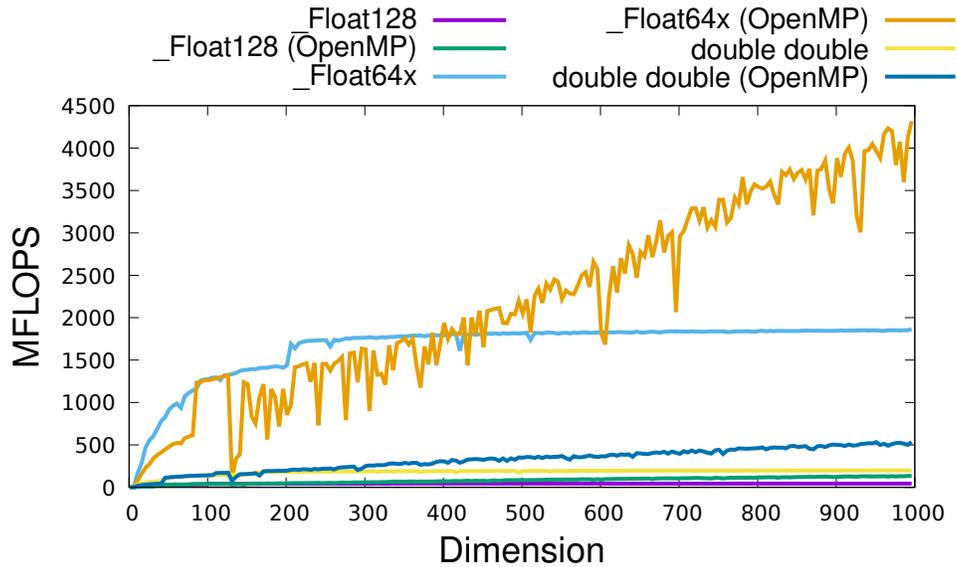}
\end{center}
\end{figure}

\begin{figure}
\caption{{\tt Rpotrf} performance on Intel Xeon E5 2623 for {\tt MPFR 512bit}, {\tt GMP 512bit} and {\tt quad-double} with/without simple OpenMP acceleration. }
\label{Rpotrf2.B}
\begin{center}
\includegraphics{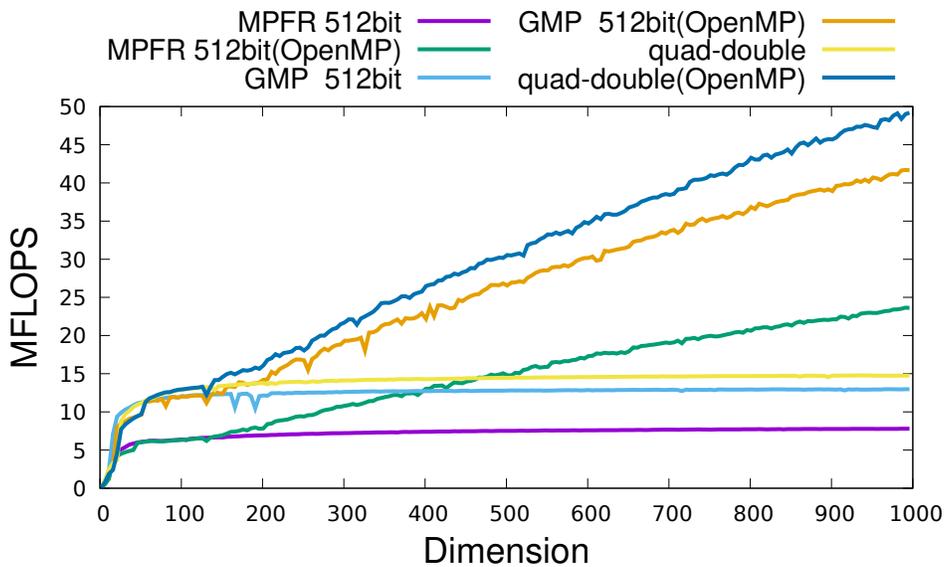}
\end{center}
\end{figure}

In Figure~\ref{Rpotrf1.B}, we show the result of {\tt Rpotrf} performance for {\tt \_Float128}, {\tt \_Float64x} and {\tt double-double}, and in Figure~\ref{Rpotrf2.B} we show the result of {\tt Rpotrf} performance for {\tt MPFR}, {\tt GMP} and {\tt quad-double} on Intel Xeon E5 2623.
The peak performances of the reference {\tt Rpotrf}s of {\tt \_Float128}, {\tt \_Float64x} and {\tt double-double} are 43.8 MFlops, 1865 MFlops, 199 MFlops, respectively.
The peak performances of simple OpenMP parallelized {\tt Rpotrf}s of {\tt \_Float128}, {\tt \_Float64x} and {\tt double-double} are 134 MFlops, 4317 MFlops, 534 MFlops, respectively.
The peak performances of the reference {\tt Rpotrf}s of {\tt MPFR 512bit}, {\tt GMP 512bit} and {\tt quad-double} are 7.8 MFlops, 13.0 MFlops, 14.8 MFlops, respectively.
The peak performances of simple OpenMP parallelized {\tt MPFR 512bit}, {\tt GMP 512bit} and {\tt quad-double} are 23.6 MFlops, 41.7 MFlops, 49.2 MFlops, respectively.\\

%%%%% Intel Core i5-8500B %%%%
\subsubsection{{\tt Rpotrf} on Intel Core i5-8500B}
\begin{figure}
\caption{ {\tt Rpotrf} performance on Intel Core i5-8500B for {\tt \_Float128}, {\tt \_Float64x} and {\tt double-double} with/without simple OpenMP acceleration. }
\label{Rpotrf1.C}
\begin{center}
\includegraphics{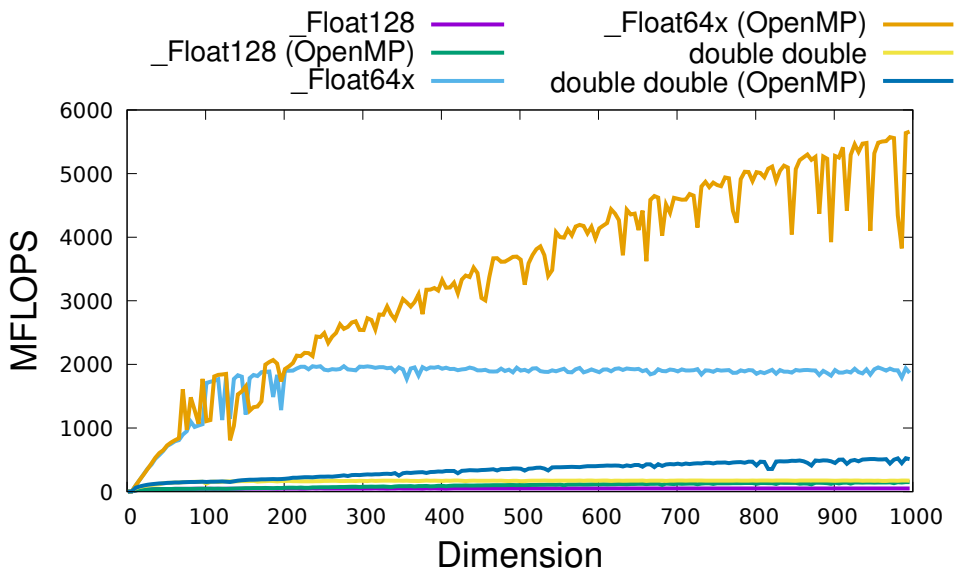}
\end{center}
\end{figure}

\begin{figure}
\caption{{\tt Rpotrf} performance on Intel Core i5-8500B for {\tt MPFR 512bit}, {\tt GMP 512bit} and {\tt quad-double} with/without simple OpenMP acceleration. }
\label{Rpotrf2.C}
\begin{center}
\includegraphics{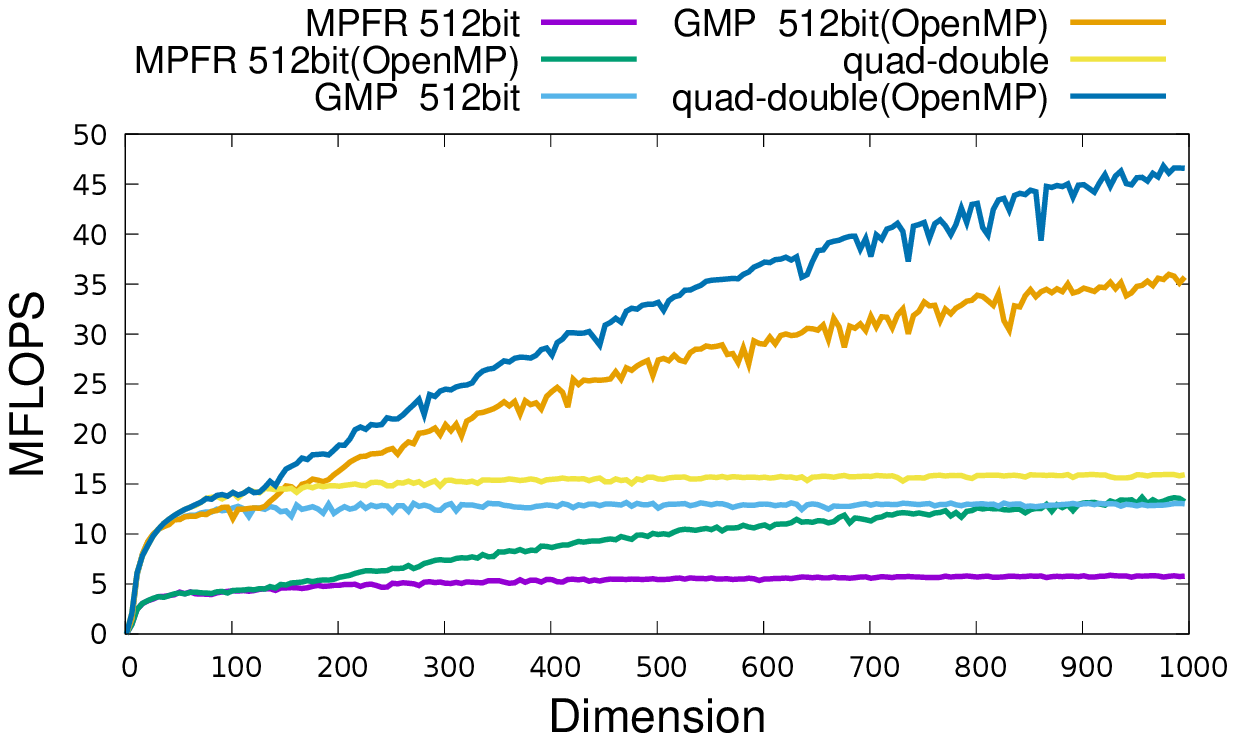}
\end{center}
\end{figure}

In Figure~\ref{Rpotrf1.C}, we show the result of {\tt Rpotrf} performance for {\tt \_Float128}, {\tt \_Float64x} and {\tt double-double}, and in Figure~\ref{Rpotrf2.C} we show the result of {\tt Rpotrf} performance for {\tt MPFR}, {\tt GMP} and {\tt quad-double} on Intel Core i5-8500B.
The peak performances of the reference {\tt Rpotrf}s of {\tt \_Float128}, {\tt \_Float64x} and {\tt double-double} are 50.5 MFlops, 1972 MFlops, 176 MFlops, respectively.
The peak performances of simple OpenMP parallelized {\tt Rpotrf}s of {\tt \_Float128}, {\tt \_Float64x} and {\tt double-double} are 150 MFlops, 5660 MFlops, 526 MFlops, respectively.
The peak performances of the reference {\tt Rpotrf}s of {\tt MPFR 512bit}, {\tt GMP 512bit} and {\tt quad-double} are 5.9 MFlops, 13.1 MFlops, 15.9 MFlops, respectively.
The peak performances of simple OpenMP parallelized {\tt MPFR 512bit}, {\tt GMP 512bit} and {\tt quad-double} are 13.7 MFlops, 36.0 MFlops, 46.8 MFlops, respectively.\\

%%%%% Raspberry Pi4 ARM Cortex A72 %%%%
\subsubsection{{\tt Rpotrf} on Raspberry Pi4 ARM Cortex A72}
\begin{figure}
\caption{ {\tt Rpotrf} performance on Raspberry Pi4 ARM Cortex A72 for {\tt \_Float128} and {\tt double-double} with/without simple OpenMP acceleration. }
\label{Rpotrf1.G}
\begin{center}
\includegraphics{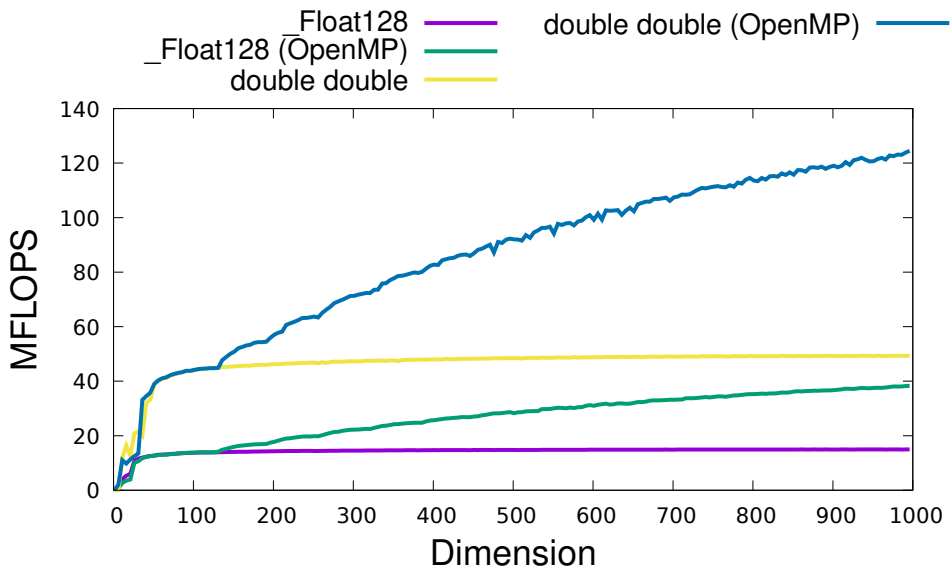}
\end{center}
\end{figure}

\begin{figure}
\caption{{\tt Rpotrf} performance on Raspberry Pi4 ARM Cortex A72 for {\tt MPFR 512bit}, {\tt GMP 512bit} and {\tt quad-double} with/without simple OpenMP acceleration. }
\label{Rpotrf2.G}
\begin{center}
\includegraphics{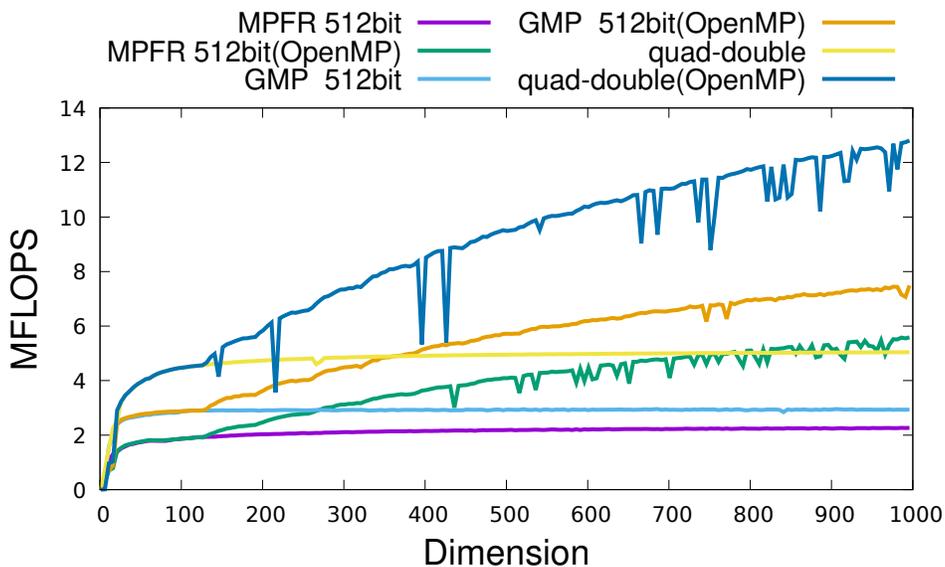}
\end{center}
\end{figure}

In Figure~\ref{Rpotrf1.G}, we show the result of {\tt Rpotrf} performance for {\tt \_Float128} and {\tt double-double}, and in Figure~\ref{Rpotrf2.G} we show the result of {\tt Rpotrf} performance for {\tt MPFR}, {\tt GMP} and {\tt quad-double} on Raspberry Pi4 ARM Cortex A72.
The peak performances of the reference {\tt Rpotrf}s of {\tt \_Float128} and {\tt double-double} are 15.0 MFlops and 49.4 MFlops, respectively.
The peak performances of simple OpenMP parallelized {\tt Rpotrf}s of {\tt \_Float128} and {\tt double-double} are 38.3 MFlops and 124 MFlops, respectively.
The peak performances of the reference {\tt Rpotrf}s of {\tt MPFR 512bit}, {\tt GMP 512bit} and {\tt quad-double} are 2.3 MFlops, 3.0 MFlops, 5.0 MFlops, respectively.
The peak performances of simple OpenMP parallelized {\tt MPFR 512bit}, {\tt GMP 512bit} and {\tt quad-double} are 5.6 MFlops, 7.5 MFlops, 12.8 MFlops, respectively.\\

\subsection{{\tt Raxpy} benchmarks}
%%%%% AMD Ryzen 3970X %%%%
\subsubsection{{\tt Raxpy} on AMD Ryzen 3970X}
\begin{figure}
\caption{ {\tt Raxpy} performance on AMD Ryzen 3970X for {\tt \_Float128}, {\tt \_Float64x} and {\tt double-double} with/without simple OpenMP acceleration. }
\label{Raxpy1.A}
\begin{center}
\includegraphics{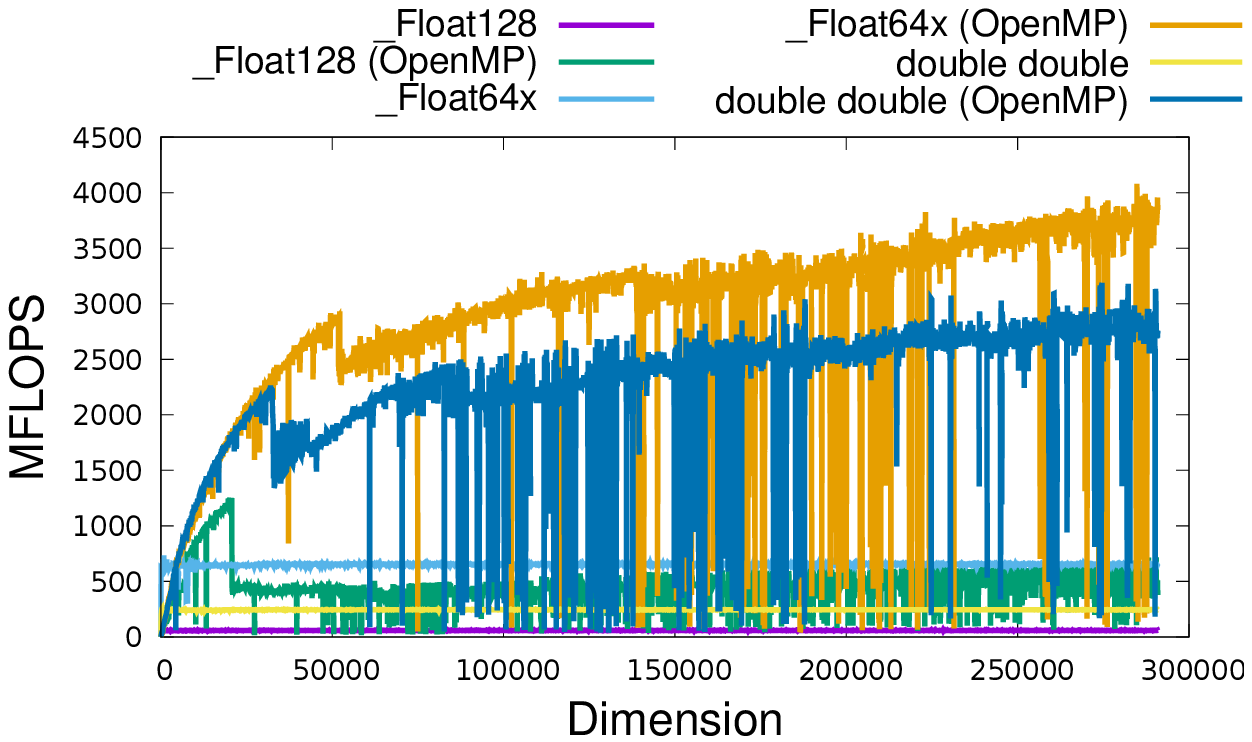}
\end{center}
\end{figure}

\begin{figure}
\caption{{\tt Raxpy} performance on AMD Ryzen 3970X for {\tt MPFR 512bit}, {\tt GMP 512bit} and {\tt quad-double} with/without simple OpenMP acceleration. }
\label{Raxpy2.A}
\begin{center}
\includegraphics{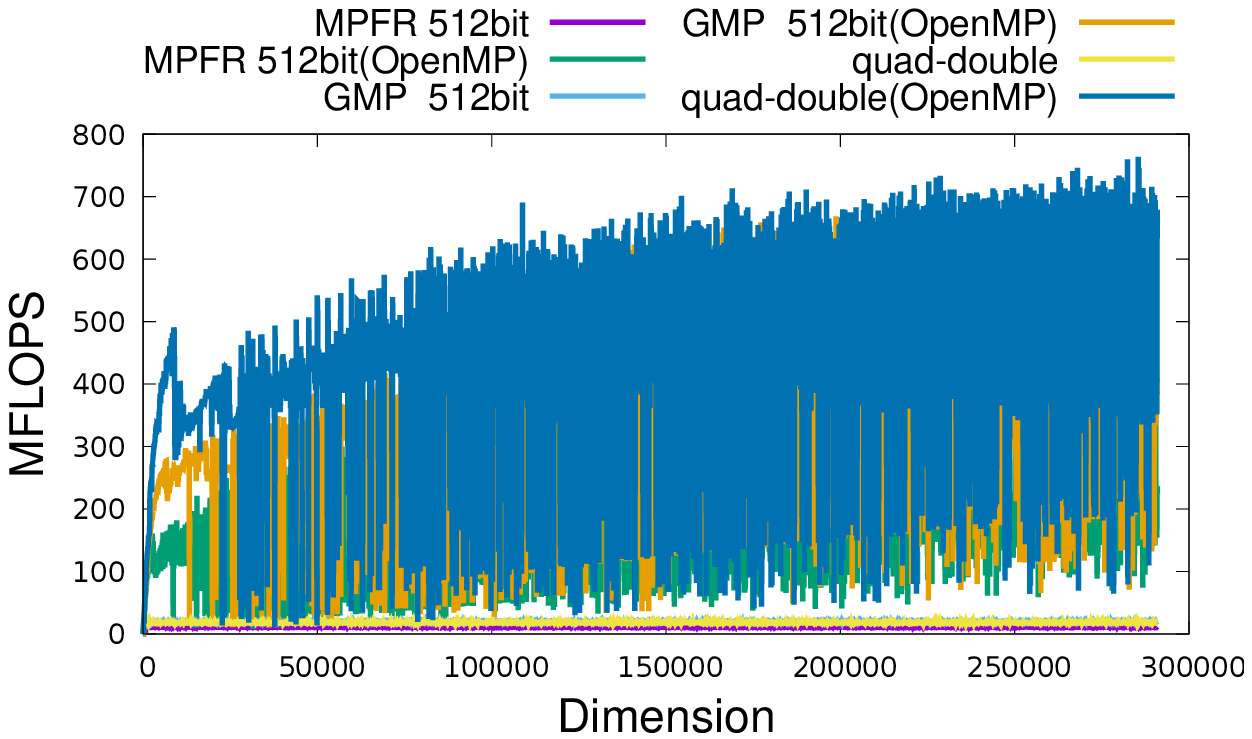}
\end{center}
\end{figure}

In Figure~\ref{Raxpy1.A}, we show the result of {\tt Raxpy} performance for {\tt \_Float128}, {\tt \_Float64x} and {\tt double-double}, and in Figure~\ref{Raxpy2.A} we show the result of {\tt Raxpy} performance for {\tt MPFR}, {\tt GMP} and {\tt quad-double} on AMD Ryzen 3970X.
The peak performances of the reference {\tt Raxpy}s of {\tt \_Float128}, {\tt \_Float64x} and {\tt double-double} are 60.9 MFlops, 733 MFlops, 258 MFlops, respectively.
The peak performances of simple OpenMP parallelized {\tt Raxpy}s of {\tt \_Float128}, {\tt \_Float64x} and {\tt double-double} are 1250 MFlops, 4079 MFlops, 3188 MFlops, respectively.
The peak performances of the reference {\tt Raxpy}s of {\tt MPFR 512bit}, {\tt GMP 512bit} and {\tt quad-double} are 11.7 MFlops, 20.9 MFlops, 20.3 MFlops, respectively.
The peak performances of simple OpenMP parallelized {\tt MPFR 512bit}, {\tt GMP 512bit} and {\tt quad-double} are 404 MFlops, 683 MFlops, 764 MFlops, respectively.\\

%%%%% Intel Xeon E5 2623 %%%%
\subsubsection{{\tt Raxpy} on Intel Xeon E5 2623}
\begin{figure}
\caption{ {\tt Raxpy} performance on Intel Xeon E5 2623 for {\tt \_Float128}, {\tt \_Float64x} and {\tt double-double} with/without simple OpenMP acceleration. }
\label{Raxpy1.B}
\begin{center}
\includegraphics{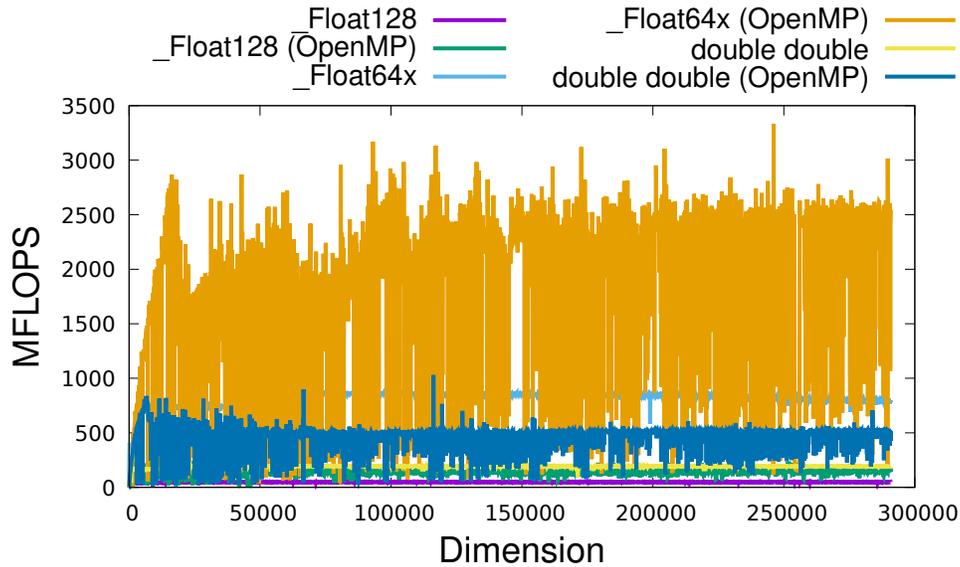}
\end{center}
\end{figure}

\begin{figure}
\caption{{\tt Raxpy} performance on Intel Xeon E5 2623 for {\tt MPFR 512bit}, {\tt GMP 512bit} and {\tt quad-double} with/without simple OpenMP acceleration. }
\label{Raxpy2.B}
\begin{center}
\includegraphics{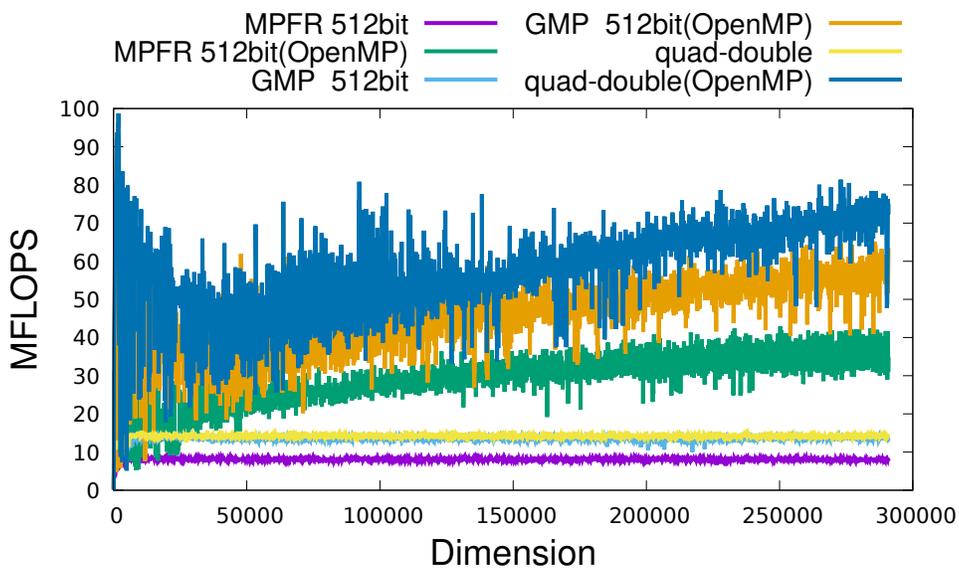}
\end{center}
\end{figure}

In Figure~\ref{Raxpy1.B}, we show the result of {\tt Raxpy} performance for {\tt \_Float128}, {\tt \_Float64x} and {\tt double-double}, and in Figure~\ref{Raxpy2.B} we show the result of {\tt Raxpy} performance for {\tt MPFR}, {\tt GMP} and {\tt quad-double} on Intel Xeon E5 2623.
The peak performances of the reference {\tt Raxpy}s of {\tt \_Float128}, {\tt \_Float64x} and {\tt double-double} are 50.3 MFlops, 866 MFlops, 195 MFlops, respectively.
The peak performances of simple OpenMP parallelized {\tt Raxpy}s of {\tt \_Float128}, {\tt \_Float64x} and {\tt double-double} are 255 MFlops, 3333 MFlops, 1030 MFlops, respectively.
The peak performances of the reference {\tt Raxpy}s of {\tt MPFR 512bit}, {\tt GMP 512bit} and {\tt quad-double} are 8.6 MFlops, 14.4 MFlops, 15.7 MFlops, respectively.
The peak performances of simple OpenMP parallelized {\tt MPFR 512bit}, {\tt GMP 512bit} and {\tt quad-double} are 43.0 MFlops, 73.0 MFlops, 98.7 MFlops, respectively.\\
%%%%% Intel Core i5-8500B %%%%
\subsubsection{{\tt Raxpy} on Intel Core i5-8500B}
\begin{figure}
\caption{ {\tt Raxpy} performance on Intel Core i5-8500B for {\tt \_Float128}, {\tt \_Float64x} and {\tt double-double} with/without simple OpenMP acceleration. }
\label{Raxpy1.C}
\begin{center}
\includegraphics{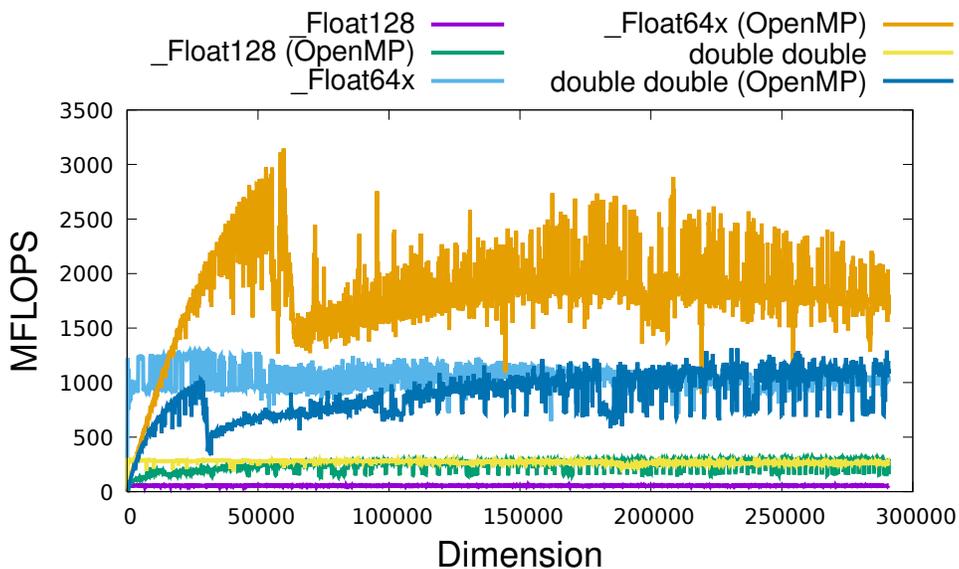}
\end{center}
\end{figure}

\begin{figure}
\caption{{\tt Raxpy} performance on Intel Core i5-8500B for {\tt MPFR 512bit}, {\tt GMP 512bit} and {\tt quad-double} with/without simple OpenMP acceleration. }
\label{Raxpy2.C}
\begin{center}
\includegraphics{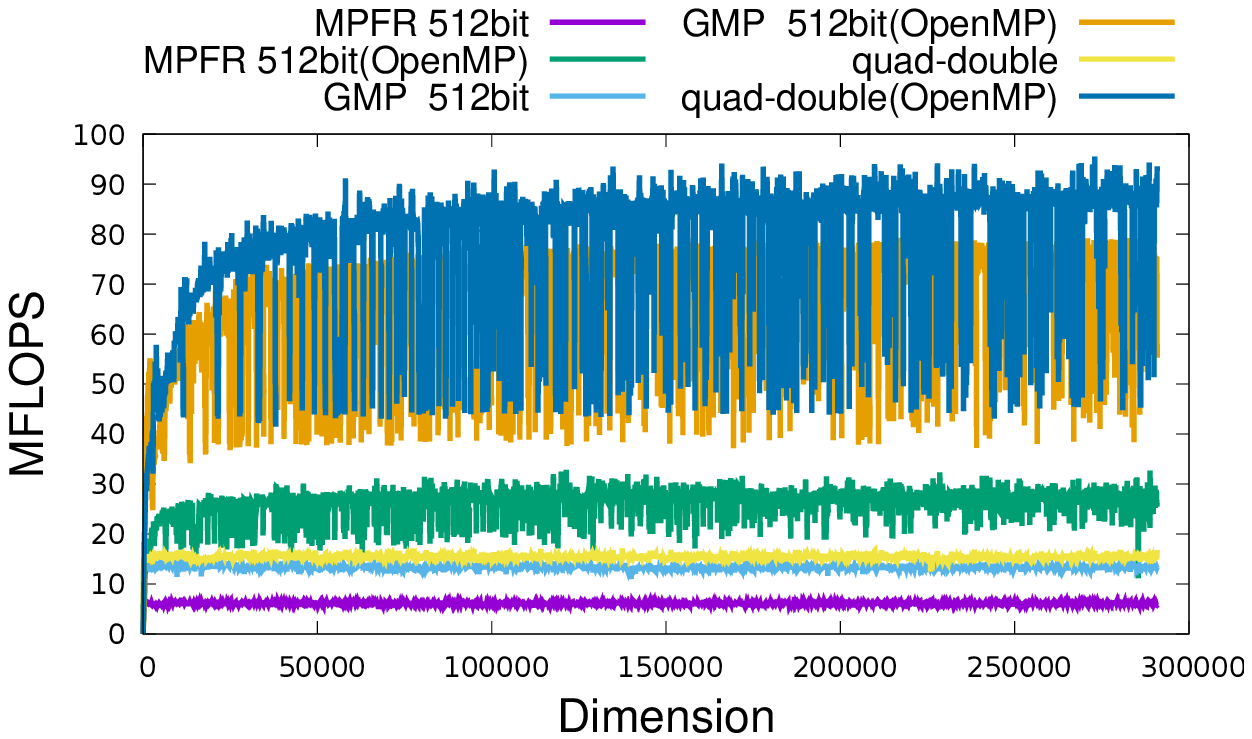}
\end{center}
\end{figure}

In Figure~\ref{Raxpy1.C}, we show the result of {\tt Raxpy} performance for {\tt \_Float128}, {\tt \_Float64x} and {\tt double-double}, and in Figure~\ref{Raxpy2.C} we show the result of {\tt Raxpy} performance for {\tt MPFR}, {\tt GMP} and {\tt quad-double} on Intel Core i5-8500B.
The peak performances of the reference {\tt Raxpy}s of {\tt \_Float128}, {\tt \_Float64x} and {\tt double-double} are 58.2 MFlops, 1303 MFlops, 308 MFlops, respectively.
The peak performances of simple OpenMP parallelized {\tt Raxpy}s of {\tt \_Float128}, {\tt \_Float64x} and {\tt double-double} are 309 MFlops, 3146 MFlops, 1316 MFlops, respectively.
The peak performances of the reference {\tt Raxpy}s of {\tt MPFR 512bit}, {\tt GMP 512bit} and {\tt quad-double} are 6.6 MFlops, 14.7 MFlops, 17.1 MFlops, respectively.
The peak performances of simple OpenMP parallelized {\tt MPFR 512bit}, {\tt GMP 512bit} and {\tt quad-double} are 32.8 MFlops, 79.4 MFlops, 95.5 MFlops, respectively.\\

%%%%% Raspberry Pi4 ARM Cortex A72 %%%%
\subsubsection{{\tt Raxpy} on Raspberry Pi4 ARM Cortex A72}
\begin{figure}
\caption{ {\tt Raxpy} performance on Raspberry Pi4 ARM Cortex A72 for {\tt \_Float128} and {\tt double-double} with/without simple OpenMP acceleration. }
\label{Raxpy1.G}
\begin{center}
\includegraphics{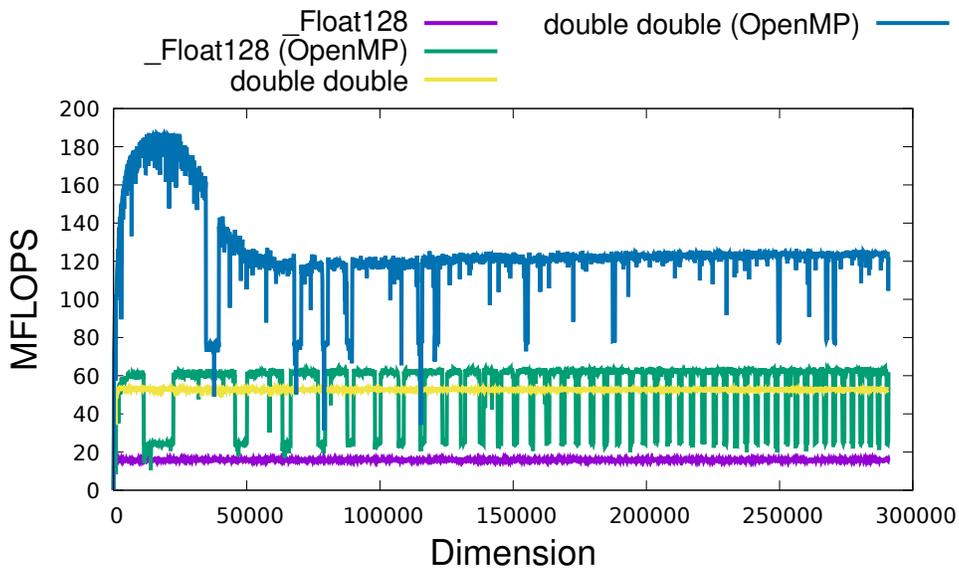}
\end{center}
\end{figure}

\begin{figure}
\caption{{\tt Raxpy} performance on Raspberry Pi4 ARM Cortex A72 for {\tt MPFR 512bit}, {\tt GMP 512bit} and {\tt quad-double} with/without simple OpenMP acceleration. }
\label{Raxpy2.G}
\begin{center}
\includegraphics{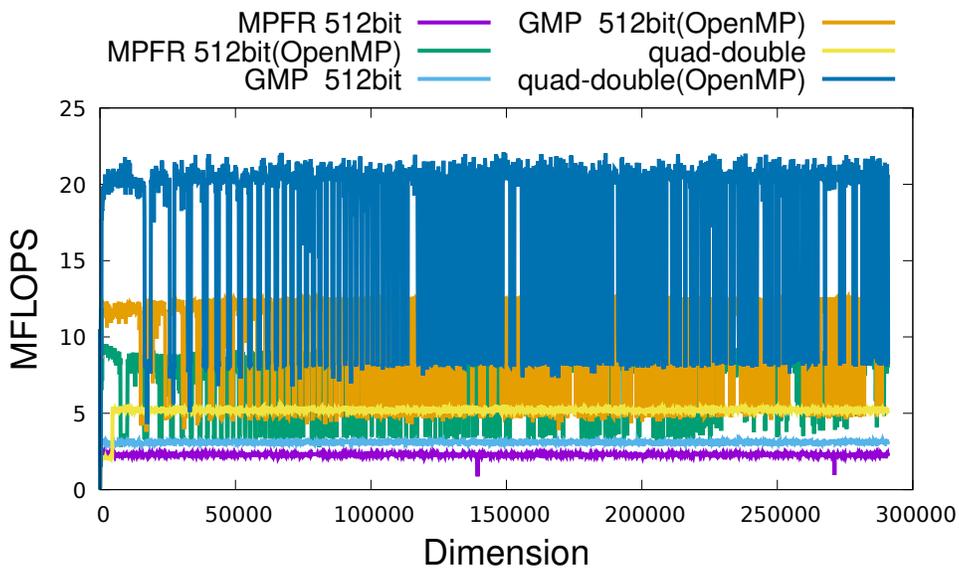}
\end{center}
\end{figure}

In Figure~\ref{Raxpy1.G}, we show the result of {\tt Raxpy} performance for {\tt \_Float128} and {\tt double-double}, and in Figure~\ref{Raxpy2.G} we show the result of {\tt Raxpy} performance for {\tt MPFR}, {\tt GMP} and {\tt quad-double} on Raspberry Pi4 ARM Cortex A72.
The peak performances of the reference {\tt Raxpy}s of {\tt \_Float128} and {\tt double-double} are 16.5 MFlops and 53.5 MFlops, respectively.
The peak performances of simple OpenMP parallelized {\tt Raxpy}s of {\tt \_Float128} and {\tt double-double} are 64.7 MFlops and 187 MFlops, respectively.
The peak performances of the reference {\tt Raxpy}s of {\tt MPFR 512bit}, {\tt GMP 512bit} and {\tt quad-double} are 2.5 MFlops, 3.3 MFlops, 5.6 MFlops, respectively.
The peak performances of simple OpenMP parallelized {\tt MPFR 512bit}, {\tt GMP 512bit} and {\tt quad-double} are 9.5 MFlops, 12.7 MFlops, 22.1 MFlops, respectively.\\

\subsection{{\tt Rdot} benchmarks}
%%%%% AMD Ryzen 3970X %%%%
\subsubsection{{\tt Rdot} on AMD Ryzen 3970X}
\begin{figure}
\caption{ {\tt Rdot} performance on AMD Ryzen 3970X for {\tt \_Float128}, {\tt \_Float64x} and {\tt double-double} with/without simple OpenMP acceleration. }
\label{Rdot1.A}
\begin{center}
\includegraphics{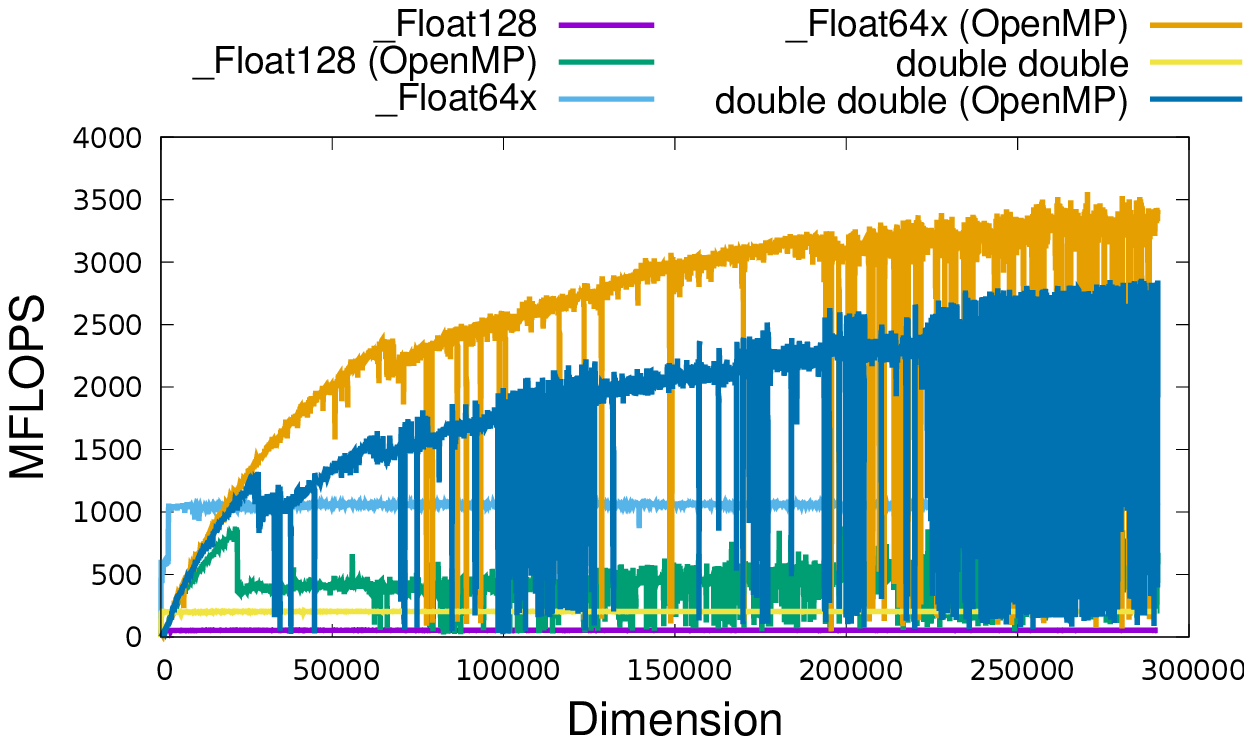}
\end{center}
\end{figure}

\begin{figure}
\caption{{\tt Rdot} performance on AMD Ryzen 3970X for {\tt MPFR 512bit}, {\tt GMP 512bit} and {\tt quad-double} with/without simple OpenMP acceleration. }
\label{Rdot2.A}
\begin{center}
\includegraphics{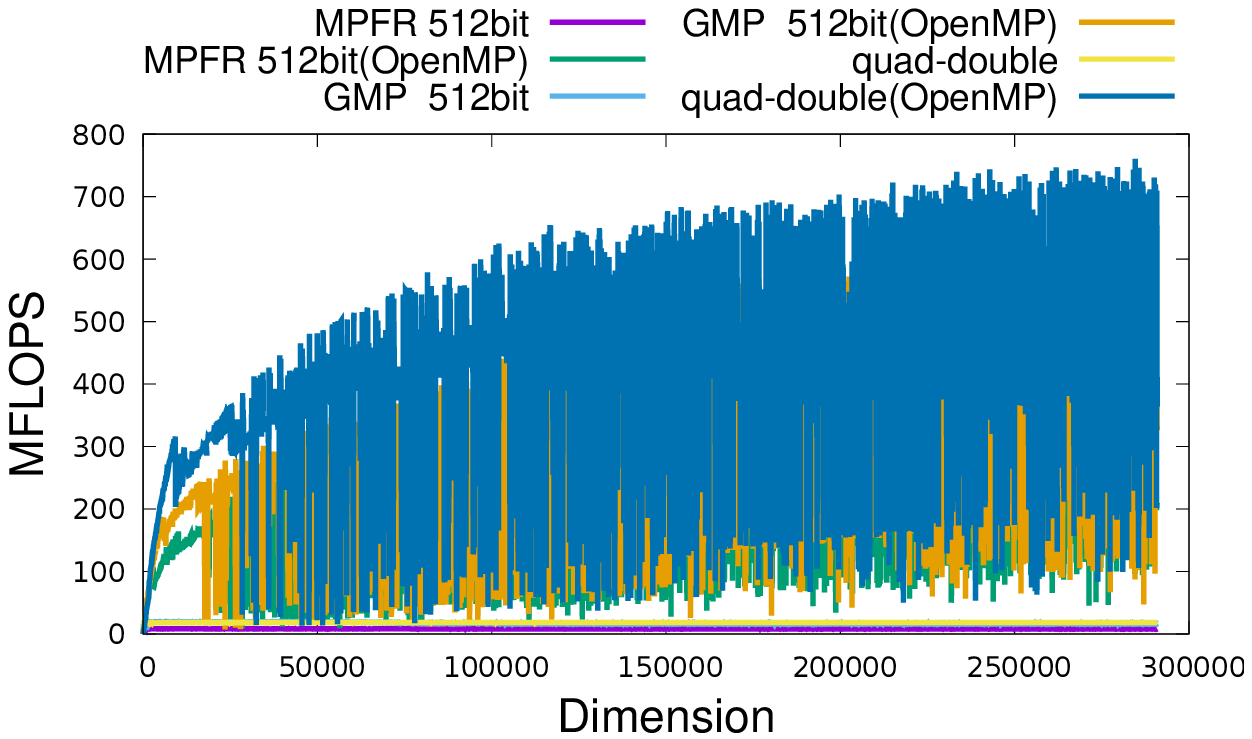}
\end{center}
\end{figure}

In Figure~\ref{Rdot1.A}, we show the result of {\tt Rdot} performance for {\tt \_Float128}, {\tt \_Float64x} and {\tt double-double}, and in Figure~\ref{Rdot2.A} we show the result of {\tt Rdot} performance for {\tt MPFR}, {\tt GMP} and {\tt quad-double} on AMD Ryzen 3970X.
The peak performances of the reference {\tt Rdot}s of {\tt \_Float128}, {\tt \_Float64x} and {\tt double-double} are 54.1 MFlops, 1077 MFlops, 207 MFlops, respectively.
The peak performances of simple OpenMP parallelized {\tt Rdot}s of {\tt \_Float128}, {\tt \_Float64x} and {\tt double-double} are 1085 MFlops, 3560 MFlops, 2866 MFlops, respectively.
The peak performances of the reference {\tt Rdot}s of {\tt MPFR 512bit}, {\tt GMP 512bit} and {\tt quad-double} are 8.5 MFlops, 19.8 MFlops, 19.4 MFlops, respectively.
The peak performances of simple OpenMP parallelized {\tt MPFR 512bit}, {\tt GMP 512bit} and {\tt quad-double} are 415 MFlops, 643 MFlops, 760 MFlops, respectively.\\

%%%%% Intel Xeon E5 2623 %%%%
\subsubsection{{\tt Rdot} on Intel Xeon E5 2623}
\begin{figure}
\caption{ {\tt Rdot} performance on Intel Xeon E5 2623 for {\tt \_Float128}, {\tt \_Float64x} and {\tt double-double} with/without simple OpenMP acceleration. }
\label{Rdot1.B}
\begin{center}
\includegraphics{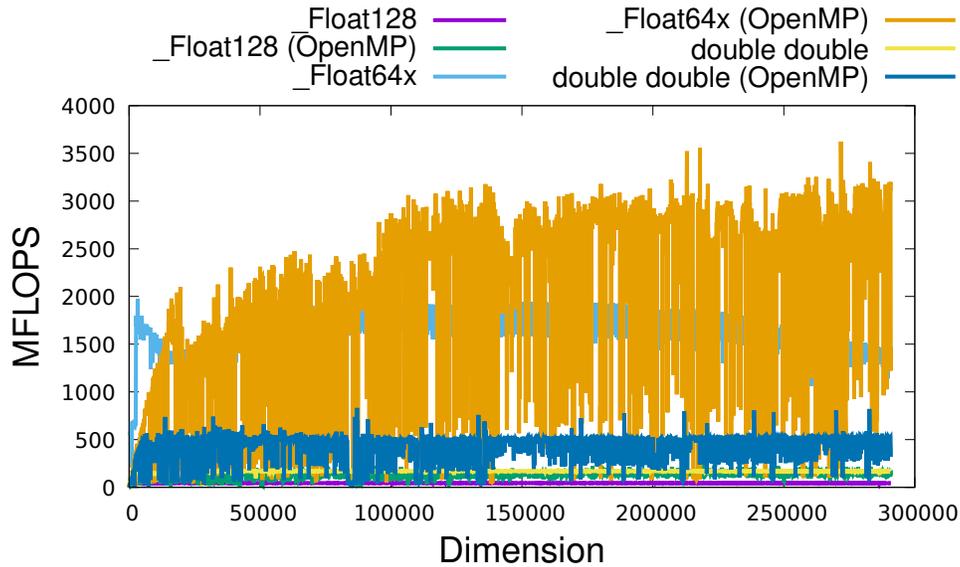}
\end{center}
\end{figure}

\begin{figure}
\caption{{\tt Rdot} performance on Intel Xeon E5 2623 for {\tt MPFR 512bit}, {\tt GMP 512bit} and {\tt quad-double} with/without simple OpenMP acceleration. }
\label{Rdot2.B}
\begin{center}
\includegraphics{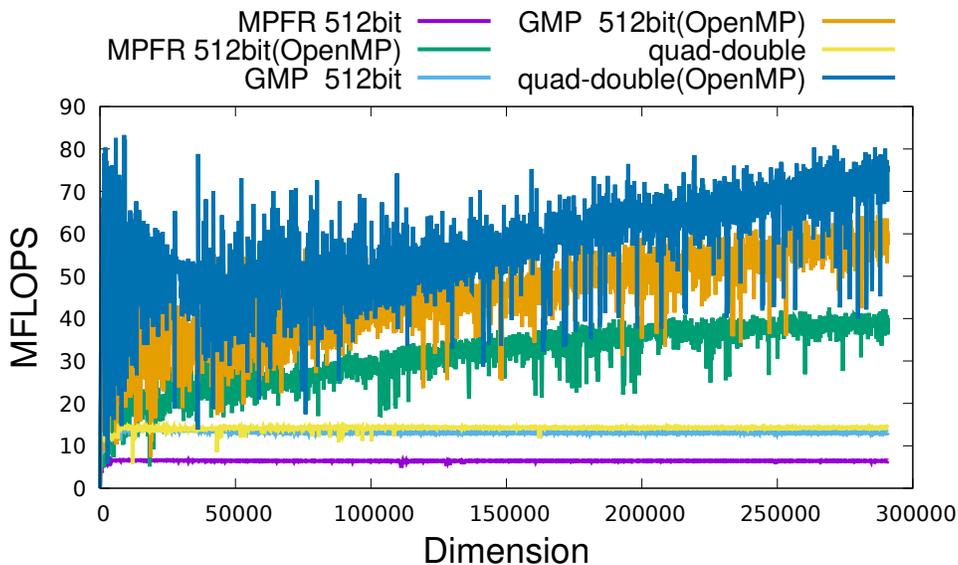}
\end{center}
\end{figure}

In Figure~\ref{Rdot1.B}, we show the result of {\tt Rdot} performance for {\tt \_Float128}, {\tt \_Float64x} and {\tt double-double}, and in Figure~\ref{Rdot2.B} we show the result of {\tt Rdot} performance for {\tt MPFR}, {\tt GMP} and {\tt quad-double} on Intel Xeon E5 2623.
The peak performances of the reference {\tt Rdot}s of {\tt \_Float128}, {\tt \_Float64x} and {\tt double-double} are 44.1 MFlops, 1977 MFlops, 171 MFlops, respectively.
The peak performances of simple OpenMP parallelized {\tt Rdot}s of {\tt \_Float128}, {\tt \_Float64x} and {\tt double-double} are 269 MFlops, 3623 MFlops, 833 MFlops, respectively.
The peak performances of the reference {\tt Rdot}s of {\tt MPFR 512bit}, {\tt GMP 512bit} and {\tt quad-double} are 6.6 MFlops, 13.9 MFlops, 14.5 MFlops, respectively.
The peak performances of simple OpenMP parallelized {\tt MPFR 512bit}, {\tt GMP 512bit} and {\tt quad-double} are 42.6 MFlops, 69.0 MFlops, 83.3 MFlops, respectively.\\

%%%%% Intel Core i5-8500B %%%%
\subsubsection{{\tt Rdot} on Intel Core i5-8500B}
\begin{figure}
\caption{ {\tt Rdot} performance on Intel Core i5-8500B for {\tt \_Float128}, {\tt \_Float64x} and {\tt double-double} with/without simple OpenMP acceleration. }
\label{Rdot1.C}
\begin{center}
\includegraphics{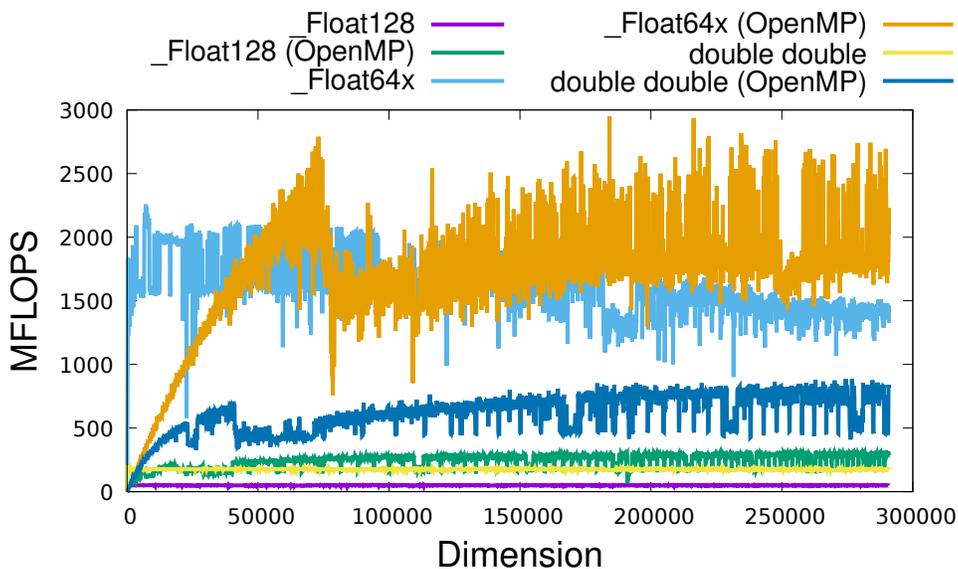}
\end{center}
\end{figure}

\begin{figure}
\caption{{\tt Rdot} performance on Intel Core i5-8500B for {\tt MPFR 512bit}, {\tt GMP 512bit} and {\tt quad-double} with/without simple OpenMP acceleration. }
\label{Rdot2.C}
\begin{center}
\includegraphics{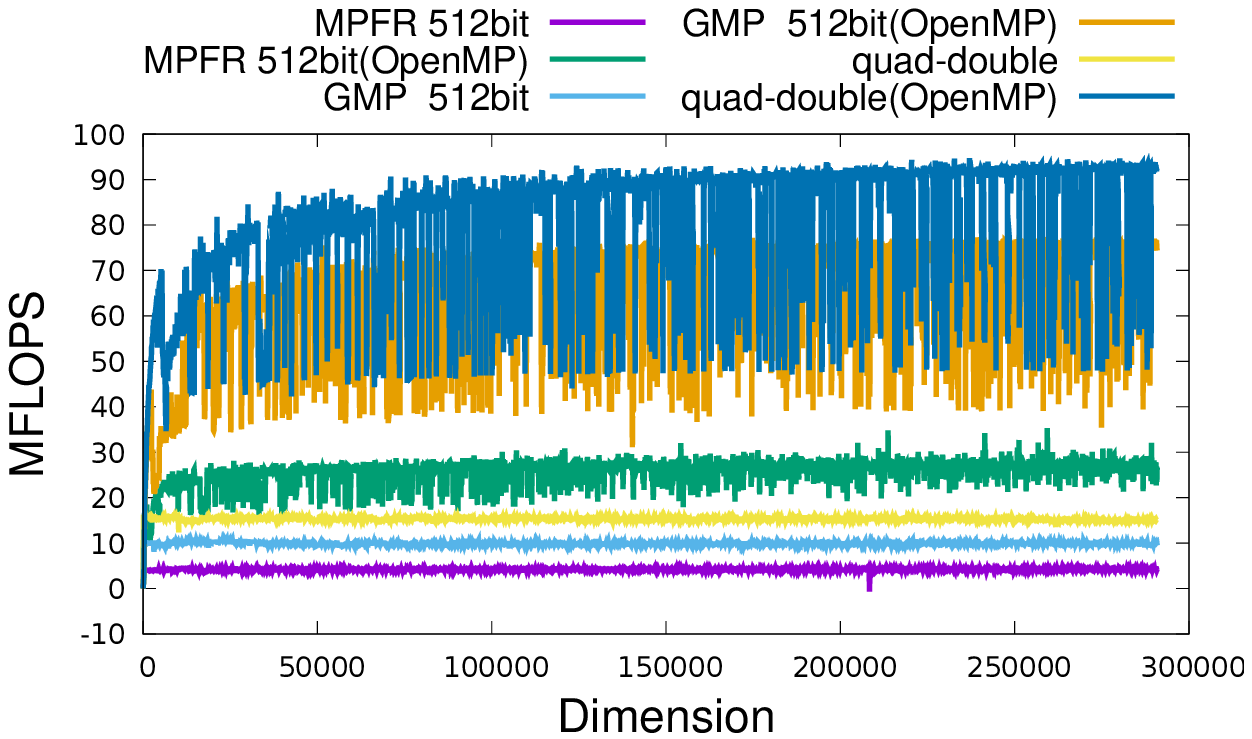}
\end{center}
\end{figure}

In Figure~\ref{Rdot1.C}, we show the result of {\tt Rdot} performance for {\tt \_Float128}, {\tt \_Float64x} and {\tt double-double}, and in Figure~\ref{Rdot2.C} we show the result of {\tt Rdot} performance for {\tt MPFR}, {\tt GMP} and {\tt quad-double} on Intel Core i5-8500B.
The peak performances of the reference {\tt Rdot}s of {\tt \_Float128}, {\tt \_Float64x} and {\tt double-double} are 52.0 MFlops, 2258 MFlops, 206 MFlops, respectively.
The peak performances of simple OpenMP parallelized {\tt Rdot}s of {\tt \_Float128}, {\tt \_Float64x} and {\tt double-double} are 320 MFlops, 2949 MFlops, 886 MFlops, respectively.
The peak performances of the reference {\tt Rdot}s of {\tt MPFR 512bit}, {\tt GMP 512bit} and {\tt quad-double} are 4.5 MFlops, 10.9 MFlops, 16.2 MFlops, respectively.
The peak performances of simple OpenMP parallelized {\tt MPFR 512bit}, {\tt GMP 512bit} and {\tt quad-double} are 35.3 MFlops, 77.5 MFlops, 94.7 MFlops, respectively.\\

%%%%% Raspberry Pi4 ARM Cortex A72 %%%%
\subsubsection{{\tt Rdot} on Raspberry Pi4 ARM Cortex A72}
\begin{figure}
\caption{ {\tt Rdot} performance on Raspberry Pi4 ARM Cortex A72 for {\tt \_Float128} and {\tt double-double} with/without simple OpenMP acceleration. }
\label{Rdot1.G}
\begin{center}
\includegraphics{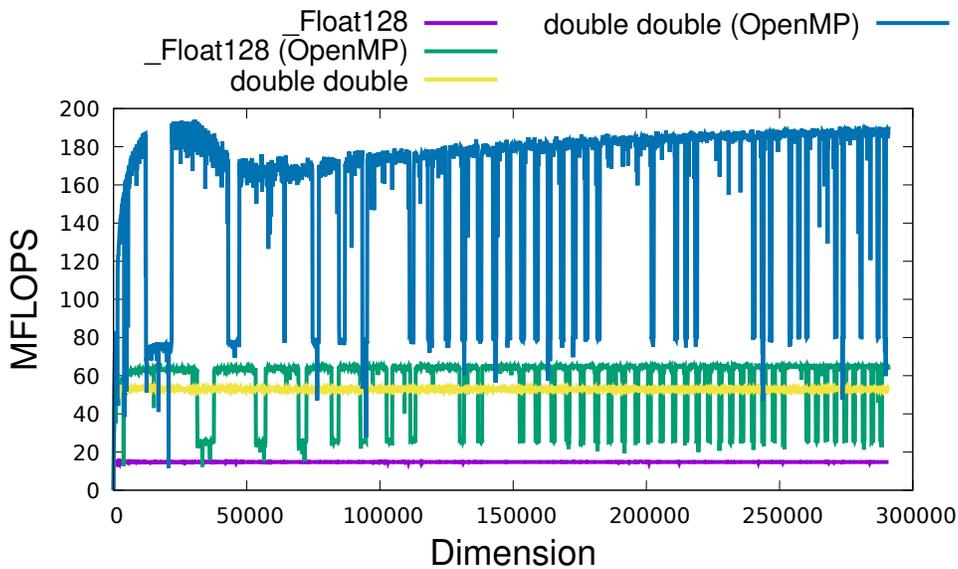}
\end{center}
\end{figure}

\begin{figure}
\caption{{\tt Rdot} performance on Raspberry Pi4 ARM Cortex A72 for {\tt MPFR 512bit}, {\tt GMP 512bit} and {\tt quad-double} with/without simple OpenMP acceleration. }
\label{Rdot2.G}
\begin{center}
\includegraphics{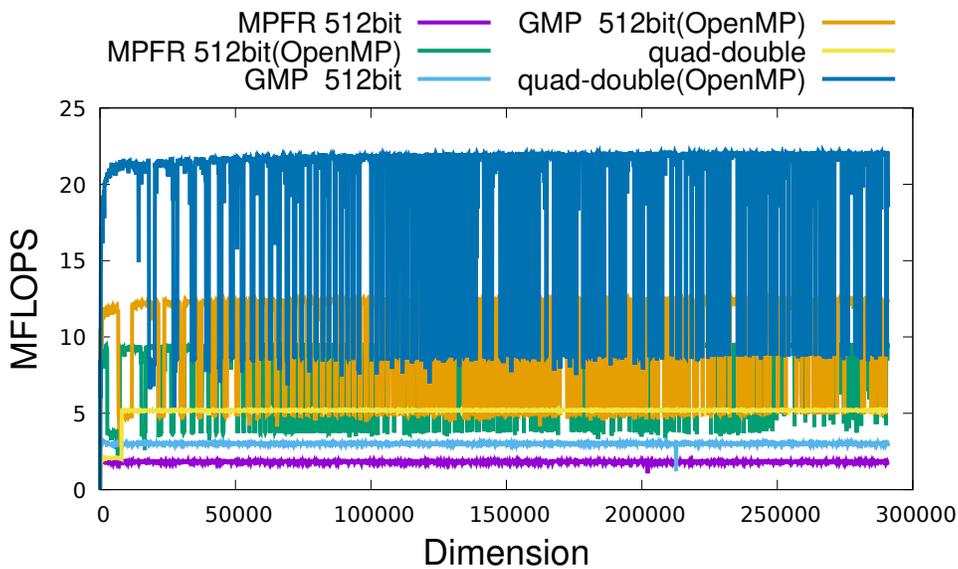}
\end{center}
\end{figure}

In Figure~\ref{Rdot1.G}, we show the result of {\tt Rdot} performance for {\tt \_Float128} and {\tt double-double}, and in Figure~\ref{Rdot2.G} we show the result of {\tt Rdot} performance for {\tt MPFR}, {\tt GMP} and {\tt quad-double} on Raspberry Pi4 ARM Cortex A72.
The peak performances of the reference {\tt Rdot}s of {\tt \_Float128} and {\tt double-double} are 15.0 MFlops and 53.8 MFlops, respectively.
The peak performances of simple OpenMP parallelized {\tt Rdot}s of {\tt \_Float128} and {\tt double-double} are 65.2 MFlops and 194 MFlops, respectively.
The peak performances of the reference {\tt Rdot}s of {\tt MPFR 512bit}, {\tt GMP 512bit} and {\tt quad-double} are 1.9 MFlops, 3.2 MFlops, 5.2 MFlops, respectively.
The peak performances of simple OpenMP parallelized {\tt MPFR 512bit}, {\tt GMP 512bit} and {\tt quad-double} are 9.7 MFlops, 12.5 MFlops, 22.2 MFlops, respectively.\\

\subsection{Raw data for the benchmarks}
At \url{https://github.com/nakatamaho/mplapack/tree/master/benchmark/results/2022}, you will find the raw data for benchmarks.

\subsection{How to take benchmarks}
Users can take performance benchmarks on their machine:
\begin{verbatim}
$ cd /home/docker/MPLAPACK/lib/*/mplapack/benchmark
$ bash -x benchall.sh
\end{verbatim}

\section{History}
\label{sec:history}
M.N started developing MPLAPACK as a by-product of an arbitrary accurate semidefinite programming solver, SDPA-GMP, around 2006~\cite{JCP2008}. The first MPLAPACK (formerly MPACK) version, 0.0.1, was released on 2008-7-15. In 2009-2-5, we released SDPA-GMP 7.1.2, and MPLAPACK supports SDPA-GMP, SDPA-QD, and SDPA-DD~\cite{SDPA-GMP,sdpa-gmpgithub,sdpa-qdgithub,sdpa-ddgithub}. On 2010-01-13, we supported Windows via mingw32. In 2010-5-21, we supported MPFR. In 2012-10-13, we released a fast implementation of a double-double version of Rgemm for NVIDIA C2050~\cite{6424545,6495966}. In 2017-3-29, we moved the web site from \url{http://mplapack.sourceforge.net/} to \url{https://github.com/nakatamaho/mplapack/}. In 2021-4-1, we renamed our project to MPLAPACK. In 2021-4-11, we supported AArch64 and released 0.9.4. In 2021-10-1, we released 1.0.0. It has huge improvements; all real subroutines are supported (except for RFP formats and mixed precision versions). In 2021, we released version 1.0.1. It fixed double-double arithmetic for Intel one API. In 2022-07-26 we released 2.0.0. It has huge improvements  now that all routines, including the RFP matrix format, are supported. Currently, only mixed precision routines are not yet supported. Since this release, we also provide all the results of tests for amd64 (aka x86\_64) (Linux/Mac/Win) and Arm64 (Linux)~\cite{lintest_mplapack,eigtest_mplapack}. We also fixed minor pieces of stuff like linking issues. On 2022-09-12, we released 2.0.1. It comes with A100 and V100 support for double-double version of {\tt Rgemm} and {\tt Rsyrk}~\cite{6424545,6495966}. Their peak performance is impressive, and approximately 400-600 GFlops~\cite{benchmark_mplapack}.

\section{Related works}
\label{sec:relatedworks}
We can relate our work in two main directions. (i) acceleration of multiple-precision extension to BLAS (especially on GPU) and (ii) development of a multiple-precision version of linear algebra packages.

For (i), In 2009, Mukunoki\etal{}~\cite{HPC-137_1} implemented a double-double version of matrix-matrix multiplication kernels for NVIDIA Tesla C1060 and evaluated the performance. In 2010, Mukunoki \etal{}~\cite{10.1007/978-3-642-28151-8_25} the quadruple precision Basic Linear Algebra Subprograms (BLAS) functions, AXPY, GEMV, and GEMM, on graphics processing units (GPUs), and evaluated their performance. On an NVIDIA Tesla C1060, their BLAS functions are approximately 30 times faster than the existing quadruple precision BLAS on an Intel Core i7 920.

In 2011, Nakasato~\cite{Nakasato2011AFG} implemented optimized dense matrix multiplication kernels for AMD Cypress GPU for various precisions. Their result includes double-double precision and attained peak performance of 30 GFlops.

In 2012, Nakata~\etal{}~\cite{6495966} implemented double-double version of {\tt Rgemm} on NVIDIA C2050. Their implementation is a complete {\tt Rgemm} implementation; thus, their implementation can be used for real applications; it can handle all the sizes in matrices and supports the transpose of each matrix. They attained 16.1GFlops with CPU-GPU transfer and 26.4GFlops (25.7GFlops with CPU-GPU transfer included). Moreover, they applied to semidefinite programming solver and attained ten times acceleration compared to CPU-only implementation.

In 2012, Yamada~\etal{} developed QPBLAS packages for CPUs~\cite{7965202}, as well as the QPBLAS-GPU package for GPUs in 2013~\cite{qpblas-gpu}. These correspond to the complete set of double-double version parts of MPBLAS, accelerating CPU and GPUs. Note that memory alignment is different from MPBLAS and written in Fortran 90.

In 2014 and 2015, Kouya~\cite{Tomonori_Kouya2014,Tomonori_Kouya2016} implemented optimized  matrix-matrix multiplications using Strassen and Winograd algorithms using MPFR, double-double, and quad-double precisions and applied them to LU factorizations. Strassen and Winograd versions are twice as fast as double-double and quad-double precisions running one core using a simple blocking algorithm. Furthermore, there is no significant performance difference between the Strassen version and Winograd. However, when we used MPFR, the Winograd version was always faster. He also found a significant loss of accuracy when performing LU factorization using Strassen and Winograd versions. 

In 2016, Joldes \etal{} implemented CAMPARY: Cuda Multiple Precision Arithmetic Library and applied it to semi-definite programming~\cite{10.1007/978-3-319-42432-3_29}. However, it was a prototype implementation and somewhat slower than our CPU implementation of ours~\cite{SDPA-GMP}. In 2017, Joldes \etal{} implemented an improved version of CAMPARY~\cite{8023060}. The significant result of this paper is that they improved {\tt Rgemm} performance on GPU for various multiple-precision versions; 1.6GFlops for triple-double precision, 976MFlops for quadruple-precision, 660MFlops for quintuple-double, 453MFlops for sextuple-double, 200MFlops for octuple-double.

In 2019, Hishinuma \etal{} implimented {\tt Rgemm} kernels for double-double precision on MIMD type acclearlator PEZY-SC2~\cite{hishinuma2019pzqd}. 
The performance of their implementation of {\tt Rgemm} the PEZY-SC2 attained 75\% of the peak performance. This value is 20 times faster than an Intel Xeon E5-2618L v3, even including the communication time between the host CPU and the PEZY-SC2.

In 2020, Isopov \etal{}~\cite{ISUPOV2020105506} took a benchmark for several level-1 multi-precision routines for MPFR, ARPREC, MPDECIMAL, MPACK, GARPREC, CUMP,, and MPRES-BLAS. MPRES-BLAS is an ongoing effort to develop a multiple-precision version of BLAS like MPBLAS on GPUS~ The\cite{10.1007/978-3-030-64616-5_4}. MPRES-BLAS defines a new arbitrary precision type suited for GPUs and fastest among CAMPARY~\cite{10.1007/978-3-319-42432-3_29}, CUMP~\cite{Nakayama2011ImplementationOM} and GARPREC~\cite{Lu:2010:SEP:1869389.1869392} when the fractions range from 424bits to 848 bits.

In 2021, Kouya~\cite{10.1007/978-3-030-86976-2_14} acceralated {\tt dd\_real}, {\tt qd\_real} and triple-double precision version of {\tt Rgemm} with AVX2 using Strassen algorithm~\cite{STRASSEN1969}. 

For (ii), Kouya has long developed C libraries for multiple-precision calculation using GMP and MPFR, including matrix operations~\cite{BNCpack}. He calls them as BNCpack and MPIBNCPack. At least, we can go back to 2003 to find there was MPIBNCpack version 0.1~\cite{MPIBNCPack01}. In 2011, version 0.7 had released, and it can solve linear equations and eigenvalue problems and many multiple-precision functions that are not directly related to linear algebra. From version 0.8 (2013-03-11), BNCpack and MPIBNCpack are integrated~\cite{BNCpack,1710.01839}.

Saito developed ZKCM~\cite{SAITOH20132005}, a C++ library for multi-precision matrix computation for quantum computer simulation. He implemented matrix inversion, singular-value decomposition of a general matrix, the diagonalization of a Hermitian matrix, and other operations.

Arb is a C library for arbitrary-precision ball arithmetic developed by Johansson~\cite{1611.02831}. It uses ball arithmetic to track
numerical errors automatically.  It can perform and solve a wide range of matrix operations, e.g.,  LU factorization, inversion, and eigenvalue problems, but it is not an extension of LAPACK to higher precision since the employed arithmetic is different. 

Multiprecision Computing Toolbox by Advanpix for MATLAB is a toolbox of MATLAB that can solve many arbitrary precision matrix operations
like diagonalization of real and complex non-symmetric matrices and singular value decomposition with
outstanding performance. They seem to employ a similar approach to ours to convert LAPACK to multiple precision versions, but details are not open and do not replace LAPACK~\cite{advanpix}. Symbolic Math Toolbox (SMT) of Matlab also provides such functionalities~\cite{matlabsymbolic}.

RalphAS has been developing GenericSchur.jl~\cite{GenericSchur}, and it calculates Schur decomposition of matrices with generic floating-point element types in Julia. However, it is not a replacement for the LAPACK library.

Johansson~\etal{} also has been developing mpmath~\cite{mpmath} can handle linear algebra (linear system solving, LU factorization, matrix inverse, matrix norms, matrix exponentials/logarithms/square roots, eigenvalues, singular values, QR factorization), written in python, and it is not a replacement of LAPACK.

Mathematica~\cite{mathematica} has multiple-precision versions of the eigenvalue problem of non-symmetric matrices, singular value decomposition problems, and other solvers.

\section{Future plans}
\label{sec:futureplans}
For version 3.0, (i) we are planning to add more optimized subroutines for MPBLAS and MPLAPACK, (ii) python and octave integration, and (iii) dropping the GMP version. Finally, for version 4.0, we plan to add a template version so we can use {\tt \_Float16} and other precisions more transparently.
Our development status can be found at \url{https://github.com/nakatamaho/mplapack#mplapack-release-process}.

\section*{Acknowledgement}

First, we are grateful to the LAPACK team for making an excellent library.
This work was supported by the Special Postdoctoral Researchers' Program of RIKEN (2008, 2009), Grant-in-Aid for Scientific Research (B) 21300017 from the Japan Society for the Promotion of Science (2009, 2010, 2011), Microsoft Research CORE6 (2010), the Japan Society for the Promotion of Science (JSPS KAKENHI Grant no. 18H03206) and TIS inc.

The author would like to thank Dr. Imamura Toshiyuki. Dr. Nakasato Naohito, Dr. Fujisawa Katsuki, Mr. Umeda Kouzo, Dr. Kouya Tomonori, Dr. Takahashi Daisuke, Dr. Goto Kazushige, Dr. Himeno Ryutaro, Dr. Hishimuna Toshiaki, Dr. Katagiri Takahiro, Dr. Ogita Takeshi, Dr. Kashiwagi Masahide, Dr. Yuasa Fukuko, Dr. Ishikawa Tadashi, Dr. Geshi Masaaki, Dr. Mukunoki Daichi and Mr. Minato Yuichiro for warm encouragement.

\bibliography{manual}

\begin{thebibliography}{10}

\bibitem{GoluVanl96}
Gene~H. Golub and Charles~F. Van~Loan.
\newblock {\em Matrix Computations}.
\newblock The Johns Hopkins University Press, third edition, 1996.

\bibitem{10.1145/567806.567807}
An updated set of basic linear algebra subprograms (blas).
\newblock {\em ACM Trans. Math. Softw.}, 28(2):135151, June 2002.

\bibitem{laug}
E.~Anderson, Z.~Bai, C.~Bischof, S.~Blackford, J.~Demmel, J.~Dongarra,
  J.~Du~Croz, A.~Greenbaum, S.~Hammarling, A.~McKenney, and D.~Sorensen.
\newblock {\em {LAPACK} Users' Guide}.
\newblock Society for Industrial and Applied Mathematics, Philadelphia, PA,
  third edition, 1999.

\bibitem{4610935}
Ieee standard for floating-point arithmetic.
\newblock {\em IEEE Std 754-2008}, pages 1--70, 2008.

\bibitem{10.1145/1356052.1356053}
Kazushige Goto and Robert A. van~de Geijn.
\newblock Anatomy of high-performance matrix multiplication.
\newblock {\em ACM Trans. Math. Softw.}, 34(3), May 2008.

\bibitem{6413635}
Zhang Xianyi, Wang Qian, and Zhang Yunquan.
\newblock Model-driven level 3 blas performance optimization on loongson 3a
  processor.
\newblock In {\em 2012 IEEE 18th International Conference on Parallel and
  Distributed Systems}, pages 684--691, 2012.

\bibitem{8519659}
Jack Dongarra, Vladimir Getov, and Kevin Walsh.
\newblock The 30th anniversary of the supercomputing conference: Bringing the
  future closer--supercomputing history and the immortality of now.
\newblock {\em Computer}, 51(10):74--85, 2018.

\bibitem{cublas}
NVIDIA.
\newblock Basic linear algebra on nvidia gpus, 2021.
\newblock \url{https://developer.nvidia.com/cublas}.

\bibitem{tdb10}
Stanimire Tomov, Jack Dongarra, and Marc Baboulin.
\newblock {Towards dense linear algebra for hybrid GPU accelerated manycore
  systems}.
\newblock {\em Parallel Computing}, 36(5-6):232--240, June 2010.

\bibitem{BAILEY201210106}
D.H. Bailey, R.~Barrio, and J.M. Borwein.
\newblock High-precision computation: Mathematical physics and dynamics.
\newblock {\em Applied Mathematics and Computation}, 218(20):10106--10121,
  2012.

\bibitem{math3020337}
David~H. Bailey and Jonathan~M. Borwein.
\newblock High-precision arithmetic in mathematical physics.
\newblock {\em Mathematics}, 3(2):337--367, 2015.

\bibitem{high:ASNA2}
Nicholas~J. Higham.
\newblock {\em Accuracy and Stability of Numerical Algorithms}.
\newblock Society for Industrial and Applied Mathematics, Philadelphia, PA,
  USA, second edition, 2002.

\bibitem{SDP}
Lieven Vandenberghe and Stephen Boyd.
\newblock Semidefinite programming.
\newblock {\em SIAM Rev.}, 38(1):4995, March 1996.

\bibitem{JCP2008}
Maho Nakata, Bastiaan~J. Braams, Katsuki Fujisawa, Mituhiro Fukuda, Jerome~K.
  Percus, Makoto Yamashita, and Zhengji Zhao.
\newblock Variational calculation of second-order reduced density matrices by
  strong n-representability conditions and an accurate semidefinite programming
  solver.
\newblock {\em The Journal of Chemical Physics}, 128(16):164113, 2008.

\bibitem{SDPA-GMP}
Maho Nakata.
\newblock A numerical evaluation of highly accurate multiple-precision
  arithmetic version of semidefinite programming solver: Sdpa-gmp, -qd and -dd.
\newblock In {\em 2010 IEEE International Symposium on Computer-Aided Control
  System Design}, pages 29--34, 2010.

\bibitem{SDPA}
Makoto Yamashita, Katsuki Fujisawa, Mituhiro Fukuda, Kazuhiro Kobayashi,
  Kazuhide Nakata, and Maho Nakata.
\newblock {\em Latest Developments in the Family for Solving Large-Scale SDPs},
  pages 687--713.
\newblock Springer US, Boston, MA, 2012.

\bibitem{sdpa-gmpgithub}
{Nakata, Maho}.
\newblock {SDPA-GMP}, retrieved September 23, 2021.
\newblock \url{https://github.com/nakatamaho/sdpa-gmp/}.

\bibitem{macosgccbug}
{GMP failure on Mac OS X 10.5}, retrieved August 29, 2022.
\newblock
  \url{https://gmplib.org/list-archives/gmp-bugs/2008-February/000930.html}.

\bibitem{merkel2014docker}
Dirk Merkel.
\newblock Docker: lightweight linux containers for consistent development and
  deployment.
\newblock {\em Linux journal}, 2014(239):2, 2014.

\bibitem{30711}
Ieee standard for binary floating-point arithmetic.
\newblock {\em ANSI/IEEE Std 754-1985}, pages 1--20, 1985.

\bibitem{18661-3}
{\em ISO/IEC TS 18661-3:2015 Information Technology - Programming languages,
  their environments, and system software interfaces - Floating-point
  extensions for C - Part 3: Interchange and extended types}.
\newblock the International Organization for Standardization, Chemin de
  Blandonnet 8 CP 401 1214 Vernier, Geneva Switzerland, 2015.

\bibitem{IBMz13}
Cedric Lichtenau, Steven Carlough, and Silvia~Melitta Mueller.
\newblock Quad precision floating point on the ibm z13.
\newblock In {\em 2016 IEEE 23nd Symposium on Computer Arithmetic (ARITH)},
  pages 87--94, 2016.

\bibitem{hida2000}
Xiaoye S.~Li Yozo~Hida and David~H. Bailey.
\newblock Quad-double arithmetic: Algorithms, implementation, and application.
\newblock In {\em Technical Report LBNL-46996}. Lawrence Berkley National
  Laboratory, 2000.

\bibitem{Knuth1997}
Donald~E. Knuth.
\newblock {\em The Art of Computer Programming, Volume 2 (3rd Ed.):
  Seminumerical Algorithms}.
\newblock Addison-Wesley Longman Publishing Co., Inc., USA, 1997.

\bibitem{Dekker1971}
T.J. Dekker.
\newblock A floating-point technique for extending the available precision.
\newblock {\em Numerische Mathematik}, 18:224--242, 1971/72.

\bibitem{ibmdd}
{6.12 Additional Floating Types}, retrieved September 7, 2022.
\newblock \url{https://gcc.gnu.org/onlinedocs/gcc/Floating-Types.html}.

\bibitem{Granlund12}
Torbj{\"o}rn Granlund and {the GMP development team}.
\newblock {\em {GNU MP}: {T}he {GNU} {M}ultiple {P}recision {A}rithmetic
  {L}ibrary}, 5.0.5 edition, 2012.
\newblock \url{http://gmplib.org/}.

\bibitem{10.1145/1236463.1236468}
Laurent Fousse, Guillaume Hanrot, Vincent Lef\`{e}vre, Patrick P\'{e}lissier,
  and Paul Zimmermann.
\newblock Mpfr: A multiple-precision binary floating-point library with correct
  rounding.
\newblock {\em ACM Trans. Math. Softw.}, 33(2):13es, June 2007.

\bibitem{mpreal}
{MPFR C++}, retrieved September 10, 2022.
\newblock \url{http://www.holoborodko.com/pavel/mpfr/}.

\bibitem{MPC}
Andreas Enge, Micka\"el Gastineau, Philippe Th\'eveny, and Paul Zimmermann.
\newblock {\em mpc --- A library for multiprecision complex arithmetic with
  exact rounding}.
\newblock INRIA, 1.1.0 edition, January 2018.
\newblock \url{http://mpc.multiprecision.org/}.

\bibitem{arm64abi}
{8 Arm C AND C++ Language Mappings}, retrieved September 10, 2022.
\newblock
  \url{https://github.com/ARM-software/abi-aa/blob/2bcab1e3b22d55170c563c3c7940134089176746/aapcs64/aapcs64.rst#811arithmetic-types}.

\bibitem{6424545}
Maho Nakata, Yasuyoshi Takao, Shigeho Noda, and Ryutaro Himeno.
\newblock A fast implementation of matrix-matrix product in double-double
  precision on nvidia c2050 and application to semidefinite programming.
\newblock In {\em 2012 Third International Conference on Networking and
  Computing}, pages 68--75, 2012.

\bibitem{Gregory1969ACO}
R.~T. Gregory and D.~Karney.
\newblock A collection of matrices for testing computational algorithms.
\newblock 1969.

\bibitem{wikipedia_svd}
{Singular value decomposition}, retrieved September 24, 2021.
\newblock \url{https://en.wikipedia.org/wiki/Singular_value_decomposition}.

\bibitem{lawn41}
Susan Blackford and Jack Dongarra.
\newblock Installation guide for {LAPACK}.
\newblock LAPACK Working Note~41, March 1992.
\newblock UT-CS-92-151, March, 1992.

\bibitem{fable}
Ralf~W. Grosse-Kunstleve, Thomas~C. Terwilliger, Nicholas~K. Sauter, and
  Paul~D. Adams.
\newblock Automatic fortran to++ conversion with.
\newblock {\em Source Code for Biology and Medicine}, 7:5, 2012.
\newblock \url{https://doi.org/10.1186/1751-0473-7-5}.

\bibitem{sdpa-qdgithub}
{Nakata, Maho}.
\newblock {SDPA-QD}, retrieved September 23, 2021.
\newblock \url{https://github.com/nakatamaho/sdpa-qd/}.

\bibitem{sdpa-ddgithub}
{Nakata, Maho}.
\newblock {SDPA-DD}, retrieved September 23, 2021.
\newblock \url{https://github.com/nakatamaho/sdpa-dd/}.

\bibitem{6495966}
Maho Nakata.
\newblock Poster: Mpack 0.7.0: Multiple precision version of blas and lapack.
\newblock In {\em 2012 SC Companion: High Performance Computing, Networking
  Storage and Analysis}, pages 1353--1353, 2012.

\bibitem{lintest_mplapack}
{Linear equation est results of MPLAPACK}, retrieved September 9, 2022.
\newblock
  \url{https://github.com/nakatamaho/mplapack/tree/master/mplapack/test/lin/results}.

\bibitem{eigtest_mplapack}
{Eigenproblem test results of MPLAPACK}, retrieved September 9, 2022.
\newblock
  \url{https://github.com/nakatamaho/mplapack/tree/master/mplapack/test/eig/results}.

\bibitem{benchmark_mplapack}
{Benchmark results of MPLAPACK and MPBLAS}, retrieved September 9, 2022.
\newblock
  \url{https://github.com/nakatamaho/mplapack/tree/master/benchmark/results/2022}.

\bibitem{HPC-137_1}
Mukunoki Daichi and Takahashi Daisuke.
\newblock Implementation and evaluation of quadruple precision blas on gpu
  ({\it in japanese}).
\newblock In {\em IPSJ SIG Technical Report}, volume 137, pages 1--6, 2009.

\bibitem{10.1007/978-3-642-28151-8_25}
Daichi Mukunoki and Daisuke Takahashi.
\newblock Implementation and evaluation of quadruple precision blas functions
  on gpus.
\newblock In Kristj{\'a}n J{\'o}nasson, editor, {\em Applied Parallel and
  Scientific Computing}, pages 249--259, Berlin, Heidelberg, 2012. Springer
  Berlin Heidelberg.

\bibitem{Nakasato2011AFG}
N.~Nakasato.
\newblock A fast gemm implementation on the cypress gpu.
\newblock {\em SIGMETRICS Perform. Evaluation Rev.}, 38:50--55, 2011.

\bibitem{7965202}
Susumu Yamada, Takuya Ina, Narimasa Sasa, Yasuhiro Idomura, Masahiko Machida,
  and Toshiyuki Imamura.
\newblock Quadruple-precision blas using bailey's arithmetic with fma
  instruction: its performance and applications.
\newblock In {\em 2017 IEEE International Parallel and Distributed Processing
  Symposium Workshops (IPDPSW)}, pages 1418--1425, 2017.

\bibitem{qpblas-gpu}
Japan Atomic~Energy Agency.
\newblock {\em Quadruple Precision BLAS Routines for GPU QPBLAS-GPU Ver.1.0
  User's Manual}, July 2013.

\bibitem{Tomonori_Kouya2014}
Tomonori Kouya.
\newblock Accelerated multiple precision matrix multiplication using strassen's
  algorithm and winograd's variant.
\newblock {\em JSIAM Letters}, 6:81--84, 2014.

\bibitem{Tomonori_Kouya2016}
Tomonori Kouya.
\newblock Performance evaluation of multiple precision matrix multiplications
  using parallelized strassen and winograd algorithms.
\newblock {\em JSIAM Letters}, 8:21--24, 2016.

\bibitem{10.1007/978-3-319-42432-3_29}
Mioara Joldes, Jean-Michel Muller, Valentina Popescu, and Warwick Tucker.
\newblock Campary: Cuda multiple precision arithmetic library and applications.
\newblock In Gert-Martin Greuel, Thorsten Koch, Peter Paule, and Andrew
  Sommese, editors, {\em Mathematical Software -- ICMS 2016}, pages 232--240,
  Cham, 2016. Springer International Publishing.

\bibitem{8023060}
Mioara Joldes, Jean-Michel Muller, and Valentina Popescu.
\newblock Implementation and performance evaluation of an extended precision
  floating-point arithmetic library for high-accuracy semidefinite programming.
\newblock In {\em 2017 IEEE 24th Symposium on Computer Arithmetic (ARITH)},
  pages 27--34, 2017.

\bibitem{hishinuma2019pzqd}
Toshiaki Hishinuma and Maho Nakata.
\newblock pzqd: Pezy-sc2 acceleration of double-double precision arithmetic
  library for high-precision blas.
\newblock In {\em International Conference on Computational \& Experimental
  Engineering and Sciences}, pages 717--736. Springer, 2019.

\bibitem{ISUPOV2020105506}
Konstantin Isupov.
\newblock Performance data of multiple-precision scalar and vector blas
  operations on cpu and gpu.
\newblock {\em Data in Brief}, 30:105506, 2020.

\bibitem{10.1007/978-3-030-64616-5_4}
Konstantin Isupov and Vladimir Knyazkov.
\newblock Multiple-precision blas library for graphics processing units.
\newblock In Vladimir Voevodin and Sergey Sobolev, editors, {\em
  Supercomputing}, pages 37--49, Cham, 2020. Springer International Publishing.

\bibitem{Nakayama2011ImplementationOM}
Takato Nakayama and D.~Takahashi.
\newblock Implementation of multiple-precision floating-point arithmetic
  library for gpu computing.
\newblock In {\em Proc. 23rd IASTED International Conference on Parallel and
  Distributed Computing and Systems (PDCS 2011)}, pages 343--349, 2011.

\bibitem{Lu:2010:SEP:1869389.1869392}
Mian Lu, Bingsheng He, and Qiong Luo.
\newblock Supporting extended precision on graphics processors.
\newblock In {\em Proceedings of the Sixth International Workshop on Data
  Management on New Hardware}, DaMoN '10, pages 19--26, New York, NY, USA,
  2010. ACM.

\bibitem{10.1007/978-3-030-86976-2_14}
Tomonori Kouya.
\newblock Acceleration of multiple precision matrix multiplication based on
  multi-component floating-point arithmetic using avx2.
\newblock In Osvaldo Gervasi, Beniamino Murgante, Sanjay Misra, Chiara Garau,
  Ivan Ble{\v{c}}i{\'{c}}, David Taniar, Bernady~O. Apduhan, Ana Maria A.~C.
  Rocha, Eufemia Tarantino, and Carmelo~Maria Torre, editors, {\em
  Computational Science and Its Applications -- ICCSA 2021}, pages 202--217,
  Cham, 2021. Springer International Publishing.

\bibitem{STRASSEN1969}
V.~STRASSEN.
\newblock Gaussian elimination is not optimal.
\newblock {\em Numerische Mathematik}, 13:354--356, 1969.

\bibitem{BNCpack}
Tomonori Kouya.
\newblock {\em BNCpack 0.7}, September 2011.
\newblock \url{http://na-inet.jp/na/bnc/}.

\bibitem{MPIBNCPack01}
Tomonori Kouya.
\newblock {\em MPIBNCpack 0.1}, September 2003.
\newblock \url{https://na-inet.jp/na/bnc/mpibncpack.pdf}.

\bibitem{1710.01839}
Tomonori Kouya.
\newblock Tuning technique for multiple precision dense matrix multiplication
  using prediction of computational time, 2017.

\bibitem{SAITOH20132005}
Akira SaiToh.
\newblock Zkcm: A c++ library for multiprecision matrix computation with
  applications in quantum information.
\newblock {\em Computer Physics Communications}, 184(8):2005--2020, 2013.

\bibitem{1611.02831}
Fredrik Johansson.
\newblock Arb: Efficient arbitrary-precision midpoint-radius interval
  arithmetic, 2016.
\newblock \url{https://arblib.org/}.

\bibitem{advanpix}
{Advanpix}.
\newblock {Multiprecision Computing Toolbox for MATLAB}, retrieved September
  23, 2021.
\newblock \url{https://www.advanpix.com/}.

\bibitem{matlabsymbolic}
Inc. The~MathWorks.
\newblock {\em Symbolic Math Toolbox}.
\newblock Natick, Massachusetts, United State, 2019.

\bibitem{GenericSchur}
{RalphAS}.
\newblock {Schur decomposition of matrices with generic floating-point element
  types in Julia}, retrieved September 23, 2021.
\newblock \url{https://github.com/RalphAS/GenericSchur.jl}.

\bibitem{mpmath}
Fredrik Johansson et~al.
\newblock {\em mpmath: a {P}ython library for arbitrary-precision
  floating-point arithmetic (version 0.18)}, December 2013.
\newblock \url{http://mpmath.org/}.

\bibitem{mathematica}
Wolfram~Research{,} Inc.
\newblock Mathematica, {V}ersion 12.3.1.
\newblock Champaign, IL, 2021.

\end{thebibliography}

\end{document}